\documentclass{article} 
\usepackage{iclr2025_conference,times}

\usepackage{amsmath,amsfonts,bm}

\def\eqref#1{equation~\ref{#1}}

\def\1{\bm{1}}

\DeclareMathAlphabet{\mathsfit}{\encodingdefault}{\sfdefault}{m}{sl}
\SetMathAlphabet{\mathsfit}{bold}{\encodingdefault}{\sfdefault}{bx}{n}

\usepackage{hyperref}
\usepackage{url}

\usepackage{booktabs}       

\usepackage{graphicx}
\usepackage{algorithm}
\usepackage{algpseudocode}
\usepackage{svg}
\usepackage{tabularx}
\usepackage{enumitem}

\usepackage{comment}

\title{HIVEX: A High-Impact Environment Suite\\for Multi-Agent Research}

\author{%
  Philipp D. Siedler \\
  Aleph Alpha Research\\
  \texttt{p.d.siedler@gmail.com} \\
}

\iclrfinalcopy 
\begin{document}

\maketitle

\begin{abstract}
Games have been vital test beds for the rapid development of Agent-based research. Remarkable progress has been achieved in the past, but it is unclear if the findings equip for real-world problems. While pressure grows, some of the most critical ecological challenges can find mitigation and prevention solutions through technology and its applications. Most real-world domains include multi-agent scenarios and require machine-machine and human-machine collaboration. Open-source environments have not advanced and are often toy scenarios, too abstract or not suitable for multi-agent research. By mimicking real-world problems and increasing the complexity of environments, we hope to advance state-of-the-art multi-agent research and inspire researchers to work on immediate real-world problems.\\
Here, we present HIVEX, an environment suite to benchmark multi-agent research focusing on ecological challenges. HIVEX includes the following environments: Wind Farm Control, Wildfire Resource Management, Drone-Based Reforestation, Ocean Plastic Collection, and Aerial Wildfire Suppression. We provide
\href{https://github.com/hivex-research/hivex-environments}{environments},
\href{https://github.com/hivex-research/hivex}{training examples}, and
\href{https://github.com/hivex-research/hivex-results}{baselines}
for the main and sub-tasks.
\footnote{GitHub Organisation: \href{https://github.com/hivex-research}{https://github.com/hivex-research}}
All trained models resulting from the experiments of this work are hosted on Hugging Face.
\footnote{Trained Models: \href{https://huggingface.co/hivex-research}{https://huggingface.co/hivex-research}}
We also provide a leaderboard on Hugging Face and encourage the community to submit models trained on our environment suite.
\footnote{Hugging Face Leaderboard: \href{https://huggingface.co/spaces/hivex-research/hivex-leaderboard}{https://huggingface.co/spaces/hivex-research/hivex-leaderboard}}
\end{abstract}

\begin{figure}[h!]
    \centering
    \includegraphics[width=\linewidth]{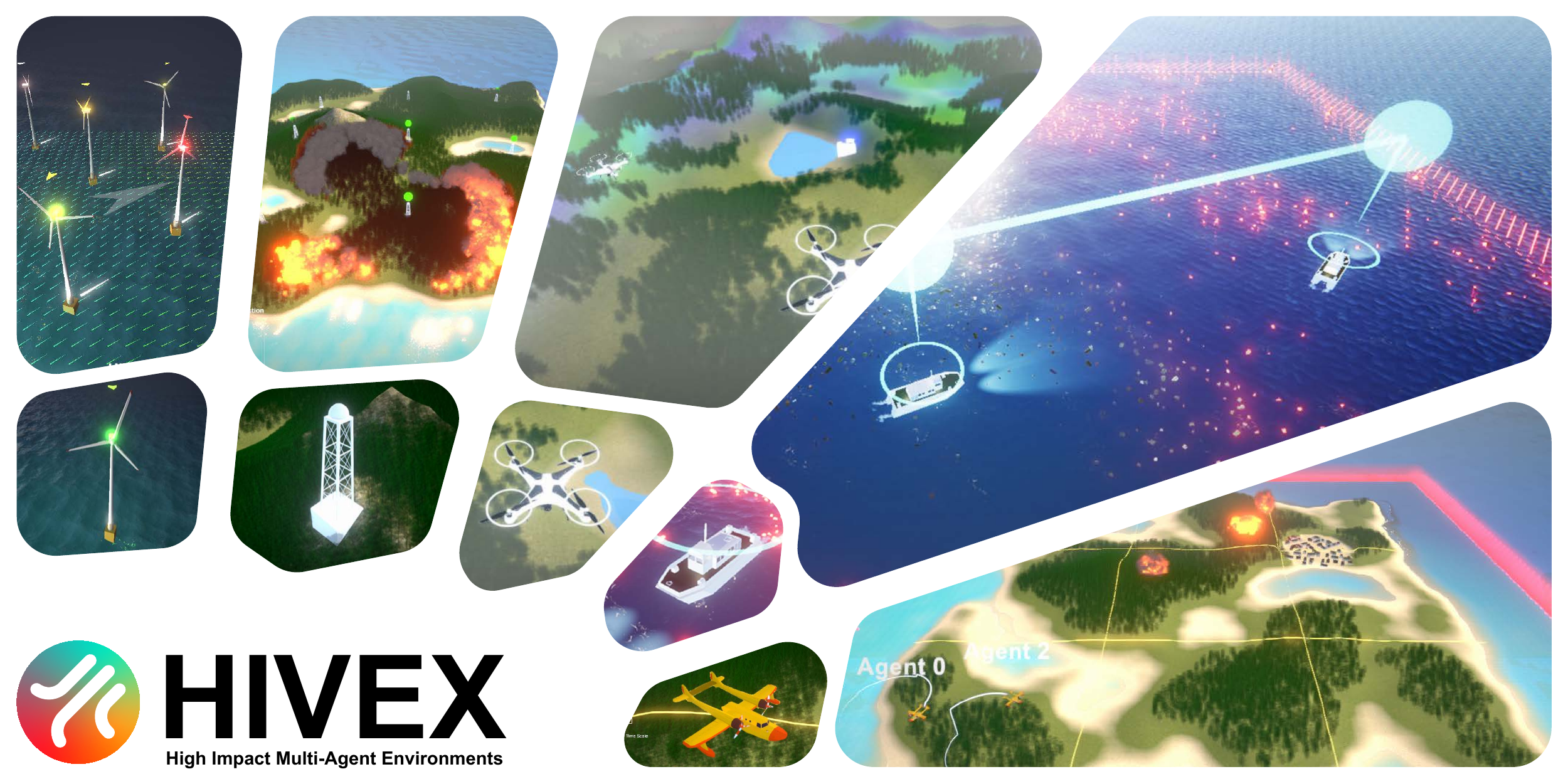}
\end{figure}

\newpage

\tableofcontents
\newpage

\section{Introduction}

Currently, no open-source benchmark for multi-agent reinforcement learning (MARL) closely mimics real-world scenarios focused on critical ecological challenges, offering sub-tasks, fine-grained terrain elevation or various layout patterns, supporting open-ended learning through procedurally generated environments and providing visual richness. Most common benchmarks with direct real-world applications are in the following domains: 1. intelligent machines and devices, 2. chemical engineering, biotechnology, and medical treatment, 3. human and society, and 4. social dilemmas \cite{ning_survey_2024}.\\
The main HIVEX environment features are either procedurally generated or sampled from a random distribution. Therefore, training and evaluation are differentiated by seed values, ensuring testing scenarios are not seen during training. We aim to assess and compare MARL algorithms, focusing on test-time evaluation with zero-shot test scenarios. If applicable, a scenario consists of an environment and a task-pattern or terrain elevation combination. Each environment has a main end-to-end task and isolated subtasks that are independent or part of the main task. Environments have between two and nine tasks, various layout patterns, or terrain elevation levels. The environments described are ordered by increasing complexity in observation size and type, action count and type, and reward granularity, including individual and collective rewards. We introduce combinations of vector and visual observations and discrete and continuous actions.

\section{Motivation: Critical Ecological Challenges}
\label{sec:motivation}

\begin{figure}[h!]
  \centering
  \includegraphics[width=\linewidth]{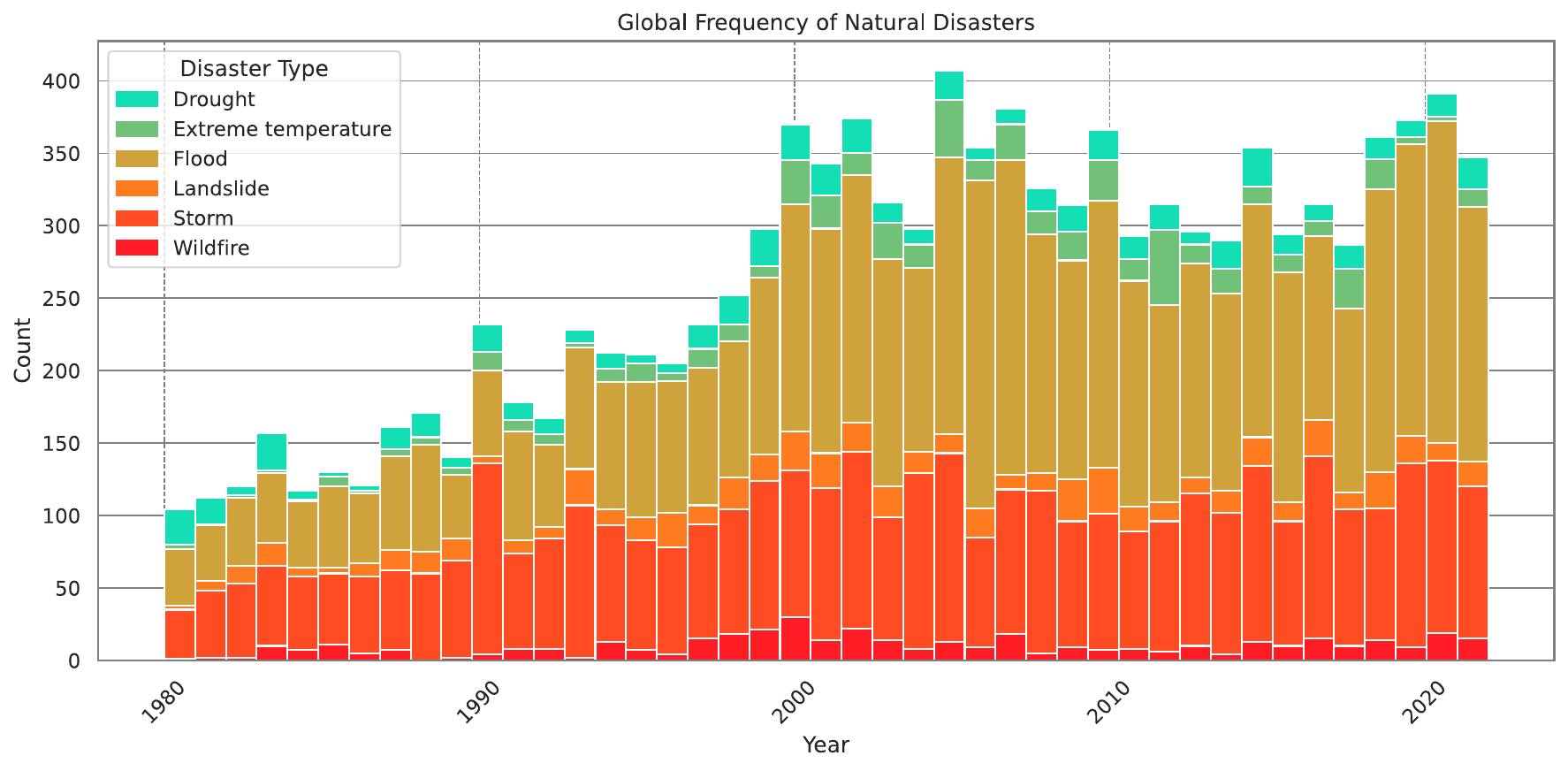}
  \vspace{-0.6cm}
  \caption{Climate-related global disasters frequency. The links between climate change and natural disasters are well documented in a wide variety of climate change literature. This graph depicts the trend in global climate-related disasters over time. Interactive plot and dataset can be explored here: \url{https://climatedata.imf.org/pages/climatechange-data}.}
  \label{fig:climate_related_disasters_frequency}
\end{figure}

Climate change is manifesting more visibly and urgently than ever \cite{archer_climate_2010, romm_climate_2022}.
We are witnessing an increase in frequent and intense weather phenomena, such as storms, droughts, fires, and floods \cite{cred__uclouvain_em-dat_2023}. Figure \ref{fig:climate_related_disasters_frequency} shows the aforementioned disaster types triple in frequency between $1980$ and $2020$.
These events are reshaping ecosystems and critically impacting agriculture and natural resources, which are vital to human survival \cite{change_managing_2012}.
A concerning report by the Intergovernmental Panel on Climate Change (IPCC) in $2022$ highlights the dire consequences of continued greenhouse gas emissions, warning that significant curbing measures are needed within the next three decades to avert catastrophic impacts. If the $1.5$ ºC degree increase in global warming cannot be negated, some impacts may be long-lasting or irreversible, such as the loss of ecosystems potentially fundamental to our existence \cite{ipcc_global_2022}.

\subsection{Mitigation, Adaptation and Disaster Response}

The battle against climate change encompasses three critical approaches: mitigation, adaptation and disaster response \cite{european_commission_how_2022}.

\begin{itemize}[noitemsep,nolistsep]
  \item Mitigation focuses on reducing emissions through transformative measures in electricity generation, transportation, building design, industry practices, and land use.
  \item Adaptation, on the other hand, is about enhancing resilience and improving disaster management strategies to prepare for the inevitable impacts of changing climate patterns.
  \item Disaster Response involves prompt and effective measures to manage emergencies caused by climate-related events. This includes providing immediate relief, medical aid, and reconstruction assistance and implementing policies for rapid response and recovery to minimize the impact on affected communities.
\end{itemize}

This tripartite approach is essential, as highlighted by the IPCC report and echoed in the research by \citet{collins_long-term_2018}, underscoring the importance of addressing both immediate and long-term aspects of climate change.

\subsection{Irreversibility}

Recent research underscores the alarming irreversibility of certain impacts of climate change.
A study at Arizona State University, published in the Proceedings of the National Academy of Sciences, explores the concept of 'rate-induced tipping' in ecological systems \cite{panahi_rate-induced_2023}.
This research is crucial in understanding when certain environmental systems, such as coral reefs, may reach a point of irreversible damage \cite{hughes_global_2018}.\\
As ocean temperatures rise due to increased carbon emissions \cite{venegas_three_2023}, corals and their symbiotic zooxanthellae (tiny cells that live within most types of coral polyps - they help the coral survive by providing it with food resulting from photosynthesis) are pushed towards a threshold beyond which severe bleaching occurs \cite{sully_global_2019}, leading to a cascade of effects on the entire reef ecosystem.
This bleaching, once initiated, cannot be reversed even if ocean temperatures were to subsequently stabilize, illustrating the permanent nature of some climate change impacts. The study emphasizes that even gradual changes in environmental parameters can suddenly trigger catastrophic system collapses, highlighting the urgency of addressing climate change proactively to prevent irreversible ecological damage \cite{panahi_rate-induced_2023}.

\subsection{Timeline and Urgency}

\begin{figure}[h!]
  \centering
  \includegraphics[width=\linewidth]{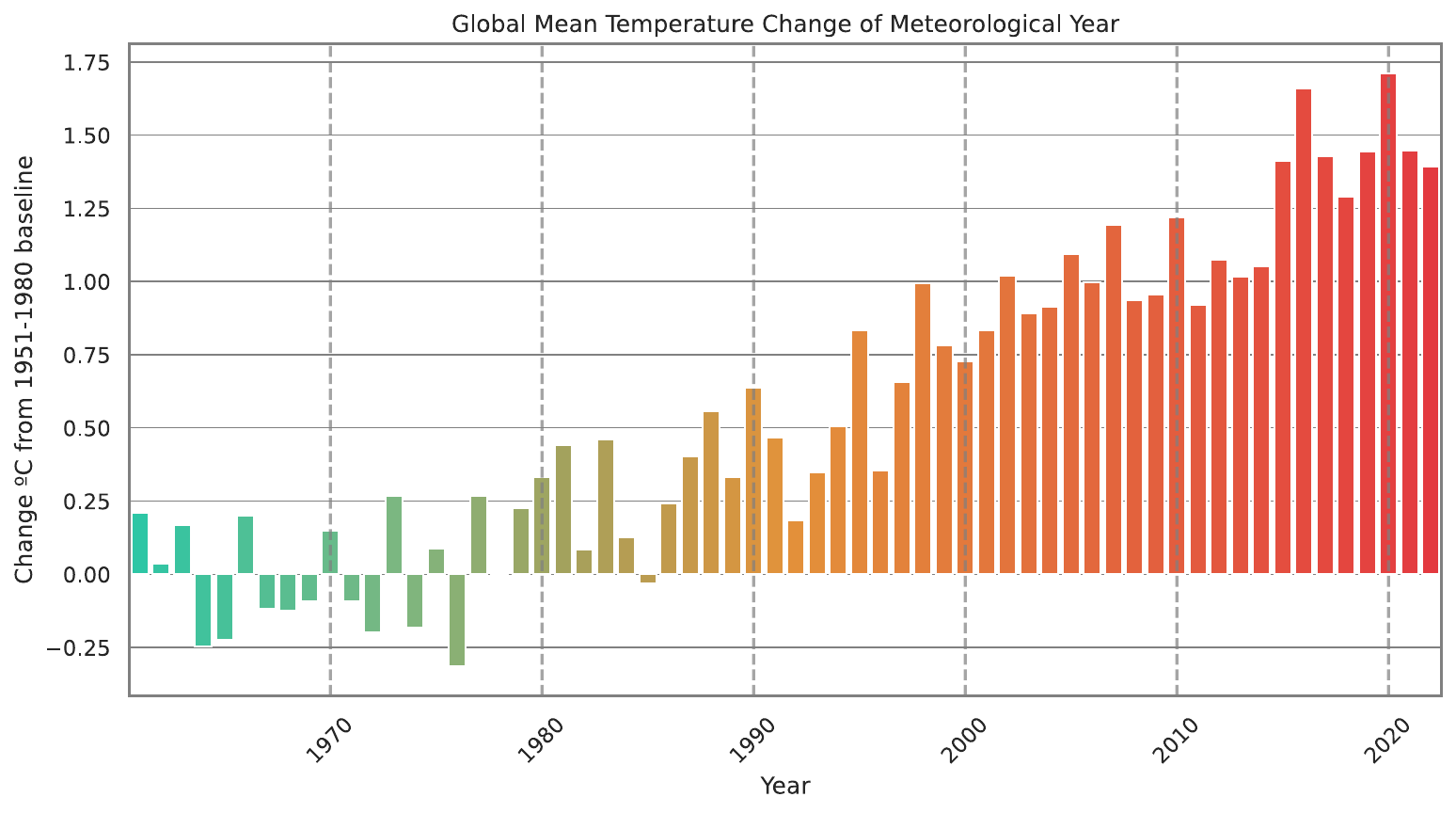}
  \vspace{-0.6cm}
  \caption{Annual Global surface temperature change. This indicator presents the global mean surface temperature change during the period 1961-2021, using temperatures between 1951 and 1980 as a baseline. This data is provided by the Food and Agriculture Organization Corporate Statistical Database (\href{https://www.fao.org/faostat/en/}{FAOSTAT}) and is based on publicly available GISTEMP data from the National Aeronautics and Space Administration Goddard Institute for Space Studies (\href{https://data.giss.nasa.gov/gistemp/}{NASA GISS}). Interactive plot and dataset can be explored here: \url{https://climatedata.imf.org/pages/climatechange-data}.}
  \label{fig:global_mean_temperature_change}
\end{figure}

The timeline for addressing climate change is critical and urgent. According to the latest insights, there's a pressing need to accelerate climate action significantly to limit global temperature rise to $1.5$ degrees Celsius.
This target requires deep, rapid, and sustained greenhouse gas emissions reductions across all sectors within this decade. Emissions need to decrease immediately to stay within these limits and be cut by nearly half by $2030$ \cite{lee_ipcc_2023}. Figure \ref{fig:global_mean_temperature_change} shows the global surface temperature change in Celsius degrees per year from the baseline temperature between $1951$ and $1980$ \cite{food_and_agriculture_organization_of_the_united_nations_faostat_1997}.

The $2023$ Yearbook of Global Climate Action, presented at the UN Climate Change Conference (COP28) \cite{hughes_global_2018}, emphasizes the urgency of scaling up climate actions. It highlights the increase in stakeholders taking climate action but also points out that the pace and scale of these actions are insufficient to meet the $1.5$-degree Celsius target. The Yearbook calls for accelerated, effective implementation of climate actions, emphasizing the critical role of governments in reducing barriers to lowering greenhouse gas emissions and the need for transformational changes in sectors like food, electricity, transport, industry, buildings, and land use.\\
A major UN report, "Climate Change 2023: Synthesis Report" by the Intergovernmental Panel on Climate Change (IPCC) \cite{lee_ipcc_2023}, underlines the significant impacts already being felt globally and the increased frequency of extreme weather events due to climate change. The report stresses the necessity of integrating adaptation to climate change with actions to reduce or avoid greenhouse gas emissions. It also points out the importance of financial and technical support for developing countries from wealthier nations to achieve these goals \cite{de-arteaga_machine_2018}.

\section{Background}

\subsection{Role of Machine Learning}

The vast array of challenges presented by climate change also opens diverse opportunities for impactful action \cite{kaack_challenges_2019, ford_big_2016}. While the situation is grave, there is immense potential for innovative solutions in areas such as renewable energy, sustainable agriculture, and resource-efficient industrial practices. The commitment to tackling these challenges is about averting disaster and harnessing the opportunity for significant environmental, economic, and social progress \cite{berendt_ai_2019, hager_artificial_2019}.

The last two years have brought climate change to the doorstep of many. Extreme heatwaves, wildfires, and floods make life increasingly difficult for animals and humans \cite{de-arteaga_machine_2018}. ML has emerged as a key tool for technological advancement in recent years. As ML and artificial intelligence (AI) use in societal and global initiatives grows, there's a pressing need to explore how these technologies can best address climate change challenges.
Many in the ML field are eager to contribute but unsure of the best approach, while various sectors are increasingly seeking ML expertise.

ML has many applications in combating climate change for various time horizons and degrees of impact \cite{rolnick_tackling_2022, ladi_applications_2022}.
Straight forward applications 
However, we think it's crucial to acknowledge its fundamental role in enhancing our understanding of climate complexities \cite{yu_computational_2013, faghmous_big_2014}. ML, with its advanced data analysis capabilities, is instrumental in deciphering the multifaceted nature of climate data. It aids scientists and researchers in identifying patterns and trends that are not immediately apparent \cite{climate_trace_-_climate_2022}, providing insights into phenomena like temperature changes, precipitation patterns, and extreme weather events. This deepened understanding is the bedrock upon which targeted solutions for climate change mitigation and adaptation are developed.

In the critical battle against climate change, ML emerges as a pivotal ally, offering a diverse array of contributions across various domains. By enabling automatic monitoring through remote sensing, ML helps in identifying key environmental changes, such as deforestation, and in assessing post-disaster damages. This technology is particularly significant in the realm of ecosystem informatics and sustainability, where it aids in understanding complex ecological dynamics and biodiversity, supporting conservation efforts and sustainable resource management \cite{dietterich_machine_2009, gomes_computational_2019, lassig_computational_2016}. ML's ability to process vast amounts of ecological data enhances our capacity to track species populations, monitor habitat changes, and predict ecological responses to various environmental stressors.

\begin{figure}[h!]
\centering
\includegraphics[width=\linewidth]{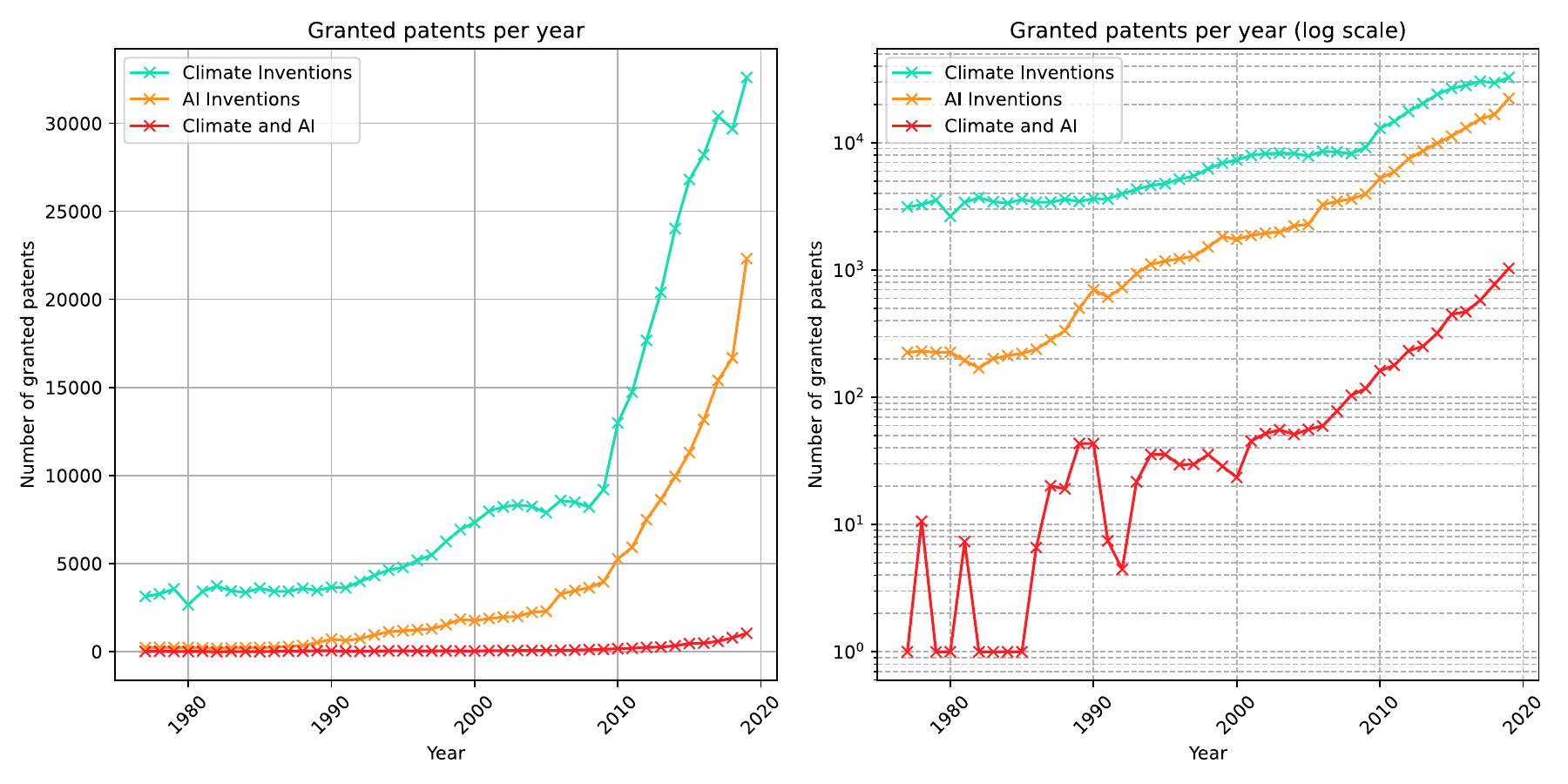}
\vspace{-0.6cm}
\caption{Left: Granted patents per year, with a steeper rise starting around 2010. Right: The rise on the left can be seen as exponential growth in climate AI patents (linear on a log scale), and this holds for climate patents and AI patents separately. Within climate patents, however, AI patents are not growing exponentially. \cite{verendel_tracking_2023,angelucci_supporting_2018}}
\label{fig:ai_climate_publication}
\end{figure}

Further, ML accelerates scientific discovery, suggesting innovative materials for batteries, construction, and carbon capture technologies. Ecosystem informatics enables the identification of patterns and relationships within ecological systems, facilitating the development of strategies to protect and sustain these vital systems. Additionally, ML optimizes systems for enhanced efficiency, evident in applications like freight consolidation, carbon market design, and reduction of food waste \cite{joppa_case_2017}. Its ability to accelerate computationally intense physical simulations, like climate and energy scheduling models, is invaluable. The integration of ML in these areas not only addresses immediate environmental concerns but also fosters long-term sustainability and resilience of ecosystems, thus playing a crucial role in mitigating the impacts of climate change. Figure \ref{fig:ai_climate_publication} shows an increase of patents granted for climate inventions, AI inventions and climate and AI between $1970$ and $2020$. This means we can directly link advancements in AI to innovation in climate-related topics.

The integration of ML in climate change mitigation not only benefits society but also propels advancements in ML itself, particularly in areas such as interpretability, causality, and uncertainty quantification. However, the challenge lies in the nature of climate-relevant data, which is often proprietary, sensitive, or not globally representative. Solutions like transfer learning and domain adaptation become crucial in addressing these data challenges. We aim to emphasize the significant potential that advancing state-of-the-art ML, utilizing real-world data and simulation environments, can go hand in hand with developing effective solutions for current pressing challenges.

\subsection{Natural Societies and Multi-Agent Research}

In MARL environments, groups of agents with baseline intelligence and ability can have a higher collective intelligence by acting together \cite{cohen_team_1997}. A shared pool of information through a collective observation space can help individual agents to learn quicker. Additionally, as a group, they can achieve objectives that would be challenging to attain individually \cite{guestrin_coordinated_2002, decker_distributed_1987, panait_cooperative_2005, mataric_using_1998}. However, acting as a collective requires collaboration. From the perspective of an individual agent, other agents in the collective and the consequences of their actions, i.e. change of the environment, can be seen as part of a dynamic environment \cite{ravula_ad_2019}. Perceiving others' actions and making sense of their intention is called intention reading, stated in the theory-of-mind (ToM) \cite{hernandez-leal_survey_2019-1}. While this is an integral part of human collaborative activities, we will assume shared intentionality \cite{tomasello_understanding_2005}.

In our quest to advance multi-agent systems and cooperative strategies, the study of animal societies like ants and meerkats offers invaluable lessons. These natural societies, characterized by intricate cooperation and complex social structures, provide a blueprint for understanding and designing efficient, self-organizing systems in human contexts.

\textbf{Ants and Cooperative Robots}: Researchers at Harvard University explored how ants cooperate to solve complex problems like transporting and building things using simple rules. They studied black carpenter ants and created a simulation to model their cooperative behaviour. This model was then used to develop robot ants (RAnts) that demonstrated similar cooperative behaviours to real ants, highlighting the potential for applying natural cooperation strategies in robotics \cite{prasath_dynamics_2022}. Recent work of ours explores distributed robotics for building architectural structures \cite{hosmer_robotic_2023}, in which robotics help each other to climb, add and remove bespoke building blocks for a dynamically changing spatial configuration.

\textbf{Ant Colonies and Social Evolution}: Certain ant species, which do not have a leader, can exhibit complex behaviours like the division of labour through self-organization. This challenges the notion that strong groups require strong leaders and suggests that even in the simplest groups, significant collaboration can occur. This research has implications for understanding the evolution of social behaviour and the early stages of complex society formation \cite{gordon_ant_2010, gordon_organization_2002}.

\textbf{Meerkats and Cooperation}: Meerkats have been studied to understand the role of testosterone in female competition and cooperative breeding. High testosterone levels in matriarch meerkats play a key role in their success and aggression, influencing the cooperative structure of the group. This study reveals that cooperation can also arise through aggressive means, shedding light on a new mechanism for the evolution of cooperative breeding \cite{clutton-brock_contributions_2001, muller_dominance_2004}.

\textbf{Meerkat Society Study}: The Kalahari Meerkat Project, led by Professor Tim Clutton-Brock, provides extensive insights into meerkat societies. The project has tracked over 3,000 meerkats, examining their life histories and the effects of climate change on their survival and development. This long-term study offers valuable data on cooperative breeding, kinship, and the resilience of meerkat groups in challenging environments \cite{komdeur_evolution_2008, newman_reproductive_2016}.

In the context of nature, Charles Darwin argues for the survival of the fittest \citep{darwin_origin_1977} and, therefore, the occurrence of competition. While in AI, the majority of significant work on MA systems consider two opposing agents only, the problems of interest of this work are cooperative MA systems, where groups of agents act together to achieve higher individual and collective goals \citep{cohen_team_1997, guestrin_coordinated_2002, decker_distributed_1987, panait_cooperative_2005, mataric_using_1998}. Just like in human society or the animal world, individuals have unique or mixtures of motives. However, we can define agents with mixed or identical motives in an MA environment simulation. Assuming shared intentionality leaves us with the question of how to collaborate. Communication can play a crucial role in collaborating successfully. Human society uses language as a communication medium \citep{baron_birchenall_animal_2016}. Agents can send signals of various types as a form of language. Nevertheless, observing others' behaviour can be a form of communication. Body language, a tail-wagging dog, or the red colour of an octopus can communicate internal states and intentions. But we can also design agents that directly share policies - state action transitions - or memory data of past experiences.

\subsection{Learning Algorithm} \label{ref:learning_algorithm}

Addressing the intricacies and challenges in multi-agent systems that operate in dynamic and complex environments requires a sophisticated blend of algorithms and methodologies. Our approach employs Proximal Policy Optimization (PPO) \cite{schulman_proximal_2017} with parameter sharing for MA training \ref{alg:2}.

At the heart of our model is the policy $\theta$, represented by a neural network with parameters that process the observations from the environment, factoring in past states and producing actions as outputs. 
Within the context of the HIVEX suite, PPO offers a stable reinforcement learning algorithm, ensuring that agents iteratively refine their strategies without drastic deviations. This is crucial given the suite's dynamic environmental events, from wildfires to ocean cleanups.
PPO is an advanced reinforcement learning algorithm that seeks to improve policy-based learning by ensuring that the updated policy does not deviate too drastically from the previous policy. This is achieved by adding a constraint or penalty to the objective function to restrict extreme policy updates \ref{alg:1}.

\textbf{Proximal Policy Optimization}: Two main concepts define the PPO \citep{schulman_proximal_2017}, a state-of-the-art, on-policy RL algorithm: 1. PPO performs the largest possible but safe gradient ascent learning step by estimating a trust region and 2. Advantage estimates how good an action in a specific state is compared to the average action. A trust region can be calculated as the quotient of the current policy to be refined $\pi_\theta(a_t|s_t)$ and the previous policy as follows $r_t(\theta) = \frac{\pi_\theta(a_t|s_t)}{\pi_{\theta_k}(a_t|s_t)} = \frac{current\ policy}{old \  policy}$. The advantage is the difference between the Q and the Value Function: $A(s,a) = Q(s,a) - V(s)$, where $s$ is the state and $a$ the action \citep{zychlinski_complete_2019}. The Q function measures the overall expected reward given state $s$, performing action $a$, and denoted as: $\mathcal{Q}(s,a) = \mathbb{E}\left[ \sum_{n=0}^{N} \gamma^n r_n \right]$. The Value Function, similar to the Q Function, measures overall expected reward, with the difference that the State Value is calculated after the action has been taken and is denoted as: $\mathcal{V}(s) = \mathbb{E}\left[ \sum_{n=0}^{N} \gamma^n r_n \right]$.

\begin{algorithm}
\caption{PPO-CLIP pseudocode \citep{openai_proximal_2021}. Multi-Agent PPO pseudocode in Appendix \ref{ref:ma_ppo}}
\begin{algorithmic}
    \item Input: initial policy parameters $\theta_0$, initial value function parameters $\phi_0$
    \For {$k=0,1,2,\ldots$}
        \State Collect set of trajectories $\mathcal{D}_k$ = \{$\tau_i$\} by running policy $\pi_k = \pi(\theta_k)$ in the environment.
        \State Compute rewards-to-go $\hat{R_t}$.
        \State Compute advantage estimates, $\hat{A_t}$ (using any method of     advantage estimation) based on the
        \State current value function $V_{\phi_k}$
        \State Update the policy by maximizing the PPO-Clip objective:
        \State $\theta_{k+1} = arg\underset{\theta}{max} \frac{1}{|\mathcal{D}_k|T} \sum_{\tau \in \mathcal{D}_k} \sum_{t = 0}^{T} \min \left( \frac{\pi_\theta(a_t|s_t)}{\pi_{\theta_k}(a_t|s_t)}A^{\pi_{\theta_k}}(s_t, a_t), g(\epsilon, A^{\pi_{\theta_k}}(s_t, a_t)) \right)$,
        \State typically via stochastic gradient ascent with Adam.
        \State Fit value function by regression on mean-squared error:
        \State $\phi_{k+1} = arg\underset{\phi}{min} \frac{1}{|\mathcal{D}_k|T} \sum_{\tau \in \mathcal{D}_k} \sum_{t = 0}^{T} \left( (V_{\phi}(s_t)-\hat{R_t} \right)$
        \State typically via some gradient descent algorithm.
    \EndFor
\end{algorithmic}
\label{alg:1}
\end{algorithm}

\clearpage

\section{The HIVEX Environment Suite} \label{ref:the_hivex_environment_suite}

HIVEX addresses ecological challenges, developed in Unity using the ML-Agents Toolkit \cite{juliani_unity_2020}. Each environment mimics a real-world scenario where multiple agents interact, collaborate, and compete, providing rich settings for multi-agent research. Scenarios include:

\begin{enumerate}
    \item \textbf{Wind Farm Control}: Agents adjust turbine orientations based on wind conditions.
    \item \textbf{Wildfire Resource Management}: Agents allocate firefighting resources during wildfires.
    \item \textbf{Drone-Based Reforestation}: Drones collaborate to plant trees in deforested areas.
    \item \textbf{Ocean Plastic Collection}: Cleanup vessels locate and retrieve plastic waste from oceans.
    \item \textbf{Aerial Wildfire Suppression}: Firefighting planes work together to extinguish wildfires and protect the village.
\end{enumerate}

Agents receive vector and visual observations from their environment and perform multi-faceted actions such as adjusting turbines, shifting resources, planting seeds, and collecting ocean plastic. Real-world constraints are imposed, such as drone battery life limitations, requiring strategic recharging to maximize efficiency.

The Hivex Environment Suite Overview table (\ref{tab:hivex_environments_fullwidth}) provides a concise summary of the five distinct HIVEX environments. Each environment includes three key metrics: the Main Task Count, representing the primary objective of the default scenario; the Sub-Task Count, indicating additional challenges within the environment; and the Terrain Elevation Levels/Patterns, which add an extra layer of environmental complexity. While WFC focuses on Patterns as its complexity dimension, WRM, DBR, and AWS incorporate Terrain Elevation Levels. In contrast, OPC does not feature an additional complexity dimension.

\begin{table}[h!]
\caption{Hivex Environment Suite Overview}
\centering
\renewcommand{\arraystretch}{1.5} 
\setlength{\tabcolsep}{8pt} 
\begin{tabular}{m{1.8cm} m{2cm} m{1.6cm} m{1.5cm} m{1.4cm} m{2.5cm}}
Screenshot & Name & Abbreviation & Main Task Count & Sub-Task Count & Terrain Elevation Levels/Patterns\\
\hline
\hline
\vspace{0.2cm}\includegraphics[width=0.15\textwidth, height=1.5cm]{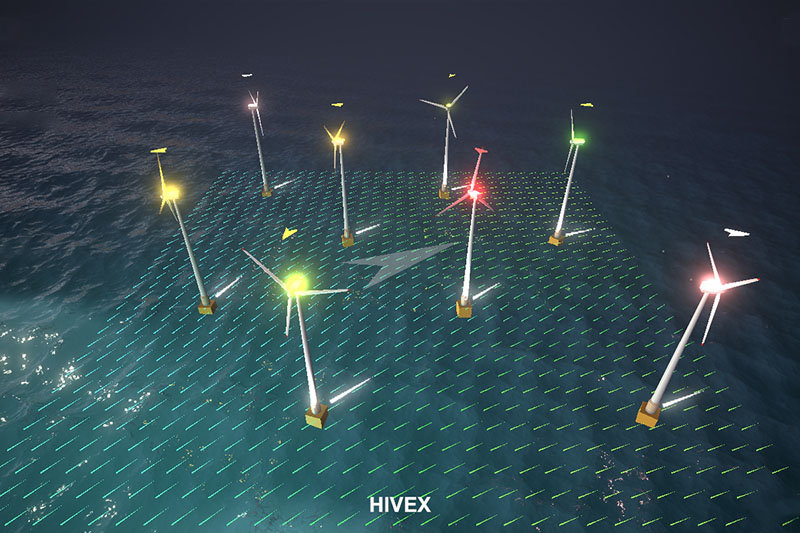}\vspace{0.0cm} 
& Wind Farm Control & WFC & 1 & 1 & 9 \\
\hline
\vspace{0.2cm}\includegraphics[width=0.15\textwidth, height=1.5cm]{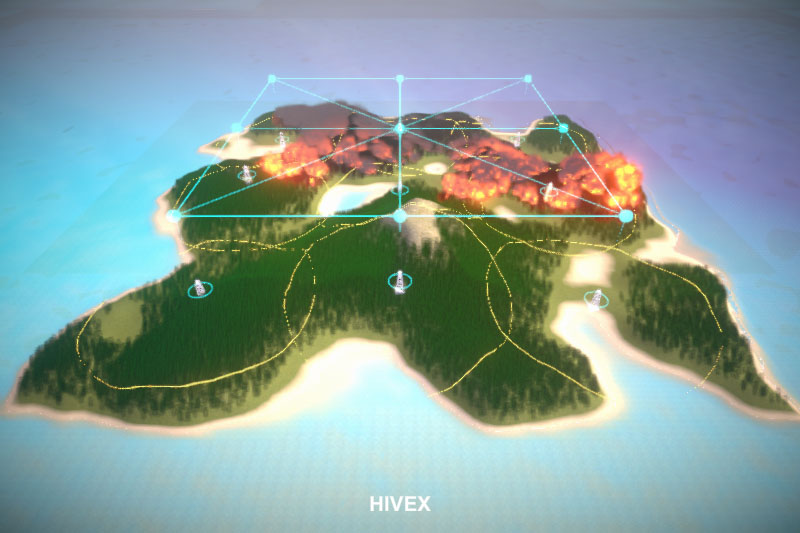}\vspace{0.0cm} 
& Wildfire Resource Management & WRM & 1 & 2 & 10 \\
\hline
\vspace{0.2cm}\includegraphics[width=0.15\textwidth, height=1.5cm]{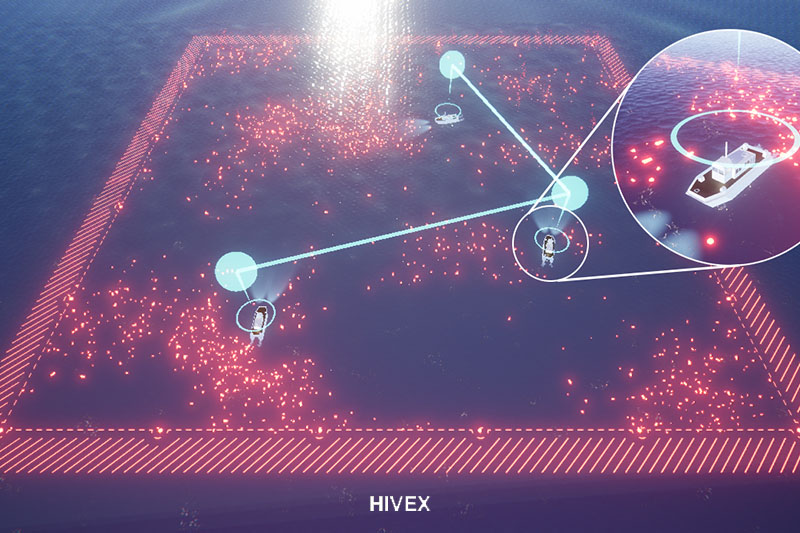}\vspace{0.0cm} 
& Ocean Plastic Collection & OPC & 1 & 3 & - \\
\hline
\vspace{0.2cm}\includegraphics[width=0.15\textwidth, height=1.5cm]{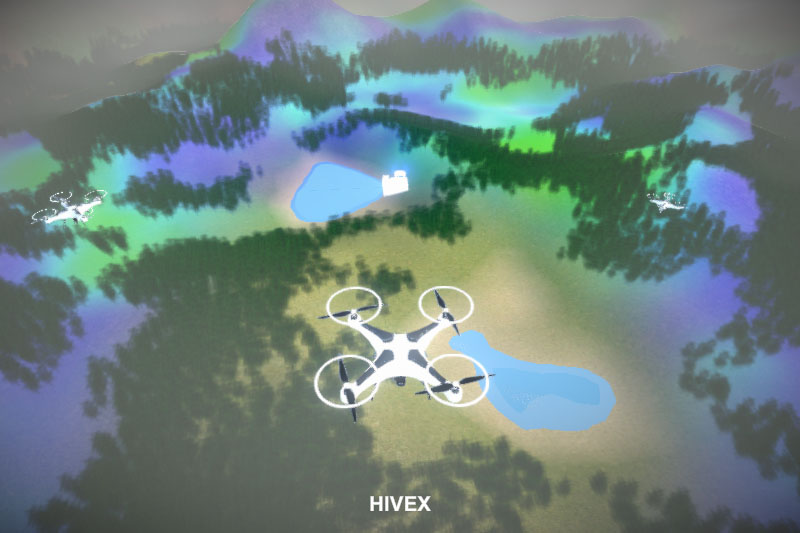}\vspace{0.0cm} 
& Drone-Based Reforestation & DBR & 1 & 6 & 10 \\
\hline
\vspace{0.2cm}\includegraphics[width=0.15\textwidth, height=1.5cm]{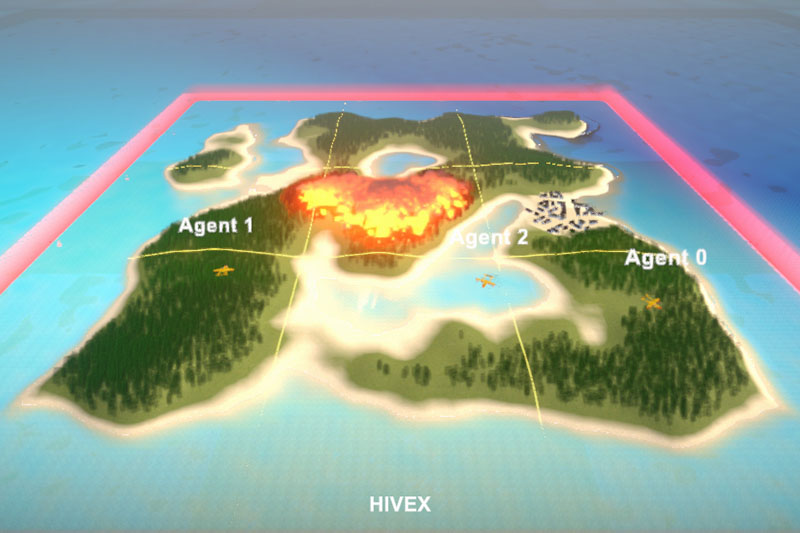}\vspace{0.0cm} 
& Aerial Wildfire Suppression & AWS & 1 & 8 & 10 \\
\hline
\end{tabular}
\label{tab:hivex_environments_fullwidth}
\end{table}

\subsection{Wind Farm Control}

\subsubsection{Environment Specification}

\begin{figure}[h!]
    \centering
    \includegraphics[width=\linewidth]{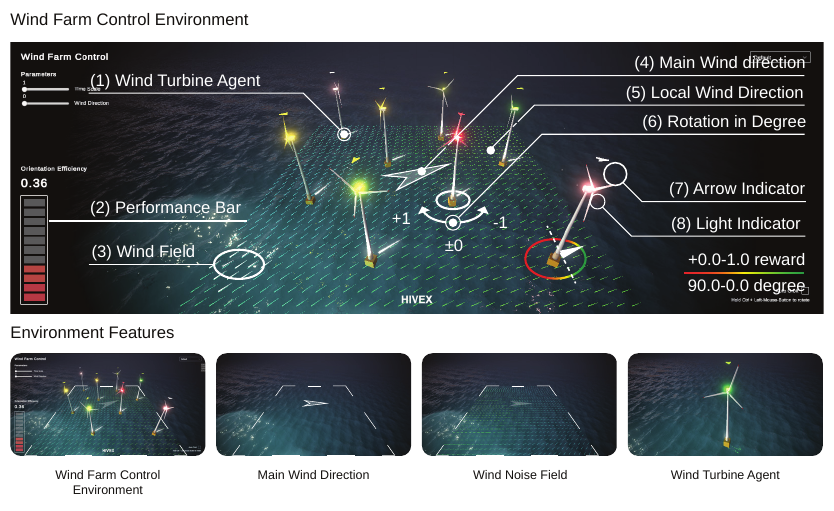}
    \vspace{-0.5cm}
    \caption{Wind Farm Control - Main environment features: Main wind direction, wind noise field sample, agent controlled wind turbine.}
    \label{fig:wind_farm_control_large}
\end{figure}

\begin{table}[h!]
\caption{Environment Specifications: Wind Farm Control}
\label{table:env_specs_WFC}
\centering 
\renewcommand{\arraystretch}{1.2} 
\begin{tabularx}{\linewidth}{l l X} 
Category & Parameter & Description/Value \\
\hline\hline
General & Episode Length & 5000 \\
        & Agent Count & 8 \\
        & Neighbour Count & 0 \\
\hline
Vector Observations (6) & Stacks & 1 \\
                         & Normalized & True \\
                         & Turbine Location (2) & \(\vec{p}(x, y)\) \\
                         & Turbine Direction (2) & \(\vec{dir}(x, y)\) \\
                         & Wind Direction (2) & \(\vec{wdir}(x, y)\) \\
\hline
Visual Observations (0) & - & - \\
\hline
Continuous Actions (0) & - & - \\
\hline
Discrete Actions (1) & Rotate Turbine & \{0: Do Nothing, 1: Rotate Left, 2: Rotate Right\} \\
\hline
\end{tabularx}
\end{table}

\subsubsection{Main Task and Rewards}

Generate Energy - The agent's goal is to rotate the wind turbine to be oriented against the wind direction and generate energy. The agent receives a positive reward in the range of $[0, 1]$ at each time step. This reward corresponds to the performance of each wind turbine and is being calculated as described in equation \ref{WFC:eq:1}. Orienting the wind turbine against the wind yields a high reward.\\
A comprehensive task list and description for the Wind Farm Control environment can be found in the Appendix \ref{task-description-and-reward-scale-wind-farm-control}. We also provide extensive reward description and calculation in the Appendix \ref{reward-description-and-calculation-wind-farm-control}.

\begin{figure}[h!]
    \centering
    \includegraphics[width=\linewidth]{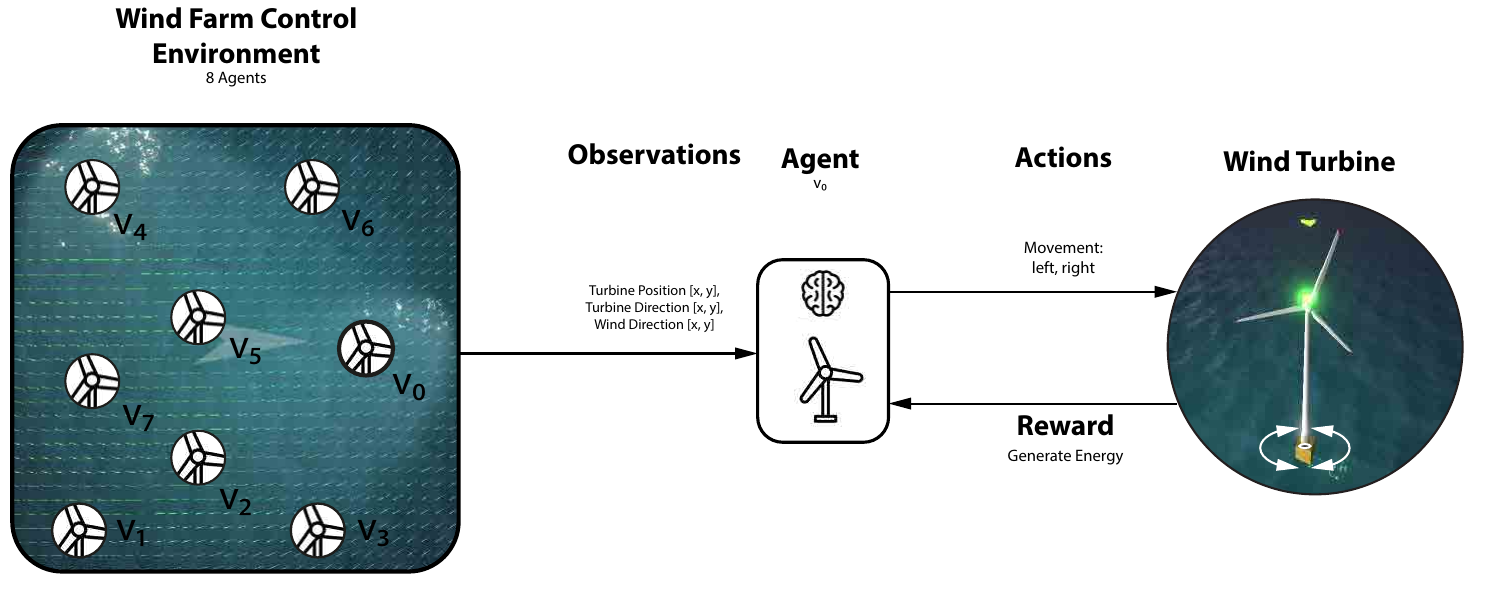}
    \vspace{-0.5cm}
    \caption{Wind Farm Control Process Diagram: The default layout of the WFC environment consists of eight wind turbines. Each turbine receives six vector inputs: its position (x, y), its orientation (x, y), and the local wind direction (x, y). The agent controlling each turbine has three discrete actions: do nothing, turn left, or turn right. The primary reward is based on the amount of wind energy generated when the turbine is optimally aligned with the wind direction.}
    \label{fig:wind_farm_control_process}
\end{figure}

\newpage

\subsection{Wildfire Resource Management}

\subsubsection{Environment Specification}

\begin{figure}[h!]
    \centering
    \includegraphics[width=\linewidth]{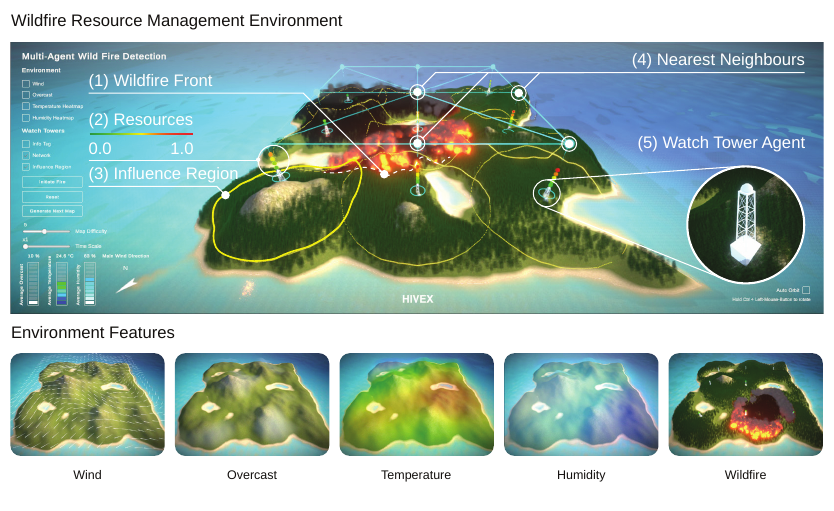}
    \vspace{-0.5cm}
    \caption{Wildfire Resource Management - Main environment features: Wind field sample, overcast field sample, temperature field sample, humidity field sample, growing wildfire.}
    \label{fig:wildfire_resource_management_large}
\end{figure}

\begin{table}[h!]
\caption{Environment Specifications: Wildfire Resource Management}
\label{table:env_specs_WRM}
\begin{center}
\begin{tabular}{lll}
Category & Parameter & Description/Value \\
\hline\hline
General & Episode Length & 500 \\
        & Agent Count & 9 \\
        & Neighbour Count & 3 \\
\hline
Vector Observations (16) & Stacks & 2 \\
                         & Normalized & True \\
                         & Closest Fire Location (3) & \(\vec{p}(x, y, z)\) \\
                         & Temperature (1) & \(t\) \\
                         & Humidity (1) & \(h\) \\
                         & Overcast (1) & \(o\) \\
                         & Total Support (1) & \(ts\) \\
\hline
Visual Observations (0) & - & - \\
\hline
Continuous Actions (0) & - & - \\
\hline
Discrete Actions (4) & Add/Sub Resource: Self & \{0: No Action, 1: Add, 2: Sub\} \\
                     & Add/Sub Resource: Neighbour 1 & \{0: No Action, 1: Add, 2: Sub\} \\
                     & Add/Sub Resource: Neighbour 2 & \{0: No Action, 1: Add, 2: Sub\} \\
                     & Add/Sub Resource: Neighbour 3 & \{0: No Action, 1: Add, 2: Sub\} \\
\hline
\end{tabular}
\end{center}
\end{table}

\subsubsection{Main Task and Rewards}

Resource Distribution - At each time step, the agent distributes a total of 1.0 resource units, in increments of 0.1, to either itself or neighbouring watchtowers. If the agent runs out of resources, it must first reallocate resources from itself or neighbouring watchtowers before redistributing. The agent's priority is to allocate resources to the watchtowers closest to and most threatened by incoming fires. The agent earns rewards based on three factors. First, it receives a positive reward corresponding to the performance of the watchtower it controls, weighted by the amount of resources allocated to itself, as described in Equation \ref{WRM:eq:1}. Second, the agent also gains a reward based on the performance of neighbouring watchtowers, which is weighted by the resources allocated to them, as outlined in Equation \ref{WRM:eq:2}. Additionally, extra rewards are given for distributing resources effectively to neighbouring watchtowers. Finally, the agent's overall reward includes a component that reflects the sum of the performance of all agent-controlled watchtowers, detailed in Equation \ref{WRM:eq:3}.\\
For more detailed information on the task descriptions and reward calculations, please refer to the Appendix (\ref{task-description-and-reward-scale-wildfire-resource-management}) and (\ref{reward-description-and-calculation-wildfire-resource-management}).

\begin{figure}[h!]
    \centering
    \includegraphics[width=\linewidth]{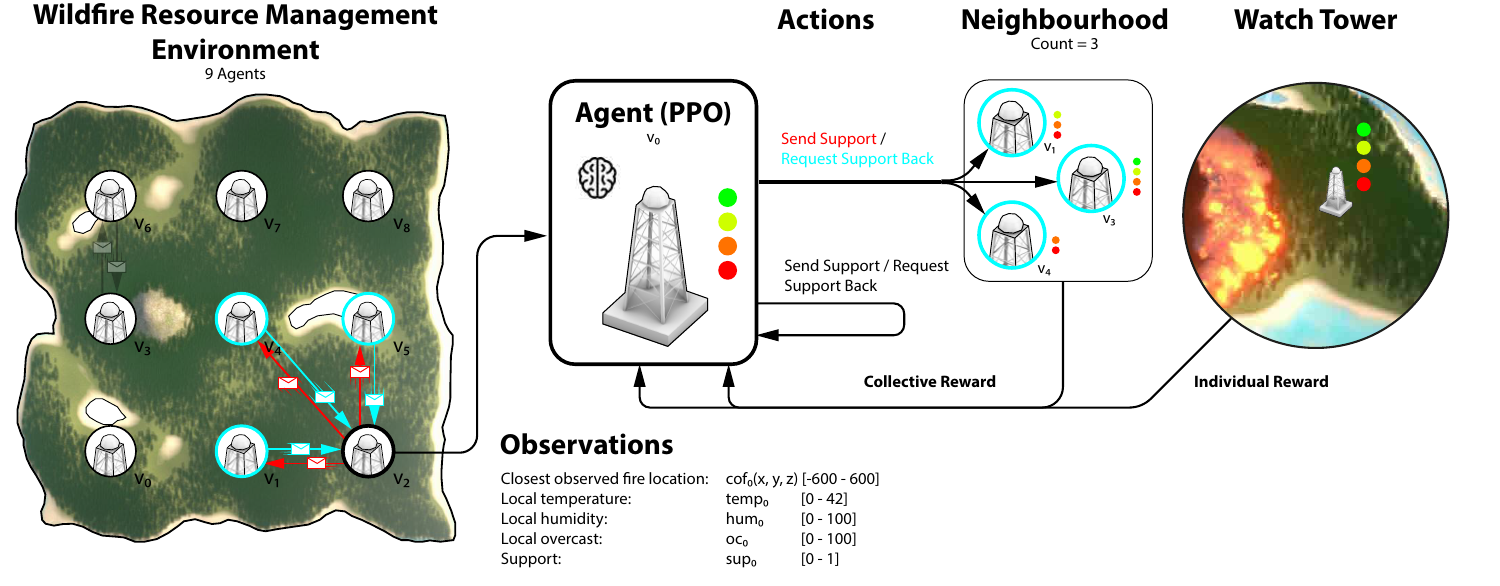}
    \vspace{-0.5cm}
    \caption{Wildfire Resource Management Process Diagram: The WRM environment consists of nine agents, each managing one of nine watchtowers. Each agent observes three environmental factors: temperature, humidity, and cloud cover, as well as whether a fire has been detected within 600 meters and the current resource level of its watchtower. Each watchtower starts with 1.0 resources, which can be allocated in 0.1 increments to either the agent's own tower or neighboring towers. Agents receive maximum rewards when their watchtower is well-resourced and a fire is approaching. For each step where the fire approaches and the watchtower is adequately prepared, the agent receives a high reward.}
    \label{fig:wildfire_resource_management_process}
\end{figure}

\newpage

\subsection{Ocean Plastic Collection}

\subsubsection{Environment Specifications}

\begin{figure}[h!]
    \centering
    \includegraphics[width=\linewidth]{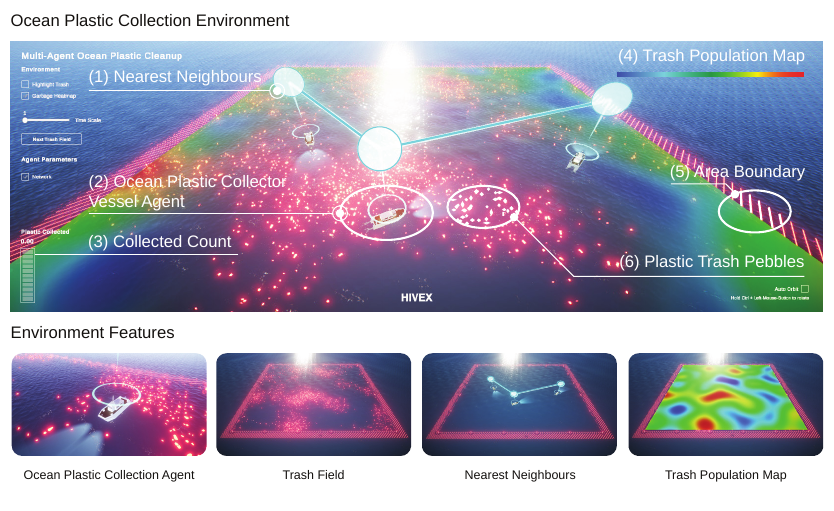}
    \vspace{-0.5cm}
    \caption{Ocean Plastic Collection - The main environment features an Agent-controlled ocean plastic collection vessel, trash field sample, nearest neighbours, and trash population map.}
    \label{fig:ocean_plastic_collection_large}
\end{figure}

\begin{table}[h!]
\caption{Environment Specifications: Ocean Plastic Collection}
\label{table:env_specs_OPC}
\begin{center}
\begin{tabular}{lll}
Category & Parameter & Description/Value \\
\hline\hline
General & Episode Length & 5000 \\
        & Agent Count & 3 \\
        & Neighbour Count & 1 \\
\hline
Vector Observations (12) & Stacks & 2 \\
                         & Normalized & True \\
                         & Local Position (2) & \(\vec{p}(x, y)\) \\
                         & Direction (2) & \(\vec{dir}(x, y)\) \\
                         & Closest Neighbouring Vessel (2) & \(\vec{np}(x, y)\) \\
\hline
Visual Observations (1250) & Resolution & 25x25x1 \\
                          & Stacks & 2 \\
                          & Normalized & True \\
                          & Trash & \(t = [0, 1]\) \\
\hline
Continuous Actions (0) & - & - \\
\hline
Discrete Actions (2) & Throttle & \{0: Do Nothing, 1: Accelerate\} \\
                     & Steer & \{0: Do Nothing, 1: Turn Right, \\
                     & & 2: Turn Left\} \\
\hline
\end{tabular}
\end{center}
\end{table}

\subsubsection{Main Task and Rewards}

Plastic Collection - The agent aims to accelerate and steer the plastic collection vessel to collect as many floating plastic pebbles as possible while avoiding crashing into other vessels and crossing the environment's border.
The agent receives a positive reward of $1$ for each floating plastic pebble collected. Furthermore, the agent receives a positive reward for the lowest collected trash count amongst all agents at each time step.
The lowest trash count is scaled by $0.01$. The steps to calculate the lowest collected trash count reward can be found in Equation \ref{OPC:eq:1}.
Finally, the agent receives a negative reward of $-100$ when the border is crossed.\\
A comprehensive task list and description for the Ocean Plastic Collection environment can be found in the Appendix \ref{task-description-and-reward-scale-ocean-plastic-collection}. We also provide extensive reward description and calculation in the Appendix \ref{reward-description-and-calculation-ocean-plastic-collection}.

\begin{figure}[h!]
    \centering
    \includegraphics[width=\linewidth]{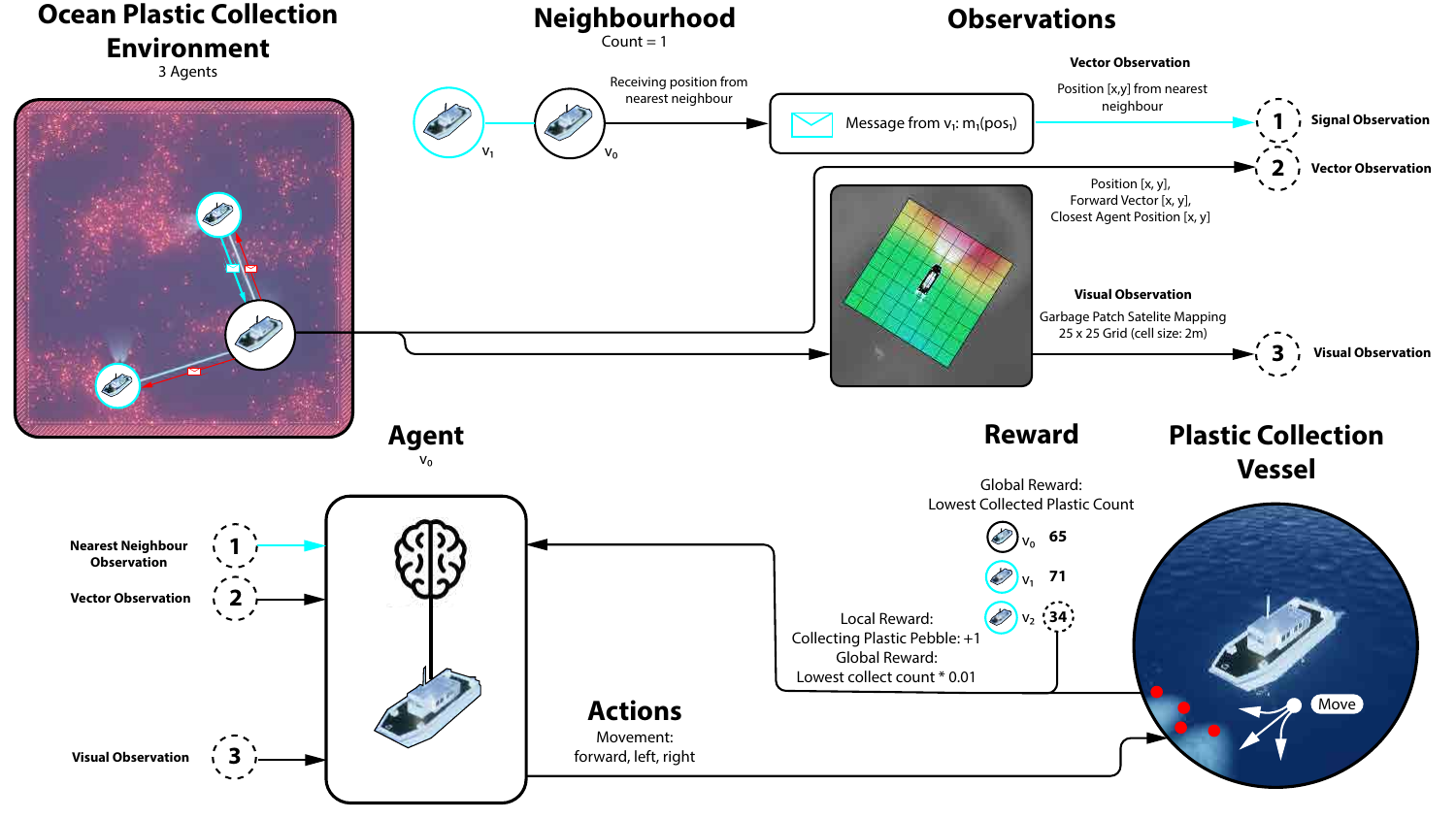}
    \vspace{-0.5cm}
    \caption{Ocean Plastic Collection Process Diagram: The default OPC environment includes three agents, each controlling a plastic collection vessel. Agents receive a 25x25 visual grid, where each cell represents 2 meters, along with vector observations such as their position (x, y), forward direction (x, y), and the position of the nearest agent (x, y). Agents can move forward, turn left, or turn right. Rewards are granted for each plastic pebble successfully collected from the ocean.}
    \label{fig:ocean_plastic_collection_process}
\end{figure}

\newpage

\subsection{Drone-Based Reforestation}

\subsubsection{Environment Specifications}

\begin{figure}[h!]
    \centering
    \includegraphics[width=\linewidth]{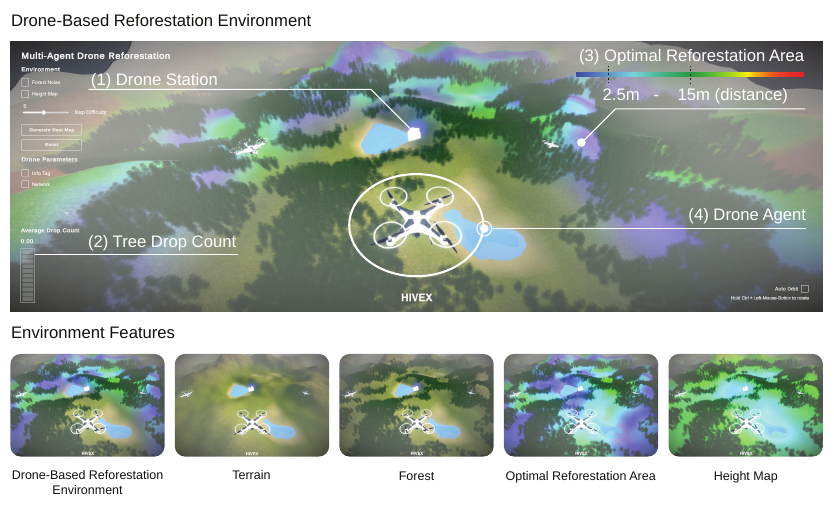}
    \vspace{-0.5cm}
    \caption{Drone-Based Reforestation - Main environment features: Terrain sample, forest sample, non-visible to agent optimal reforestation area, non-visible to agent height map.}
    \label{fig:drone_based_reforestation_large}
\end{figure}

\begin{table}[h!]
\caption{Environment Specifications: Drone-Based Reforestation}
\label{table:env_specs_DBR}
\begin{center}
\begin{tabular}{lll}
Category & Parameter & Description/Value \\
\hline\hline
General & Episode Length & 2000 \\
        & Agent Count & 3 \\
        & Neighbour Count & 0 \\
\hline
Vector Observations (20) & Stacks & 2 \\
                         & Normalized & True \\
                         & Distance to Ground (1) & \(dg\) \\
                         & Local Position (3) & \(\vec{p}(x, y, z)\) \\
                         & Direction (3) & \(\vec{dir}(x, y, z)\) \\
                         & Drone Station Height (1) & \(dsh\) \\
                         & Holding Seed (1) & \(hs = [0, 1]\) \\
                         & Energy Level (1) & \(el\) \\
\hline
Visual Observations (256) & Resolution & 16x16x1 \\
                    & Stacks & 1 \\
                    & Normalized & True \\
                    & Downward Pointing Camera & Grayscale (256), \(t = [0, 1]\) \\
\hline
Continuous Actions (3) & Throttle & \([-1, 1]\) \\
        & Steer & \([-1, 1]\) \\
        & Up/Down & \([-1, 1]\) \\
\hline
Discrete Actions (1) &  Drop Seed & \{0: Do Nothing, 1: Drop Seed\} \\
\hline
\end{tabular}
\end{center}
\end{table}

\subsubsection{Main Task and Rewards}

Maximizing Collective Tree Count - The agent's primary objective is to pick up seeds and recharge at the drone station, explore fertile ground near existing trees, and drop seeds while ensuring sufficient battery charge to return to the station. For each successful seed drop, the agent receives a reward based on two components: the quality of the drop location and its proximity to other seeds and trees. The seed quality reward ranges from 0 to 20, while the distance reward ranges from 0 to 10, giving a total possible reward of 0 to 30 for each drop. These calculations are detailed in Equation \ref{DBR:eq:3}. When carrying a seed, the agent incurs a time-step penalty of $-1 / (episode * length / 2)$, with energy depletion penalties being higher when a seed is carried. If the drone is not carrying a seed, the penalty is $-1 / episode * length$. The episode length is 2000 time steps. Additionally, the agent can receive a bonus for returning to the drone station. After a seed drop, the agent is also rewarded incrementally for reducing the distance to the station, with steps of 2.5. The incremental return reward ranges from 0 to 20 and is adjusted by a multiplier based on the seed drop quality. For example, if a seed is dropped 50 meters from the station, up to 20 incremental rewards may be received. The calculation of this reward is described in Equation \ref{DBR:eq:6}.\\
Detailed descriptions of tasks and rewards for the Drone-Based Reforestation environment are available in the Appendix \ref{task-description-and-reward-scale-drone-based-reforestation} and \ref{reward-description-and-calculation-drone-based-reforestation}.

\begin{figure}[h!]
    \centering
    \includegraphics[width=\linewidth]{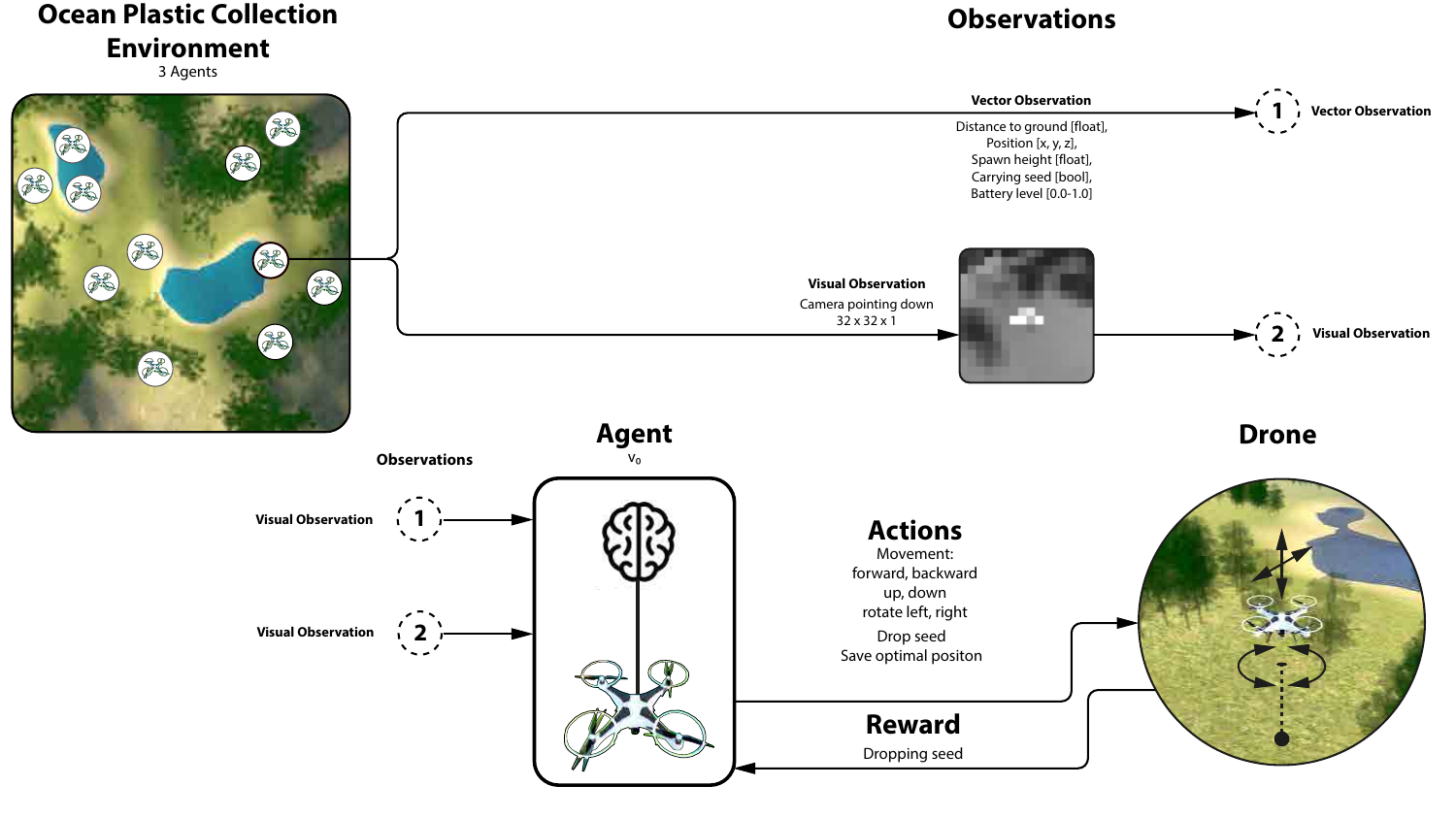}
    \vspace{-0.5cm}
    \caption{Drone-Based Reforestation Process Diagram: The default DBR environment features three agents, each controlling a drone. Each agent's observations include a vector with data such as the drone's distance to the ground, position (x, y, z), spawn height, whether it's carrying a seed, battery levels, and terrain, forest, and height maps. Additionally, agents receive a 32x32 grayscale visual observation. Agents can perform actions such as moving forward, backward, up, down, rotating left or right, saving optimal positions, and dropping a seed if carrying one. Rewards are given for successful seed drops, with bonuses for drops in highly fertile areas.}
    \label{fig:drone_based_reforestation_process}
\end{figure}

\newpage

\subsection{Aerial Wildfire Suppression}

\subsubsection{Environment Specifications}

\begin{figure}[h!]
    \centering
    \includegraphics[width=\linewidth]{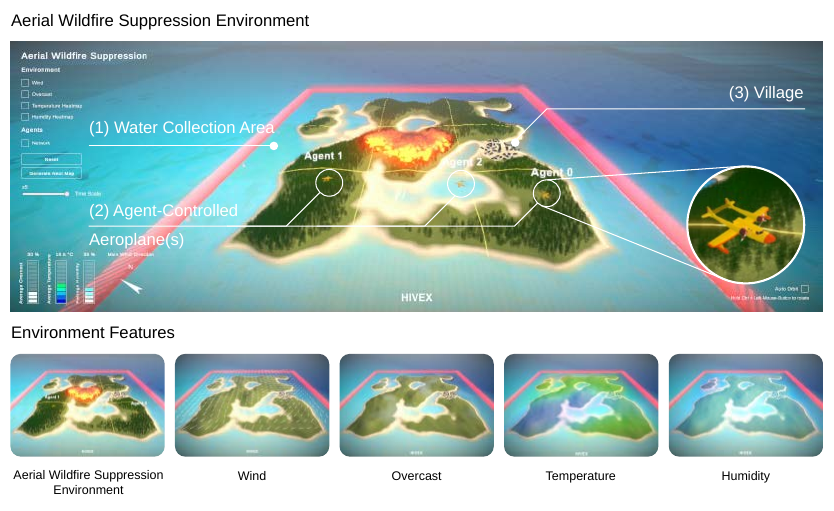}
    \vspace{-0.5cm}
    \caption{Aerial Wildfire Suppression Environment: (1) Water Collection Area, (2) Agent-controlled Wildfire Suppression Aeroplanes, (3) Village. Environment Features: Wind field sample, overcast field sample, temperature field sample, humidity field sample..}
    \label{fig:aerial_wildfire_suppression_large}
\end{figure}

\begin{table}[h!]
\caption{Environment Specifications: Aerial Wildfire Suppression}
\label{table:env_specs_AWS}
\begin{center}
\begin{tabular}{lll}
Category & Parameter & Description/Value \\
\hline\hline
General & Episode Length & 3000 \\
        & Agent Count & 3 \\
        & Neighbour Count & 0 \\
\hline
Vector Observations (8) & Stacks & 1 \\
                         & Normalized & True \\
                         & Local Position (2) & \(\vec{p}(x, y)\) \\
                         & Direction (2) & \(\vec{dir}(x, y)\) \\
                         & Holding Water (1) & \(hw = [0, 1]\) \\
                         & Closest Tree Location (2) & \(\vec{ct}(x, y)\) \\
                         & Closest Tree Burning (1) & \(ctb = [0, 1]\) \\
\hline
Visual Observations (1764) & Resolution & 42x42x3 \\
                           & Stacks & 1 \\
                           & Normalized & True \\
                           & Downward Pointing Camera & RGB, \([r, g, b] = [[0, 1], [0, 1], [0, 1]]\) \\
\hline
Continuous Actions (1) & Steer Left/Right & \([-1, 1]\) \\
\hline
Discrete Actions (1) & Drop Water & \{0: Do Nothing, 1: Drop Water\} \\
\hline
\end{tabular}
\end{center}
\end{table}

\subsubsection{Main Task and Rewards}

Minimize Fire Duration and Protect the Village - The agent’s primary goal is to pick up water and extinguish as many burning trees as possible or prepare unburned forest areas to prevent the spread of fire. A secondary goal is to protect the village by preventing fire from getting too close, either by extinguishing burning trees or redirecting the fire through tree preparation. Crossing the environment’s boundary (a $1500$x$1500$ square surrounding a $1200$x$1200$ island) results in a negative reward of $-100$. Steering the aeroplane towards the surrounding water girdle ($300$ units wide) earns a positive reward of $100$. There is also a small time-step penalty of $-1 / MaxStep$. If the fire across the entire island is extinguished, with or without agent intervention, a positive reward of 10 is given. If the fire reaches within $150$ units of the village centre, the agent receives a penalty of $-50$.\\
A detailed task list and reward breakdown for the Aerial Wildfire Suppression environment is provided in the Appendix (\ref{task-description-and-reward-scale-aerial-wildfire-suppressionl}), along with further information on reward calculations in the Appendix (\ref{reward-description-and-calculation-aerial-wildfire-suppression}).

\begin{figure}[h!]
    \centering
    \includegraphics[width=\linewidth]{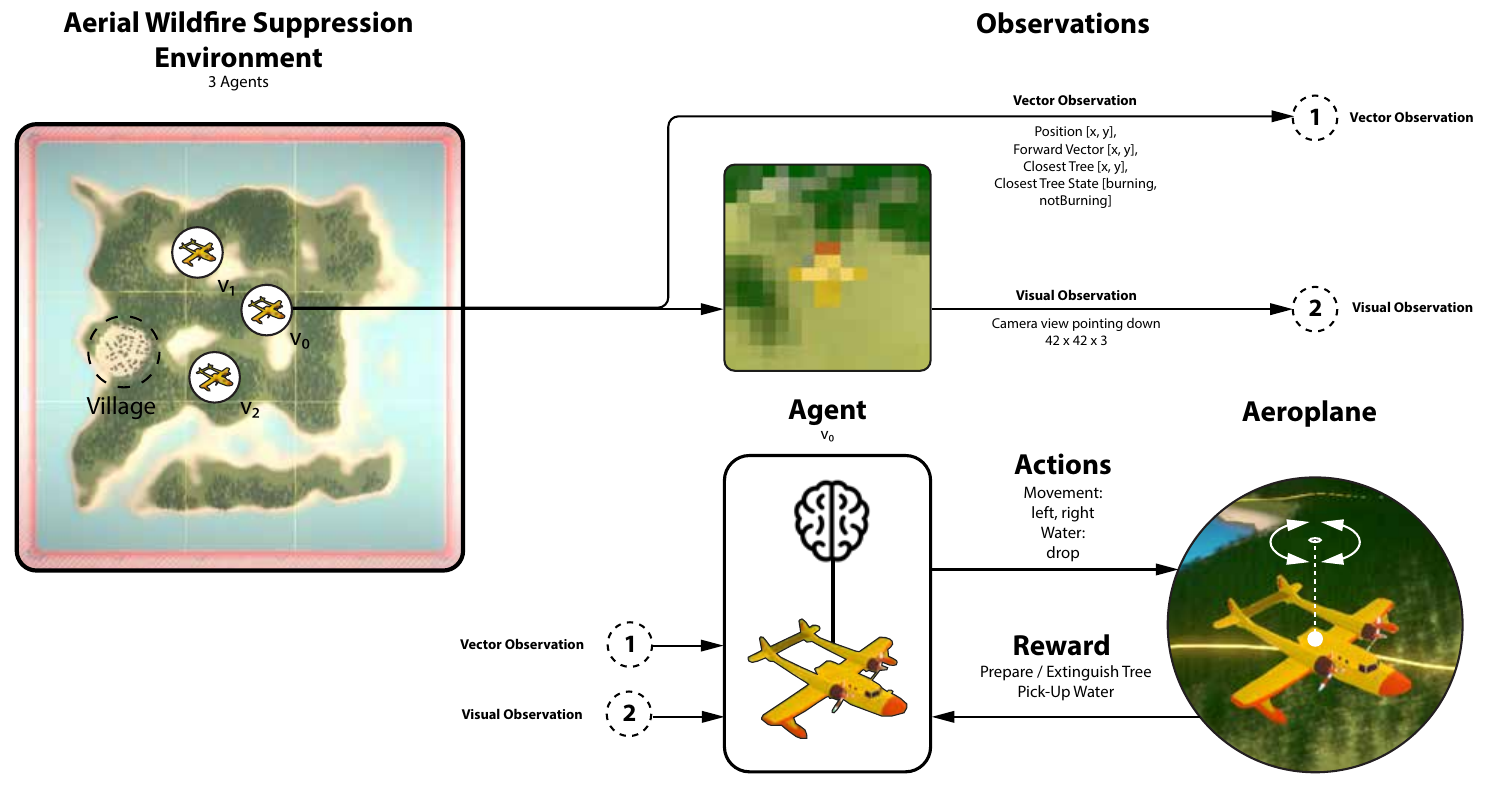}
    \vspace{-0.5cm}
    \caption{Aerial Wildfire Suppression Process Diagram: The default AWS environment consists of three agents, each controlling an airplane. Each agent receives both vector and visual observations. The vector observations include position (x, y), forward direction (x, y), the position of the nearest tree (x, y), and the tree's state: either [burning] or [not burning]. The visual observation is a 42x42x3 rgb grid. Agents can steer left, steer right, or release water. Rewards are given for extinguishing burning trees, with smaller rewards for preparing non-burning but alive trees. A small reward is also granted for picking up water.}
    \label{fig:aerial_wildfire_suppression_process}
\end{figure}

\newpage

\section{Related Work}

While the HIVEX environments can be situated close to some existing MARL benchmarks in the domain of UAVs \cite{lv_multi-agent_2023,cui_multi-agent_2020,qie_joint_2019,pham_cooperative_2018}, energy supply \cite{riedmiller_reinforcement_2001} and resource handling \cite{han_budget_2021,perolat_multi-agent_2017,ben_noureddine_multi-agent_2017}, we believe there is a gap for critical ecological challenges such as wildfires \cite{maccarthy_new_2022, tyukavina_global_2022}, pollution \cite{wef_new_2016} and deforestation \cite{dow_goldman_estimating_2020}.

Many environment suits available are grid-based and have very simple 2D visual representations such as Level-Based Foraging \cite{christianos_shared_2021}, PressurePlate, Multi-Robot Warehouse (RWARE) \cite{papoudakis_benchmarking_2021}, Pommerman \cite{resnick_pommerman_2022}, or Overcooked \cite{carroll_utility_2020} and many more. By enriching the visual representation of these environments and reducing the level of abstraction, we believe we can attract a broader range of disciplines to engage with the HIVEX environments suite.

Procedurally generating environment features, such as level design, tasks \cite{vinyals_grandmaster_2019, berner_dota_2019}, and agent populations have been adopted in various environment suits, such as Meltingpot \cite{leibo_scalable_2021}, Neural MMO \cite{suarez_neural_2019} and Capture the Flag \cite{jaderberg_human-level_2019}. We procedurally generate terrains in various terrain elevation levels for Wildfire Resource Management, Drone-Based Reforestation and Aerial Wildfire Suppression environments \ref{fig:drone_based_reforestation_env_scenarios}. The environments Wind Farm Control and Ocean Plastic Collection utilize noise maps and random sampling \ref{fig:wind_farm_control_env_scenarios}, \ref{fig:wildfire_resource_management_env_scenarios}, \ref{fig:drone_based_reforestation_env_scenarios}, \ref{fig:ocean_plastic_collection_env_scenarios}, \ref{fig:aerial_wildfire_suppression_env_scenarios}.

DeepMind's work Melting Pot is a suite of test scenarios for multi-agent reinforcement learning emphasising social situations \cite{leibo_scalable_2021}. While we do not directly target social aspects in our environments, our previous work has shown significant performance improvements when introducing communication mechanisms in earlier versions of HIVEX environments \cite{siedler_power_2021, siedler_dynamic_2022, siedler_collaborative_2022, siedler_learning_2023}. However, Melting Pot, with its 50 substrates (environments) and 256 unique scenarios (tasks), has influenced the structural design of our environment suite.

Work such as Neural MMO or LUX \cite{chen_emergent_2023} focuses on efficient large agent number environments. However, we believe that this is not as important for our work, as the scenarios we have presented do not require large amounts of agents. Nevertheless, we have shown that our environments scale well across increasing numbers of agents.

There is a trade-off between simulated environments and experience samples from the real world. While The latter might be expensive, mixtures of both can lead to success \cite{shashua_sim_2021}. HIVEX focuses on simulated environments. However, we would like to shorten the sim-to-real gap in future work.

\section{Experiments and Results} \label{ref:results}

\begin{figure}[h!]
\centering
\includegraphics[width=\linewidth]{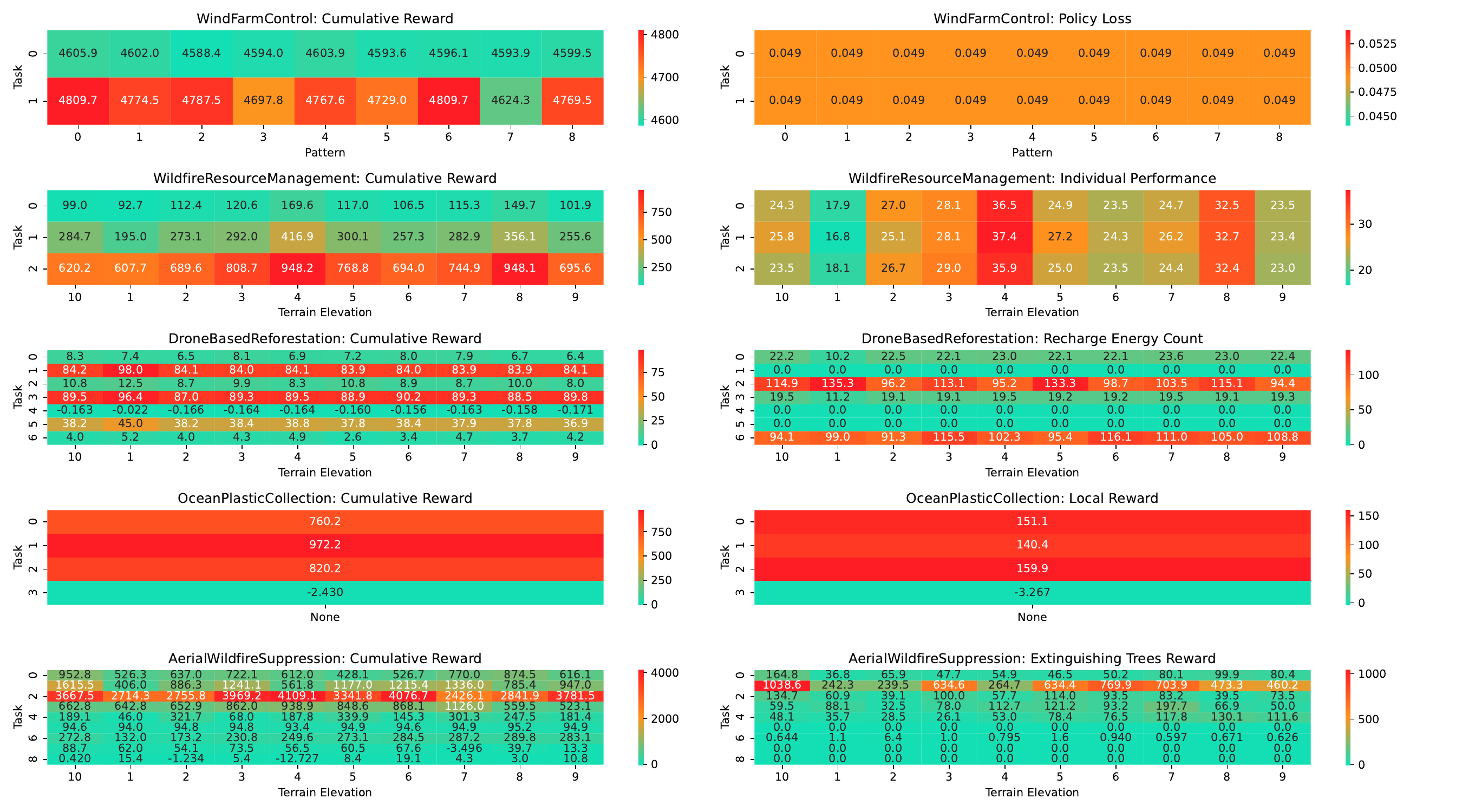}
\caption{Average test results for all environments for Cumulative Reward and environment-specific metrics such as 1. Wind Farm Control: Policy Loss, 2. Wildfire Resource Management: Individual Performance is the isolated individual performance, 3. Drone-Based Reforestation: Recharge Energy Count, which indicates how often a drone returned to the drone station to recharge energy and pick up a new seed; 4. Ocean Plastic Collection: Local Reward, which is the reward for collecting plastic pebbles, 5. Aerial Wildfire Suppression: Extinguishing Trees Reward.}
\label{fig:results}
\end{figure}

We have trained and tested all environments across all tasks and terrain elevation levels or patterns three times and report the average and the error margin \ref{fig:results}. The test runs represent the baseline for the HIVEX environment suite. Extensive results can be found in the Appendix in the section Additional Results \ref{ref:additional_results}. Furthermore, all checkpoints and logs can be found in the \href{https://github.com/hivex-research/hivex-results}{hivex-results} repository. We have used Proximal Policy Optimization (PPO) \cite{schulman_proximal_2017} for all train and test runs (Appendix: Learning Algorithm \ref{sec:ppo}). We provide hyperparameters for training in the Hyperparameters section \ref{ref:hyperparameters}.

\begin{figure}[h!]
\centering
\includegraphics[width=\linewidth]{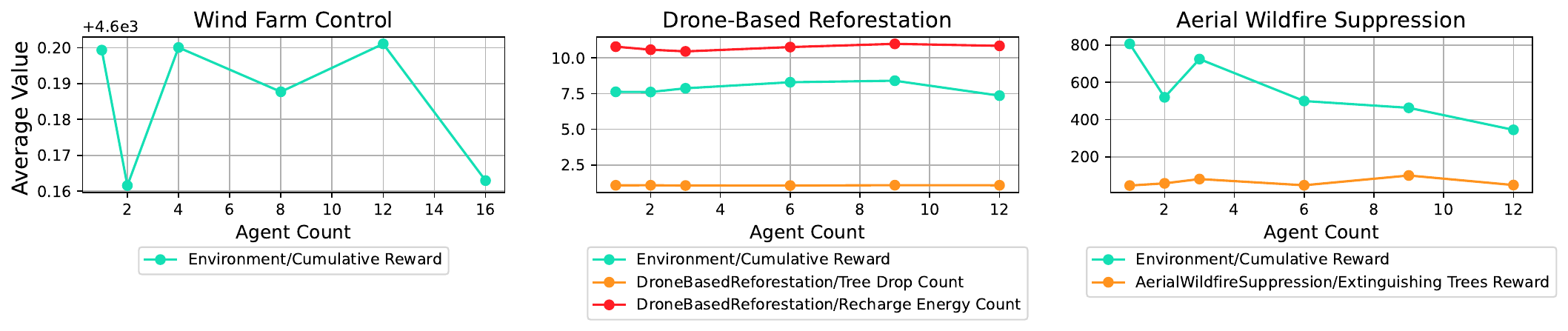}
\caption{Agent Number Scalability Test of Wind Farm Control, Drone-Based Reforestation, and Aerial Wildfire Suppression environments.}
\label{fig:scale-test}
\end{figure}

We tested the scalability of selected HIVEX environments with larger agent numbers, including Wind Farm Control, Drone-Based Reforestation, and Aerial Wildfire Suppression. Wildfire Resource Management and Ocean Plastic Collection were excluded from scalability tests: the former has a fixed layout and agent count, while the latter's fixed amount of floating plastic would reduce per-agent performance with an increased agent count. Wind Farm Control has been tested on $[1, 2, 4, 8, 12, 16]$, Drone-Based Reforestation and Aerial Wildfire Suppression on $[1, 2, 3, 6, 9, 12]$ agent counts \ref{fig:scale-test}.

\section{Discussion}

The cumulative reward performance in Wind Farm Control exhibits a stable trajectory across various layout patterns, indicating a well-optimized policy that effectively manages changing wind conditions. Despite minor fluctuations, the overall trend remains consistent across different tasks.

In Wildfire Resource Management, cumulative rewards show greater variability as task difficulty increases. Although rewards initially rise with terrain elevation levels, they plateau and fluctuate at higher levels, such as 4 and 8, marking the highest recorded reward. A higher terrain elevation level has steeper mountains and a more structured but sparse distribution of forest volume along mountain ranges. This suggests the model struggles in open fields where fire behaviour is less predictable. Nevertheless, the model performs reasonably in most scenarios, demonstrating its adaptability in real-world wildfire resource allocation. This trend is further evident in the individual performance data.

The Drone-Based Reforestation task demonstrates relatively stable but declining cumulative rewards, indicating the model's efficiency in reforestation efforts despite struggling in more challenging scenarios involving steep terrain and sparse forest areas. The "Recharge Energy Count" metric remains steady, even as terrain elevation increases, suggesting that while the agent struggles to find optimal drop locations, it maintains consistent drop and recharge activity. This metric's stability across tasks suggests potential for improvement, such as testing more energy-demanding tasks or introducing tighter energy consumption constraints.

In Aerial Wildfire Suppression, task performance appears highly sensitive to terrain elevation, with rewards dropping as complexity increases. While the model performs well in scenarios with sparse forest volume and limited fire spread, it struggles in scenarios with denser forests where fires can spread in all directions. As in other tasks, higher terrain elevation reflects steeper terrain and sparser forest distribution, requiring more frequent water drops as fires spread more unpredictably. The "Extinguishing Trees Reward" metric also reflects this variability, emphasizing the need for refined strategies, such as pre-wetting trees to direct the fire in lower-terrain elevation scenarios.

Overall, the baseline model demonstrates varying success across difficulties and environments. The baseline results indicate that the model efficiently learns routine conditions, but its performance declines as the complexity of the tasks increases. This indicates that the environments effectively introduce new challenges across scenarios, patterns, or terrain elevation levels. Future work should focus on adding even more difficult scenarios and edge cases.

The scalability analysis reveals that multi-agent systems in all three environments - Wind Farm Control, Drone-Based Reforestation, and Aerial Wildfire Suppression - exhibit stable and positive performance trends as agent counts increase. In Wind Farm Control, the cumulative reward remains stable across all tested agent counts, indicating that the system scales effectively without significant performance degradation.\\
In Drone-Based Reforestation, the cumulative reward scales well, with only a minor decrease beyond 9 agents. Tree drop counts remain stable, reflecting consistent performance, while energy consumption shows a slight upward trend, demonstrating good scalability with manageable resource trade-offs.\\
For Aerial Wildfire Suppression, the cumulative reward is generally stable as agent numbers increase, with a slight dip before recovering toward 12 agents. The extinguishing reward follows a similar pattern, showing an upward trend as agents increase, indicating that the system scales well despite minor fluctuations. Overall, these environments demonstrate good scalability across agent counts with only minor trade-offs in specific metrics \ref{fig:scale-test}.

\section{Limitations and Potential Impacts}

While our simulations provide a valuable foundation for MARL research in addressing critical ecological challenges, several limitations may affect their generalizability and real-world applicability. One major limitation is how accurately these simulations represent real-world scenarios. Despite efforts to closely model actual environments, simulations inevitably simplify complex conditions, often failing to capture unexpected environmental variables and interactions with dynamic objects.\\
For instance, turbines in the Wind Farm Control environment can be turned much faster than in reality, and wind directions shift too quickly and randomly. In contrast, real-world wind tends to have a predominant direction in specific regions. In the Ocean Plastic Collection environment, vessel turning and acceleration speeds are significantly exaggerated. Similarly, in the Reforestation environment, agents can pick up seeds simply by being near the drone station, which does not reflect real-world conditions. Fire spreads much faster in the Wildfire Resource Management and Aerial Wildfire Suppression environments. Specifically, resources are distributed too quickly in the Wildfire Resource Management environment, while the claim is that the scenarios are in remote areas.\\
Additionally, water-carrying planes turn much faster than would be possible in reality, even when fully loaded. Furthermore, the camera feed resolution in the Drone-Based Reforestation and Aerial Wildfire Suppression environments is lower than what would be needed in practice. Although the simulations perform well with low resolution, we anticipate more challenges with diverse objects in real-world scenarios.

These discrepancies could impact the real-world applicability of our findings, but there are still promising areas for implementation. For instance, algorithms developed in the Wind Farm Control environment, despite their simplified wind patterns, could contribute to optimizing wind farm layouts and improving maintenance strategies, as seen in efforts by companies like Siemens Gamesa, which integrates AI for predictive maintenance in real wind farms \cite{su_optimal_2023}. Similarly, wildfire management strategies derived from simulations, though faster than real-world conditions, could assist in resource distribution planning and suppression tactics, akin to systems used by CAL FIRE in the United States \cite{hernandez_machine_2024}. Lastly, despite its simplified nature, our reforestation environment could enhance large-scale efforts such as the Great Green Wall initiative in Africa, which seeks to restore degraded lands using new technologies \cite{gravesen_great_2022}. These applications demonstrate the potential utility of our simulations when combined with real-world data and in-field validation.

A key limitation of the current environment design is its potential for bias, as the terrains and landscapes are generated within a single climate zone. This restricts the diversity of environmental conditions, excluding deserts, rocky regions, and other ecosystems with distinct flora and fauna. To address this, future work could incorporate real geographic data from diverse global regions, including terrain, forest structure, and environmental variables like wind speed, precipitation, temperature, and cloud cover. Collaboration with companies and research labs will also be necessary to adjust agent-controlled objects to align with real-world capabilities. However, for specific applications such as wildfire or reforestation simulations, only certain areas of the world are particularly relevant, which naturally limits the range of applicable environments. For instance, wildfire simulations are most pertinent in regions such as Russia, Canada, and the United States, which experience the highest tree cover loss due to fires \cite{tyukavina_global_2022}. Conversely, reforestation efforts are more urgent in areas like the Sahara, the Zinder and Maradi regions \cite{pausata_greening_2020}, and the Amazon Rainforest \cite{dow_goldman_estimating_2020}. Thus, while the HIVEX environment suite offers a promising starting point, fine-tuning based on real-world data is essential to achieve meaningful real-world applications.

The HIVEX environment suite is designed for training and testing on accessible end-user hardware. Our simulations have been successfully executed on systems with an NVIDIA GeForce RTX 3090, an AMD Ryzen 9 7950X 16-Core Processor, and 64 GB of RAM specifications within the range of many gaming laptops and desktop computers. As such, researchers and practitioners do not need specialized, large-scale computational clusters, making our approach accessible to those with mid-range to high-end consumer hardware. Future optimizations could further reduce these requirements for even broader accessibility.

\section{Conclusion}

The HIVEX suite is a novel open-source benchmark that simulates real-world critical ecological challenges. It supports multi-agent and open-ended research across diverse tasks and scenarios by offering procedurally generated environments, as well as adjustable layout patterns and terrain elevation levels. The wide range of environments, tasks, and scenarios provides a broad spectrum of challenges, making HIVEX a valuable tool for testing new algorithms. In conclusion, while addressing critical ecological challenges remains the primary focus, it is equally important to highlight the multi-agent nature of the HIVEX suite. This characteristic plays a central role in enabling diverse, open-ended research across a variety of tasks and scenarios.

Future work aims to narrow the sim-to-real gap by incorporating real-world data, such as terrain and weather conditions. Additionally, key research directions include exploring whether a single policy can generalize across terrain levels, patterns, sub-tasks, and environments in the HIVEX suite, as well as investigating how effectively knowledge can transfer between environments and tasks. Another important question is whether modular architectures can scale more effectively than end-to-end approaches in these scenarios. Finally, future research will also focus on understanding social behavior within these environments, particularly by leveraging communication dynamics. These directions will help guide future exploration and ensure that HIVEX continues to serve as a robust platform for advancing research in multi-agent reinforcement learning.

\clearpage


\bibliography{references}

\begin{thebibliography}{83}
\providecommand{\natexlab}[1]{#1}
\providecommand{\url}[1]{\texttt{#1}}
\expandafter\ifx\csname urlstyle\endcsname\relax
  \providecommand{\doi}[1]{doi: #1}\else
  \providecommand{\doi}{doi: \begingroup \urlstyle{rm}\Url}\fi

\bibitem[Angelucci et~al.(2018)Angelucci, Hurtado-Albir, and Volpe]{angelucci_supporting_2018}
Stefano Angelucci, F.~Javier Hurtado-Albir, and Alessia Volpe.
\newblock Supporting global initiatives on climate change: {The} {EPO}'s “{Y02}-{Y04S}” tagging scheme.
\newblock \emph{World Patent Information}, 54:\penalty0 S85--S92, September 2018.
\newblock ISSN 0172-2190.
\newblock \doi{10.1016/j.wpi.2017.04.006}.
\newblock URL \url{https://www.sciencedirect.com/science/article/pii/S0172219016300618}.

\bibitem[Archer \& Rahmstorf(2010)Archer and Rahmstorf]{archer_climate_2010}
David Archer and Stefan Rahmstorf.
\newblock \emph{The {Climate} {Crisis}: {An} {Introductory} {Guide} to {Climate} {Change}}.
\newblock Cambridge University Press, 2010.
\newblock ISBN 978-0-521-40744-1.
\newblock Google-Books-ID: CH5V1Bq9ZnQC.

\bibitem[Barón~Birchenall(2016)]{baron_birchenall_animal_2016}
Leonardo Barón~Birchenall.
\newblock Animal {Communication} and {Human} {Language}: {An} overview.
\newblock \emph{International Journal of Comparative Psychology}, 29\penalty0 (1), 2016.
\newblock ISSN 0889-3675.
\newblock URL \url{https://escholarship.org/uc/item/3b7977qr}.

\bibitem[Ben~Noureddine et~al.(2017)Ben~Noureddine, Gharbi, and Ben~Ahmed]{ben_noureddine_multi-agent_2017}
Dhouha Ben~Noureddine, Atef Gharbi, and Samir Ben~Ahmed.
\newblock Multi-agent {Deep} {Reinforcement} {Learning} for {Task} {Allocation} in {Dynamic} {Environment}:.
\newblock In \emph{Proceedings of the 12th {International} {Conference} on {Software} {Technologies}}, pp.\  17--26, Madrid, Spain, 2017. SCITEPRESS - Science and Technology Publications.
\newblock ISBN 978-989-758-262-2.
\newblock \doi{10.5220/0006393400170026}.
\newblock URL \url{http://www.scitepress.org/DigitalLibrary/Link.aspx?doi=10.5220/0006393400170026}.

\bibitem[Berendt(2019)]{berendt_ai_2019}
Bettina Berendt.
\newblock {AI} for the {Common} {Good}?! {Pitfalls}, challenges, and ethics pen-testing.
\newblock \emph{Paladyn, Journal of Behavioral Robotics}, 10\penalty0 (1):\penalty0 44--65, January 2019.
\newblock ISSN 2081-4836.
\newblock \doi{10.1515/pjbr-2019-0004}.
\newblock URL \url{https://www.degruyter.com/document/doi/10.1515/pjbr-2019-0004/html}.
\newblock Publisher: De Gruyter Open Access Section: Paladyn.

\bibitem[Berner et~al.(2019)Berner, Brockman, Chan, Cheung, Dennison, Farhi, Fischer, Hashme, Hesse, Józefowicz, Gray, Olsson, Pachocki, Petrov, Salimans, Schlatter, Schneider, Sidor, Sutskever, Tang, Wolski, and Zhang]{berner_dota_2019}
Christopher Berner, Greg Brockman, Brooke Chan, Vicki Cheung, Christy Dennison, David Farhi, Quirin Fischer, Shariq Hashme, Chris Hesse, Rafal Józefowicz, Scott Gray, Catherine Olsson, Jakub Pachocki, Michael Petrov, Tim Salimans, Jeremy Schlatter, Jonas Schneider, Szymon Sidor, Ilya Sutskever, Jie Tang, Filip Wolski, and Susan Zhang.
\newblock Dota 2 with {Large} {Scale} {Deep} {Reinforcement} {Learning}.
\newblock \emph{arXiv.org}, pp.\ ~66, December 2019.

\bibitem[Calvin et~al.(2023)Calvin, Dasgupta, Krinner, Mukherji, Thorne, Trisos, Romero, Aldunce, Barrett, Blanco, Cheung, Connors, Denton, Diongue-Niang, Dodman, Garschagen, Geden, Hayward, Jones, Jotzo, Krug, Lasco, Lee, Masson-Delmotte, Meinshausen, Mintenbeck, Mokssit, Otto, Pathak, Pirani, Poloczanska, Pörtner, Revi, Roberts, Roy, Ruane, Skea, Shukla, Slade, Slangen, Sokona, Sörensson, Tignor, Van~Vuuren, Wei, Winkler, Zhai, Zommers, Hourcade, Johnson, Pachauri, Simpson, Singh, Thomas, Totin, Arias, Bustamante, Elgizouli, Flato, Howden, Méndez-Vallejo, Pereira, Pichs-Madruga, Rose, Saheb, Sánchez~Rodríguez, Ürge Vorsatz, Xiao, Yassaa, Alegría, Armour, Bednar-Friedl, Blok, Cissé, Dentener, Eriksen, Fischer, Garner, Guivarch, Haasnoot, Hansen, Hauser, Hawkins, Hermans, Kopp, Leprince-Ringuet, Lewis, Ley, Ludden, Niamir, Nicholls, Some, Szopa, Trewin, Van Der~Wijst, Winter, Witting, Birt, Ha, Romero, Kim, Haites, Jung, Stavins, Birt, Ha, Orendain, Ignon, Park, Park, Reisinger, Cammaramo, Fischlin,
  Fuglestvedt, Hansen, Ludden, Masson-Delmotte, Matthews, Mintenbeck, Pirani, Poloczanska, Leprince-Ringuet, and Péan]{lee_ipcc_2023}
Katherine Calvin, Dipak Dasgupta, Gerhard Krinner, Aditi Mukherji, Peter~W. Thorne, Christopher Trisos, José Romero, Paulina Aldunce, Ko~Barrett, Gabriel Blanco, William~W.L. Cheung, Sarah Connors, Fatima Denton, Aïda Diongue-Niang, David Dodman, Matthias Garschagen, Oliver Geden, Bronwyn Hayward, Christopher Jones, Frank Jotzo, Thelma Krug, Rodel Lasco, Yune-Yi Lee, Valérie Masson-Delmotte, Malte Meinshausen, Katja Mintenbeck, Abdalah Mokssit, Friederike~E.L. Otto, Minal Pathak, Anna Pirani, Elvira Poloczanska, Hans-Otto Pörtner, Aromar Revi, Debra~C. Roberts, Joyashree Roy, Alex~C. Ruane, Jim Skea, Priyadarshi~R. Shukla, Raphael Slade, Aimée Slangen, Youba Sokona, Anna~A. Sörensson, Melinda Tignor, Detlef Van~Vuuren, Yi-Ming Wei, Harald Winkler, Panmao Zhai, Zinta Zommers, Jean-Charles Hourcade, Francis~X. Johnson, Shonali Pachauri, Nicholas~P. Simpson, Chandni Singh, Adelle Thomas, Edmond Totin, Paola Arias, Mercedes Bustamante, Ismail Elgizouli, Gregory Flato, Mark Howden, Carlos Méndez-Vallejo,
  Joy~Jacqueline Pereira, Ramón Pichs-Madruga, Steven~K. Rose, Yamina Saheb, Roberto Sánchez~Rodríguez, Diana Ürge Vorsatz, Cunde Xiao, Noureddine Yassaa, Andrés Alegría, Kyle Armour, Birgit Bednar-Friedl, Kornelis Blok, Guéladio Cissé, Frank Dentener, Siri Eriksen, Erich Fischer, Gregory Garner, Céline Guivarch, Marjolijn Haasnoot, Gerrit Hansen, Mathias Hauser, Ed~Hawkins, Tim Hermans, Robert Kopp, Noëmie Leprince-Ringuet, Jared Lewis, Debora Ley, Chloé Ludden, Leila Niamir, Zebedee Nicholls, Shreya Some, Sophie Szopa, Blair Trewin, Kaj-Ivar Van Der~Wijst, Gundula Winter, Maximilian Witting, Arlene Birt, Meeyoung Ha, José Romero, Jinmi Kim, Erik~F. Haites, Yonghun Jung, Robert Stavins, Arlene Birt, Meeyoung Ha, Dan Jezreel~A. Orendain, Lance Ignon, Semin Park, Youngin Park, Andy Reisinger, Diego Cammaramo, Andreas Fischlin, Jan~S. Fuglestvedt, Gerrit Hansen, Chloé Ludden, Valérie Masson-Delmotte, J.B.~Robin Matthews, Katja Mintenbeck, Anna Pirani, Elvira Poloczanska, Noëmie Leprince-Ringuet,
  and Clotilde Péan.
\newblock {IPCC}, 2023: {Climate} {Change} 2023: {Synthesis} {Report}. {Contribution} of {Working} {Groups} {I}, {II} and {III} to the {Sixth} {Assessment} {Report} of the {Intergovernmental} {Panel} on {Climate} {Change} [{Core} {Writing} {Team}, {H}. {Lee} and {J}. {Romero} (eds.)]. {IPCC}, {Geneva}, {Switzerland}.
\newblock Technical report, Intergovernmental Panel on Climate Change (IPCC), July 2023.
\newblock URL \url{https://www.ipcc.ch/report/ar6/syr/}.
\newblock Edition: First.

\bibitem[Carroll et~al.(2020)Carroll, Shah, Ho, Griffiths, Seshia, Abbeel, and Dragan]{carroll_utility_2020}
Micah Carroll, Rohin Shah, Mark~K. Ho, Thomas~L. Griffiths, Sanjit~A. Seshia, Pieter Abbeel, and Anca Dragan.
\newblock On the {Utility} of {Learning} about {Humans} for {Human}-{AI} {Coordination}, January 2020.
\newblock URL \url{http://arxiv.org/abs/1910.05789}.
\newblock arXiv:1910.05789 [cs, stat].

\bibitem[Change(2012)]{change_managing_2012}
Intergovernmental Panel on~Climate Change.
\newblock \emph{Managing the {Risks} of {Extreme} {Events} and {Disasters} to {Advance} {Climate} {Change} {Adaptation}: {Special} {Report} of the {Intergovernmental} {Panel} on {Climate} {Change}}.
\newblock Cambridge University Press, May 2012.
\newblock ISBN 978-1-107-02506-6.
\newblock Google-Books-ID: nQg3SJtkOGwC.

\bibitem[Chen et~al.(2023)Chen, Tao, Chen, Shen, Li, Yu, Cheng, Zhu, and Li]{chen_emergent_2023}
Hanmo Chen, Stone Tao, Jiaxin Chen, Weihan Shen, Xihui Li, Chenghui Yu, Sikai Cheng, Xiaolong Zhu, and Xiu Li.
\newblock Emergent collective intelligence from massive-agent cooperation and competition, January 2023.
\newblock URL \url{http://arxiv.org/abs/2301.01609}.
\newblock arXiv:2301.01609 [cs].

\bibitem[Christianos et~al.(2021)Christianos, Schäfer, and Albrecht]{christianos_shared_2021}
Filippos Christianos, Lukas Schäfer, and Stefano~V. Albrecht.
\newblock Shared {Experience} {Actor}-{Critic} for {Multi}-{Agent} {Reinforcement} {Learning}, May 2021.
\newblock URL \url{http://arxiv.org/abs/2006.07169}.
\newblock arXiv:2006.07169 [cs].

\bibitem[{Climate TRACE -}(2022)]{climate_trace_-_climate_2022}
{Climate TRACE -}.
\newblock Climate {TRACE}, 2022.
\newblock URL \url{https://climatetrace.org/}.

\bibitem[Clutton-Brock et~al.(2001)Clutton-Brock, Brotherton, O'Riain, Griffin, Gaynor, Kansky, Sharpe, and McIlrath]{clutton-brock_contributions_2001}
T.~H. Clutton-Brock, P.~N.~M. Brotherton, M.~J. O'Riain, A.~S. Griffin, D.~Gaynor, R.~Kansky, L.~Sharpe, and G.~M. McIlrath.
\newblock Contributions to cooperative rearing in meerkats.
\newblock \emph{Animal Behaviour}, 61\penalty0 (4):\penalty0 705--710, April 2001.
\newblock ISSN 0003-3472.
\newblock \doi{10.1006/anbe.2000.1631}.
\newblock URL \url{https://www.sciencedirect.com/science/article/pii/S0003347200916312}.

\bibitem[Cohen et~al.(1997)Cohen, Levesque, and Smith]{cohen_team_1997}
Philip Cohen, Hector Levesque, and Ira Smith.
\newblock On {Team} {Formation}.
\newblock \emph{Synthese Library}, pp.\  87--114, 1997.

\bibitem[Collins et~al.(2018)Collins, Knutti, Arblaster, Dufresne, Fichefet, Gao, Jr, Johns, Krinner, Shongwe, Weaver, Wehner, Allen, Andrews, Beyerle, Bitz, Bony, Booth, Brooks, Brovkin, Browne, Brutel-Vuilmet, Cane, Chadwick, Cook, Cook, Eby, Fasullo, Forest, Forster, Good, Goosse, Gregory, Hegerl, Hezel, Hodges, Holland, Huber, Joshi, Kharin, Kushnir, Lawrence, Lee, Liddicoat, Lucas, Lucht, Marotzke, Massonnet, Matthews, Meinshausen, Morice, Otto, Patricola, Philippon, Rahmstorf, Riley, Saenko, Seager, Sedláček, Shaffrey, Shindell, Sillmann, Stevens, Stott, Webb, Zappa, Zickfeld, Joussaume, Mokssit, Taylor, and Tett]{collins_long-term_2018}
Matthew Collins, Reto Knutti, Julie Arblaster, Jean-Louis Dufresne, Thierry Fichefet, Xuejie Gao, William J~Gutowski Jr, Tim Johns, Gerhard Krinner, Mxolisi Shongwe, Andrew~J Weaver, Michael Wehner, Myles~R Allen, Tim Andrews, Urs Beyerle, Cecilia~M Bitz, Sandrine Bony, Ben B~B Booth, Harold~E Brooks, Victor Brovkin, Oliver Browne, Claire Brutel-Vuilmet, Mark Cane, Robin Chadwick, Ed~Cook, Kerry~H Cook, Michael Eby, John Fasullo, Chris~E Forest, Piers Forster, Peter Good, Hugues Goosse, Jonathan~M Gregory, Gabriele~C Hegerl, Paul~J Hezel, Kevin~I Hodges, Marika~M Holland, Markus Huber, Manoj Joshi, Viatcheslav Kharin, Yochanan Kushnir, David~M Lawrence, Robert~W Lee, Spencer Liddicoat, Christopher Lucas, Wolfgang Lucht, Jochem Marotzke, François Massonnet, H~Damon Matthews, Malte Meinshausen, Colin Morice, Alexander Otto, Christina~M Patricola, Gwenaëlle Philippon, Stefan Rahmstorf, William~J Riley, Oleg Saenko, Richard Seager, Jan Sedláček, Len~C Shaffrey, Drew Shindell, Jana Sillmann, Bjorn Stevens,
  Peter~A Stott, Robert Webb, Giuseppe Zappa, Kirsten Zickfeld, Sylvie Joussaume, Abdalah Mokssit, Karl Taylor, and Simon Tett.
\newblock Long-term {Climate} {Change}: {Projections}, {Commitments} and {Irreversibility}.
\newblock \emph{The Intergovernmental Panel on Climate Change}, 2018.

\bibitem[Commission(2022)]{european_commission_how_2022}
European Commission.
\newblock How the {EU} is helping partner countries fight climate change - {European} {Commission}, 2022.
\newblock URL \url{https://climate.ec.europa.eu/news-your-voice/stories/how-eu-helping-partner-countries-fight-climate-change_en}.

\bibitem[Cui et~al.(2020)Cui, Liu, and Nallanathan]{cui_multi-agent_2020}
Jingjing Cui, Yuanwei Liu, and Arumugam Nallanathan.
\newblock Multi-{Agent} {Reinforcement} {Learning}-{Based} {Resource} {Allocation} for {UAV} {Networks}.
\newblock \emph{IEEE Transactions on Wireless Communications}, 19\penalty0 (2):\penalty0 729--743, February 2020.
\newblock ISSN 1558-2248.
\newblock \doi{10.1109/TWC.2019.2935201}.
\newblock URL \url{https://ieeexplore.ieee.org/document/8807386}.
\newblock Conference Name: IEEE Transactions on Wireless Communications.

\bibitem[Darwin(1977)]{darwin_origin_1977}
Charles Darwin.
\newblock On the origin of species by means of natural selection, or, {The} preservation of favoured races in the struggle for life., 1977.
\newblock URL \url{https://www.loc.gov/item/06017473/}.

\bibitem[De-Arteaga et~al.(2018)De-Arteaga, Herlands, Neill, and Dubrawski]{de-arteaga_machine_2018}
Maria De-Arteaga, William Herlands, Daniel~B. Neill, and Artur Dubrawski.
\newblock Machine {Learning} for the {Developing} {World}.
\newblock \emph{ACM Transactions on Management Information Systems}, 9\penalty0 (2):\penalty0 9:1--9:14, August 2018.
\newblock ISSN 2158-656X.
\newblock \doi{10.1145/3210548}.
\newblock URL \url{https://dl.acm.org/doi/10.1145/3210548}.

\bibitem[Decker(1987)]{decker_distributed_1987}
Keith~S. Decker.
\newblock Distributed problem-solving techniques: {A} survey.
\newblock \emph{IEEE Transactions on Systems, Man, \& Cybernetics}, 17\penalty0 (5):\penalty0 729--740, 1987.
\newblock ISSN 0018-9472(Print).
\newblock \doi{10.1109/TSMC.1987.6499280}.
\newblock Place: US Publisher: Institute of Electrical \& Electronics Engineers Inc.

\bibitem[Dietterich(2009)]{dietterich_machine_2009}
Thomas~G Dietterich.
\newblock Machine {Learning} in {Ecosystem} {Informatics} and {Sustainability}.
\newblock \emph{Proceedings of the Twenty-First International Joint Conference on Artificial Intelligence}, 2009.

\bibitem[Dow~Goldman et~al.(2020)Dow~Goldman, Weisse, Harris, and Schneider]{dow_goldman_estimating_2020}
Elizabeth Dow~Goldman, Mikaela Weisse, Nancy Harris, and Martina Schneider.
\newblock Estimating the {Role} of {Seven} {Commodities} in {Agriculture}-{Linked} {Deforestation}: {Oil} {Palm}, {Soy}, {Cattle}, {Wood} {Fiber}, {Cocoa}, {Coffee}, and {Rubber}.
\newblock \emph{World Resources Institute}, 2020.
\newblock \doi{10.46830/writn.na.00001}.
\newblock URL \url{https://www.wri.org/research/}.

\bibitem[Faghmous \& Kumar(2014)Faghmous and Kumar]{faghmous_big_2014}
James~H. Faghmous and Vipin Kumar.
\newblock A {Big} {Data} {Guide} to {Understanding} {Climate} {Change}: {The} {Case} for {Theory}-{Guided} {Data} {Science}.
\newblock \emph{Big Data}, 2\penalty0 (3):\penalty0 155--163, September 2014.
\newblock ISSN 2167-6461.
\newblock \doi{10.1089/big.2014.0026}.
\newblock URL \url{https://www.liebertpub.com/doi/full/10.1089/big.2014.0026}.
\newblock Publisher: Mary Ann Liebert, Inc., publishers.

\bibitem[Ford et~al.(2016)Ford, Tilleard, Berrang-Ford, Araos, Biesbroek, Lesnikowski, MacDonald, Hsu, Chen, and Bizikova]{ford_big_2016}
James~D. Ford, Simon~E. Tilleard, Lea Berrang-Ford, Malcolm Araos, Robbert Biesbroek, Alexandra~C. Lesnikowski, Graham~K. MacDonald, Angel Hsu, Chen Chen, and Livia Bizikova.
\newblock Big data has big potential for applications to climate change adaptation.
\newblock \emph{Proceedings of the National Academy of Sciences}, 113\penalty0 (39):\penalty0 10729--10732, September 2016.
\newblock \doi{10.1073/pnas.1614023113}.
\newblock URL \url{https://www.pnas.org/doi/abs/10.1073/pnas.1614023113}.
\newblock Publisher: Proceedings of the National Academy of Sciences.

\bibitem[Gomes et~al.(2019)Gomes, Dietterich, Barrett, Conrad, Dilkina, Ermon, Fang, Farnsworth, Fern, Fern, Fink, Fisher, Flecker, Freund, Fuller, Gregoire, Hopcroft, Kelling, Kolter, Powell, Sintov, Selker, Selman, Sheldon, Shmoys, Tambe, Wong, Wood, Wu, Xue, Yadav, Yakubu, and Zeeman]{gomes_computational_2019}
Carla Gomes, Thomas Dietterich, Christopher Barrett, Jon Conrad, Bistra Dilkina, Stefano Ermon, Fei Fang, Andrew Farnsworth, Alan Fern, Xiaoli Fern, Daniel Fink, Douglas Fisher, Alexander Flecker, Daniel Freund, Angela Fuller, John Gregoire, John Hopcroft, Steve Kelling, Zico Kolter, Warren Powell, Nicole Sintov, John Selker, Bart Selman, Daniel Sheldon, David Shmoys, Milind Tambe, Weng-Keen Wong, Christopher Wood, Xiaojian Wu, Yexiang Xue, Amulya Yadav, Abdul-Aziz Yakubu, and Mary~Lou Zeeman.
\newblock Computational sustainability: computing for a better world and a sustainable future.
\newblock \emph{Communications of the ACM}, 62\penalty0 (9):\penalty0 56--65, August 2019.
\newblock ISSN 0001-0782.
\newblock \doi{10.1145/3339399}.
\newblock URL \url{https://dl.acm.org/doi/10.1145/3339399}.

\bibitem[Gordon(2002)]{gordon_organization_2002}
Deborah~M. Gordon.
\newblock The organization of work in social insect colonies.
\newblock \emph{Complexity}, 8\penalty0 (1):\penalty0 43--46, 2002.
\newblock ISSN 1099-0526.
\newblock \doi{10.1002/cplx.10048}.
\newblock URL \url{https://onlinelibrary.wiley.com/doi/abs/10.1002/cplx.10048}.
\newblock \_eprint: https://onlinelibrary.wiley.com/doi/pdf/10.1002/cplx.10048.

\bibitem[Gordon(2010)]{gordon_ant_2010}
Deborah~M. Gordon.
\newblock \emph{Ant {Encounters}: {Interaction} {Networks} and {Colony} {Behavior}}.
\newblock Princeton University Press, 2010.
\newblock ISBN 978-0-691-13879-4.
\newblock URL \url{https://www.jstor.org/stable/j.ctt7rpzh}.

\bibitem[Gravesen \& Funder(2022)Gravesen and Funder]{gravesen_great_2022}
Marie Gravesen and Mikkel Funder.
\newblock \emph{The {Great} {Green} {Wall}: {An} {Overview} and {Lessons} {Learnt}}.
\newblock ResearchGate, February 2022.
\newblock \doi{10.13140/RG.2.2.35246.18241}.

\bibitem[Guestrin et~al.(2002)Guestrin, Lagoudakis, and Parr]{guestrin_coordinated_2002}
Carlos Guestrin, Michail Lagoudakis, and Ronald Parr.
\newblock Coordinated {Reinforcement} {Learning}.
\newblock In \emph{In {Proceedings} of the {ICML}-2002 {The} {Nineteenth} {International} {Conference} on {Machine} {Learning}}, pp.\  227--234, 2002.

\bibitem[Hager et~al.(2019)Hager, Drobnis, Fang, Ghani, Greenwald, Lyons, Parkes, Schultz, Saria, Smith, and Tambe]{hager_artificial_2019}
Gregory~D. Hager, Ann Drobnis, Fei Fang, Rayid Ghani, Amy Greenwald, Terah Lyons, David~C. Parkes, Jason Schultz, Suchi Saria, Stephen~F. Smith, and Milind Tambe.
\newblock Artificial {Intelligence} for {Social} {Good}, January 2019.
\newblock URL \url{http://arxiv.org/abs/1901.05406}.
\newblock arXiv:1901.05406 [cs].

\bibitem[Han \& Arndt(2021)Han and Arndt]{han_budget_2021}
Benjamin Han and Carl Arndt.
\newblock Budget {Allocation} as a {Multi}-{Agent} {System} of {Contextual} \& {Continuous} {Bandits}.
\newblock In \emph{Proceedings of the 27th {ACM} {SIGKDD} {Conference} on {Knowledge} {Discovery} \& {Data} {Mining}}, {KDD} '21, pp.\  2937--2945, New York, NY, USA, August 2021. Association for Computing Machinery.
\newblock ISBN 978-1-4503-8332-5.
\newblock \doi{10.1145/3447548.3467124}.
\newblock URL \url{https://dl.acm.org/doi/10.1145/3447548.3467124}.

\bibitem[Hernandez \& Hoskins(2024)Hernandez and Hoskins]{hernandez_machine_2024}
Kassandra Hernandez and Aaron~B. Hoskins.
\newblock Machine learning algorithms applied to wildfire data in {California}'s central valley.
\newblock \emph{Trees, Forests and People}, 15:\penalty0 100516, March 2024.
\newblock ISSN 2666-7193.
\newblock \doi{10.1016/j.tfp.2024.100516}.
\newblock URL \url{https://www.sciencedirect.com/science/article/pii/S2666719324000244}.

\bibitem[Hernandez-Leal et~al.(2019)Hernandez-Leal, Kartal, and Taylor]{hernandez-leal_survey_2019-1}
Pablo Hernandez-Leal, Bilal Kartal, and Matthew~E. Taylor.
\newblock A {Survey} and {Critique} of {Multiagent} {Deep} {Reinforcement} {Learning}.
\newblock \emph{Autonomous Agents and Multi-Agent Systems}, 33\penalty0 (6):\penalty0 750--797, November 2019.
\newblock ISSN 1387-2532, 1573-7454.
\newblock \doi{10.1007/s10458-019-09421-1}.
\newblock URL \url{http://arxiv.org/abs/1810.05587}.
\newblock arXiv: 1810.05587.

\bibitem[Hosmer et~al.(2023)Hosmer, Mutis, Hughes, He, and Siedler]{hosmer_robotic_2023}
Tyson Hosmer, Sergio Mutis, Eric Hughes, Ziming He, and Philipp Siedler.
\newblock Robotic {Reconfiguration} with {Deep} {Multi}-{Agent} {Reinforcement} {Learning}.
\newblock \emph{ACADIA}, 2023.
\newblock URL \url{https://papers.cumincad.org/data/works/att/acadia23_v2_72.pdf}.

\bibitem[Hughes et~al.(2018)Hughes, Kerry, Baird, Connolly, Dietzel, Eakin, Heron, Hoey, Hoogenboom, Liu, McWilliam, Pears, Pratchett, Skirving, Stella, and Torda]{hughes_global_2018}
Terry~P. Hughes, James~T. Kerry, Andrew~H. Baird, Sean~R. Connolly, Andreas Dietzel, C.~Mark Eakin, Scott~F. Heron, Andrew~S. Hoey, Mia~O. Hoogenboom, Gang Liu, Michael~J. McWilliam, Rachel~J. Pears, Morgan~S. Pratchett, William~J. Skirving, Jessica~S. Stella, and Gergely Torda.
\newblock Global warming transforms coral reef assemblages.
\newblock \emph{Nature}, 556\penalty0 (7702):\penalty0 492--496, April 2018.
\newblock ISSN 1476-4687.
\newblock \doi{10.1038/s41586-018-0041-2}.
\newblock URL \url{https://www.nature.com/articles/s41586-018-0041-2}.
\newblock Number: 7702 Publisher: Nature Publishing Group.

\bibitem[{Ipcc}(2022)]{ipcc_global_2022}
{Ipcc}.
\newblock \emph{Global {Warming} of 1.5°{C}: {IPCC} {Special} {Report} on {Impacts} of {Global} {Warming} of 1.5°{C} above {Pre}-industrial {Levels} in {Context} of {Strengthening} {Response} to {Climate} {Change}, {Sustainable} {Development}, and {Efforts} to {Eradicate} {Poverty}}.
\newblock Cambridge University Press, 1 edition, June 2022.
\newblock ISBN 978-1-00-915794-0 978-1-00-915795-7.
\newblock \doi{10.1017/9781009157940}.
\newblock URL \url{https://www.cambridge.org/core/product/identifier/9781009157940/type/book}.

\bibitem[Jaderberg et~al.(2019)Jaderberg, Czarnecki, Dunning, Marris, Lever, Castaneda, Beattie, Rabinowitz, Morcos, Ruderman, Sonnerat, Green, Deason, Leibo, Silver, Hassabis, Kavukcuoglu, and Graepel]{jaderberg_human-level_2019}
Max Jaderberg, Wojciech~M. Czarnecki, Iain Dunning, Luke Marris, Guy Lever, Antonio~Garcia Castaneda, Charles Beattie, Neil~C. Rabinowitz, Ari~S. Morcos, Avraham Ruderman, Nicolas Sonnerat, Tim Green, Louise Deason, Joel~Z. Leibo, David Silver, Demis Hassabis, Koray Kavukcuoglu, and Thore Graepel.
\newblock Human-level performance in first-person multiplayer games with population-based deep reinforcement learning.
\newblock \emph{Science}, 364\penalty0 (6443):\penalty0 859--865, May 2019.
\newblock ISSN 0036-8075, 1095-9203.
\newblock \doi{10.1126/science.aau6249}.
\newblock URL \url{http://arxiv.org/abs/1807.01281}.
\newblock arXiv:1807.01281 [cs, stat].

\bibitem[Joppa(2017)]{joppa_case_2017}
Lucas~N. Joppa.
\newblock The case for technology investments in the environment.
\newblock \emph{Nature}, 552\penalty0 (7685):\penalty0 325--328, December 2017.
\newblock \doi{10.1038/d41586-017-08675-7}.
\newblock URL \url{https://www.nature.com/articles/d41586-017-08675-7}.
\newblock Bandiera\_abtest: a Cg\_type: Comment Number: 7685 Publisher: Nature Publishing Group Subject\_term: Climate change, Biodiversity.

\bibitem[Juliani et~al.(2020)Juliani, Berges, Teng, Cohen, Harper, Elion, Goy, Gao, Henry, Mattar, and Lange]{juliani_unity_2020}
Arthur Juliani, Vincent-Pierre Berges, Ervin Teng, Andrew Cohen, Jonathan Harper, Chris Elion, Chris Goy, Yuan Gao, Hunter Henry, Marwan Mattar, and Danny Lange.
\newblock Unity: {A} {General} {Platform} for {Intelligent} {Agents}, May 2020.
\newblock URL \url{http://arxiv.org/abs/1809.02627}.
\newblock arXiv:1809.02627 [cs, stat].

\bibitem[Kaack(2019)]{kaack_challenges_2019}
Lynn~Helena Kaack.
\newblock \emph{Challenges and {Prospects} for {Data}-{Driven} {Climate} {Change} {Mitigation}}.
\newblock Ph.{D}., Carnegie Mellon University, United States -- Pennsylvania, 2019.
\newblock URL \url{https://www.proquest.com/docview/2194980972/abstract/C1EE164709C34F0EPQ/1}.
\newblock ISBN: 9780438963153.

\bibitem[Komdeur et~al.(2008)Komdeur, Eikenaar, Brouwer, and Richardson]{komdeur_evolution_2008}
Jan Komdeur, Cas Eikenaar, Lyanne Brouwer, and David~S Richardson.
\newblock The {Evolution} and {Ecology} of {Cooperative} {Breeding} in {Vertebrates}.
\newblock In \emph{Encyclopedia of {Life} {Sciences}}. John Wiley \& Sons, Ltd, 2008.
\newblock ISBN 978-0-470-01590-2.
\newblock \doi{10.1002/9780470015902.a0021218}.
\newblock URL \url{https://onlinelibrary.wiley.com/doi/abs/10.1002/9780470015902.a0021218}.
\newblock \_eprint: https://onlinelibrary.wiley.com/doi/pdf/10.1002/9780470015902.a0021218.

\bibitem[Ladi et~al.(2022)Ladi, Jabalameli, and Sharifi]{ladi_applications_2022}
Tahmineh Ladi, Shaghayegh Jabalameli, and Ayyoob Sharifi.
\newblock Applications of machine learning and deep learning methods for climate change mitigation and adaptation.
\newblock \emph{Environment and Planning B: Urban Analytics and City Science}, 49\penalty0 (4):\penalty0 1314--1330, May 2022.
\newblock ISSN 2399-8083.
\newblock \doi{10.1177/23998083221085281}.
\newblock URL \url{https://doi.org/10.1177/23998083221085281}.
\newblock Publisher: SAGE Publications Ltd STM.

\bibitem[Leibo et~al.(2021)Leibo, Duéñez-Guzmán, Vezhnevets, Agapiou, Sunehag, Koster, Matyas, Beattie, Mordatch, and Graepel]{leibo_scalable_2021}
Joel~Z. Leibo, Edgar Duéñez-Guzmán, Alexander~Sasha Vezhnevets, John~P. Agapiou, Peter Sunehag, Raphael Koster, Jayd Matyas, Charles Beattie, Igor Mordatch, and Thore Graepel.
\newblock Scalable {Evaluation} of {Multi}-{Agent} {Reinforcement} {Learning} with {Melting} {Pot}, July 2021.
\newblock URL \url{http://arxiv.org/abs/2107.06857}.
\newblock arXiv:2107.06857 [cs].

\bibitem[Lv et~al.(2023)Lv, Xiao, Du, Niu, Xing, and Xu]{lv_multi-agent_2023}
Zefang Lv, Liang Xiao, Yousong Du, Guohang Niu, Chengwen Xing, and Wenyuan Xu.
\newblock Multi-{Agent} {Reinforcement} {Learning} {Based} {UAV} {Swarm} {Communications} {Against} {Jamming}.
\newblock \emph{IEEE Transactions on Wireless Communications}, 22\penalty0 (12):\penalty0 9063--9075, December 2023.
\newblock ISSN 1558-2248.
\newblock \doi{10.1109/TWC.2023.3268082}.
\newblock URL \url{https://ieeexplore.ieee.org/document/10107729}.
\newblock Conference Name: IEEE Transactions on Wireless Communications.

\bibitem[Lässig et~al.(2016)Lässig, Kersting, and Morik]{lassig_computational_2016}
Jörg Lässig, Kristian Kersting, and Katharina Morik.
\newblock Computational {Sustainability} {\textbar} {SpringerLink}, 2016.
\newblock URL \url{https://link.springer.com/book/10.1007/978-3-319-31858-5}.

\bibitem[MacCarthy et~al.(2022)MacCarthy, Tyukavina, Weisse, and Harris]{maccarthy_new_2022}
James MacCarthy, Sasha Tyukavina, Mikaela Weisse, and Nancy Harris.
\newblock New {Data} {Confirms}: {Forest} {Fires} {Are} {Getting} {Worse}, August 2022.
\newblock URL \url{https://www.wri.org/insights/global-trends-forest-fires}.

\bibitem[MATARIC(1998)]{mataric_using_1998}
MAJA~J. MATARIC.
\newblock Using communication to reduce locality in distributed multiagent learning.
\newblock \emph{Journal of Experimental \& Theoretical Artificial Intelligence}, 10\penalty0 (3):\penalty0 357--369, July 1998.
\newblock ISSN 0952-813X.
\newblock \doi{10.1080/095281398146806}.
\newblock URL \url{https://doi.org/10.1080/095281398146806}.
\newblock Publisher: Taylor \& Francis \_eprint: https://doi.org/10.1080/095281398146806.

\bibitem[Muller \& Wrangham(2004)Muller and Wrangham]{muller_dominance_2004}
Martin~N Muller and Richard~W Wrangham.
\newblock Dominance, aggression and testosterone in wild chimpanzees: a test of the ‘challenge hypothesis’.
\newblock \emph{Animal Behaviour}, 67\penalty0 (1):\penalty0 113--123, January 2004.
\newblock ISSN 0003-3472.
\newblock \doi{10.1016/j.anbehav.2003.03.013}.
\newblock URL \url{https://www.sciencedirect.com/science/article/pii/S0003347203003981}.

\bibitem[Newman et~al.(2016)Newman, Eyre, Kimble, Arcos-Burgos, Hogg, and Easteal]{newman_reproductive_2016}
Saul~J. Newman, Simon Eyre, Catherine~H. Kimble, Mauricio Arcos-Burgos, Carolyn~J. Hogg, and Simon Easteal.
\newblock Reproductive success is predicted by social dynamics and kinship in managed animal populations, May 2016.
\newblock URL \url{https://f1000research.com/articles/5-870}.

\bibitem[Ning \& Xie(2024)Ning and Xie]{ning_survey_2024}
Zepeng Ning and Lihua Xie.
\newblock A survey on multi-agent reinforcement learning and its application.
\newblock \emph{Journal of Automation and Intelligence}, February 2024.
\newblock ISSN 2949-8554.
\newblock \doi{10.1016/j.jai.2024.02.003}.
\newblock URL \url{https://www.sciencedirect.com/science/article/pii/S2949855424000042}.

\bibitem[of~the United~Nations(1997)]{food_and_agriculture_organization_of_the_united_nations_faostat_1997}
Food {and} Agriculture~Organization of~the United~Nations.
\newblock {FAOSTAT} {Statistical} {Database}: {Temperature} change statistics 1961–2022, 1997.

\bibitem[OpenAI(2021)]{openai_proximal_2021}
Spinning~Up OpenAI.
\newblock Proximal {Policy} {Optimization} — {Spinning} {Up} documentation, 2021.
\newblock URL \url{https://spinningup.openai.com/en/latest/algorithms/ppo.html}.

\bibitem[Panahi et~al.(2023)Panahi, Do, Hastings, and Lai]{panahi_rate-induced_2023}
Shirin Panahi, Younghae Do, Alan Hastings, and Ying-Cheng Lai.
\newblock Rate-induced tipping in complex high-dimensional ecological networks.
\newblock \emph{Proceedings of the National Academy of Sciences}, 120\penalty0 (51):\penalty0 e2308820120, December 2023.
\newblock \doi{10.1073/pnas.2308820120}.
\newblock URL \url{https://www.pnas.org/doi/full/10.1073/pnas.2308820120}.
\newblock Publisher: Proceedings of the National Academy of Sciences.

\bibitem[Panait \& Luke(2005)Panait and Luke]{panait_cooperative_2005}
Liviu Panait and Sean Luke.
\newblock Cooperative {Multi}-{Agent} {Learning}: {The} {State} of the {Art}.
\newblock \emph{Autonomous Agents and Multi-Agent Systems}, 11\penalty0 (3):\penalty0 387--434, November 2005.
\newblock ISSN 1387-2532, 1573-7454.
\newblock \doi{10.1007/s10458-005-2631-2}.
\newblock URL \url{http://link.springer.com/10.1007/s10458-005-2631-2}.

\bibitem[Papoudakis et~al.(2021)Papoudakis, Christianos, Schäfer, and Albrecht]{papoudakis_benchmarking_2021}
Georgios Papoudakis, Filippos Christianos, Lukas Schäfer, and Stefano~V. Albrecht.
\newblock Benchmarking {Multi}-{Agent} {Deep} {Reinforcement} {Learning} {Algorithms} in {Cooperative} {Tasks}, November 2021.
\newblock URL \url{http://arxiv.org/abs/2006.07869}.
\newblock arXiv:2006.07869 [cs, stat].

\bibitem[Pausata et~al.(2020)Pausata, Gaetani, Messori, Berg, Maia~de Souza, Sage, and deMenocal]{pausata_greening_2020}
Francesco S.~R. Pausata, Marco Gaetani, Gabriele Messori, Alexis Berg, Danielle Maia~de Souza, Rowan~F. Sage, and Peter~B. deMenocal.
\newblock The {Greening} of the {Sahara}: {Past} {Changes} and {Future} {Implications}.
\newblock \emph{One Earth}, 2\penalty0 (3):\penalty0 235--250, March 2020.
\newblock ISSN 2590-3322.
\newblock \doi{10.1016/j.oneear.2020.03.002}.
\newblock URL \url{https://www.sciencedirect.com/science/article/pii/S2590332220301007}.

\bibitem[Perolat et~al.(2017)Perolat, Leibo, Zambaldi, Beattie, Tuyls, and Graepel]{perolat_multi-agent_2017}
Julien Perolat, Joel~Z. Leibo, Vinicius Zambaldi, Charles Beattie, Karl Tuyls, and Thore Graepel.
\newblock A multi-agent reinforcement learning model of common-pool resource appropriation.
\newblock \emph{arXiv:1707.06600 [cs, q-bio]}, September 2017.
\newblock URL \url{http://arxiv.org/abs/1707.06600}.
\newblock arXiv: 1707.06600.

\bibitem[Pham et~al.(2018)Pham, La, Feil-Seifer, and Nefian]{pham_cooperative_2018}
Huy~Xuan Pham, Hung~Manh La, David Feil-Seifer, and Aria Nefian.
\newblock Cooperative and {Distributed} {Reinforcement} {Learning} of {Drones} for {Field} {Coverage}, September 2018.
\newblock URL \url{http://arxiv.org/abs/1803.07250}.
\newblock arXiv:1803.07250 [cs] version: 2.

\bibitem[Prasath et~al.(2022)Prasath, Mandal, Giardina, Kennedy, Murthy, and Mahadevan]{prasath_dynamics_2022}
S~Ganga Prasath, Souvik Mandal, Fabio Giardina, Jordan Kennedy, Venkatesh~N Murthy, and L~Mahadevan.
\newblock Dynamics of cooperative excavation in ant and robot collectives.
\newblock \emph{eLife}, 11:\penalty0 e79638, October 2022.
\newblock ISSN 2050-084X.
\newblock \doi{10.7554/eLife.79638}.
\newblock URL \url{https://doi.org/10.7554/eLife.79638}.
\newblock Publisher: eLife Sciences Publications, Ltd.

\bibitem[Qie et~al.(2019)Qie, Shi, Shen, Xu, Li, and Wang]{qie_joint_2019}
Han Qie, Dianxi Shi, Tianlong Shen, Xinhai Xu, Yuan Li, and Liujing Wang.
\newblock Joint {Optimization} of {Multi}-{UAV} {Target} {Assignment} and {Path} {Planning} {Based} on {Multi}-{Agent} {Reinforcement} {Learning}.
\newblock \emph{IEEE Access}, 7:\penalty0 146264--146272, 2019.
\newblock ISSN 2169-3536.
\newblock \doi{10.1109/ACCESS.2019.2943253}.
\newblock URL \url{https://ieeexplore.ieee.org/document/8846699/}.

\bibitem[Ravula et~al.(2019)Ravula, Alkoby, and Stone]{ravula_ad_2019}
Manish Ravula, Shani Alkoby, and Peter Stone.
\newblock Ad {Hoc} {Teamwork} {With} {Behavior} {Switching} {Agents}.
\newblock In \emph{Proceedings of the {Twenty}-{Eighth} {International} {Joint} {Conference} on {Artificial} {Intelligence}}, pp.\  550--556, Macao, July 2019. International Joint Conferences on Artificial Intelligence Organization.
\newblock \doi{10.24963/ijcai.2019/78}.
\newblock URL \url{https://doi.org/10.24963/ijcai.2019/78}.

\bibitem[Resnick et~al.(2022)Resnick, Eldridge, Ha, Britz, Foerster, Togelius, Cho, and Bruna]{resnick_pommerman_2022}
Cinjon Resnick, Wes Eldridge, David Ha, Denny Britz, Jakob Foerster, Julian Togelius, Kyunghyun Cho, and Joan Bruna.
\newblock Pommerman: {A} {Multi}-{Agent} {Playground}, April 2022.
\newblock URL \url{http://arxiv.org/abs/1809.07124}.
\newblock arXiv:1809.07124 [cs].

\bibitem[Riedmiller et~al.(2001)Riedmiller, Moore, and Schneider]{riedmiller_reinforcement_2001}
Martin Riedmiller, Andrew Moore, and Jeff Schneider.
\newblock Reinforcement {Learning} for {Cooperating} and {Communicating} {Reactive} {Agents} in {Electrical} {Power} {Grids}.
\newblock In Markus Hannebauer, Jan Wendler, and Enrico Pagello (eds.), \emph{Balancing {Reactivity} and {Social} {Deliberation} in {Multi}-{Agent} {Systems}}, pp.\  137--149, Berlin, Heidelberg, 2001. Springer.
\newblock ISBN 978-3-540-44568-5.
\newblock \doi{10.1007/3-540-44568-4_9}.

\bibitem[Rolnick et~al.(2022)Rolnick, Donti, Kaack, Kochanski, Lacoste, Sankaran, Ross, Milojevic-Dupont, Jaques, Waldman-Brown, Luccioni, Maharaj, Sherwin, Mukkavilli, Kording, Gomes, Ng, Hassabis, Platt, Creutzig, Chayes, and Bengio]{rolnick_tackling_2022}
David Rolnick, Priya~L. Donti, Lynn~H. Kaack, Kelly Kochanski, Alexandre Lacoste, Kris Sankaran, Andrew~Slavin Ross, Nikola Milojevic-Dupont, Natasha Jaques, Anna Waldman-Brown, Alexandra~Sasha Luccioni, Tegan Maharaj, Evan~D. Sherwin, S.~Karthik Mukkavilli, Konrad~P. Kording, Carla~P. Gomes, Andrew~Y. Ng, Demis Hassabis, John~C. Platt, Felix Creutzig, Jennifer Chayes, and Yoshua Bengio.
\newblock Tackling {Climate} {Change} with {Machine} {Learning}.
\newblock \emph{ACM Computing Surveys}, 55\penalty0 (2):\penalty0 42:1--42:96, February 2022.
\newblock ISSN 0360-0300.
\newblock \doi{10.1145/3485128}.
\newblock URL \url{https://dl.acm.org/doi/10.1145/3485128}.

\bibitem[Romm(2022)]{romm_climate_2022}
Joseph~J. Romm.
\newblock \emph{Climate {Change}: {What} {Everyone} {Needs} to {Know}}.
\newblock Oxford University Press, 2022.
\newblock ISBN 978-0-19-764712-7.

\bibitem[Schulman et~al.(2017)Schulman, Wolski, Dhariwal, Radford, and Klimov]{schulman_proximal_2017}
John Schulman, Filip Wolski, Prafulla Dhariwal, Alec Radford, and Oleg Klimov.
\newblock Proximal {Policy} {Optimization} {Algorithms}.
\newblock \emph{arXiv:1707.06347 [cs]}, August 2017.
\newblock URL \url{http://arxiv.org/abs/1707.06347}.
\newblock arXiv: 1707.06347.

\bibitem[Shashua et~al.(2021)Shashua, Di~Castro, and Mannor]{shashua_sim_2021}
Shirli Di~Castro Shashua, Dotan Di~Castro, and Shie Mannor.
\newblock Sim and {Real}: {Better} {Together}, October 2021.
\newblock URL \url{http://arxiv.org/abs/2110.00445}.
\newblock arXiv:2110.00445 [cs, stat].

\bibitem[Siedler(2021)]{siedler_power_2021}
Philipp~Dominic Siedler.
\newblock The {Power} of {Communication} in a {Distributed} {Multi}-{Agent} {System}.
\newblock \emph{arXiv:2111.15611 [cs]}, December 2021.
\newblock URL \url{http://arxiv.org/abs/2111.15611}.
\newblock arXiv: 2111.15611.

\bibitem[Siedler(2022{\natexlab{a}})]{siedler_collaborative_2022}
Philipp~Dominic Siedler.
\newblock Collaborative {Auto}-{Curricula} {Multi}-{Agent} {Reinforcement} {Learning} with {Graph} {Neural} {Network} {Communication} {Layer} for {Open}-ended {Wildfire}-{Management} {Resource} {Distribution}, April 2022{\natexlab{a}}.
\newblock URL \url{http://arxiv.org/abs/2204.11350}.
\newblock arXiv:2204.11350 [cs].

\bibitem[Siedler(2022{\natexlab{b}})]{siedler_dynamic_2022}
Philipp~Dominic Siedler.
\newblock Dynamic {Collaborative} {Multi}-{Agent} {Reinforcement} {Learning} {Communication} for {Autonomous} {Drone} {Reforestation}, November 2022{\natexlab{b}}.
\newblock URL \url{http://arxiv.org/abs/2211.15414}.
\newblock arXiv:2211.15414 [cs].

\bibitem[Siedler(2023)]{siedler_learning_2023}
Philipp~Dominic Siedler.
\newblock Learning to {Communicate} and {Collaborate} in a {Competitive} {Multi}-{Agent} {Setup} to {Clean} the {Ocean} from {Macroplastics}, April 2023.
\newblock URL \url{http://arxiv.org/abs/2304.05872}.
\newblock arXiv:2304.05872 [cs].

\bibitem[Su et~al.(2023)Su, Weng, Yu, and Wu]{su_optimal_2023}
Tai-Sheng Su, Xin-Yu Weng, Vincent~F. Yu, and Chin-Chun Wu.
\newblock Optimal maintenance planning for offshore wind farms under an uncertain environment.
\newblock \emph{Ocean Engineering}, 283:\penalty0 115033, September 2023.
\newblock ISSN 0029-8018.
\newblock \doi{10.1016/j.oceaneng.2023.115033}.
\newblock URL \url{https://www.sciencedirect.com/science/article/pii/S0029801823014178}.

\bibitem[Suarez et~al.(2019)Suarez, Du, Isola, and Mordatch]{suarez_neural_2019}
Joseph Suarez, Yilun Du, Phillip Isola, and Igor Mordatch.
\newblock Neural {MMO}: {A} {Massively} {Multiagent} {Game} {Environment} for {Training} and {Evaluating} {Intelligent} {Agents}, March 2019.
\newblock URL \url{http://arxiv.org/abs/1903.00784}.
\newblock arXiv:1903.00784 [cs, stat].

\bibitem[Sully et~al.(2019)Sully, Burkepile, Donovan, Hodgson, and van Woesik]{sully_global_2019}
S.~Sully, D.~E. Burkepile, M.~K. Donovan, G.~Hodgson, and R.~van Woesik.
\newblock A global analysis of coral bleaching over the past two decades.
\newblock \emph{Nature Communications}, 10\penalty0 (1):\penalty0 1264, March 2019.
\newblock ISSN 2041-1723.
\newblock \doi{10.1038/s41467-019-09238-2}.
\newblock URL \url{https://www.nature.com/articles/s41467-019-09238-2}.
\newblock Publisher: Nature Publishing Group.

\bibitem[Tomasello et~al.(2005)Tomasello, Carpenter, Call, Behne, and Moll]{tomasello_understanding_2005}
Michael Tomasello, Malinda Carpenter, Josep Call, Tanya Behne, and Henrike Moll.
\newblock Understanding and sharing intentions: the origins of cultural cognition.
\newblock \emph{The Behavioral and Brain Sciences}, 28\penalty0 (5):\penalty0 675--691; discussion 691--735, October 2005.
\newblock ISSN 0140-525X.
\newblock \doi{10.1017/S0140525X05000129}.

\bibitem[Tyukavina et~al.(2022)Tyukavina, Potapov, Hansen, Pickens, Stehman, Turubanova, Parker, Zalles, Lima, Kommareddy, Song, Wang, and Harris]{tyukavina_global_2022}
Alexandra Tyukavina, Peter Potapov, Matthew~C. Hansen, Amy~H. Pickens, Stephen~V. Stehman, Svetlana Turubanova, Diana Parker, Viviana Zalles, André Lima, Indrani Kommareddy, Xiao-Peng Song, Lei Wang, and Nancy Harris.
\newblock Global {Trends} of {Forest} {Loss} {Due} to {Fire} {From} 2001 to 2019.
\newblock \emph{Frontiers in Remote Sensing}, 3, March 2022.
\newblock ISSN 2673-6187.
\newblock \doi{10.3389/frsen.2022.825190}.
\newblock URL \url{https://www.frontiersin.org/journals/remote-sensing/articles/10.3389/frsen.2022.825190/full}.
\newblock Publisher: Frontiers.

\bibitem[UCLouvain(2023)]{cred__uclouvain_em-dat_2023}
CRED~/ UCLouvain.
\newblock {EM}-{DAT} - {The} international disaster database, 2023.
\newblock URL \url{https://www.emdat.be/}.

\bibitem[Venegas et~al.(2023)Venegas, Acevedo, and Treml]{venegas_three_2023}
Roberto~M. Venegas, Jorge Acevedo, and Eric~A. Treml.
\newblock Three decades of ocean warming impacts on marine ecosystems: {A} review and perspective.
\newblock \emph{Deep Sea Research Part II: Topical Studies in Oceanography}, 212:\penalty0 105318, December 2023.
\newblock ISSN 0967-0645.
\newblock \doi{10.1016/j.dsr2.2023.105318}.
\newblock URL \url{https://www.sciencedirect.com/science/article/pii/S0967064523000681}.

\bibitem[Verendel(2023)]{verendel_tracking_2023}
Vilhelm Verendel.
\newblock Tracking artificial intelligence in climate inventions with patent data.
\newblock \emph{Nature Climate Change}, 13\penalty0 (1):\penalty0 40--47, January 2023.
\newblock ISSN 1758-6798.
\newblock \doi{10.1038/s41558-022-01536-w}.
\newblock URL \url{https://www.nature.com/articles/s41558-022-01536-w}.
\newblock Publisher: Nature Publishing Group.

\bibitem[Vinyals et~al.(2019)Vinyals, Babuschkin, Czarnecki, Mathieu, Dudzik, Chung, Choi, Powell, Ewalds, Georgiev, Oh, Horgan, Kroiss, Danihelka, Huang, Sifre, Cai, Agapiou, Jaderberg, Vezhnevets, Leblond, Pohlen, Dalibard, Budden, Sulsky, Molloy, Paine, Gulcehre, Wang, Pfaff, Wu, Ring, Yogatama, Wünsch, McKinney, Smith, Schaul, Lillicrap, Kavukcuoglu, Hassabis, Apps, and Silver]{vinyals_grandmaster_2019}
Oriol Vinyals, Igor Babuschkin, Wojciech~M. Czarnecki, Michaël Mathieu, Andrew Dudzik, Junyoung Chung, David~H. Choi, Richard Powell, Timo Ewalds, Petko Georgiev, Junhyuk Oh, Dan Horgan, Manuel Kroiss, Ivo Danihelka, Aja Huang, Laurent Sifre, Trevor Cai, John~P. Agapiou, Max Jaderberg, Alexander~S. Vezhnevets, Rémi Leblond, Tobias Pohlen, Valentin Dalibard, David Budden, Yury Sulsky, James Molloy, Tom~L. Paine, Caglar Gulcehre, Ziyu Wang, Tobias Pfaff, Yuhuai Wu, Roman Ring, Dani Yogatama, Dario Wünsch, Katrina McKinney, Oliver Smith, Tom Schaul, Timothy Lillicrap, Koray Kavukcuoglu, Demis Hassabis, Chris Apps, and David Silver.
\newblock Grandmaster level in {StarCraft} {II} using multi-agent reinforcement learning.
\newblock \emph{Nature}, 575\penalty0 (7782):\penalty0 350--354, November 2019.
\newblock ISSN 1476-4687.
\newblock \doi{10.1038/s41586-019-1724-z}.
\newblock URL \url{https://www.nature.com/articles/s41586-019-1724-z}.
\newblock Number: 7782 Publisher: Nature Publishing Group.

\bibitem[WEF(2016)]{wef_new_2016}
WEF.
\newblock The {New} {Plastics} {Economy}: {Rethinking} the future of plastics, 2016.
\newblock URL \url{https://www.weforum.org/reports/the-new-plastics-economy-rethinking-the-future-of-plastics/}.

\bibitem[Yu et~al.(2013)Yu, Chawla, and Simoff]{yu_computational_2013}
Ting Yu, Nitesh Chawla, and Simeon Simoff.
\newblock Computational {Intelligent} {Data} {Analysis} for {Sustainable} {Development}, 2013.
\newblock URL \url{https://www.routledge.com/Computational-Intelligent-Data-Analysis-for-Sustainable-Development/Yu-Chawla-Simoff/p/book/9781138198692}.

\bibitem[Zychlinski(2019)]{zychlinski_complete_2019}
Shaked Zychlinski.
\newblock The {Complete} {Reinforcement} {Learning} {Dictionary}, November 2019.
\newblock URL \url{https://towardsdatascience.com/the-complete-reinforcement-learning-dictionary-e16230b7d24e}.

\end{thebibliography}
\bibliographystyle{iclr2025_conference}

\clearpage

\appendix

\section{Appendix}

\subsection{Resources}
\label{sec:resources}

\begin{itemize}[noitemsep,nolistsep]
    \item NVIDIA GeForce RTX 3090
    \item Driver version 536.23
    \item AMD Ryzen 9 7950X 16-Core Processor
    \item 64 GB RAM
\end{itemize}

\subsection{Multi-Agent PPO Pseudocode}\label{ref:ma_ppo}

\begin{algorithm}
\caption{Multi-Agent PPO Pseudocode}
\begin{algorithmic}
    \For {$iteration=1,2,\ldots$}
        \For {$actor=1,2,\ldots,N$}
            \State Run policy $\pi_{\theta_{old}}$ in environment for $T$ time steps
            \State Compute advantage estimates $\hat{A}_{1},\ldots,\hat{A}_{T}$
        \EndFor
        \State Optimize surrogate $L$ wrt. $\theta$, with $K$ epochs and minibatch size $M\leq NT$
        \State $\theta_{old}\leftarrow\theta$
    \EndFor
\end{algorithmic}
\label{alg:2}
\end{algorithm}

\clearpage
\subsection{Hyperparameters}\label{ref:hyperparameters}
\subsubsection{Hyperparameter Description}
\label{Appendix:hyper}
\begin{table}[h!]
\centering
    \begin{tabular}{p{0.3\textwidth}p{0.3\textwidth}p{0.3\textwidth}}
    \toprule
    Hyperparameter & Typical Range & Description\\
    \midrule
    Gamma & $0.8-0.995$ & discount factor for future rewards\\
    Lambda & $0.9-0.95$ & used when calculating the Generalized Advantage Estimate (GAE)\\
    Buffer Size & $2048-409600$ & how many experiences should be collected before updating the model\\
    Batch Size & $512-5120$ (continuous), $32-512$ (discrete) & number of experiences used for one iteration of a gradient descent update.\\\
    Number of Epochs & $3-10$ & number of passes through the experience buffer during gradient descent\\
    Learning Rate & $1e-5-1e-3$ & strength of each gradient descent update step\\
    Time Horizon & $32-2048$ & number of steps of experience to collect per-agent before adding it to the experience buffer\\
    Max Steps & $5e5-1e7$ & number of steps of the simulation (multiplied by frame-skip) during the training process\\
    Beta & $1e-4-1e-2$ & strength of the entropy regularization, which makes the policy "more random"\\
    Epsilon & $0.1-0.3$ & acceptable threshold of divergence between the old and new policies during gradient descent updating\\
    Normalize & $true/false$ & weather normalization is applied to the vector observation inputs\\
    Number of Layers & $1-3$ & number of hidden layers present after the observation input\\
    Hidden Units & $32-512$ & number of units in each fully connected layer of the neural network\\
    \midrule
    Intrinsic Curiosity Module\\
    \midrule
    Curiosity Encoding Size & $64-256$ & size of hidden layer used to encode the observations within the intrinsic curiosity module\\
    Curiosity Strength & $0.1-0.001$ & magnitude of the intrinsic reward generated by the intrinsic curiosity module\\
    \bottomrule
    \end{tabular}
\caption{Hyperparameters Description: \url{https://github.com/Unity-Technologies/ml-agents/blob/main/docs/Training-Configuration-File.md}}
\label{sample-table}
\end{table}

\clearpage
\subsubsection{Train and Test Hyperparameters: Wind Farm Control}
\begin{verbatim}
behaviors:
  Agent:
    trainer_type: ppo
    hyperparameters:
      batch_size: 256
      buffer_size: 2048
      learning_rate: 0.0003 # testing: 0.0
      beta: 0.005
      epsilon: 0.2
      lambd: 0.95
      num_epoch: 3
      learning_rate_schedule: linear # testing: constant
    network_settings:
      normalize: false
      hidden_units: 64
      num_layers: 2
    reward_signals:
      extrinsic:
        gamma: 0.9
        strength: 1.0
    keep_checkpoints: 5
    max_steps: 8000000 # testing: 8000000
    time_horizon: 2048
    summary_freq: 40000 # testing: 40000
    threaded: true

engine_settings:
  no_graphics: true

env_settings:
  env_path: /dev_environments/Hivex_WindFarmControl_win
  seed: 5000 # testing: 6000

environment_parameters:
  # Pattern: 0 Default, 1 Grid, 2 Chain, 3 Circle, 4 Square, 5 Cross,
  # 6 Two_Rows, 7 Field, 8 Random
  pattern: [0, 1, 2, 3, 4, 5, 6, 7, 8]
  task: [0, 1] # Generate Energy: 0, Avoid Damage: 1
\end{verbatim}

\clearpage
\subsubsection{Train and Test Hyperparameters: Wildfire Resource Management}
\begin{verbatim}
behaviors:
  Agent:
    trainer_type: ppo
    hyperparameters:
      batch_size: 128
      buffer_size: 2048
      learning_rate: 0.0003 # testing: 0.0
      beta: 0.01
      epsilon: 0.2
      lambd: 0.95
      num_epoch: 3
      learning_rate_schedule: linear # testing: constant
    network_settings:
      normalize: false
      hidden_units: 512
      num_layers: 2
      vis_encode_type: simple
    reward_signals:
      extrinsic:
        gamma: 0.99
        strength: 1.0
      curiosity:
        gamma: 0.99
        strength: 0.02
        encoding_size: 256
        learning_rate: 0.0003 # testing: 0.0
    keep_checkpoints: 5
    max_steps: 4500000 # testing: 450000
    time_horizon: 2048
    summary_freq: 4500 # testing: 4500
    threaded: true

engine_settings:
  no_graphics: true

env_settings:
  env_path: /dev_environments/Hivex_WildfireResourceManagement_win
  seed: 5000 # testing: 6000

environment_parameters:
  terrain_level: [1, 2, 3, 4, 5, 6, 7, 8, 9, 10]
  task: [0, 1, 2] # Main: 0, Distribute All: 1, Keep All: 2
\end{verbatim}

\clearpage
\subsubsection{Training Hyperparameters: Drone-Based Reforestation}
\begin{verbatim}
behaviors:
  Agent:
    trainer_type: ppo
    hyperparameters:
      batch_size: 1024
      buffer_size: 10240
      learning_rate: 0.0003 # testing: 0.0
      beta: 0.005
      epsilon: 0.2
      lambd: 0.95
      num_epoch: 3
      learning_rate_schedule: linear # testing: constant
    network_settings:
      normalize: false
      hidden_units: 128
      num_layers: 2
      vis_encode_type: resnet
    reward_signals:
      extrinsic:
        gamma: 0.99
        strength: 0.9
        network_settings:
          vis_encode_type: resnet
      curiosity:
        gamma: 0.99
        strength: 0.1
        encoding_size: 256
        learning_rate: 0.0003 # testing: 0.0
        network_settings:
          vis_encode_type: resnet
    keep_checkpoints: 5
    max_steps: 2000000 # testing: 2000000
    time_horizon: 10240
    summary_freq: 10000 # testing: 10000
    threaded: true

engine_settings:
  no_graphics: true

env_settings:
  env_path: /dev_environments/Hivex_DroneBasedReforestation_win
  seed: 5000 # testing: 6000

environment_parameters:
  terrain_level: [1, 2, 3, 4, 5, 6, 7, 8, 9, 10]
  task: [0, 1, 2, 3, 4, 5, 6, 7]
  # Main: 0, Find Closest Tree: 1, Group Up: 2, Pick Up Seed: 3,
  # Drop Seed: 4, Find High Potential Area: 5,
  # Find High Terrain: 6, Explore Furthest: 7
\end{verbatim}

\clearpage
\subsubsection{Training Hyperparameters: Ocean Plastic Collection}
\begin{verbatim}
behaviors:
  Agent:
    trainer_type: ppo
    hyperparameters:
      batch_size: 1024
      buffer_size: 10240
      learning_rate: 0.0003 # testing: 0.0
      beta: 0.005
      epsilon: 0.2
      lambd: 0.95
      num_epoch: 3
      learning_rate_schedule: linear # testing: constant
    network_settings:
      normalize: false
      hidden_units: 128
      num_layers: 2
      vis_encode_type: resnet
    reward_signals:
      extrinsic:
        gamma: 0.99
        strength: 0.9
        network_settings:
          vis_encode_type: resnet
      curiosity:
        gamma: 0.99
        strength: 0.1
        encoding_size: 256
        learning_rate: 0.0003 # testing: 0.0
        network_settings:
          vis_encode_type: resnet
    keep_checkpoints: 5
    max_steps: 3000000 # testing: 150000
    time_horizon: 10240
    summary_freq: 15000 # testing: 15000
    threaded: true

engine_settings:
  no_graphics: true

env_settings:
  env_path: /dev_environments/Hivex_OceanPlasticCollection_win
  seed: 5000 # testing: 6000

environment_parameters:
  task: [0, 1, 2, 3]
  # Main: 0, Find High Pollution Area: 1,
  # Group up: 2, Avoid Plastic: 3
\end{verbatim}

\clearpage
\subsubsection{Training Hyperparameters: Aerial Wildfire Suppression}
\begin{verbatim}
behaviors:
  Agent:
    trainer_type: ppo
    hyperparameters:
      batch_size: 256
      buffer_size: 4096
      learning_rate: 0.0003
      beta: 0.005
      epsilon: 0.2
      lambd: 0.95
      num_epoch: 3
      learning_rate_schedule: linear
    network_settings:
      normalize: false
      hidden_units: 256
      num_layers: 2
      vis_encode_type: simple
    reward_signals:
      extrinsic:
        gamma: 0.995
        strength: 1.0
    keep_checkpoints: 5
    max_steps: 1800000 # testing: 180000
    time_horizon: 4096
    summary_freq: 9000 # testing: 9000
    threaded: true

engine_settings:
  no_graphics: true

env_settings:
  env_path: /dev_environments/Hivex_AerialWildfireSuppression_win
  num_envs: 12
  seed: 5000 # testing: 6000

environment_parameters:
  terrain_level: [1, 2, 3, 4, 5, 6, 7, 8, 9, 10]
  task: [0, 1, 2, 3, 4, 5, 6, 7, 8]
  # Main Task: 0, Maximize Extinguishing Trees: 1,
  # Maximize Preparing Trees: 2, Minimze Time of Fire Burning: 3,
  # Protect Village: 4, Pick Up Water: 5, Drop Water: 6,
  # Find Fire: 7, Find Village: 8
\end{verbatim}

\clearpage

\subsection{Environment Scenario Samples}

\subsubsection{Wind Farm Control}

\begin{figure}[h!]
    \centering
    \includegraphics[width=\linewidth]{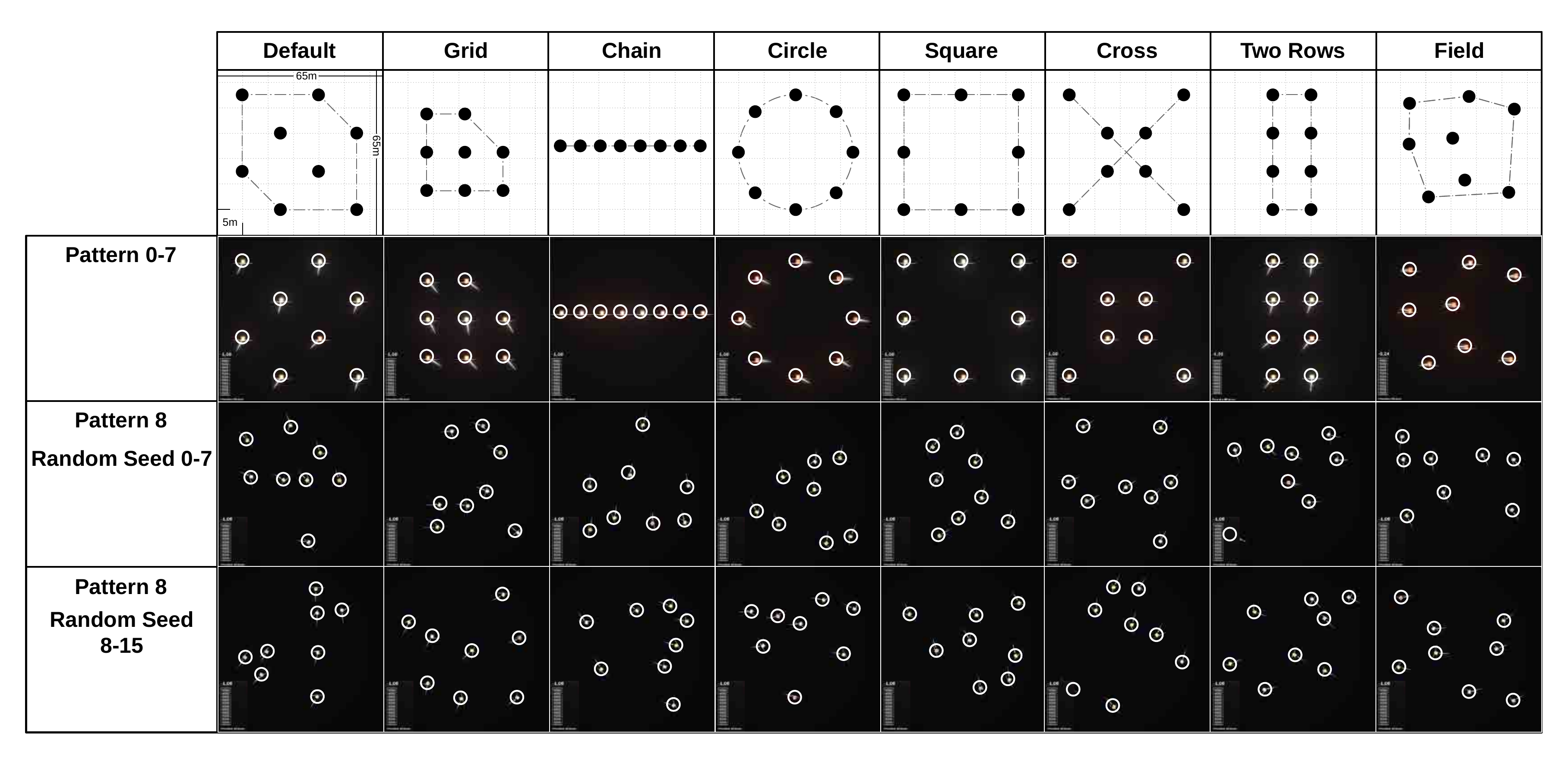}
    \caption{Wind Farm turbine layout patterns 0-7 [Default, Grid, Chain, Circle, Square, Cross, Two Rows, Field] and various seeds for the layout pattern 8 [Random].}
    \label{fig:wind_farm_control_env_scenarios}
\end{figure}

\clearpage

\subsubsection{Wildfire Resource Management}

\begin{figure}[h!]
    \centering
    \includegraphics[width=\linewidth]{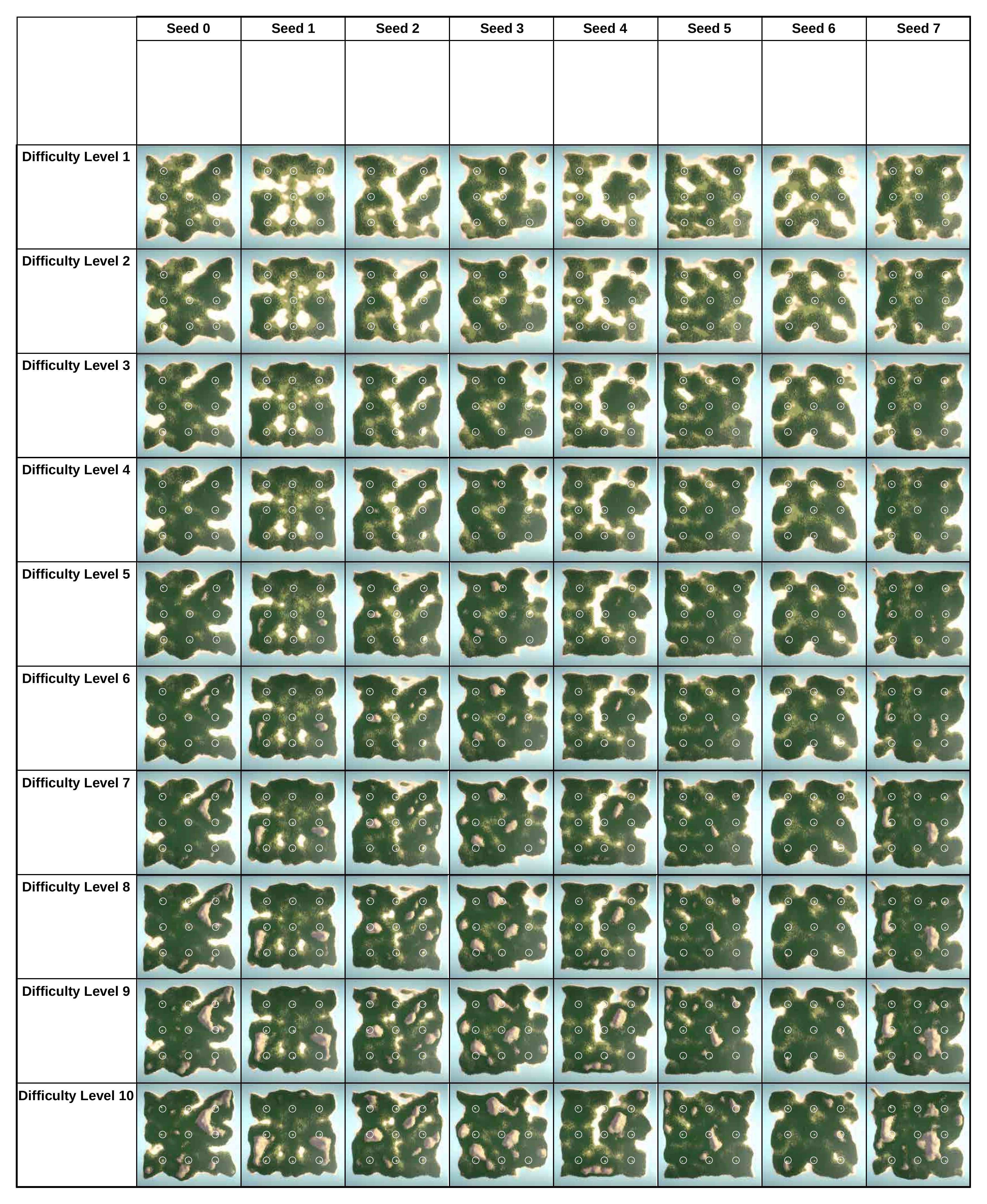}
    \caption{Wildfire Resource Management environment samples showing terrain elevation levels 1-10, top to bottom, and random seeds 0-7, left to right.}
    \label{fig:wildfire_resource_management_env_scenarios}
\end{figure}

\clearpage

\subsubsection{Drone-Based Reforestation}

\begin{figure}[h!]
    \centering
    \includegraphics[width=\linewidth]{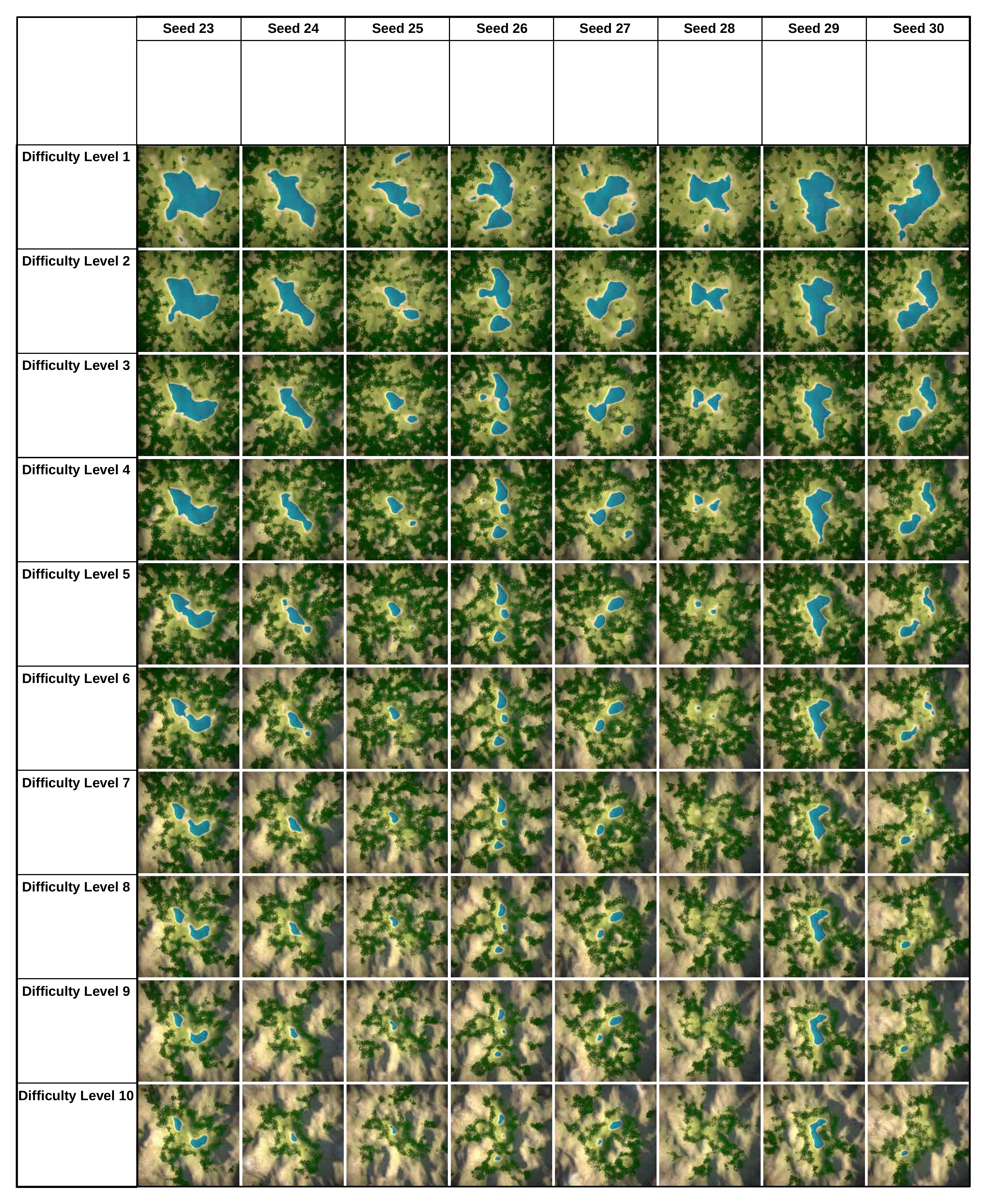}
    \caption{Drone-Based Reforestation environment samples showing terrain elevation levels 1-10, top to bottom, and random seeds 23-30, left to right.}
    \label{fig:drone_based_reforestation_env_scenarios}
\end{figure}

\clearpage

\subsubsection{Ocean Plastic Collection}

\begin{figure}[h!]
    \centering
    \includegraphics[width=0.96\linewidth]{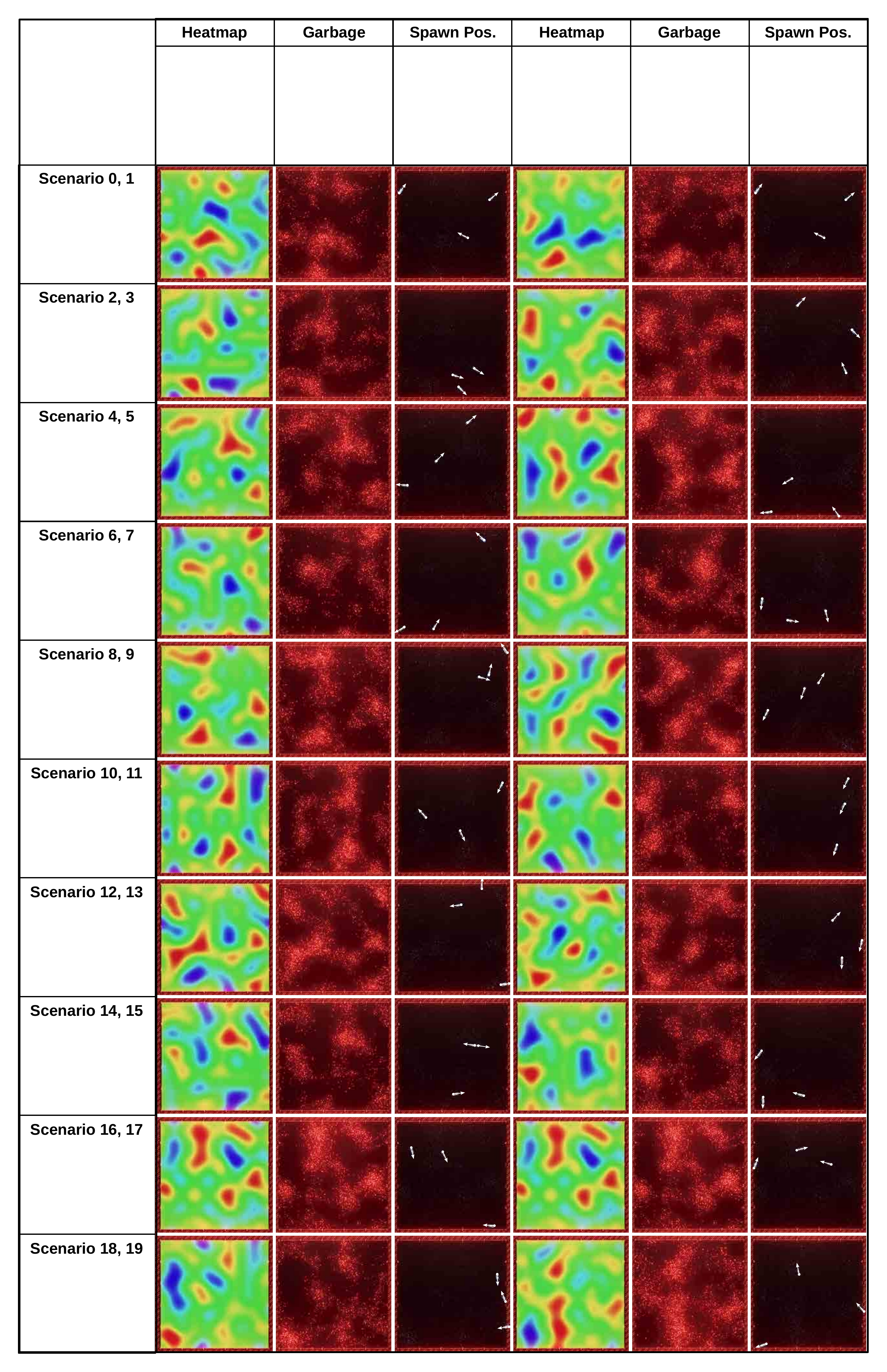}
    \caption{Ocean Plastic Collection environment samples seeds 0-19 with pollution heatmap and spawn positions for agent-controlled vessels.}
    \label{fig:ocean_plastic_collection_env_scenarios}
\end{figure}

\clearpage

\subsubsection{Aerial Wildfire Suppression}

\begin{figure}[h!]
    \centering
    \includegraphics[width=\linewidth]{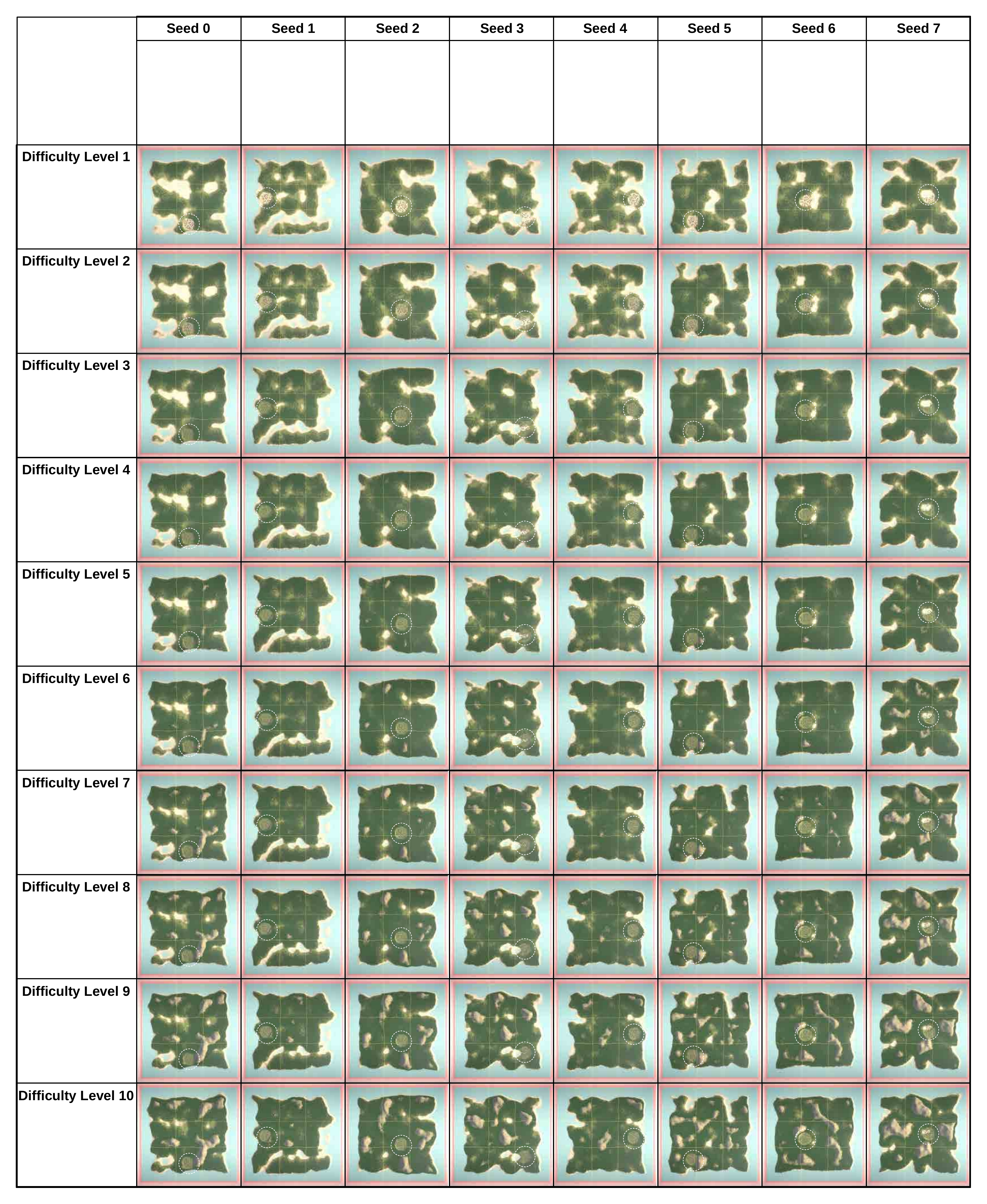}
    \caption{Aerial Wildfire Suppression environment samples showing terrain elevation levels 1-10 , top to bottom, and random seeds 0-7, left to right.}
    \label{fig:aerial_wildfire_suppression_env_scenarios}
\end{figure}

\clearpage

\subsection{Reward Description and Calculation}

\subsubsection{Wind Farm Control}\label{reward-description-and-calculation-wind-farm-control}

\textbf{Reward Description}

\begin{enumerate}[noitemsep,nolistsep]
    \item \textbf{Generate Energy} - This is a positive reward given at each time-step, in the range $[0, 1]$. This reward corresponds to the performance of each wind turbine and is being calculated as described in equation \ref{WFC:eq:1}. Orienting the wind turbine against the wind yields a high reward.
    \item \textbf{Avoid Damage} - This is a positive reward given at each time-step, in the range $[0, 1]$. We remap the angle between the wind direction and the turbine's orientation linearly from $[0, 90]$ degrees to $[0, 1]$ reward and from $[90, 180]$ degrees to $[1, 0]$ reward. Orienting the wind turbine so that the rotor blades are parallel to the wind direction yields high reward.
\end{enumerate}
\vspace{1cm}
\textbf{Reward Calculation}

\textbf{1. Generate Energy} - First, we need to describe how performance is calculated for each wind turbine:

Let us define:
\begin{itemize}[noitemsep,nolistsep]
    \item $a_{\text{turbine}} = 0.1$ --- Acceleration of the turbine motor.
    \item $\theta$ --- The angle between the wind turbine orientation and the wind direction at the turbine.
    \item $P(\theta) = 0.0$ --- Performance, initialized as $0.0$, dependent on angle $\theta$.
    \item $W(\theta)$ --- Wind force at wind turbine, dependent on angle $\theta$.
    \item $d$ --- Wind turbine drag.
\end{itemize}

Calculation steps:
\begin{enumerate}[noitemsep,nolistsep]
    \item Calculation of wind force $W$ based on angle $\theta$:
    \begin{equation}
    W(\theta) = 
    \begin{cases} 
    0 & \text{if } \theta < 0.5 \\
    \text{Map}(\theta, 0.5, 1, 0, 1) & \text{if } 0.5 \leq \theta \leq 1 
    \end{cases}
    \end{equation} 
    The "Map" function linearly interpolates the value of force from 0 to 1 as angle increases from 0.5 to 1.
    \item Calculation of drag:
    \begin{equation}
    d = -0.1 \times P(\theta)
    \end{equation}
    \item Updating performance $P$ with drag and wind force:
    \begin{equation} 
    P(\theta) = P(\theta) + d + W(\theta) \times a_{\text{turbine}}
    \end{equation}
    \item Clamping performance $P(\theta)$ between 0 and 1 is the reward $R(\theta)$:
    \begin{equation} 
    R(\theta) = \max(0, \min(1, P(\theta)))
    \end{equation}  \label{WFC:eq:1}
    Here, \(\max(0, \min(1, P(\theta)))\) limits \(P(\theta)\) within the interval [0, 1], ensuring it neither falls below 0 nor exceeds 1.
\end{enumerate}

\textbf{2. Avoid Damage}
The avoid damage reward $R(\theta)$ can be calculated as follows:

Let us define:
\begin{itemize}[noitemsep,nolistsep]
    \item $\theta$ --- The angle between the wind turbine orientation and the wind direction at the turbine.
\end{itemize}

Calculation steps:
\begin{enumerate}[noitemsep,nolistsep]
    \item Calculation of avoid damage reward based on angle $\theta$:
    \begin{equation}
        R(\theta) = \begin{cases} 
        \frac{\theta}{90} & \text{if } 0 \leq \theta \leq 90 \\
        2 - \frac{\theta}{90} & \text{if } 90 < \theta \leq 180 
        \end{cases}
    \end{equation}
\end{enumerate}

\subsubsection{Wildfire Resource Management}\label{reward-description-and-calculation-wildfire-resource-management}

\textbf{Reward Description}

\begin{enumerate}[noitemsep,nolistsep]
    \item \textbf{Watch Tower Performance} - This is a positive reward given at each time step, corresponding to the performance of the agent-controlled watch tower only. This reward is weighted by the resources distributed by self to self. Equation \ref{WRM:eq:1} describes how the individual performance and reward are calculated.
    
    \item \textbf{Neighbour Performance} - This is a positive reward given at each time step, corresponding to the sum of the performance of the neighbouring agent-controlled watch towers. This reward is weighted by the resources distributed by self to neighbouring watch towers. Equation \ref{WRM:eq:2} describes how the neighbour performance and reward are calculated. Agents receive additional rewards if they distribute useful resources to neighbouring watch towers.
    
    \item \textbf{Collective Performance} - This is a positive reward given at each time step, corresponding to the sum of the performance of all agent-controlled watch towers. Equation \ref{WRM:eq:3} describes how the collective performance and reward are calculated.
\end{enumerate}
\vspace{1cm}
\textbf{Reward Calculation}

\textbf{1. Watch Tower Performance} - First, we need to calculate the performance of each watch tower agent.
Let us define:
\begin{itemize}[noitemsep,nolistsep]
    \item $d_{\text{thresh}} = 200$ --- Threshold distance to a fire, used for normalization.
    \item $\vec{x}_0$ and $\vec{x}_1$ --- 3D vector positions of the closest observed fire at timesteps 0 and 1, respectively.
    \item $C(\vec{x})$ --- Function calculating the distance from the current watch tower to the closest observed fire at position $\vec{x}$.
    \item $d_0 = C(\vec{x}_0)$ and $d_1 = C(\vec{x}_1)$ --- Distances to the closest fire at timesteps 0 and 1.
    \item $d_{1 normalized}$ --- Normalized distance at timestep 1: $d_{1 normalized} = \frac{d_1}{d_{\text{thresh}}}$.
    \item $m$ --- Indicates whether the fire is moving towards the tower: $m = (C(\vec{x}_1) < C(\vec{x}_0))$.
    \item $s = 270$ and $a = 5$ --- Constants for the broken power law.
\end{itemize}

Calculation steps:
\begin{enumerate}[noitemsep,nolistsep]
    \item  The remapped distance factor based on the direction of movement is given by:
    \begin{equation}
    d_{1 normalized}' = \begin{cases} 
    0.5 - 0.5 \times d_{1 normalized} & \text{if } m \\
    0.5 + 0.5 \times d_{1 normalized} & \text{if not } m
    \end{cases}
    \end{equation}

    \item The adjusted distance factor using the broken power law is:
    \begin{equation}
    d_{1 normalized}'' = \left(1 + \left(d_{1 normalized}' \times \frac{1000}{s}\right)^a\right)^{-\frac{1}{2}}
    \end{equation}

    \item This is the watch tower performance metric $p$:
    \begin{equation}
    P = d_{1 normalized}''
    \end{equation}
\end{enumerate}

Now, we can calculate the reward $R(p, r_{\text{distributed}}, r_{\text{supporting}})$ by defining:

\begin{itemize}[noitemsep,nolistsep]
    \item $r_{\text{distributed}}$ --- Total supporting resources distributed from self and others.
    \item $r_{\text{supporting}}$ --- The amount of supporting resources from self only.
    \item $P$ --- Performance metric calculated as outlined above.
\end{itemize} 
Calculation steps:
\begin{equation} \label{WRM:eq:1}
R(p, r_{\text{distributed}}, r_{\text{supporting}}) = P \times r_{\text{supporting}} \times r_{\text{supporting}}
\end{equation}

\textbf{2. Neighbour Performance Reward}: We now describe how the neighbour reward is calculated.


Let us define the following:
\begin{itemize}[noitemsep,nolistsep]
    \item $p_i$ --- Represents the performance metric for the $i$-th watch tower.
    \item $n$ --- The number of neighbouring watch towers is 3.
    \item $R_{\text{neighbourhood}}$ --- The neighbour reward across neighbouring watch towers $n \in N$.
\end{itemize}

The neighbour performance reward $R(n, p_i)$ calculation involves the following steps:

\begin{enumerate}[noitemsep,nolistsep]
    \item Sum over neighbouring watch towers individual performance:
    \begin{equation} \label{WRM:eq:2}
    R_{\text{neighbourhood}}(n, p_i) = \sum_{i=1}^n p_i
    \end{equation}
\end{enumerate}

\textbf{3. Collective Performance Reward}: We now describe how the collective reward is calculated.


Let us define the following:
\begin{itemize}[noitemsep,nolistsep]
    \item $p_i$ --- Represents the performance metric for the $i$-th watch tower.
    \item $n$ --- The total number of watch towers.
    \item $R_{\text{collective}}$ --- The collective reward across all watch towers $n \in N$.
\end{itemize}

The collective performance reward $R_{\text{collective}}$ calculation involves the following steps:

\begin{enumerate}[noitemsep,nolistsep]
    \item Compute the Mean Squared Error MSE$(n, p_i)$ of watch tower performances:
    \begin{equation}
    \text{MSE}(n, p_i) = \frac{1}{n} \sum_{i=1}^n p_i^2
    \end{equation}
    
    \item Calculate the collective reward:
    \begin{equation} \label{WRM:eq:3}
    R_{\text{collective}}(n, p_i) = 1 - \left|1 - \sqrt{\text{MSE}(n, p_i)}\right|
    \end{equation}
\end{enumerate}

\subsubsection{Ocean Plastic Collection}\label{reward-description-and-calculation-ocean-plastic-collection}

\textbf{Reward Description}

\begin{enumerate}[noitemsep,nolistsep]
    \item \textbf{Collect Trash} - This is a positive reward of $1$ given for each floating plastic pebble collected.
    
    \item \textbf{Lowest Collected Trash Count} - This is a positive reward given at each time step for the lowest collected trash count amongst all agents. The lowest trash count is scaled by $0.01$. The steps to calculate the lowest collected trash count reward can be found in Equation \ref{OPC:eq:1}.
    
    \item \textbf{Crossed Border} - This is a negative reward of $-100$ given when the border is crossed.
    
    \item \textbf{Collided with Other Vessel} - This is a negative reward of $-100$ given when colliding with other vessel.
    
    \item \textbf{Close to Other Vessel} - This is a positive reward of $1$ given at each time step when the distance to the other vessel is smaller than or equal to $10$. The steps to calculate the close to other vessel reward can be found in Equation \ref{OPC:eq:2}.
    
    \item \textbf{Nearby Trash Count Delta} - This is a positive reward given when the nearby trash field population is higher than it has been until this time step. The reward given is the delta between the previous nearby trash field population count and the current. A nearby trash field population count is calculated by finding all floating plastic pebbles around a vessel with a radius smaller than or equal to $25$. The steps to calculate the nearby trash count delta reward can be found in Equation \ref{OPC:eq:3}.
    
    \item \textbf{Collide with Trash} - This is a negative reward of $-1$ given when the agent-controlled vessel is colliding with a floating plastic pebble.  
\end{enumerate}
\vspace{1cm}
\textbf{Reward Calculation}

\textbf{1. Collect Trash} - To calculate the  Collect Trash reward, let us define the following:

\begin{itemize}[noitemsep,nolistsep]
    \item $r_t$ --- Reward for each trash pebble collected.
\end{itemize}

Calculation steps:

\begin{enumerate}[noitemsep,nolistsep]
    \item Get the number of collected trash pebbles:
    \begin{equation}
    r_t = \sum_{i=1}^{N} \mathbb{I}(p_i \text{ is collected})
    \end{equation}
\end{enumerate}

\textbf{2. Lowest Collected Trash Count} - To calculate the lowest collected trash count reward, let us define the following:

\begin{itemize}[noitemsep,nolistsep]
    \item $a = 0.01$ --- Lowest trash count factor.
    \item $T$ --- Set of all agents lowest collected trash count.
\end{itemize}

Calculation steps:

\begin{enumerate}[noitemsep,nolistsep]
    \item Get the lowest trash count from all agents:
    \begin{equation}
    M(T) = \min(t_1, t_2, \dots, t_n) \text{, where $t \in T$}
    \end{equation}
    
    \item Calculate the lowest collected trash count reward $R(T)$:
    \begin{equation} \label{OPC:eq:1}
    R(T) = M(T) \times a
    \end{equation}
\end{enumerate}

\textbf{3. Crossed Border} - To calculate the Crossed Border reward, let us define the following:
\begin{itemize}[noitemsep,nolistsep]
    \item $eh = 200$ --- The environment half extend.
    \item $\vec{p}$ --- The vessel position.
    \item $r_{cb}$ --- Crossed boundary reward.
\end{itemize}

Calculation steps:

\begin{enumerate}[noitemsep,nolistsep]
    \item We can now calculate the Crossed Border reward:
    \begin{equation}
    r_{cb} =
    \begin{cases}
    -100 & \text{if } \left( p_x > eh \text{ or } p_x < -eh \text{ or } p_y > eh \text{ or } p_y < -eh \right) \\
    0 & \text{otherwise}
    \end{cases}
    \end{equation}
\end{enumerate}

\textbf{4. Collided with Other Vessel} - To calculate the Collided with Other Vessel reward, let us define the following:

\begin{itemize}[noitemsep,nolistsep]
    \item $\vec{p}$ --- The vessel position.
    \item $N_p$ --- Neighbouring vessel positions.
    \item $r_{c}$ --- Collision reward.
\end{itemize}

Calculation steps:

\begin{enumerate}[noitemsep,nolistsep]
    \item We can now calculate the Collided with Other Vessel reward:
    \begin{equation}
    r_{c} = 
    \begin{cases}
    -100 & \text{if } \exists \vec{n} \in N_p \text{ such that } \vec{p} \text{ collides with } \vec{n} \\
    0 & \text{otherwise}
    \end{cases}
    \end{equation}
\end{enumerate}

\textbf{5. Close to Other Vessel} - To calculate the lowest collected trash count reward, let us define the following:

\begin{itemize}[noitemsep,nolistsep]
    \item $d$ --- Distance to closest neighbouring vessel.
    \item $d_{\text{thresh}}$ --- Distance threshold to closest neighbouring vessel.    
\end{itemize}

Calculation steps:

\begin{enumerate}[noitemsep,nolistsep]
    \item Calculate close to other vessel reward $r$.
    \begin{equation} \label{OPC:eq:2}
    r = 
    \begin{cases}
    10 & \text{if } d < d_{\text{thresh}} \\
    0 & \text{otherwise}
    \end{cases}
    \end{equation}
\end{enumerate}

\textbf{6. Nearby Trash Count Delta} - To calculate the nearby trash count delta reward, let us define the following:

\begin{itemize}[noitemsep,nolistsep]
    \item $d_{\text{threshold}} = 25$ --- Trash count nearby distance threshold.
    \item $P$ --- All floating plastic pebble positions, $\{\vec{p_1}, \vec{p_2}, \ldots, \vec{p_n}\} \in P$.
    \item $ntc_{\text{old}} = 0$ --- Old nearby trash count.
    \item $ntc_{\text{current}}$ --- Current nearby trash count.
    \item $ntc_{\text{difference}}$ --- Difference between the old and current nearby trash count.
\end{itemize}

Calculation steps:

\begin{enumerate}[noitemsep,nolistsep]
    \item The nearby trash count is calculated by considering only floating plastic pebbles with a distance below $d_{\text{threshold}}$:
    \begin{equation}
    ntc_{\text{current}} = \sum_{i=1}^n [ \text{dist}(p_i) < d_{\text{threshold}} ]
    \end{equation}

    \item If the current nearby trash count $ntc_{\text{current}}$ is larger than the old nearby trash count $ntc_{\text{old}}$, the difference between the two is the reward $r(ntc_{\text{difference}})$:
    \begin{equation}
    ntc_{\text{difference}} = ntc_{\text{current}} - ntc_{\text{old}}
    \end{equation}
    \begin{equation} \label{OPC:eq:3}
    r(ntc_{\text{difference}}) = \max(0, ntc_{\text{difference}})    
    \end{equation}

    \item Finally the old nearby trash count $ntc_{\text{old}}$ is updated with the current nearby trash count $ntc_{\text{current}}$:
    \begin{equation}
    ntc_{\text{old}} = ntc_{\text{current}}
    \end{equation}
\end{enumerate}

\textbf{7. Collide with Trash} - To calculate the Collide with Trash reward, let us define the following:

\begin{itemize}[noitemsep,nolistsep]
    \item $\vec{p}$ --- The vessel position.
    \item $P_t$ --- All trash pebble positions.
    \item $r_{p}$ --- Collision reward.
\end{itemize}

Calculation steps:

\begin{enumerate}[noitemsep,nolistsep]
    \item We can now calculate the Collide with Trash reward:
    \begin{equation}
    r_{c} = 
    \begin{cases}
    -100 & \text{if } \exists \vec{n} \in P_t \text{ such that } \vec{p} \text{ collides with } \vec{p_t} \\
    0 & \text{otherwise}
    \end{cases}
    \end{equation}
\end{enumerate}

\subsubsection{Drone-Based Reforestation}\label{reward-description-and-calculation-drone-based-reforestation}

\textbf{Reward Description}

\begin{enumerate}[noitemsep,nolistsep]
    \item \textbf{Drop Seed} - This is a positive reward given at each seed drop. The drop seed reward consists of the quality of the drop location, a seed reward, in the range of $[0, 20]$ and distance to other seeds and existing trees, a distance reward, in the range of $[0, 10]$. Therefore, the resulting total drop seed reward is in the range of $[0, 30]$. The steps to calculate the total drop seed reward can be found in Equation \ref{DBR:eq:3}.
    
    \item \textbf{Deplete Energy Holding Seed} - This is a negative reward of $-1 / (\text{episode length} / 2)$ given at each time step if the drone is carrying a seed. The deplete energy reward at each time step is higher when carrying a seed than if not carrying a seed. The episode length is $2000$.
    
    \item \textbf{Deplete Energy No Seed} - This is a negative reward of $-1 / (\text{episode length})$ given at each time step if the drone is not carrying a seed. The episode length is $2000$.
    
    \item \textbf{Pick-up Seed} - This is an optional positive reward given when a drone is returned to the drone station. There are two tasks in which this reward is given. In "Subtask: Pick-up Seed at Base" a reward of $100$ is given and in "Subtask: Explore Furthest Distance and Return to Base" the reward is the furthest distance that has been explored and can be in the range of $[0, 200]$.
    
    \item \textbf{Incremental Running Back} - After a seed has been dropped, this reward is given incrementally when flying back to the drone station. If the distance to the drone station at time-step $t_{-1}$ is larger than the current distance, this reward is given at incremental steps of $2.5$. The range of the incremental running back reward is $[0, 20]$, which can be modified by the running back multiplier, depending on the seed drop quality. If the Seed has been dropped 50 meters away from the drone station, an incremental running back reward can be received $20$ times. The steps to calculate the incremental running back reward can be found in Equation \ref{DBR:eq:6}.
    
    \item \textbf{Group-up} - This is a positive reward of $10$, given at each time-step, if the distance to any neighbouring drone is smaller than $5$. The steps to calculate the group-up reward can be found in Equation \ref{DBR:eq:7}.
    
    \item \textbf{High Fertility Location Delta} - This is a reward given every time a higher fertility potential seed drop location has been found. The reward is the delta between the old and the new potential, if the new potential is higher than the old. The range of the reward is $[0, 1]$. The steps to calculate the high fertility location delta reward can be found in Equation \ref{DBR:eq:8}.
    
    \item \textbf{High Landscape Point Delta} - This is a reward given every time a higher point on the terrain landscape has been found. The reward is the delta between the old and the new height, if the new height is higher than the old. The reward range is $[0, 40]$, as $40$ is the environment's height boundary. The steps to calculate the high landscape point delta reward can be found in Equation \ref{DBR:eq:9}.
    
    \item \textbf{Far Distance Explored Delta} - This is a reward given every time a further distance has been explored. The reward is the delta between the old distance and the new, if the new distance is further than the old. The reward range is $[0, 200]$, as $200$ is the environment's half extend. The steps to calculate the far distance explored delta reward can be found in Equation \ref{DBR:eq:10}.
    
    \item \textbf{Find Close Tree} - This is a reward given when a tree has been found within a $20$ meter radius. The reward given is $100$. The steps to calculate the find close tree reward can be found in Equation \ref{DBR:eq:11}.
\end{enumerate}
\vspace{1cm}
\textbf{Reward Calculation}

\textbf{1. Drop Seed} - To calculate the drop seed reward, we need to calculate the actual seed drop reward and a distance reward. To calculate the seed drop reward, let us define the following:

\begin{itemize}[noitemsep,nolistsep]
    \item $dot_{\text{max}} = 75$ --- Maximum distance to other trees.
    \item $dot_{\text{min}} = 2.5$ -- Minimum distance to other trees.
    \item $dnt$ --- Closest distance to new trees.
    \item $det$ --- Closest distance to existing trees.
    \item $sdrm = 20$ --- Seed drop reward multiplier. 
    \item $r_s(det ,dot_{\text{min}}, dot_{\text{max}})$ --- Seed drop reward.
\end{itemize}

Calculation steps:

\begin{enumerate}[noitemsep,nolistsep]
    \item The following condition needs to hold true for this reward to be larger than $0$. This ensures that the newly dropped seed is far enough from existing and seeds dropped in the past, but also that the seed is not too far away from the existing forest.
    \begin{equation} \label{DBR:eq:11}
    (dot_{\text{min}} \leq det \leq dot_{\text{max}}) \text{ and } (dnt \geq dot_{\text{min}})
    \end{equation}

    \item First, we remap the distance to existing and new trees to $[1, 0]$ so that a high reward is given when the seed is dropped close to existing or new trees.
    \begin{equation}
    r_s(det ,dot_{\text{min}}, dot_{\text{max}}) = \text{{Remap}}(det, dot_{\text{min}}, dot_{\text{max}}, 1, 0)
    \end{equation}
    
    \item Applying Multiplier:
    \begin{equation}
    r_s(det ,dot_{\text{min}}, dot_{\text{max}}) = r_s(det ,dot_{\text{min}}, dot_{\text{max}}) \times \text{sdrm}
    \end{equation}
\end{enumerate}

We now describe how the distance reward is calculated. Let us define:

\begin{itemize}[noitemsep,nolistsep]
    \item $sdd$ --- Seed drop distance to drone station.
    \item $ew = 200$ --- Environment half extend.
    \item $drm = 10$ --- Distance reward multiplier.
    \item $r_{d}(sdd_{\text{normalized}}, drm)$ --- Distance reward.
\end{itemize}

Calculation steps:

\begin{enumerate}[noitemsep,nolistsep]
    \item The seed drop reward needs to be larger than $0$ for the distance reward to be applied.
    \begin{equation} \label{DBR:eq:12}
    0 < r_s(det ,dot_{\text{min}}, dot_{\text{max}})
    \end{equation}

    \item Calculate the distance reward using the normalized seed drop distance to the drone station.
    \begin{equation}
    sdd_{\text{normalized}} = sdd / ew
    \end{equation}
    \begin{equation}
    r_{d}(sdd_{\text{normalized}}, drm) = sdd_{\text{normalized}} \times drm
    \end{equation}
\end{enumerate}

The total reward for dropping a seed consists of the drop seed reward \ref{DBR:eq:11} and the distance reward \ref{DBR:eq:12}.

\begin{itemize}[noitemsep,nolistsep]
    \item $r_s$ --- Seed drop reward, calculated as described above.
    \item $r_d$ --- Distance reward, calculated as described above.
    \item $r_{sd}(r_s, r_d)$ --- The total seed drop reward.
\end{itemize}

\begin{equation} \label{DBR:eq:13}
r_{sd}(r_s, r_d) = r_s + r_d
\end{equation}

\textbf{2. Deplete Energy Holding Seed} - To calculate the deplete energy holding seed reward, let us define the following:

\begin{itemize}[noitemsep,nolistsep]
    \item $\text{episode length}_\text{max} = 2000$ --- Max episode length.
    \item $der_{\text{holding seed}}(\text{episode length}_\text{max})$ --- Deplete energy reward while holding a seed.    
\end{itemize}

\begin{equation} \label{DBR:eq:2}
der_{\text{holding seed}}(\text{episode length}_\text{max}) = -1 / (\text{episode length}_\text{max} / 2)
\end{equation}

\textbf{3. Deplete Energy No Seed} - To calculate the deplete energy no seed reward, let us define the following:

\begin{itemize}[noitemsep,nolistsep]
    \item $\text{episode length}_\text{max} = 2000$ --- Max episode length.
    \item $der_{\text{no seed}}(\text{episode length}_\text{max})$ --- Deplete energy reward without holding a seed.    
\end{itemize}

\begin{equation} \label{DBR:eq:3}
der_{\text{no seed}}(\text{episode length}_\text{max}) = -1 / (\text{episode length}_\text{max})
\end{equation}

\textbf{4. Pick-up Seed} - To calculate the Pick-up Seed reward, let us define the following:

\begin{itemize}[noitemsep,nolistsep]
    \item $p$ --- Drone position.
    \item $d$ --- Drone station position.
    \item $r_{ps}$ --- Pick-up seed reward.
\end{itemize}

Calculation steps:

\begin{enumerate}[noitemsep,nolistsep]
    \item We can now calculate the Pick-up Seed reward:
    \begin{equation}
    r_{ps} = 
    \begin{cases}
    1 & \text{if } \text{distance}(p, d) = 0 \\
    0 & \text{otherwise}
    \end{cases}
    \end{equation}
\end{enumerate}

\textbf{5. Incremental Running Back} - To calculate the incremental running back reward we need to calculate the seed drop reward \ref{DBR:eq:11} and distance reward \ref{DBR:eq:12}. Let us define the following:


\begin{itemize}[noitemsep,nolistsep]
    \item $d_0$ --- Current distance to drone station at time-step 0 in incremental steps.
    \item $\vec{p}_0$ --- Current position at time-step 0.
    \item $\vec{dp}$ --- Drone station position.
    \item $s = 2.5$ --- Incremental step size towards drone station.
    \item $r_s$ --- Seed drop reward, calculated as described above.
    \item $r_d$ --- Distance reward, calculated as described above.
    \item $r_p = 20$ --- Possible intermediate reward for running back to the drone station.
    \item $sdrm = 20$ --- Seed drop reward multiplier. 
    \item $drm = 10$ --- Distance reward multiplier.
    \item $rbm$ --- Running back multiplier.
    \item $r_{sd}(r_s, r_d)$ --- The total seed drop reward.
    \item $r_{rb}$ --- Reward for running back, given incrementally at step $s$ sized increments.
    \item $d_{\text{init}}$ --- Initial distance to drone station, this is assigned when a seed has been dropped.
    \item $d_{\text{charge}} = 7.5$ --- Distance to drone station to charge and pick-up seed. 
\end{itemize}

Calculation steps:

\begin{enumerate}[noitemsep,nolistsep]
    \item The condition for the reward to be given is that the current distance from the drone to the drone station is smaller than in time-step $t_{-1}$. The current distance $d_0$ is calculated as follows:
    \begin{equation}
    d_0 = \left\lfloor \sqrt{\sum_{i=1}^n (p_{0i} - dp_i)^2} / s \right\rfloor
    \end{equation}
    \begin{equation}
    \text{If } d_0 < d_{-1} \text{ continue with next step.}     
    \end{equation}
    
    \item Let us first calculate the running back multiplier $rbm$ by normalizing the sum of seed drop and distance rewards.
    \begin{equation}
    rbm = (r_s + r_d) / (sdrm + drm)
    \end{equation}

    \item We can now calculate the reward for running back to the drone station:
    \begin{equation}
    r_{rb} = (r_p \times rbm) / (d_{\text{init}} - d_{\text{charge}} / s)
    \end{equation}

    \item Finally, we need to ensure that the reward $r_{rb}$ is equal to or above $0$ and equal to or below $r_p$:
    \begin{equation} \label{DBR:eq:5}
    r_{rb} = 
    \begin{cases}
    0 & \text{if } r_{rb} \leq 0 \\
    r_p & \text{if } r_{rb} > r_p \\
    r_{rb} & \text{otherwise}
    \end{cases}
    \end{equation}
\end{enumerate}

\textbf{6. Group-up} - To calculate the group-up reward we need to define the following:

\begin{itemize}[noitemsep,nolistsep]
    \item $n_{c}$ --- Closest neighbour.
    \item $d_{\text{thresh}} = 5$ --- Distance threshold to closest drone.
    \item $\vec{p}$ --- Current local drone position.
    \item $d_{cn}$ --- Distance to closest neighbour
    \item $r_{gu}$ --- Reward for grouping up. 
\end{itemize}

Calculation steps:

\begin{enumerate}[noitemsep,nolistsep]
    \item Let us calculate the distance to the closest neighbour:
    \begin{equation}
    d_{cn} = \sqrt{\sum_{i=1}^n (p_{i} - n_{ci})^2}
    \end{equation}

    \item We can now calculate the reward for grouping up:
    \begin{equation} \label{DBR:eq:6}
    r_{gu} = 
    \begin{cases}
    0 & \text{if } d_{\text{thresh}} \leq d_{cn} \\
    10 & \text{otherwise}
    \end{cases}
    \end{equation}
\end{enumerate}

\textbf{7. High Fertility Location Delta} - To calculate the high fertility location delta reward, let us define the following:

\begin{itemize}[noitemsep,nolistsep]
    \item $dot_{\text{max}} = 75$ --- Maximum distance to other trees.
    \item $dot_{\text{min}} = 2.5$ -- Minimum distance to other trees.
    \item $dnt$ --- Closest distance to new trees.
    \item $det$ --- Closest distance to existing trees.
    \item $\vec{p}$ --- Current local drone position.
    \item $d_{cet}$ --- Distance to closest existing tree.
    \item $d_{cds}$ --- Distance to closest dropped seed.
    \item $\text{pot}_{\text{old}} = 0$ --- Old potential seed drop fertility, initialized as $0$.
    \item $\text{pot}_{\text{current}}$ --- Current potential seed drop fertility.
    \item $r_{fl}$ --- High fertility location delta reward.
    
\end{itemize}

Calculation steps:

\begin{enumerate}[noitemsep,nolistsep]
    \item If $det$ is smaller or equal to $dot_{\text{max}}$, $det$ is larger or equal to $dot_{\text{min}}$ and $dnt$ is larger or equal to $dot_{\text{min}}$, then follow the next calculation step, otherwise the reward $r_{fl}$ is $0$.

    \item Calculate the current potential:
    \begin{equation}
    \text{pot}_{\text{current}} = \text{Map}(det, dot_{\text{min}}, dot_{\text{max}}, 1, 0)
    \end{equation}

    \item We can now calculate the high fertility location reward:
    \begin{equation} \label{DBR:eq:7}
    r_{fl} =
    \begin{cases}
    \text{pot}_{\text{current}} - \text{pot}_{\text{old}} & \text{if } \text{pot}_{\text{old}} < \text{pot}_{\text{current}}, \text{delta of current and old potential} \\
    0 & \text{otherwise}
    \end{cases}
    \end{equation}
\end{enumerate}

\textbf{8. High Landscape Point Delta} - To calculate the high landscape point delta reward, let us define the following:

\begin{itemize}[noitemsep,nolistsep]
    \item $\vec{p}$ --- Current local drone position.
    \item $h_{\text{old}} = 0$ --- Old height, initialized as $0$.
    \item $h_{\text{current}}$ --- Current height.
    \item $h_(\vec{x})$ --- Get height at position $\vec{x}$.
    \item $r_h$ --- Height delta reward.
    
\end{itemize}

Calculation steps:

\begin{enumerate}[noitemsep,nolistsep]
    \item Calculate the current height:
    \begin{equation}
    h_{\text{current}} = h_(\vec{p})
    \end{equation}

    \item We can now calculate the hight landscape point delta reward:
    \begin{equation} \label{DBR:eq:8}
    r_{fl} =
    \begin{cases}
    h_{\text{current}} - h_{\text{old}} & \text{if } h_{\text{old}} < h_{\text{current}}, \text{delta of current and old height} \\
    0 & \text{otherwise}
    \end{cases}
    \end{equation}
\end{enumerate}

\textbf{9. Far Distance Explored Delta} - To calculate the far distance explored delta reward, let us define the following:

\begin{itemize}[noitemsep,nolistsep]
    \item $\vec{p}$ --- Current local drone position.
    \item $d_{\text{old}} = 0$ --- Old furthest distance to drone station, initialized as $0$.
    \item $d_{\text{current}}$ --- Current furthest distance to drone station.
    \item $d_(\vec{x})$ --- Get distance to drone station at position $\vec{x}$.
    \item $r_{fd}$ --- Far distance delta reward.
\end{itemize}

Calculation steps:

\begin{enumerate}[noitemsep,nolistsep]
    \item Calculate the current furthest distance:
    \begin{equation}
    d_{\text{current}} =
    \begin{cases}
    d_(\vec{p}) & \text{if } d_(\vec{p}) > d_{\text{old}} \\
    d_{\text{old}} & \text{otherwise}
    \end{cases}    
    \end{equation}

    \item We can now calculate the far distance delta reward:
    \begin{equation} \label{DBR:eq:9}
    r_{fd} =
    \begin{cases}
    d_{\text{current}} - d_{\text{old}} & \text{if } d_{\text{old}} < d_{\text{current}}, \text{delta of current and old furthest distance} \\
    0 & \text{otherwise}
    \end{cases}
    \end{equation}
\end{enumerate}

\textbf{10. Find Close Tree} - To calculate the find close tree reward, let us define the following:

\begin{itemize}[noitemsep,nolistsep]
    \item $\vec{p}$ --- Current local drone position.
    \item $ew = 200$ --- Environment half extend.
    \item $d_{cet}$ --- Distance to closest existing tree.
    \item $cet(\vec{x})$ --- Get closest existing tree given a location.
    \item $r_{ct}$ --- Find close tree reward.
\end{itemize}

Calculation steps:

\begin{enumerate}[noitemsep,nolistsep]
    \item Let us calculate the distance to the closest existing tree and normalize using the environment half extend:
    \begin{equation}
    d_{cet} = cet(\vec{p}) / ew
    \end{equation}

    \item If $d_{cet} < 20$ a reward of $100$ is given:
    \begin{equation} \label{DBR:eq:10}
    r_{ct} = 
    \begin{cases}
    100 & \text{if } d_{cet} \leq 20 \\
    0 & \text{otherwise}
    \end{cases}
    \end{equation}
\end{enumerate}

\subsubsection{Aerial Wildfire Suppression}\label{reward-description-and-calculation-aerial-wildfire-suppression}

\textbf{Reward Description}

\begin{enumerate}[noitemsep,nolistsep]
    \item \textbf{Crossed Border} - This is a negative reward of $-100$ given when the border of the environment is crossed. The border is a square around the island in the size of $1500$ by $1500$. The island is $1200$ by $1200$.
    
    \item \textbf{Pick-up Water} - This is a positive reward of $1$ given when the agent steers the aeroplane towards the water. The island is $1200$ by $1200$ and there is a girdle of water around the island with a width of $300$.
    
    \item \textbf{Fire Out} - This is a positive reward of $10$ given when the fire on the whole island dies out, with or without the active assistance of the agent.
    
    \item \textbf{Too Close to Village} - This is a negative reward of $-50$ given when the fire is closer than $150$ to the centre of the village.
    
    \item \textbf{Time Step Burning} - This is a negative reward of $-0.01$ given at each time-step, while the fire is burning.
    
    \item \textbf{Find Fire} - This is a positive reward of $100$ given when a burning tree has been found.
    
    \item \textbf{Find Village} - This is a positive reward of $100$ given when the village has been found, and the distance between the current local aeroplane position and the village is less than $150$.
    
    \item \textbf{Extinguishing Tree} - This is a positive reward in the range of $[0, 5]$ given for each tree that has been in the state burning in time-step $t_{-1}$ and is now extinguished by dropping water at its location.
    
    \item \textbf{Preparing Tree} - This is a positive reward in the range of $[0, 1]$ given for each tree that has been in the state not burning in time-step $t_{-1}$ and is now wet by dropping water at its location.
\end{enumerate}
\vspace{1cm}
\textbf{Reward Calculation}

\textbf{1. Crossed Border} - To calculate the Crossed Border reward, let us define the following:
\begin{itemize}[noitemsep,nolistsep]
    \item $eh = 750$ --- The environment half extend.
    \item $\vec{p}$ --- The drone position.
    \item $r_{cb}$ --- Crossed boundary reward.
\end{itemize}

Calculation steps:

\begin{enumerate}[noitemsep,nolistsep]
    \item We can now calculate the Crossed Border reward:
    \begin{equation} \label{AWS:eq:1}
    r_{cb} =
    \begin{cases}
    -100 & \text{if } \left( p_x > eh \text{ or } p_x < -eh \text{ or } p_y > eh \text{ or } p_y < -eh \right) \\
    0 & \text{otherwise}
    \end{cases}
    \end{equation}
\end{enumerate}

\textbf{2. Pick-up Water} - To calculate the Pick-up Water reward, let us define the following:

\begin{itemize}[noitemsep,nolistsep]
    \item $eh = 750$ --- The environment half extend.
    \item $ih = 600$ --- Island half extend.
    \item $\vec{p}$ --- The drone position.
    \item $r_{pw}$ --- Pick-up Water reward.
\end{itemize}

Calculation steps:

\begin{enumerate}[noitemsep,nolistsep]
    \item We can now calculate the Pick-up Water reward:
    \begin{equation} \label{DBR:eq:14}
    r_{pw} = 
    \begin{cases}
    1 & \text{if } \left( p_x < eh \text{ or } p_x > -eh \text{ or } p_y < eh \text{ or } p_y > -eh \right) \\
      & \text{and } \left( p_x > ih \text{ or } p_x < -ih \text{ or } p_y > ih \text{ or } p_y < -ih \right) \\
    0 & \text{otherwise}
    \end{cases}
    \end{equation}
\end{enumerate}

\textbf{3. Fire Out} - To calculate the Fire Out reward, let us define the following:

\begin{itemize}[noitemsep,nolistsep]
    \item $T$ --- All tree states.
    \item $r_{nb}$ --- No burning tree reward.
\end{itemize}

Calculation steps:

\begin{enumerate}[noitemsep,nolistsep]
    \item We can now calculate the Fire Out reward:
    \begin{equation}
    r_{nb} =
    \begin{cases}
    10 & \text{if } \forall t \in T, \, t \neq \text{"burning"} \\ 
    0 & \text{otherwise} \\
    \end{cases}
    \end{equation}
\end{enumerate}

\textbf{4. Too Close to Village} - To calculate the Too Close to Village reward, let us define the following:

\begin{itemize}[noitemsep,nolistsep]
    \item $T_c$ --- All tree states, closer to or equal to $150$ to the village.
    \item $r_{cv}$ --- Too Close to Village reward.
\end{itemize}

Calculation steps:

\begin{enumerate}[noitemsep,nolistsep]
    \item We can now calculate the Fire Out reward:
    \begin{equation}
    r_{cc} =
    \begin{cases}
    -50 & \text{if } \exists t \in T_c, \, t = \text{"burning"} \\ 
    0 & \text{otherwise}
    \end{cases}
    \end{equation}
\end{enumerate}

\textbf{5. Time Step Burning} - To calculate the Time Step Burning reward, let us define the following:

\begin{itemize}[noitemsep,nolistsep]
    \item $T$ --- All tree states.
    \item $r_{tsb}$ --- Time Step Burning reward.
\end{itemize}

Calculation steps:

\begin{enumerate}[noitemsep,nolistsep]
    \item We can now calculate the Time Step Burning reward:
    \begin{equation}
    r_{tsb} =
    \begin{cases}
    -0.01 & \text{if } \forall t \in T, \, t = \text{"burning"} \\
    0 & \text{otherwise} \\
    \end{cases}
    \end{equation}
\end{enumerate}

\textbf{6. Find Fire} - To calculate the Find Fire reward, let us define the following:

\begin{itemize}[noitemsep,nolistsep]
    \item $\vec{p}$ --- The drone position.
    \item $d_{t} = 150$ --- Distance threshold.
    \item $T$ --- All tree states.
    \item $r_{f}$ --- Find Fire reward.
\end{itemize}

Calculation steps:

\begin{enumerate}[noitemsep,nolistsep]
    \item We can now calculate the Find Fire reward:
    \begin{equation}
    r_{f} =
    \begin{cases}
    100 & \text{if } \exists t \in T \text{ such that } \text{distance}(p) < d_{t} \text{ meters and } t = \text{"burning"} \\
    0 & \text{otherwise} \\
    \end{cases}
    \end{equation}
\end{enumerate}

\textbf{7. Find Village} - To calculate the Find Village reward, let us define the following:

\begin{itemize}[noitemsep,nolistsep]
    \item $\vec{p}$ --- The drone position.
    \item $d_{t} = 150$ --- Distance threshold.
    \item $r_{v}$ --- Find Village reward.
\end{itemize}

Calculation steps:

\begin{enumerate}[noitemsep,nolistsep]
    \item We can now calculate the Find Village reward:
    \begin{equation}
    r_{v} =
    \begin{cases}
    100 & \text{if } \text{distance}(\vec{p}) \leq d_{t} \text{ meters} \\
    0 & \text{otherwise}
    \end{cases}
    \end{equation}
\end{enumerate}

\textbf{8. Extinguishing Tree} - To calculate the Extinguish Tree reward, let us define the following:

\begin{itemize}[noitemsep,nolistsep]
    \item $T$ --- All tree states.
    \item $r_{e}$ --- Extinguish Tree reward.
\end{itemize}

Calculation steps:

\begin{enumerate}[noitemsep,nolistsep]
    \item We can now calculate the Extinguish Tree reward:
    \begin{equation}
    r_{e} = 5 \sum_{t \in T} \mathbb{I}(t_{\text{previous}} = \text{"burning"} \text{ and } t_{\text{current}} = \text{"extinguished"})
    \end{equation}
\end{enumerate}

\textbf{9. Preparing Tree} - To calculate the Preparing Tree reward, let us define the following:

\begin{itemize}[noitemsep,nolistsep]
    \item $T$ --- All tree states.
    \item $r_{p}$ --- Preparing Tree reward.
\end{itemize}

Calculation steps:

\begin{enumerate}[noitemsep,nolistsep]
    \item We can now calculate the Preparing Tree reward:
    \begin{equation}
    r_{e} = \sum_{t \in T} \mathbb{I}(t_{\text{previous}} = \text{"not Burning"} \text{ and } t_{\text{current}} = \text{"wet"})
    \end{equation}
\end{enumerate}

\subsection{Task Description and Reward Scale}

\subsubsection{Wind Farm Control}\label{task-description-and-reward-scale-wind-farm-control}

\textbf{Task Description}

\begin{enumerate}[noitemsep,nolistsep]
  \item \textbf{Main Task: Generate Energy } - This is the main task of the environment. The agent's goal is to rotate the wind turbine to be oriented against the wind direction and hence generate energy.
  \item \textbf{Subtask: Avoid Damage} - This is a subtask to turn the wind turbine 90 degrees away so that the wind turbine rotor blades are parallel to the wind direction, avoiding damage to the wind turbine's rotor blades.
\end{enumerate}
\vspace{1cm}
\textbf{Reward Scale}

\begin{table}[h!]
  \caption{Main- and Sub-Task Reward Scale}
  \label{wfc-task-rewards}
  \centering
  \begin{tabularx}{\textwidth}{p{3cm} p{0.77cm} p{0.77cm}}
    \toprule
     & \multicolumn{2}{c}{Task} \\
    \cmidrule(r){2-3}
    Reward              & 1. & 2. \\
    \midrule
    1. Generate Energy  & 1 & 0 \\
    2. Avoid Damage     & 0 & 1 \\
    \bottomrule
  \end{tabularx}
\end{table}

\subsubsection{Wildfire Resource Management}\label{task-description-and-reward-scale-wildfire-resource-management}

\textbf{Task Descriptions}

\begin{enumerate}[noitemsep,nolistsep]
    \item \textbf{Main Task: Distribute Resources} - This is the main task of the environment. The goal of the agent is to distribute a total of 1.0 resources at each time step to self or neighbouring watch towers. If the agent is out of resources, it has to remove resources from self or neighbouring watch towers before re-distribution. The resources should be distributed to the watch towers where the fire is closest and incoming.
    \item \textbf{Subtask: Keep All} - This is a subtask with the same goal as the main task, however distributing resources to self yields higher rewards than distributing them to neighbouring watch towers.
    \item \textbf{Subtask: Distribute All} - This is a subtask with the same goal as the main task, however distributing resources to neighbouring watch towers yields higher rewards than distributing them to self.
\end{enumerate}
\vspace{1cm}
\textbf{Reward Scale}

\begin{table}[h!]
  \caption{Main- and Sub-Task Reward Scale}
  \label{wrm-task-rewards}
  \centering
  \begin{tabularx}{\textwidth}{p{4.6cm} p{0.77cm} p{0.77cm} p{0.77cm}}
    \toprule
     & \multicolumn{3}{c}{Task} \\
    \cmidrule(r){2-4}
    Reward                          & 1.    & 2.    & 3. \\
    \midrule
    1. Watch Tower Performance      & 1     & 10    & 1 \\
    2. Neighbourhood Performance    & 1     & 1     & 10  \\
    2. Collective Performance       & 1     & 1     & 1  \\
    \bottomrule
  \end{tabularx}
\end{table}

\subsubsection{Ocean Plastic Collection}\label{task-description-and-reward-scale-ocean-plastic-collection}

\textbf{Task Description}

\begin{enumerate}[noitemsep,nolistsep]
    \item \textbf{Main Task: Plastic Collection} - This is the main task of the environment. The goal for the agent is to accelerate and steer the plastic collection vessel to collect as many floating plastic pebbles as possible while avoiding crashing into other vessels and crossing the environments border.
    \item \textbf{Subtask: Find Highest Polluted Area} - This is a subtask with the goal of finding the highest trash population area in a given scenario.
    \item \textbf{Subtask: Group Up} - This is a subtask with the goal of finding other vessels and staying close to other vessels while collecting as many floating plastic pebbles as possible.
    \item \textbf{Subtask: Avoid Plastic} - This is a subtask with the goal of avoiding floating plastic pebbles.
\end{enumerate}
\vspace{1cm}
\textbf{Reward Scale}

\begin{table}[h!]
  \caption{Main- and Sub-Task Reward Scale}
  \label{opc-task-rewards}
  \centering
  \begin{tabularx}{\textwidth}{p{4.7cm} p{0.77cm} p{0.77cm} p{0.77cm} p{0.77cm}}
    \toprule
     & \multicolumn{4}{c}{Task} \\
    \cmidrule(r){2-5}
    Reward                              & 1.    & 2.    & 3.    & 4. \\
    \midrule
    1. Collect Trash                    & 1     & 1     & 1     & -1 \\ %
    2. Global Lowest Trash Collected    & 1     & 1     & 1     & 0 \\ %
    3. Crossed Border                   & 1     & 1     & 1     & 1 \\ %
    4. Collided with Other Vessel       & 1     & 1     & 1     & 1 \\ %
    5. Close to Other Vessel            & 0     & 0     & 1     & 0 \\ %
    6. Nearby Trash Count Delta         & 0     & 1     & 0     & 0 \\ %
    7. Collide with Trash               & 0     & 0     & 0     & 1 \\ %
    \bottomrule
  \end{tabularx}
\end{table}

\clearpage 

\subsubsection{Drone-Based Reforestation}\label{task-description-and-reward-scale-drone-based-reforestation}

\textbf{Task Description}

\begin{enumerate}[noitemsep,nolistsep]
  \item \textbf{Main Task: Maximize Collective Planted Tree Count} - This is the main task of the environment. The goal for the agent is to pick up a seed and re-charge batteries at the drone station, explore to find fertile ground for the seed, that is, a location that is close to existing trees, and drop the seed while maintaining enough battery charge to return to the drone station.
  \item \textbf{Subtask: Find Closest Forest Perimeter} - This is a subtask with the goal of finding the closest forest perimeter.
  \item \textbf{Subtask: Pick-up Seed at Base} - This is a subtask with the goal of going back to the drone station, picking up a seed, and recharging the battery. In this subtask, the initial position of drones is random instead of at the drone station.
  \item \textbf{Subtask: Drop Seed} - This is a subtask with the goal of finding the most fertile soil and dropping a seed.
  \item \textbf{Subtask: Find Highest Potential Seed Drop Location} - This is a subtask with the goal of finding soil with the highest fertility.
  \item \textbf{Subtask: Find Highest Point on Landscape} - This is a subtask with the goal of finding the highest point on the landscape.
  \item \textbf{Subtask: Explore Furthest Distance and Return to Base} - This is a subtask with the goal of exploring the furthest from the drone station and returning.
\end{enumerate}
\vspace{1cm}
\textbf{Reward Scale}

\begin{table}[h!]
  \caption{Main- and Sub-Task Reward Scale}
  \label{dbr-task-rewards}
  \centering
  \begin{tabularx}{\textwidth}{p{4.5cm} p{0.6cm} p{0.6cm} p{0.6cm} p{0.6cm} p{0.6cm} p{0.6cm} p{0.6cm} p{0.9cm}}
    \toprule
     & \multicolumn{8}{c}{Task} \\
    \cmidrule(r){2-9}
    Reward                          & 1.    & 2.    & 3.    & 4.    & 5.    & 6.    & 7. & 8.\\
    \midrule
    1. Drop Seed                    & 1     & 0     & 0     & 0     & 1     & 0     & 0  & 0\\
    2. Deplete Energy Holding Seed  & 1     & 1     & 1     & 1     & 1     & 1     & 1  & 1\\
    3. Deplete Energy No Seed       & 1     & 1     & 1     & 1     & 1     & 1     & 1  & 1\\
    4. Pick-up Seed                 & 1     & 0     & 100     & 1   & 1     & 0     & 0  & 0-200\\
    5. Incremental Running Back     & 1     & 0     & 0     & 1     & 1     & 0     & 0  & 1\\
    6. High Fertility Location Delta& 0     & 0     & 0     & 0     & 0     & 1     & 0  & 0\\
    7. High Landscape Point Delta   & 0     & 0     & 0     & 0     & 0     & 0     & 1  & 0\\
    8. Far Distance Explored Delta  & 0     & 0     & 0     & 0     & 0     & 0     & 0  & 1\\
    9. Find Close Tree             & 0     & 1     & 0     & 0     & 0     & 0     & 0  & 0\\
    \bottomrule
  \end{tabularx}
\end{table}

\subsubsection{Aerial Wildfire Suppression}\label{task-description-and-reward-scale-aerial-wildfire-suppressionl}

\textbf{Task Description}

\begin{enumerate}[noitemsep,nolistsep]
  \item \textbf{Main Task: Minimize Time Fire Burning and Prevent Fire From Moving Towards Village} - This is the main task of the environment. The goal for the agent is to pick up water and extinguish as many burning trees as possible or prepare a forest that is not yet burning. A secondary goal is to protect the village from approaching fire by extinguishing burning trees before they get too close to the village or redirecting the fire by preparing trees.
  \item \textbf{Subtask: Maximize Extinguished Burning Trees} - This is a subtask with the goal of extinguishing as many burning trees as possible.  
  \item \textbf{Subtask: Maximize Preparing Non-Burning Trees} - This is a subtask with the goal of preparing as many non-burning trees as possible.
  \item \textbf{Subtask: Minimize Time Fire Burning} - This is a subtask with the goal of minimizing the time of trees burning.
  \item \textbf{Subtask: Protect Village} - This is a subtask with the goal of protecting the village from approaching fire.
  \item \textbf{Subtask: Pick Up water} - This is a subtask with the goal of picking up water.
  \item \textbf{Subtask: Drop Water} - This is a subtask with the goal of dropping water anywhere.
  \item \textbf{Subtask: Find Fire} - This is a subtask with the goal of finding a burning tree.
  \item \textbf{Subtask: Find Village} - This is a subtask with the goal of finding the village.
\end{enumerate}
\vspace{1cm}
\textbf{Reward Scale}

\begin{table}[h!]
  \caption{Main- and Sub-Task Reward Scale}
  \label{aws-task-rewards}
  \centering
  \begin{tabularx}{\textwidth}{p{3cm} p{0.77cm} p{0.77cm} p{0.77cm} p{0.77cm} p{0.77cm} p{0.77cm} p{0.77cm} p{0.77cm} p{0.77cm}}
    \toprule
     & \multicolumn{9}{c}{Task} \\
    \cmidrule(r){2-10}
    Reward                      & 1. & 2. & 3. & 4. & 5. & 6. & 7. & 8. & 9. \\
    \midrule
    1. Crossed Border           & 1 & 1 & 1 & 1 & 1 & 1 & 1 & 1 & 1 \\
    2. Pick-up Water            & 1 & 1 & 1 & 1 & 1 & 100 & 1 & 0 & 0 \\
    3. Fire Out                 & 1 & 1 & 1 & 1 & 1 & 0 & 0 & 0 & 0 \\
    4. Too Close to Village     & 1 & 1 & 1 & 1 & 10 & 0 & 0 & 0 & 0 \\
    5. Time Step Burning        & 0 & 0 & 0 & 1 & 0 & 0 & 0 & 0 & 0 \\
    6. Find Fire                & 0 & 0 & 0 & 0 & 0 & 0 & 0 & 1 & 0 \\
    7. Find Village             & 0 & 0 & 0 & 0 & 0 & 0 & 0 & 0 & 1 \\
    Drop Water &  &  &  &  &  &  &  & \\
    \midrule
    8. Extinguishing Tree       & 1 & 10 & 1 & 1 & 1 & 1 & 1 & 0 & 0 \\
    9. Preparing Tree           & 1 & 1  & 5 & 1 & 1 & 1 & 1 & 0 & 0 \\
    \bottomrule
  \end{tabularx}
\end{table}

\clearpage

\subsection{Additional Results} \label{ref:additional_results}

\subsubsection{Wind Farm Control: Train \& Test Metrics}

\begin{figure}[h!]
\centering
\includegraphics[width=\linewidth]{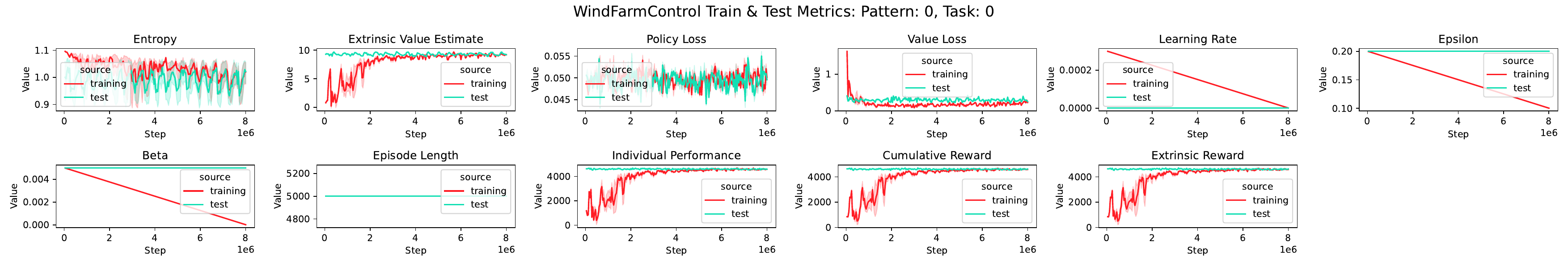}
\vspace{-0.6cm}
\caption{Wind Farm Control: Train \& Test Metrics: Pattern 0, Task 0.}
\end{figure}

\begin{figure}[h!]
\centering
\includegraphics[width=\linewidth]{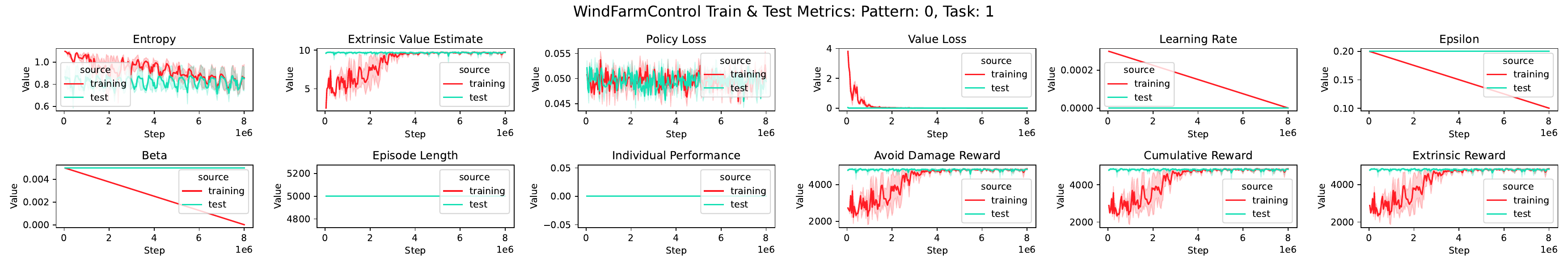}
\vspace{-0.6cm}
\caption{Wind Farm Control: Train \& Test Metrics: Pattern 0, Task 1.}
\end{figure}

\begin{figure}[h!]
\centering
\includegraphics[width=\linewidth]{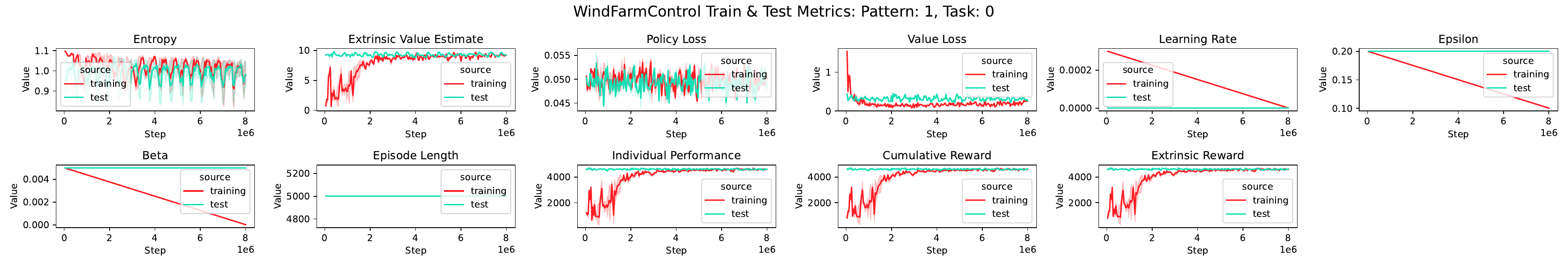}
\vspace{-0.6cm}
\caption{Wind Farm Control: Train \& Test Metrics: Pattern 1, Task 0.}
\end{figure}

\begin{figure}[h!]
\centering
\includegraphics[width=\linewidth]{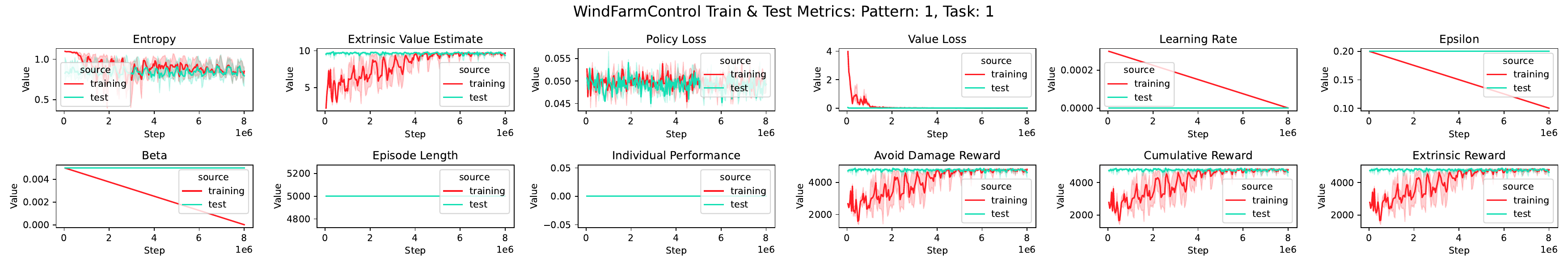}
\vspace{-0.6cm}
\caption{Wind Farm Control: Train \& Test Metrics: Pattern 1, Task 1.}
\end{figure}

\begin{figure}[h!]
\centering
\includegraphics[width=\linewidth]{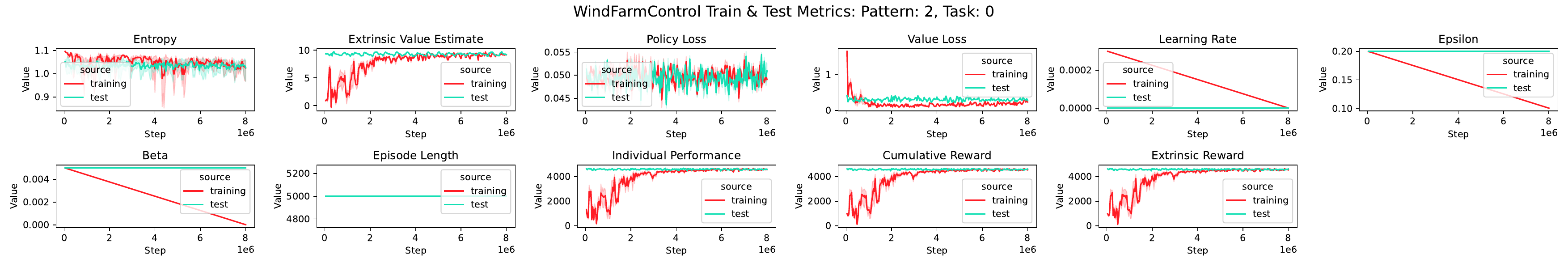}
\vspace{-0.6cm}
\caption{Wind Farm Control: Train \& Test Metrics: Pattern 2, Task 0.}
\end{figure}

\begin{figure}[h!]
\centering
\includegraphics[width=\linewidth]{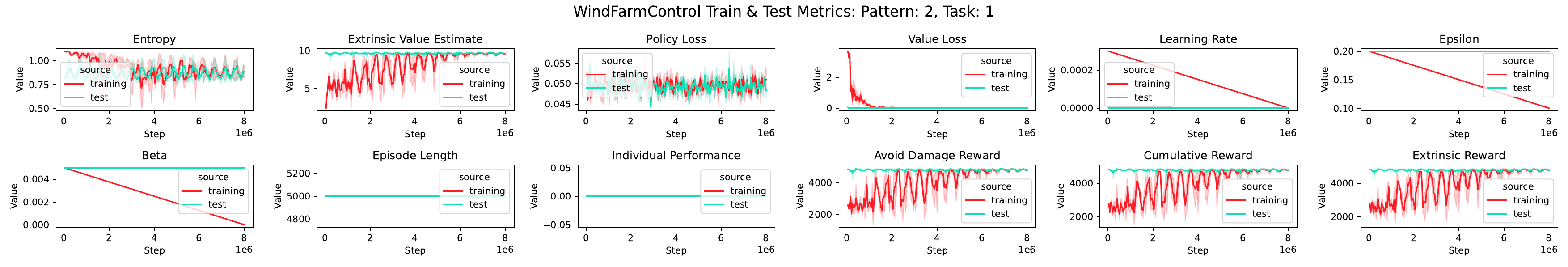}
\vspace{-0.6cm}
\caption{Wind Farm Control: Train \& Test Metrics: Pattern 2, Task 1.}
\end{figure}

\clearpage
\begin{figure}[h!]
\centering
\includegraphics[width=\linewidth]{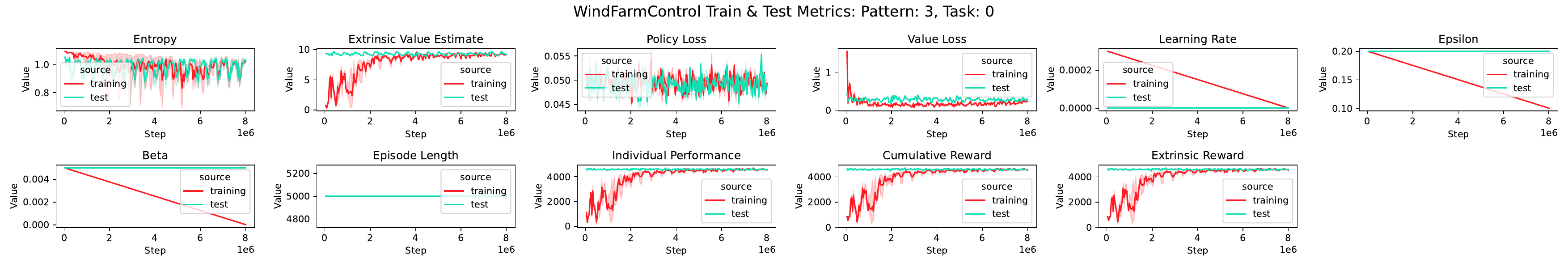}
\vspace{-0.6cm}
\caption{Wind Farm Control: Train \& Test Metrics: Pattern 3, Task 0.}
\end{figure}

\begin{figure}[h!]
\centering
\includegraphics[width=\linewidth]{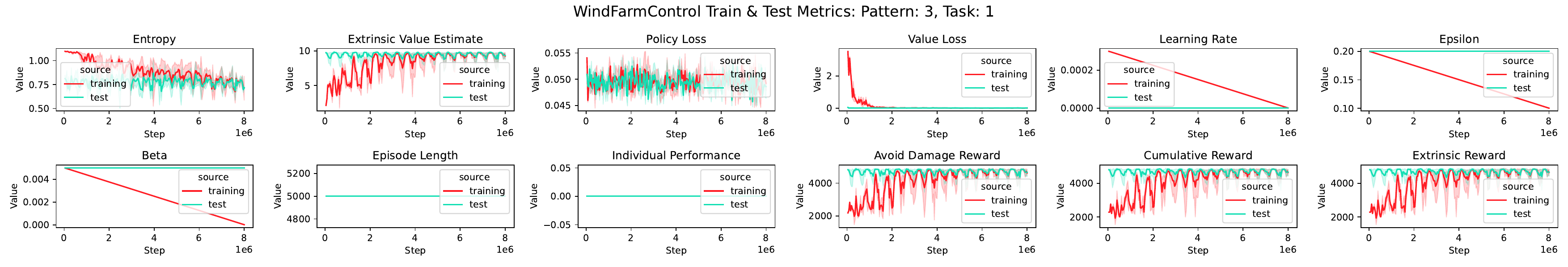}
\vspace{-0.6cm}
\caption{Wind Farm Control: Train \& Test Metrics: Pattern 3, Task 1.}
\end{figure}

\begin{figure}[h!]
\centering
\includegraphics[width=\linewidth]{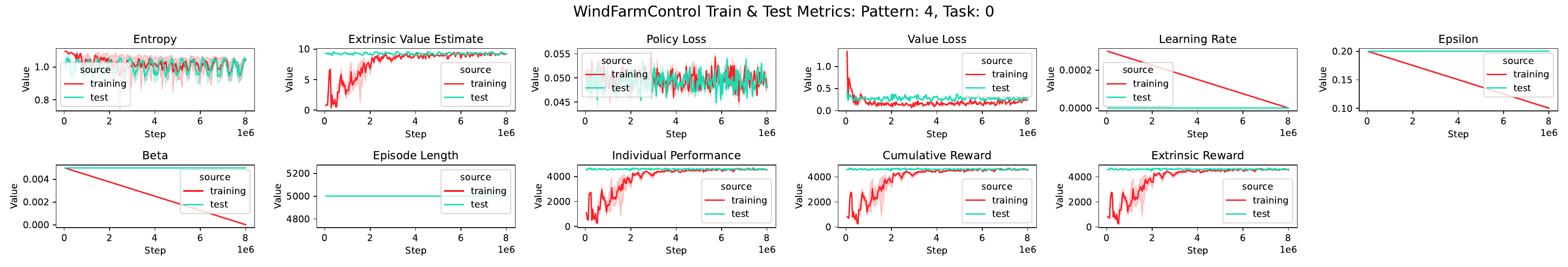}
\vspace{-0.6cm}
\caption{Wind Farm Control: Train \& Test Metrics: Pattern 4, Task 0.}
\end{figure}

\begin{figure}[h!]
\centering
\includegraphics[width=\linewidth]{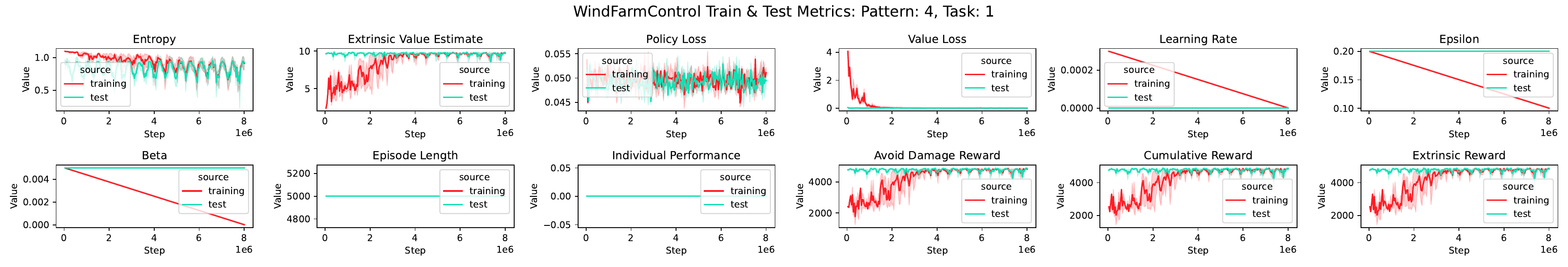}
\vspace{-0.6cm}
\caption{Wind Farm Control: Train \& Test Metrics: Pattern 4, Task 1.}
\end{figure}

\begin{figure}[h!]
\centering
\includegraphics[width=\linewidth]{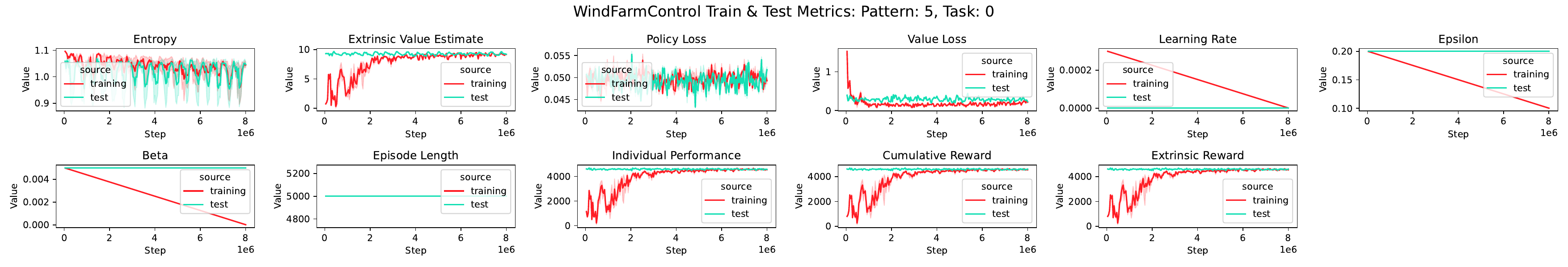}
\vspace{-0.6cm}
\caption{Wind Farm Control: Train \& Test Metrics: Pattern 5, Task 0.}
\end{figure}

\begin{figure}[h!]
\centering
\includegraphics[width=\linewidth]{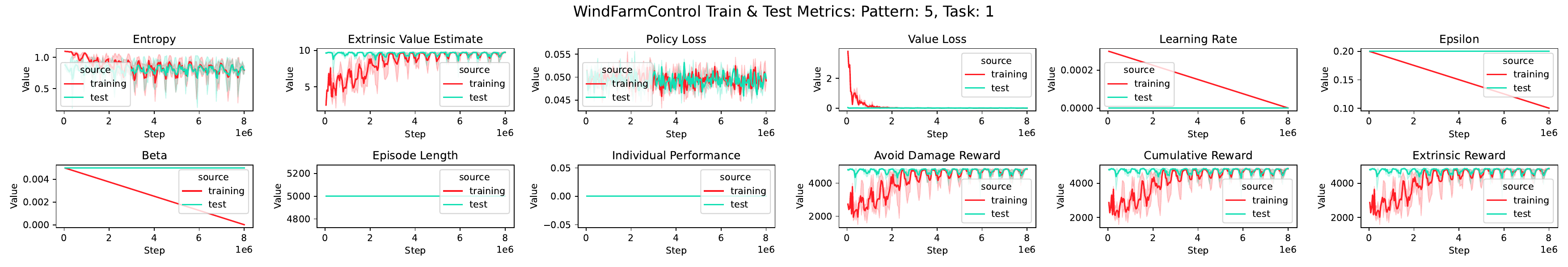}
\vspace{-0.6cm}
\caption{Wind Farm Control: Train \& Test Metrics: Pattern 5, Task 1.}
\end{figure}

\clearpage
\begin{figure}[h!]
\centering
\includegraphics[width=\linewidth]{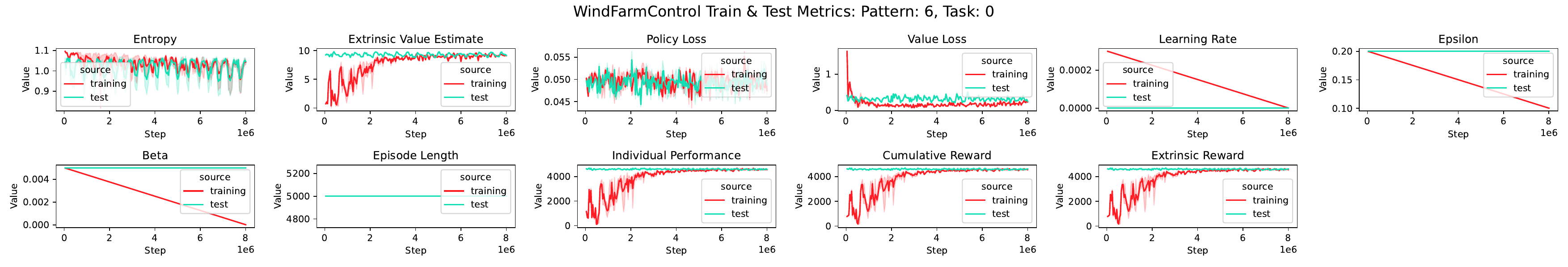}
\vspace{-0.6cm}
\caption{Wind Farm Control: Train \& Test Metrics: Pattern 6, Task 0.}
\end{figure}

\begin{figure}[h!]
\centering
\includegraphics[width=\linewidth]{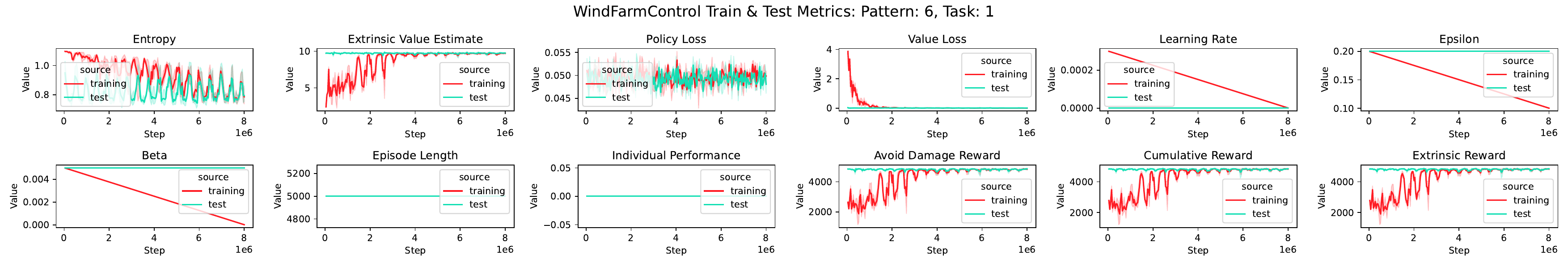}
\vspace{-0.6cm}
\caption{Wind Farm Control: Train \& Test Metrics: Pattern 6, Task 1.}
\end{figure}

\begin{figure}[h!]
\centering
\includegraphics[width=\linewidth]{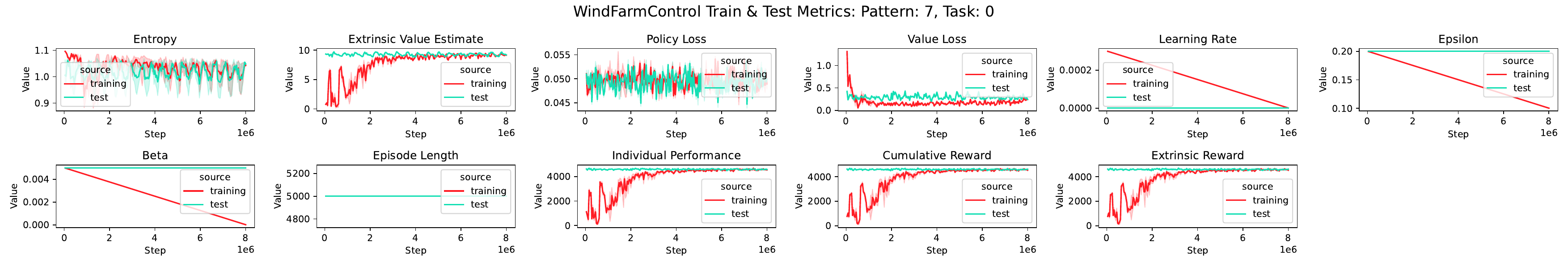}
\vspace{-0.6cm}
\caption{Wind Farm Control: Train \& Test Metrics: Pattern 7, Task 0.}
\end{figure}

\begin{figure}[h!]
\centering
\includegraphics[width=\linewidth]{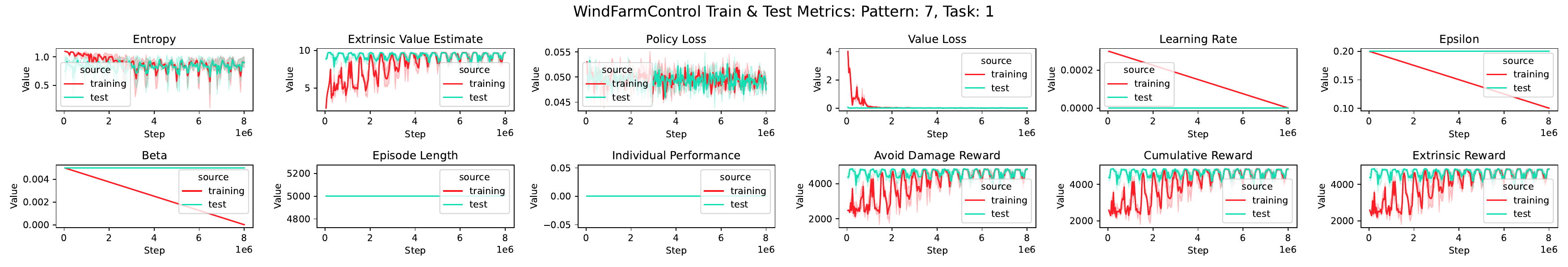}
\vspace{-0.6cm}
\caption{Wind Farm Control: Train \& Test Metrics: Pattern 7, Task 1.}
\end{figure}

\begin{figure}[h!]
\centering
\includegraphics[width=\linewidth]{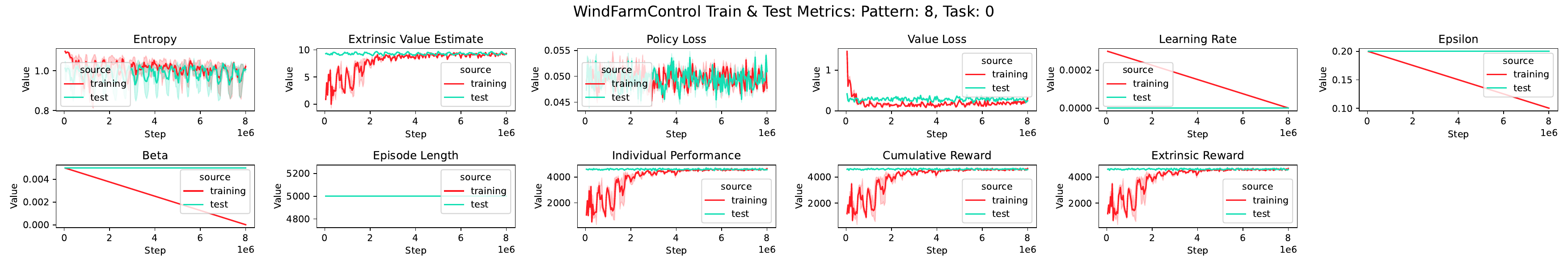}
\vspace{-0.6cm}
\caption{Wind Farm Control: Train \& Test Metrics: Pattern 8, Task 0.}
\end{figure}

\begin{figure}[h!]
\centering
\includegraphics[width=\linewidth]{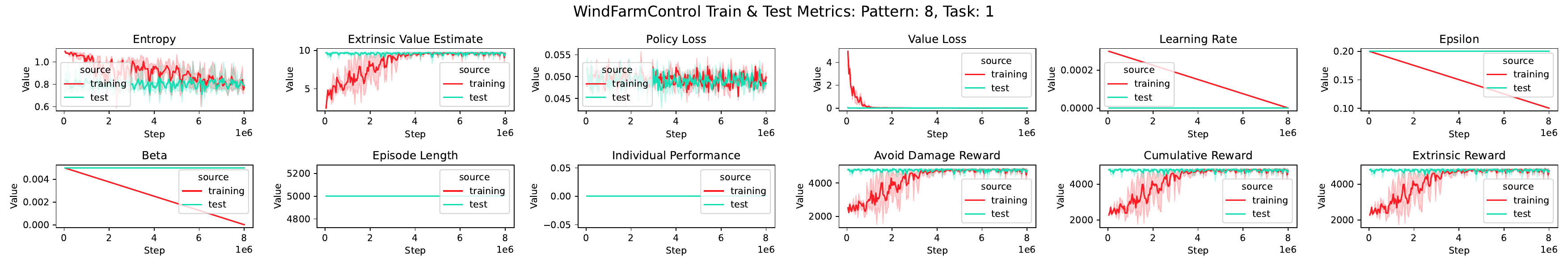}
\vspace{-0.6cm}
\caption{Wind Farm Control: Train \& Test Metrics: Pattern 8, Task 1.}
\end{figure}

\clearpage

\subsubsection{Wind Farm Control: Average Test Metric - Task VS Pattern}

\begin{figure}[h!]
\centering[h!]
\includegraphics[width=\linewidth]{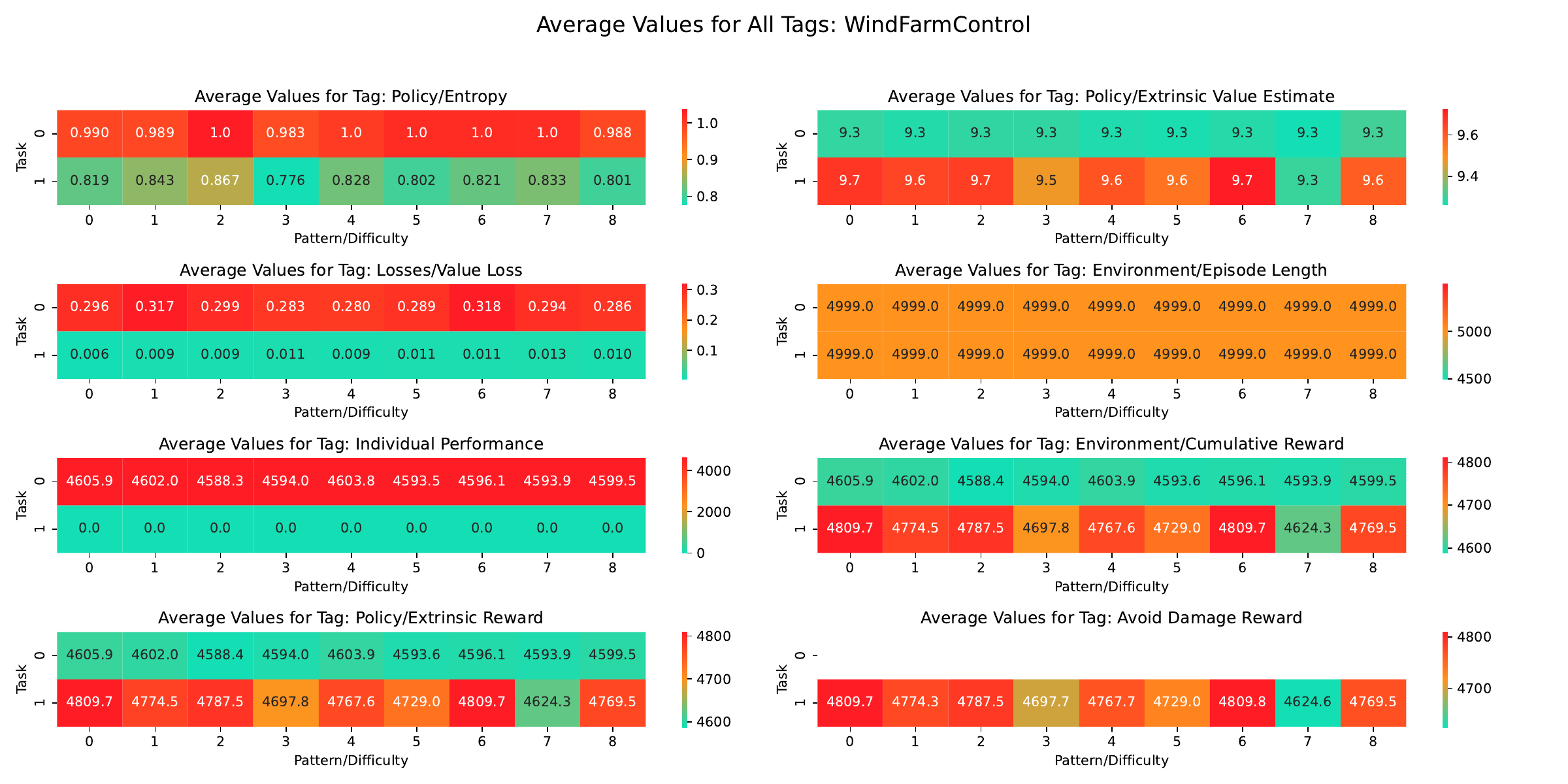}
\caption{Wind Farm Control: Average Train \& Test Metrics.}
\end{figure}

\clearpage

\subsubsection{Wildfire Resource Management: Train \& Test Metrics}
\begin{figure}[h!]
\centering
\includegraphics[width=\linewidth]{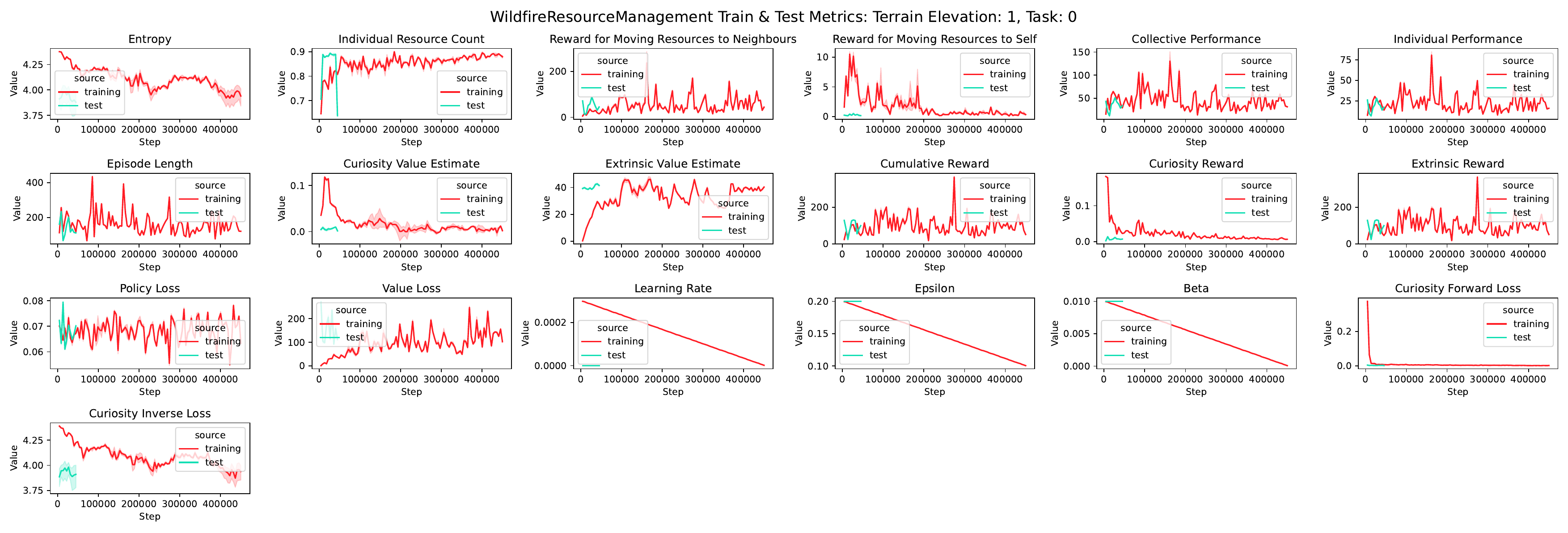}
\vspace{-0.6cm}
\caption{Wildfire Resource Management: Train \& Test Metrics: Terrain Elevation 1, Task 0.}
\end{figure}

\begin{figure}[h!]
\centering
\includegraphics[width=\linewidth]{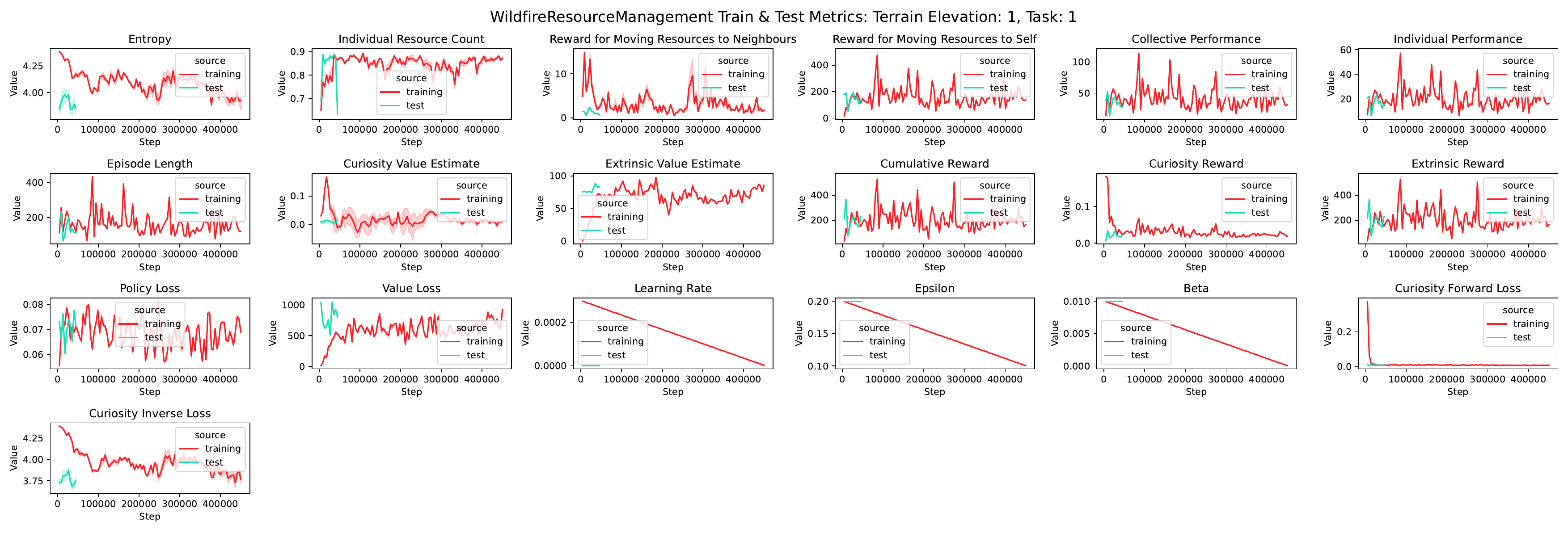}
\vspace{-0.6cm}
\caption{Wildfire Resource Management: Train \& Test Metrics: Terrain Elevation 1, Task 1.}
\end{figure}

\begin{figure}[h!]
\centering
\includegraphics[width=\linewidth]{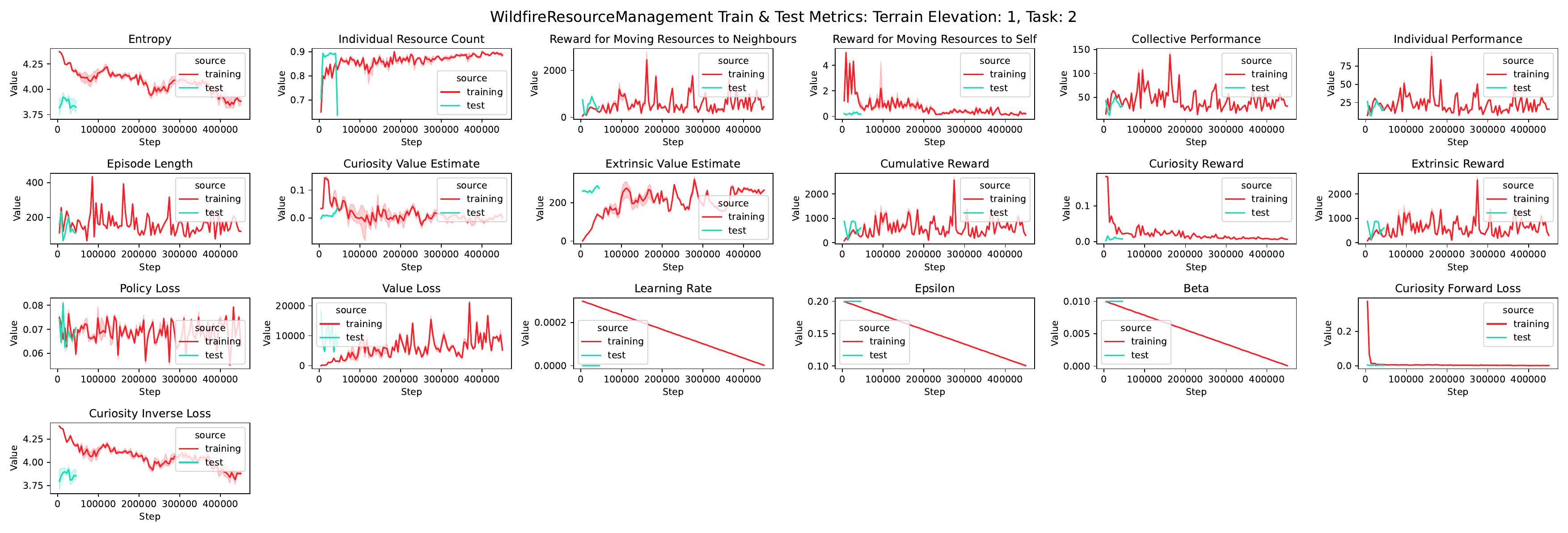}
\vspace{-0.6cm}
\caption{Wildfire Resource Management: Train \& Test Metrics: Terrain Elevation 1, Task 2.}
\end{figure}

\clearpage

\begin{figure}[h!]
\centering
\includegraphics[width=\linewidth]{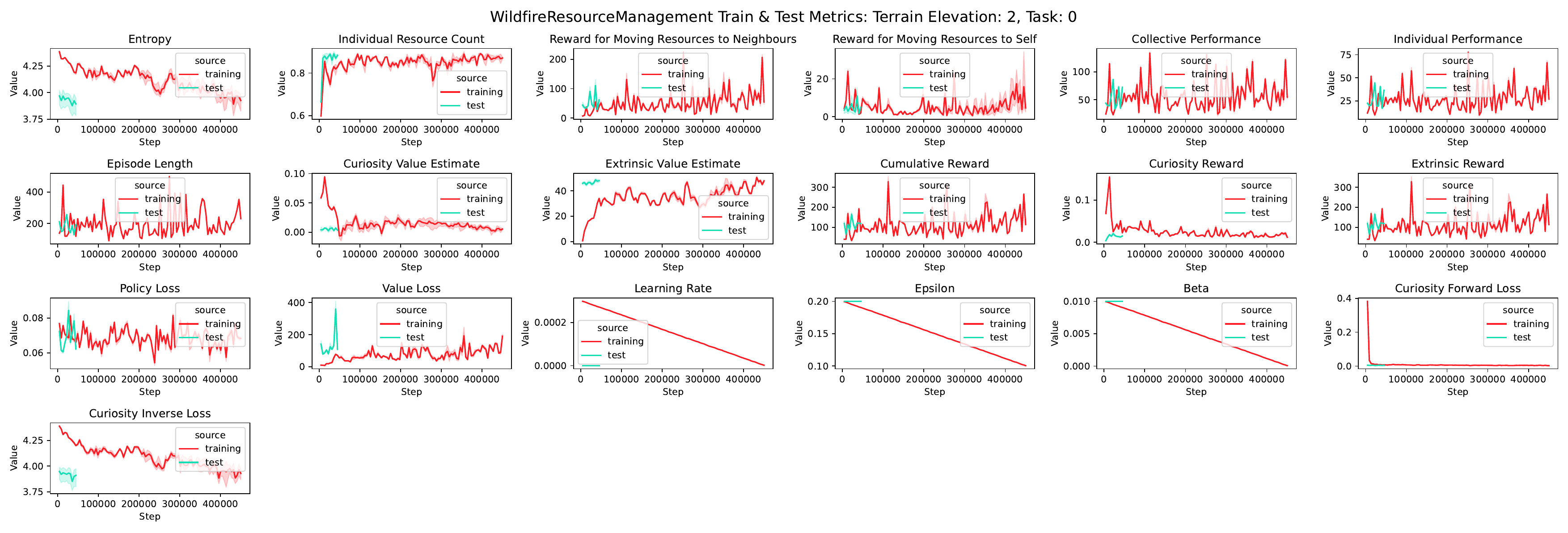}
\vspace{-0.6cm}
\caption{Wildfire Resource Management: Train \& Test Metrics: Terrain Elevation 2, Task 0.}
\end{figure}

\begin{figure}[h!]
\centering
\includegraphics[width=\linewidth]{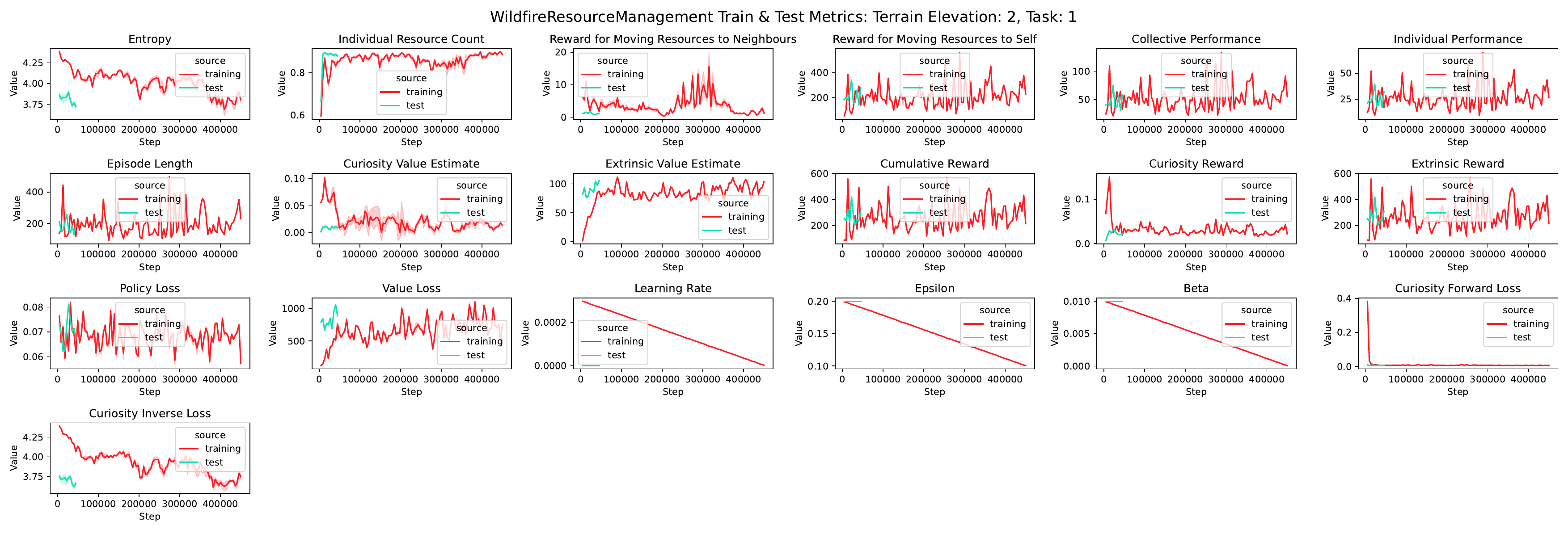}
\vspace{-0.6cm}
\caption{Wildfire Resource Management: Train \& Test Metrics: Terrain Elevation 2, Task 1.}
\end{figure}

\begin{figure}[h!]
\centering
\includegraphics[width=\linewidth]{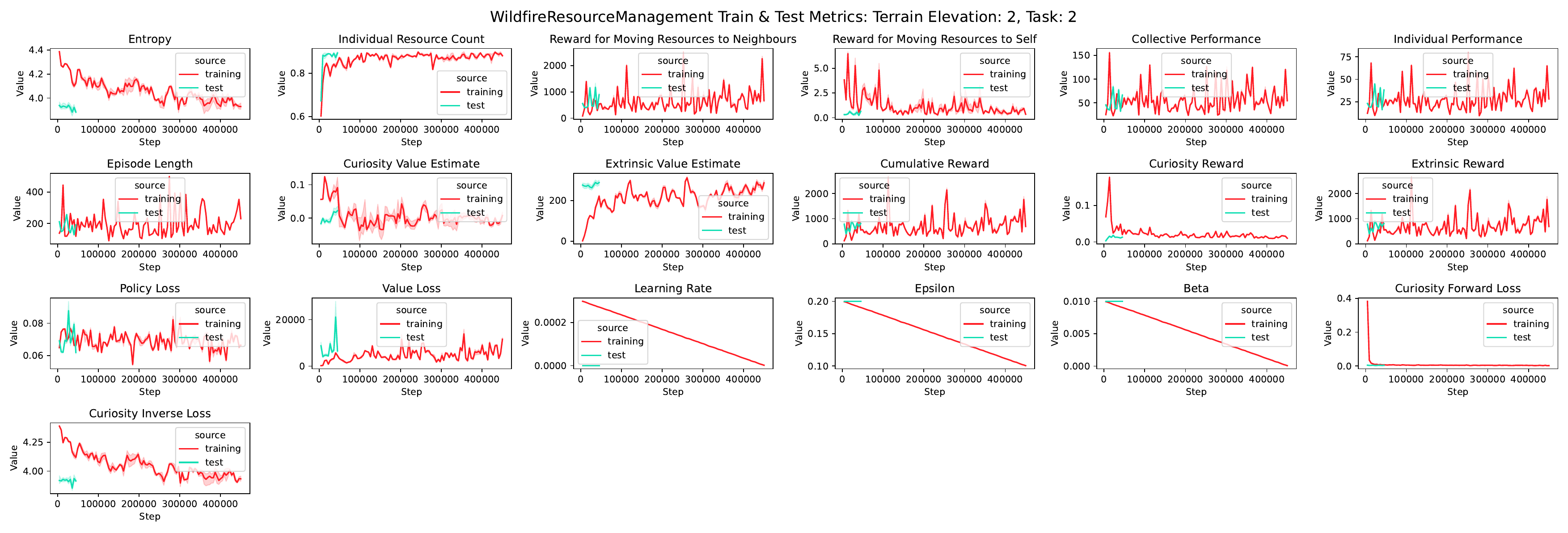}
\vspace{-0.6cm}
\caption{Wildfire Resource Management: Train \& Test Metrics: Terrain Elevation 2, Task 2.}
\end{figure}

\clearpage

\begin{figure}[h!]
\centering
\includegraphics[width=\linewidth]{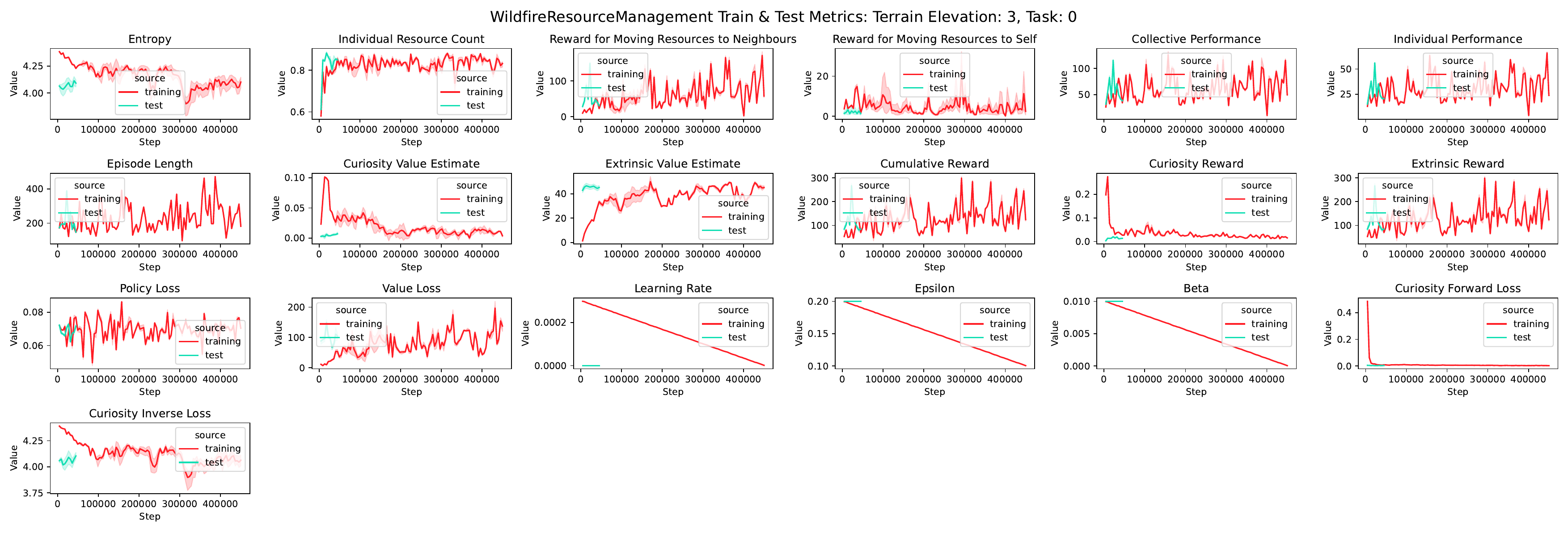}
\vspace{-0.6cm}
\caption{Wildfire Resource Management: Train \& Test Metrics: Terrain Elevation 3, Task 0.}
\end{figure}

\begin{figure}[h!]
\centering
\includegraphics[width=\linewidth]{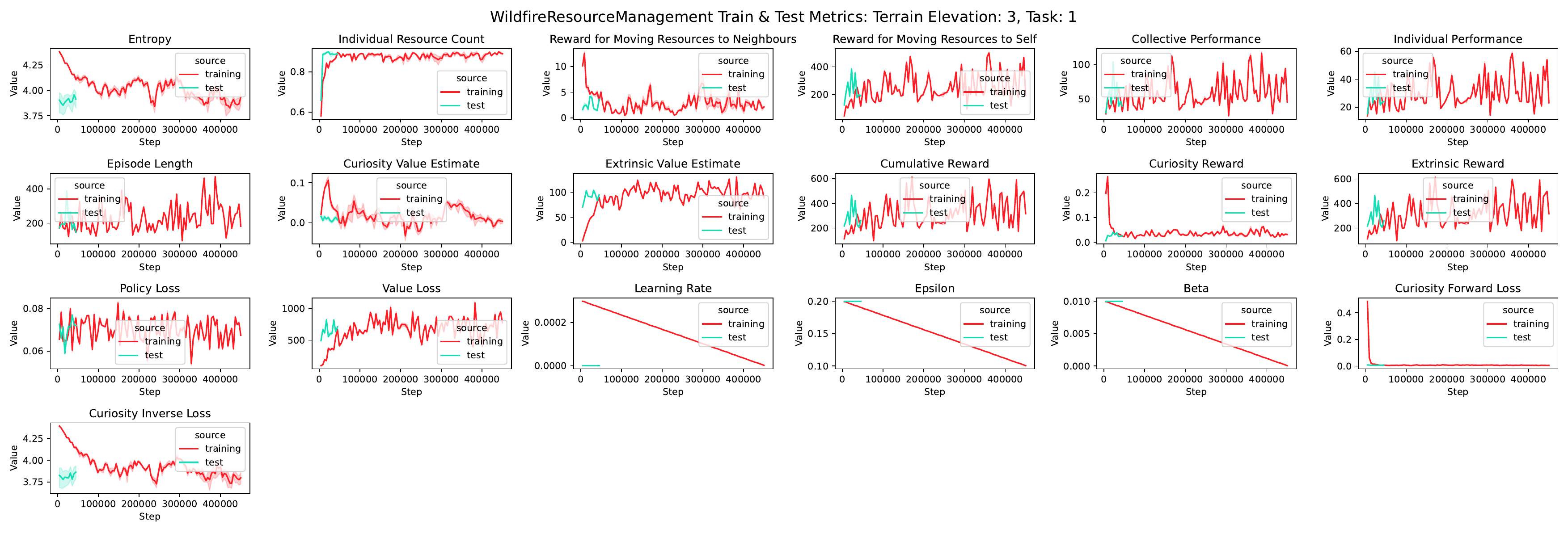}
\vspace{-0.6cm}
\caption{Wildfire Resource Management: Train \& Test Metrics: Terrain Elevation 3, Task 1.}
\end{figure}

\begin{figure}[h!]
\centering
\includegraphics[width=\linewidth]{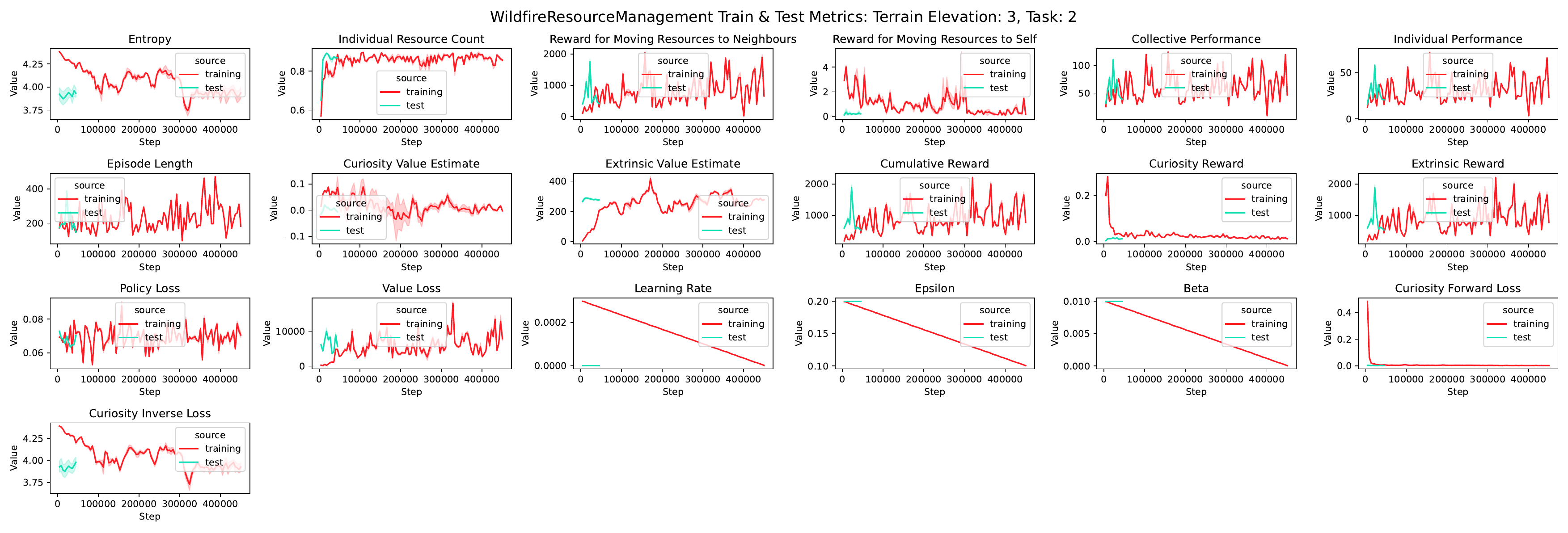}
\vspace{-0.6cm}
\caption{Wildfire Resource Management: Train \& Test Metrics: Terrain Elevation 3, Task 2.}
\end{figure}

\clearpage

\begin{figure}[h!]
\centering
\includegraphics[width=\linewidth]{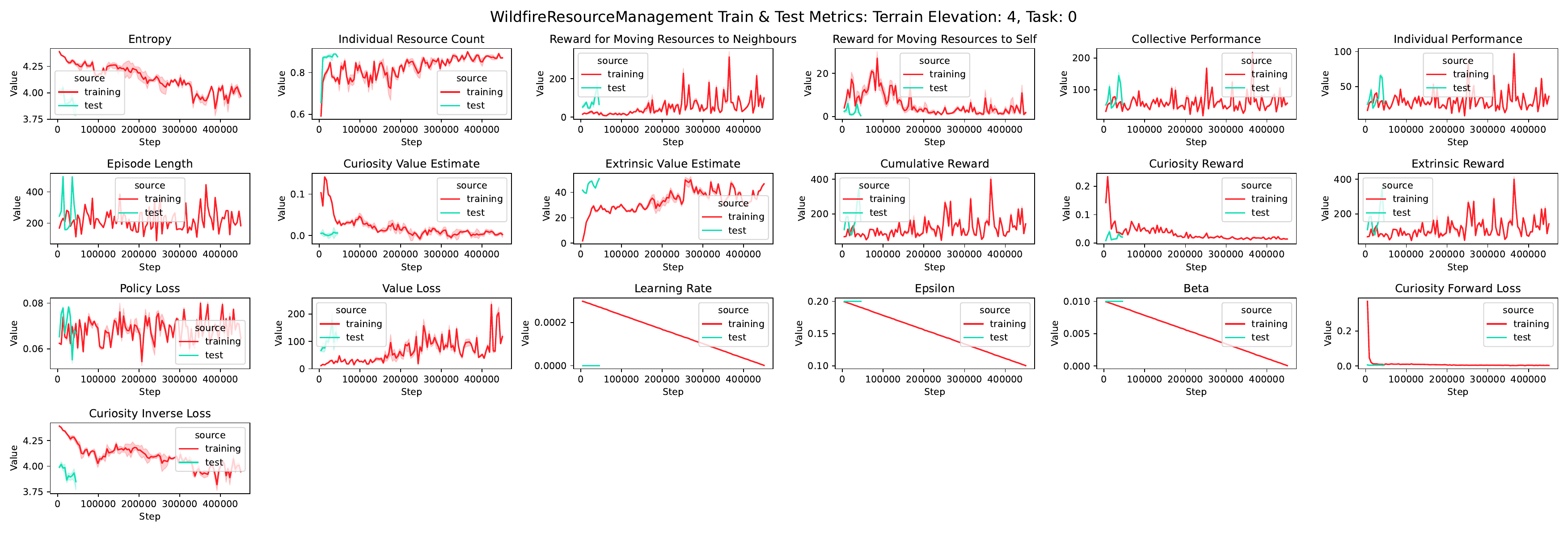}
\vspace{-0.6cm}
\caption{Wildfire Resource Management: Train \& Test Metrics: Terrain Elevation 4, Task 0.}
\end{figure}

\begin{figure}[h!]
\centering
\includegraphics[width=\linewidth]{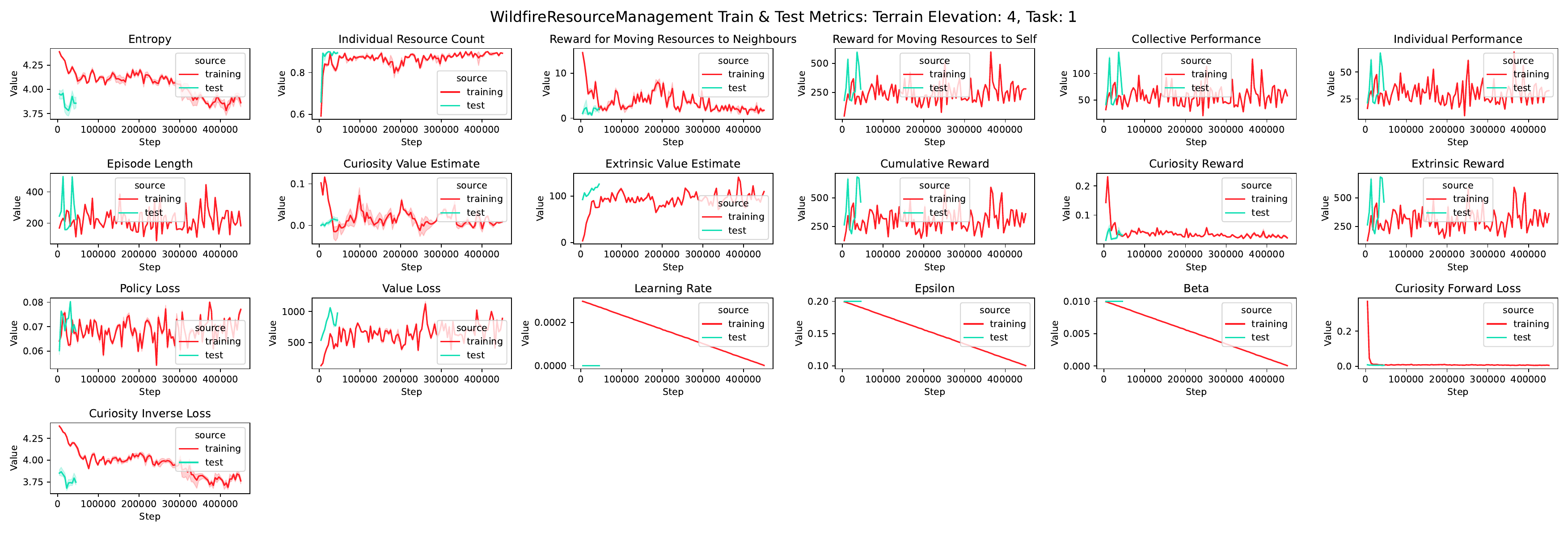}
\vspace{-0.6cm}
\caption{Wildfire Resource Management: Train \& Test Metrics: Terrain Elevation 4, Task 1.}
\end{figure}

\begin{figure}[h!]
\centering
\includegraphics[width=\linewidth]{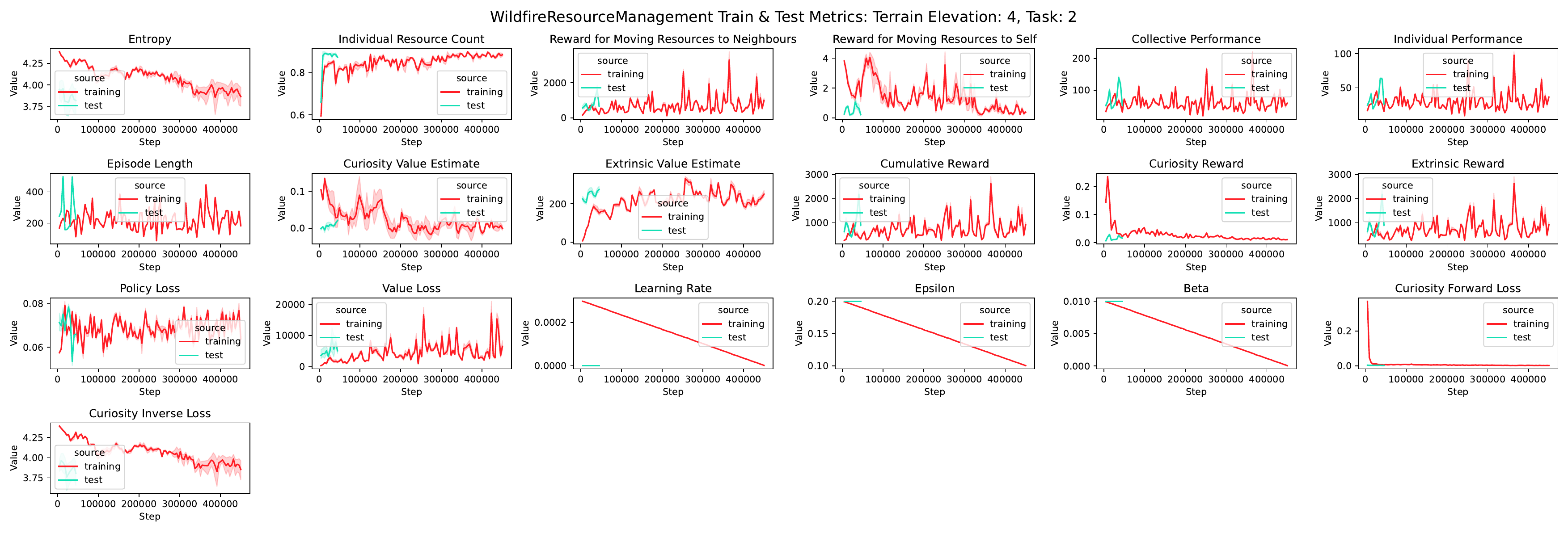}
\vspace{-0.6cm}
\caption{Wildfire Resource Management: Train \& Test Metrics: Terrain Elevation 4, Task 2.}
\end{figure}

\clearpage

\begin{figure}[h!]
\centering
\includegraphics[width=\linewidth]{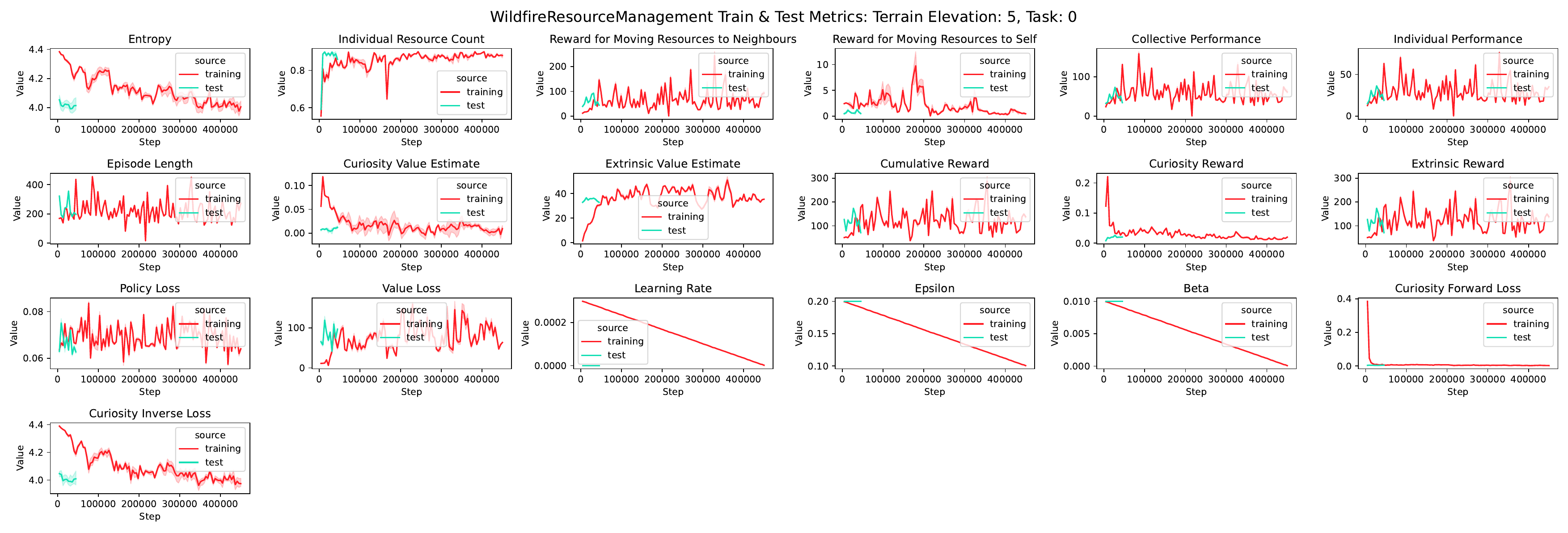}
\vspace{-0.6cm}
\caption{Wildfire Resource Management: Train \& Test Metrics: Terrain Elevation 5, Task 0.}
\end{figure}

\begin{figure}[h!]
\centering
\includegraphics[width=\linewidth]{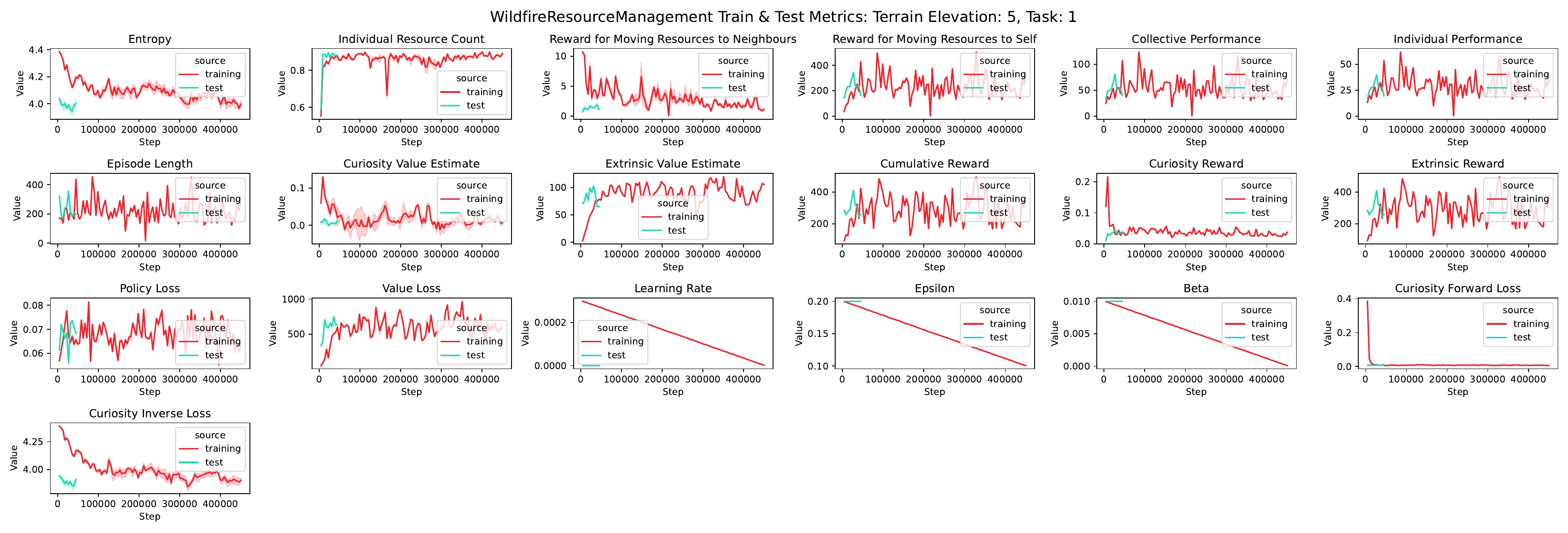}
\vspace{-0.6cm}
\caption{Wildfire Resource Management: Train \& Test Metrics: Terrain Elevation 5, Task 1.}
\end{figure}

\begin{figure}[h!]
\centering
\includegraphics[width=\linewidth]{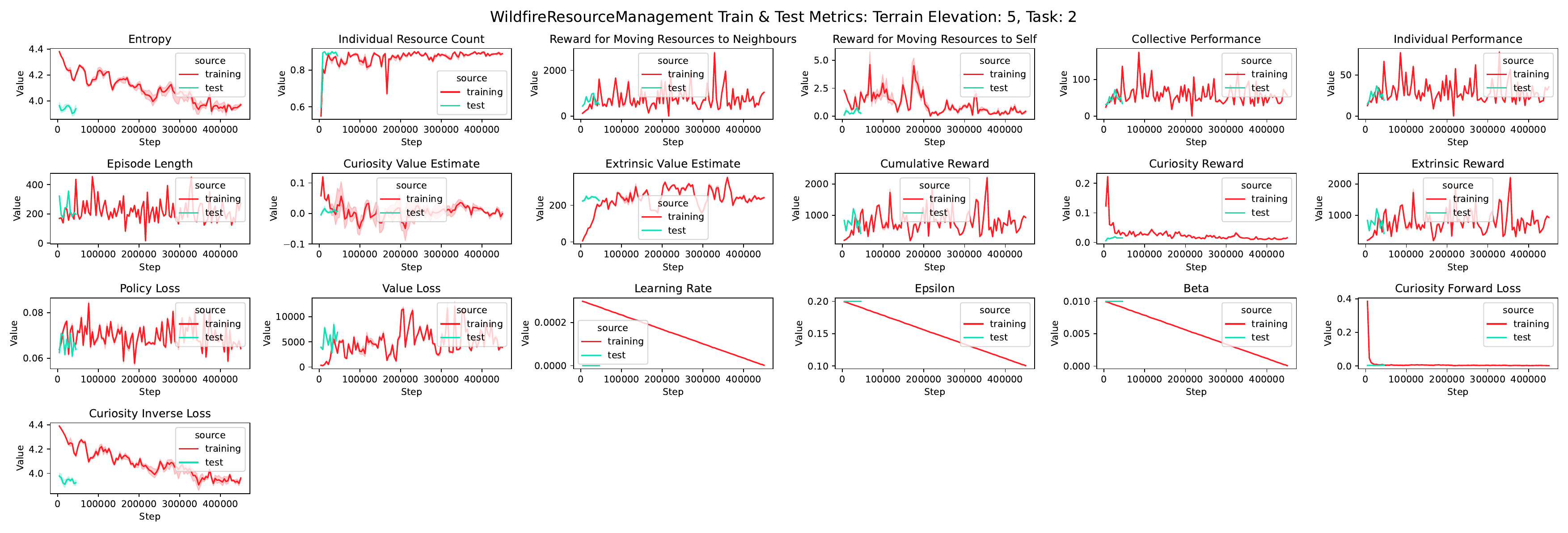}
\vspace{-0.6cm}
\caption{Wildfire Resource Management: Train \& Test Metrics: Terrain Elevation 5, Task 2.}
\end{figure}

\clearpage

\begin{figure}[h!]
\centering
\includegraphics[width=\linewidth]{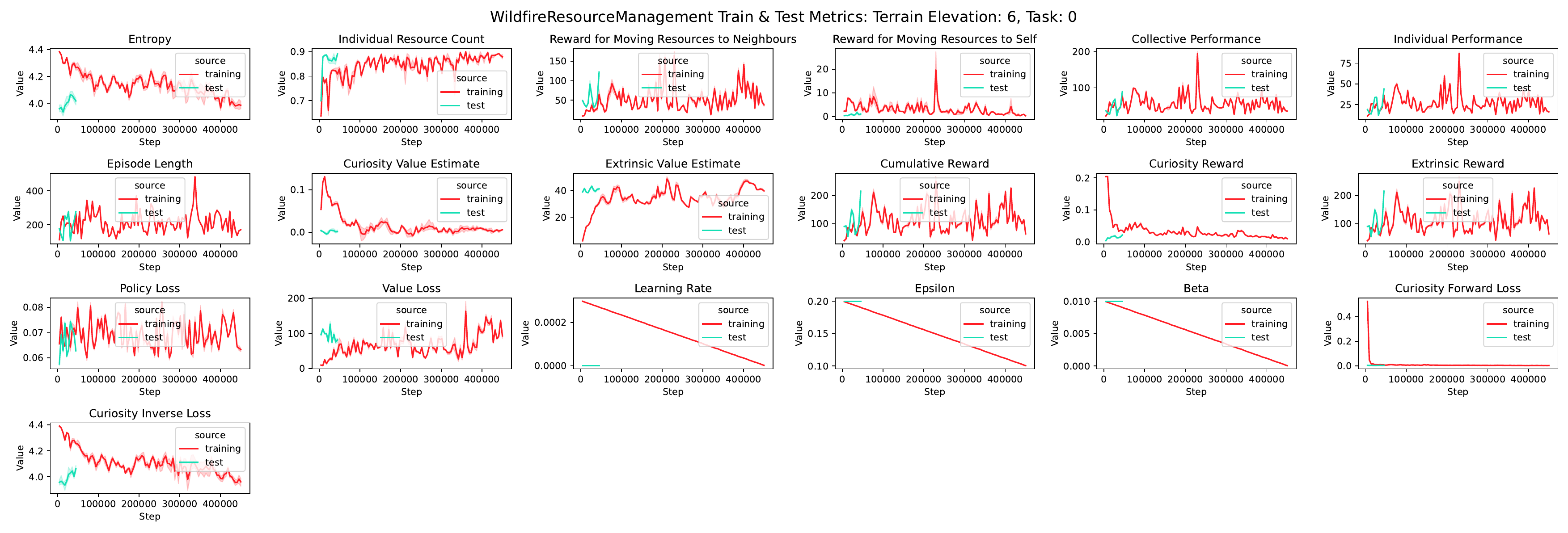}
\vspace{-0.6cm}
\caption{Wildfire Resource Management: Train \& Test Metrics: Terrain Elevation 6, Task 0.}
\end{figure}

\begin{figure}[h!]
\centering
\includegraphics[width=\linewidth]{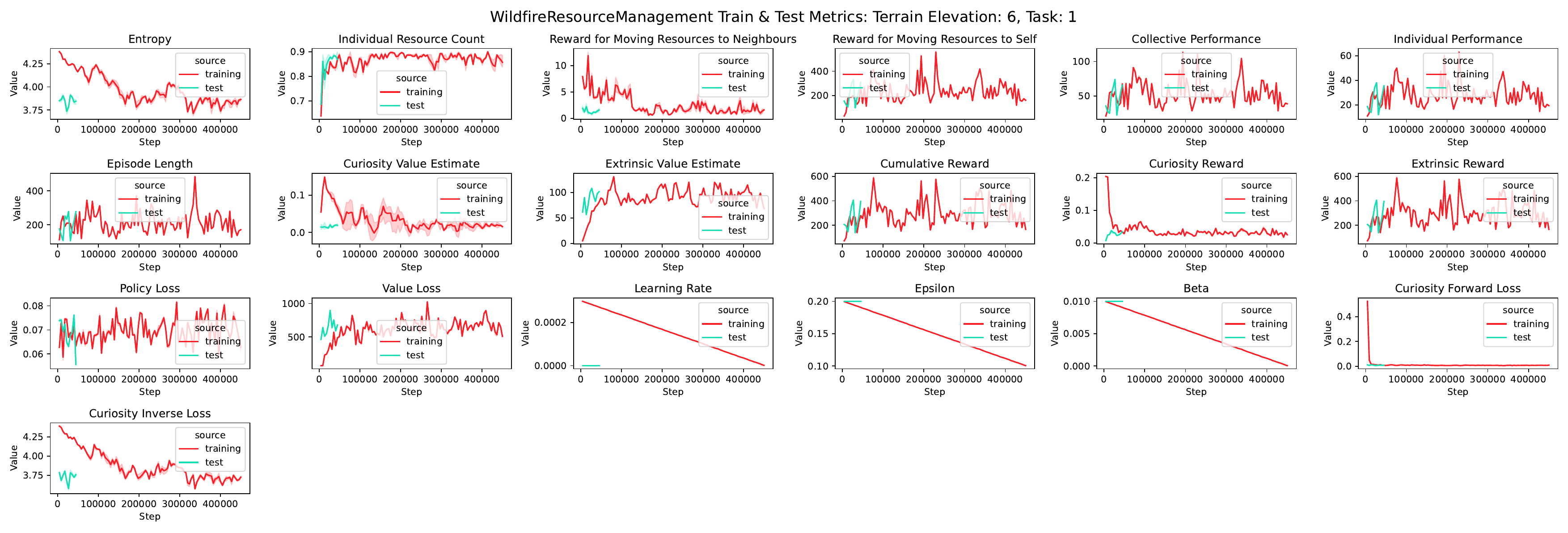}
\vspace{-0.6cm}
\caption{Wildfire Resource Management: Train \& Test Metrics: Terrain Elevation 6, Task 1.}
\end{figure}

\begin{figure}[h!]
\centering
\includegraphics[width=\linewidth]{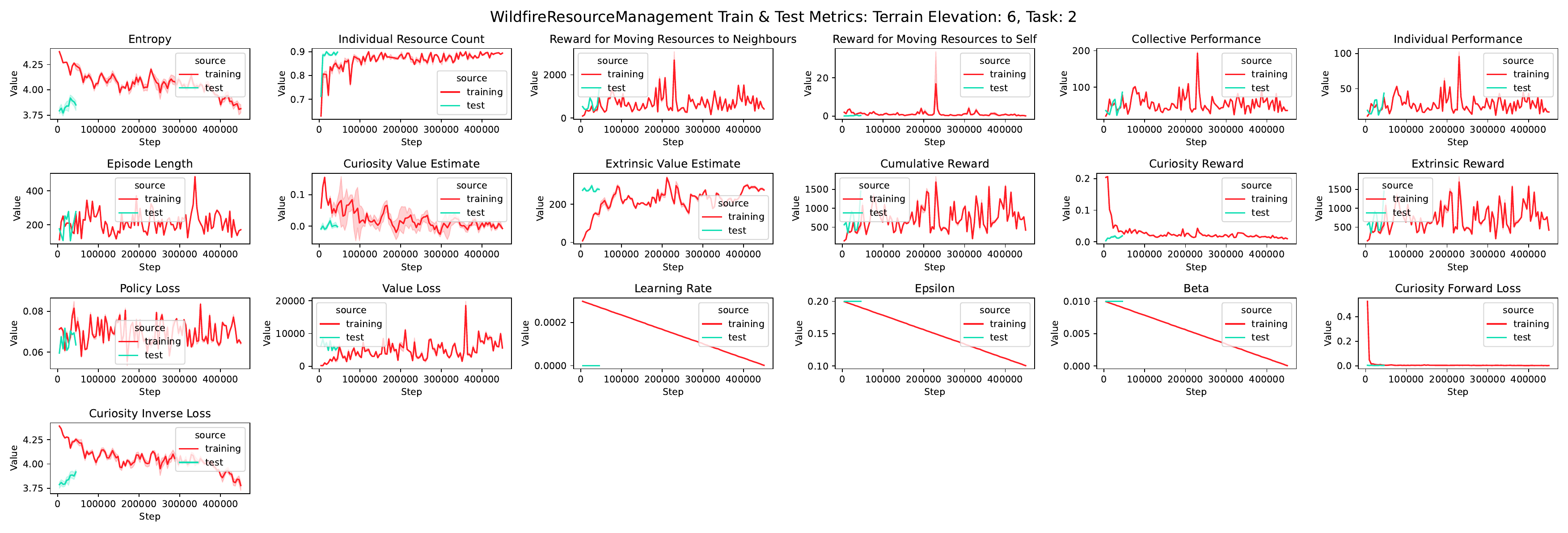}
\vspace{-0.6cm}
\caption{Wildfire Resource Management: Train \& Test Metrics: Terrain Elevation 6, Task 2.}
\end{figure}

\clearpage

\begin{figure}[h!]
\centering
\includegraphics[width=\linewidth]{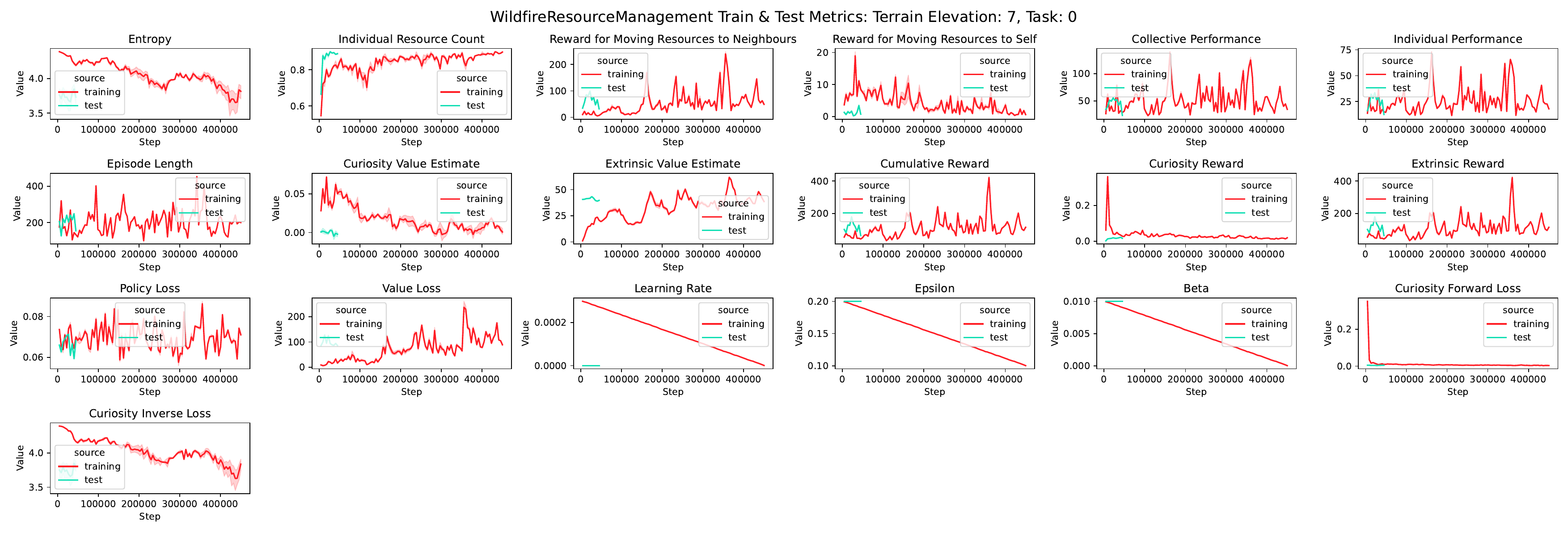}
\vspace{-0.6cm}
\caption{Wildfire Resource Management: Train \& Test Metrics: Terrain Elevation 7, Task 0.}
\end{figure}

\begin{figure}[h!]
\centering
\includegraphics[width=\linewidth]{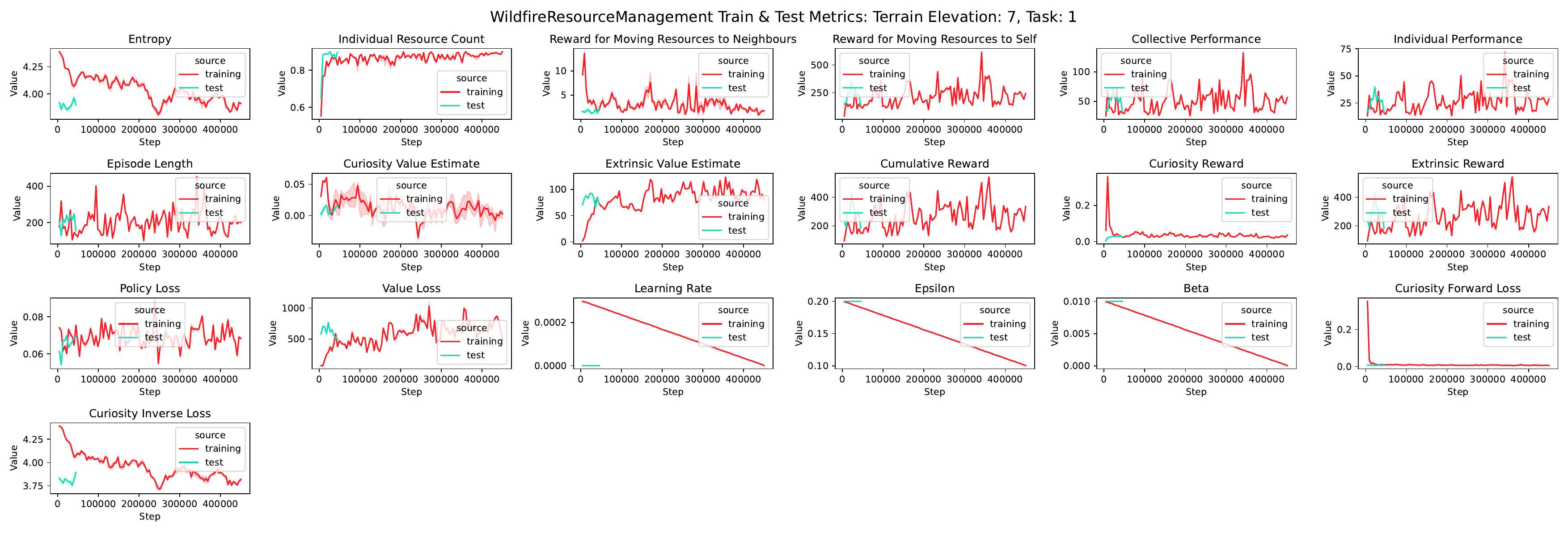}
\vspace{-0.6cm}
\caption{Wildfire Resource Management: Train \& Test Metrics: Terrain Elevation 7, Task 1.}
\end{figure}

\begin{figure}[h!]
\centering
\includegraphics[width=\linewidth]{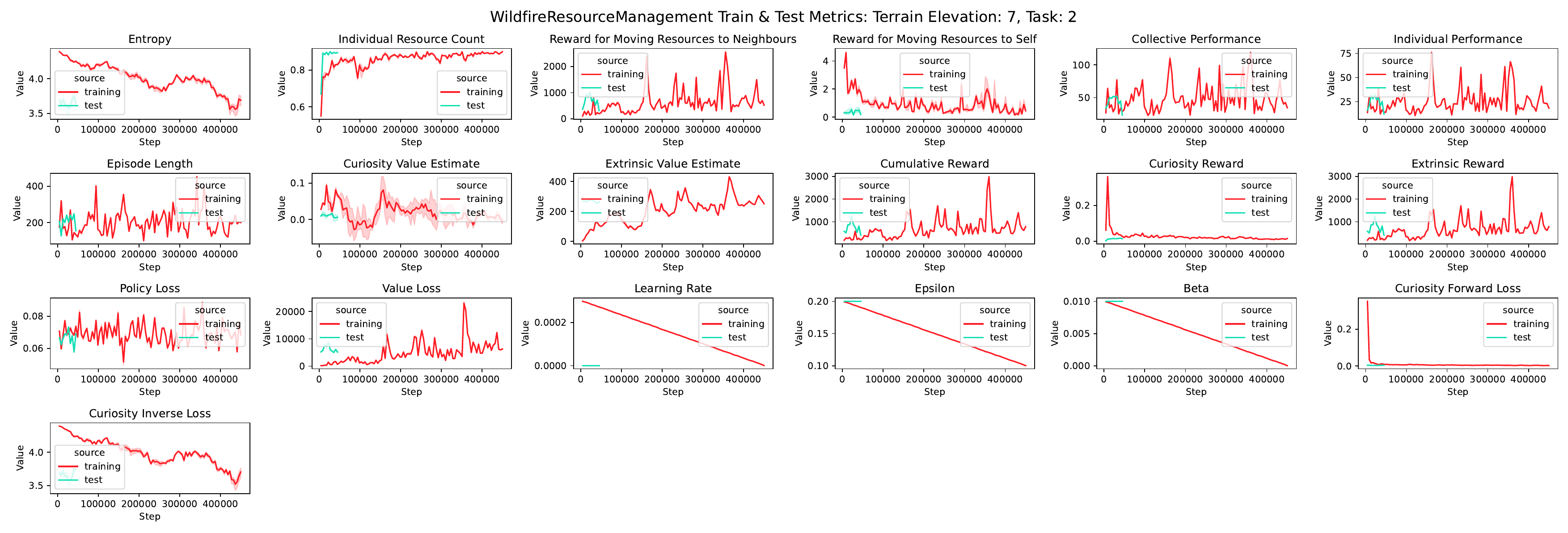}
\vspace{-0.6cm}
\caption{Wildfire Resource Management: Train \& Test Metrics: Terrain Elevation 7, Task 2.}
\end{figure}

\clearpage

\begin{figure}[h!]
\centering
\includegraphics[width=\linewidth]{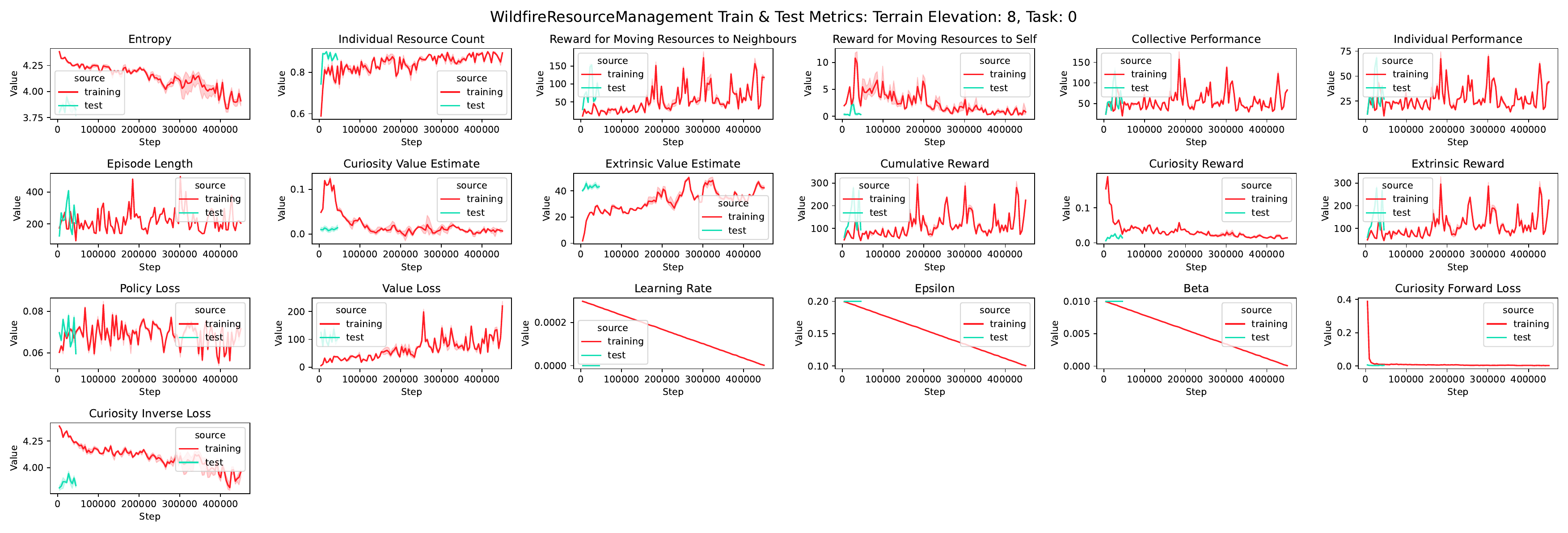}
\vspace{-0.6cm}
\caption{Wildfire Resource Management: Train \& Test Metrics: Terrain Elevation 8, Task 0.}
\end{figure}

\begin{figure}[h!]
\centering
\includegraphics[width=\linewidth]{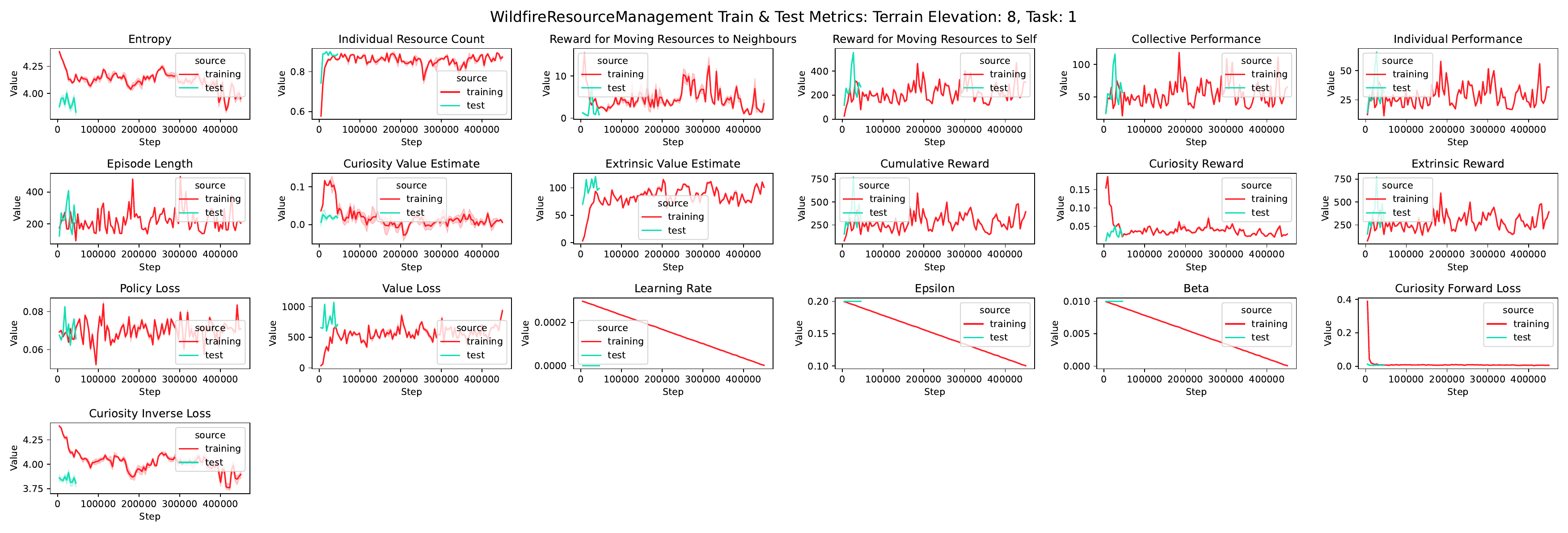}
\vspace{-0.6cm}
\caption{Wildfire Resource Management: Train \& Test Metrics: Terrain Elevation 8, Task 1.}
\end{figure}

\begin{figure}[h!]
\centering
\includegraphics[width=\linewidth]{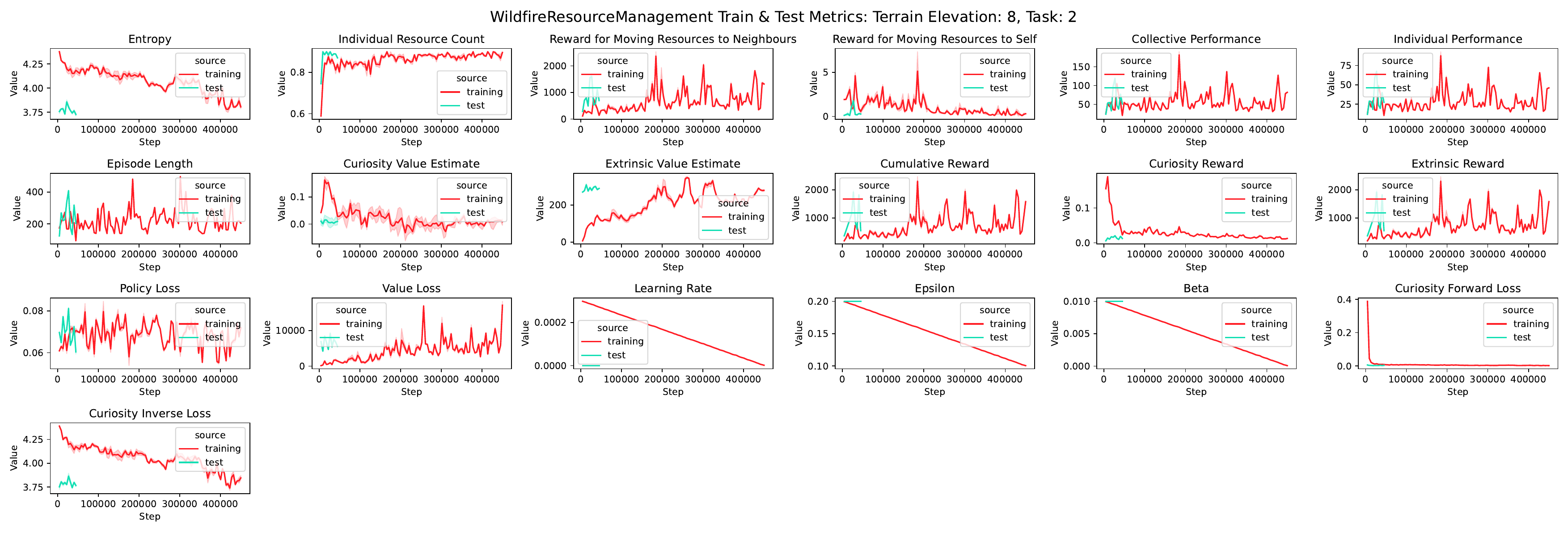}
\vspace{-0.6cm}
\caption{Wildfire Resource Management: Train \& Test Metrics: Terrain Elevation 8, Task 2.}
\end{figure}

\clearpage

\begin{figure}[h!]
\centering
\includegraphics[width=\linewidth]{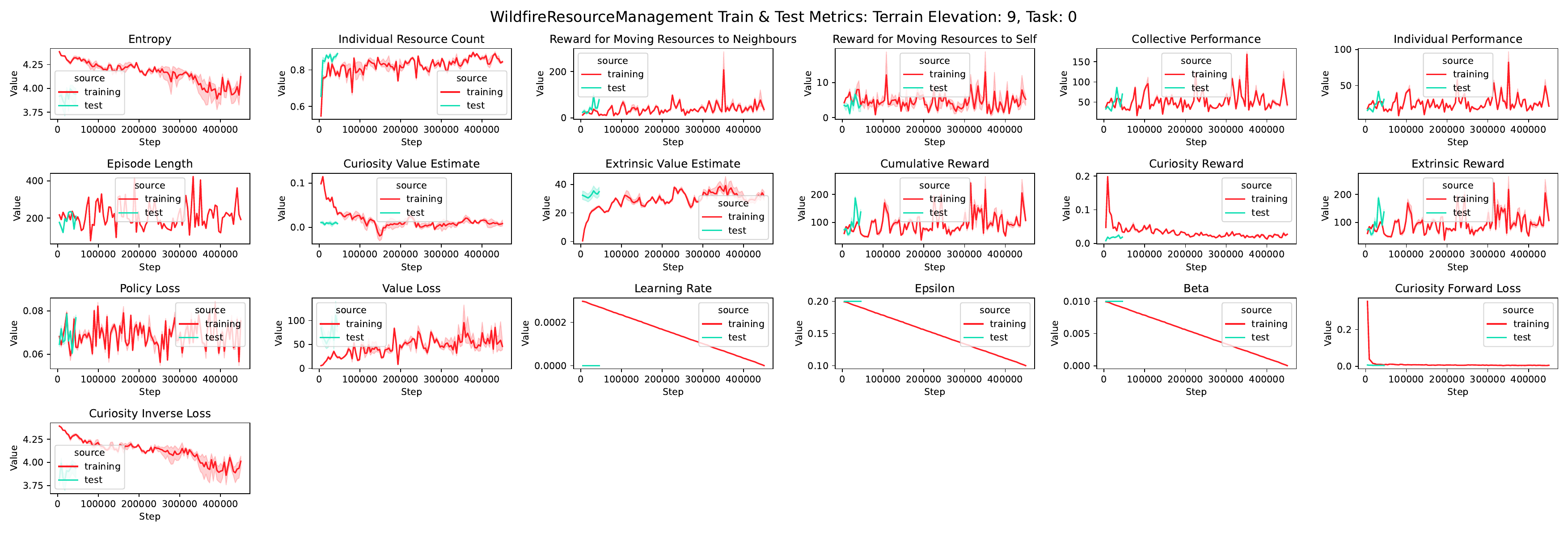}
\vspace{-0.6cm}
\caption{Wildfire Resource Management: Train \& Test Metrics: Terrain Elevation 9, Task 0.}
\end{figure}

\begin{figure}[h!]
\centering
\includegraphics[width=\linewidth]{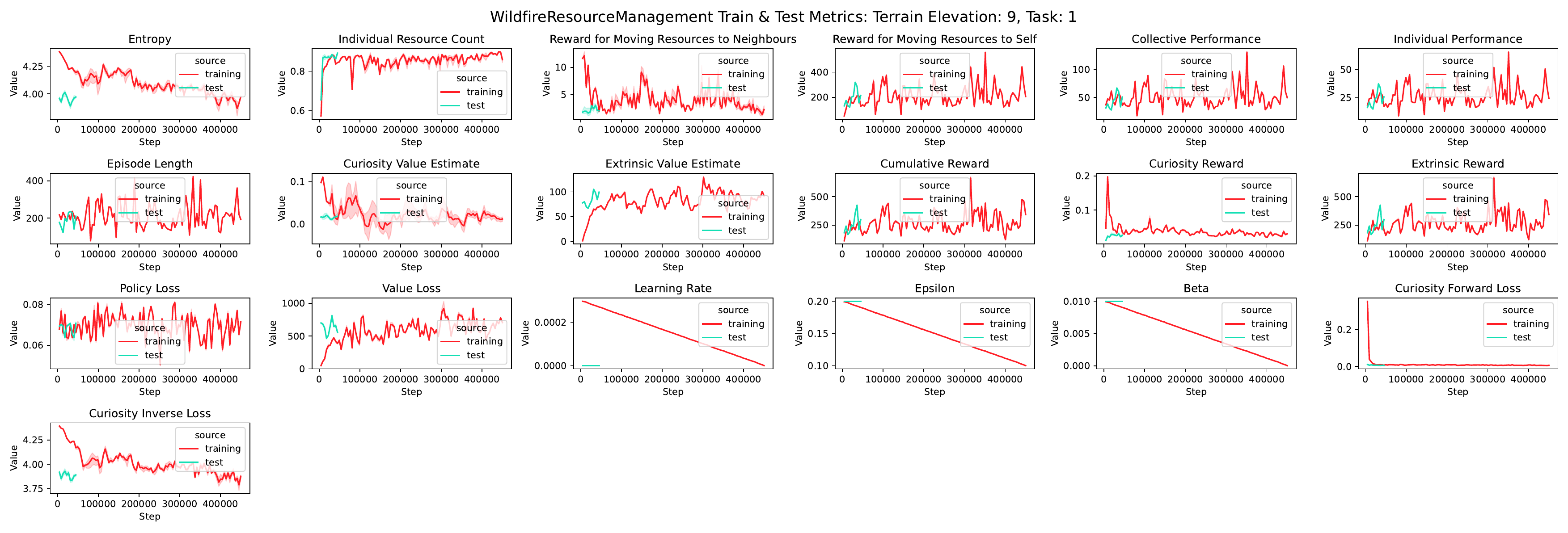}
\vspace{-0.6cm}
\caption{Wildfire Resource Management: Train \& Test Metrics: Terrain Elevation 9, Task 1.}
\end{figure}

\begin{figure}[h!]
\centering
\includegraphics[width=\linewidth]{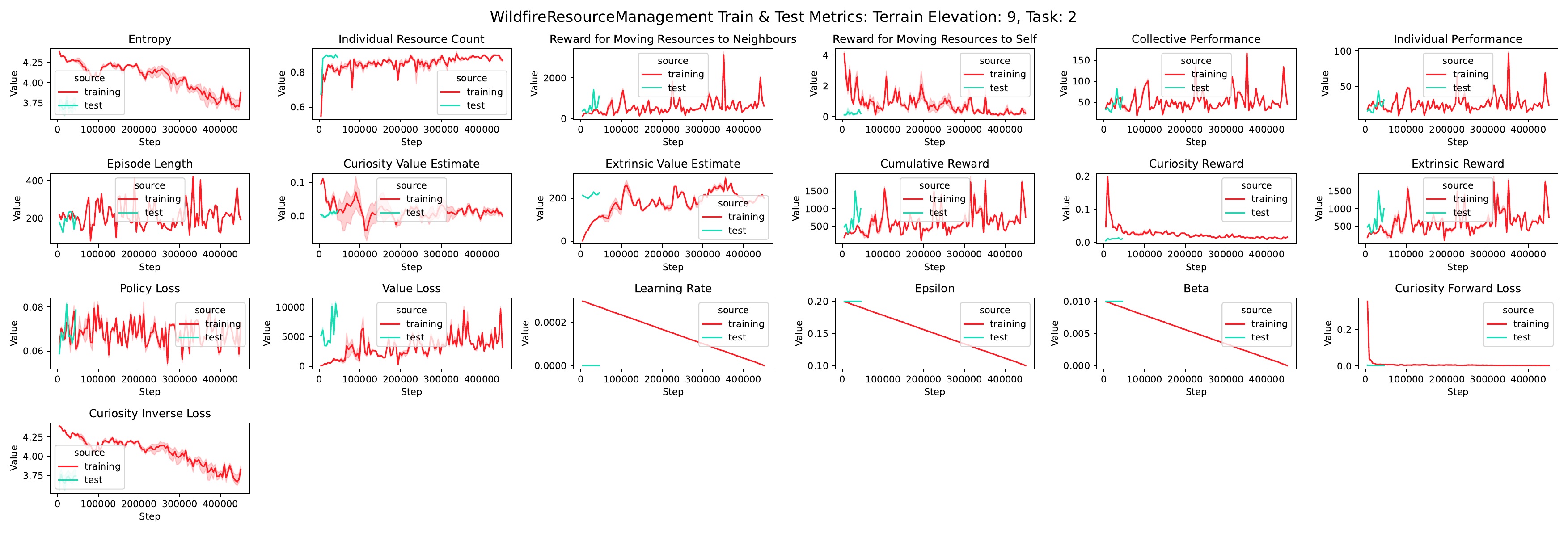}
\vspace{-0.6cm}
\caption{Wildfire Resource Management: Train \& Test Metrics: Terrain Elevation 9, Task 2.}
\end{figure}

\clearpage

\begin{figure}[h!]
\centering
\includegraphics[width=\linewidth]{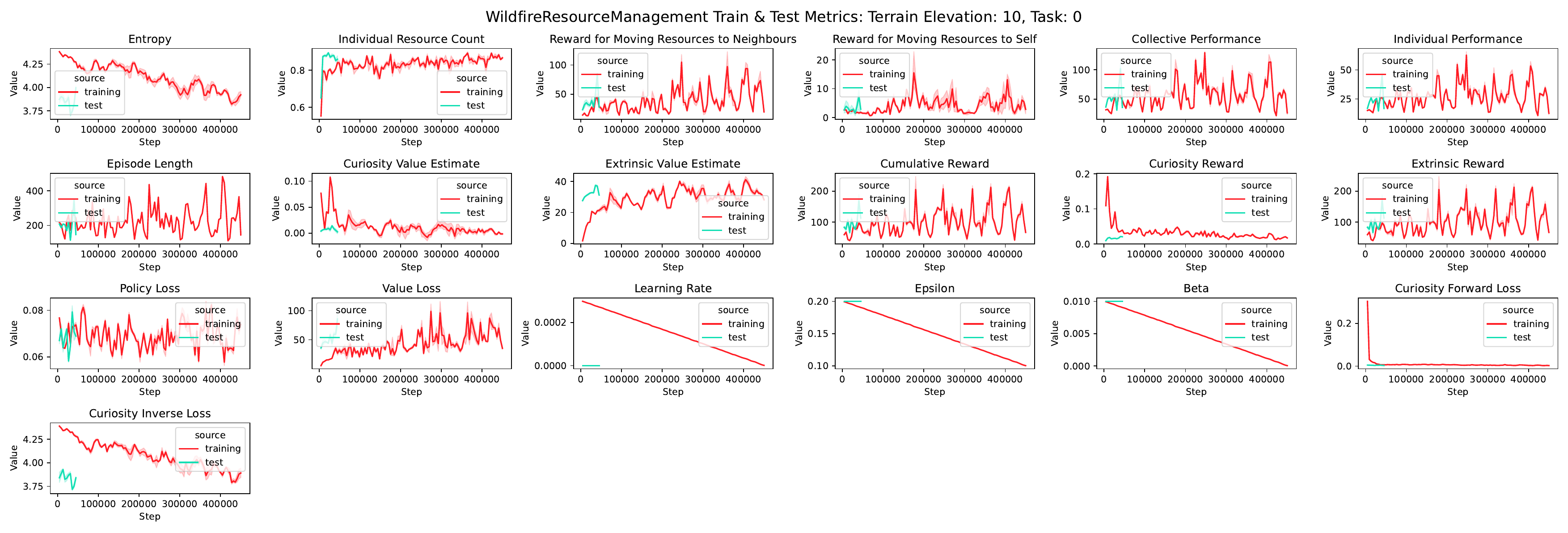}
\vspace{-0.6cm}
\caption{Wildfire Resource Management: Train \& Test Metrics: Terrain Elevation 10, Task 0.}
\end{figure}

\begin{figure}[h!]
\centering
\includegraphics[width=\linewidth]{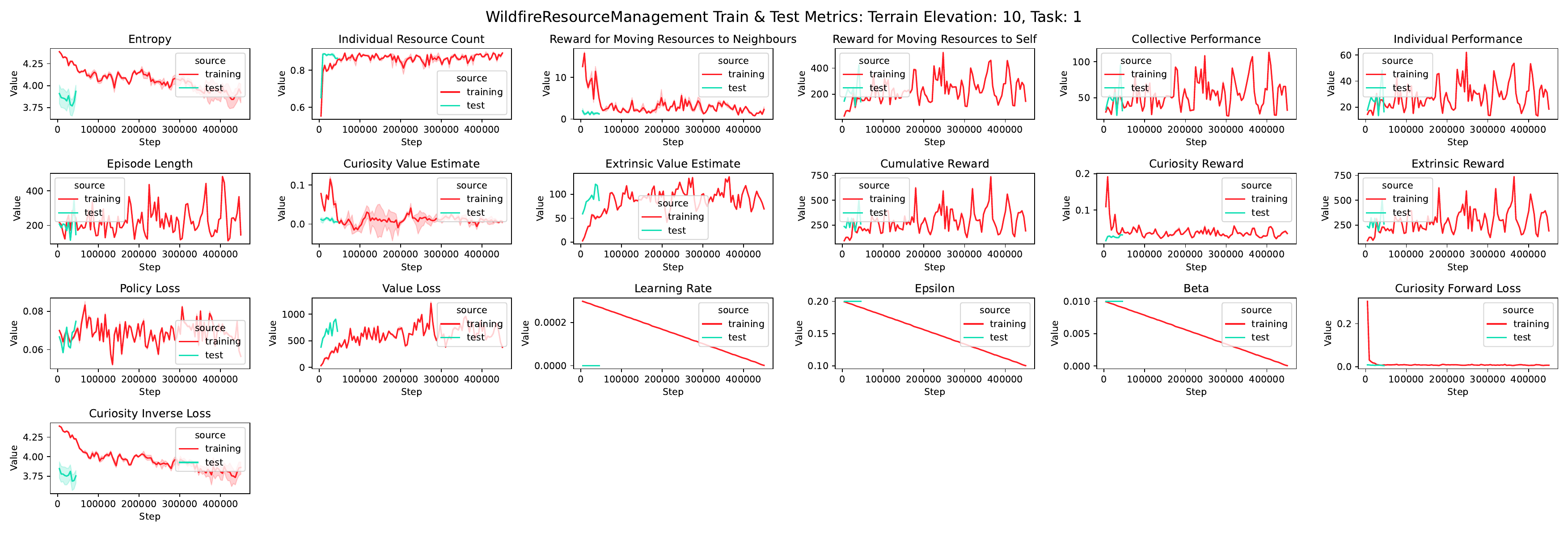}
\vspace{-0.6cm}
\caption{Wildfire Resource Management: Train \& Test Metrics: Terrain Elevation 10, Task 1.}
\end{figure}

\begin{figure}[h!]
\centering
\includegraphics[width=\linewidth]{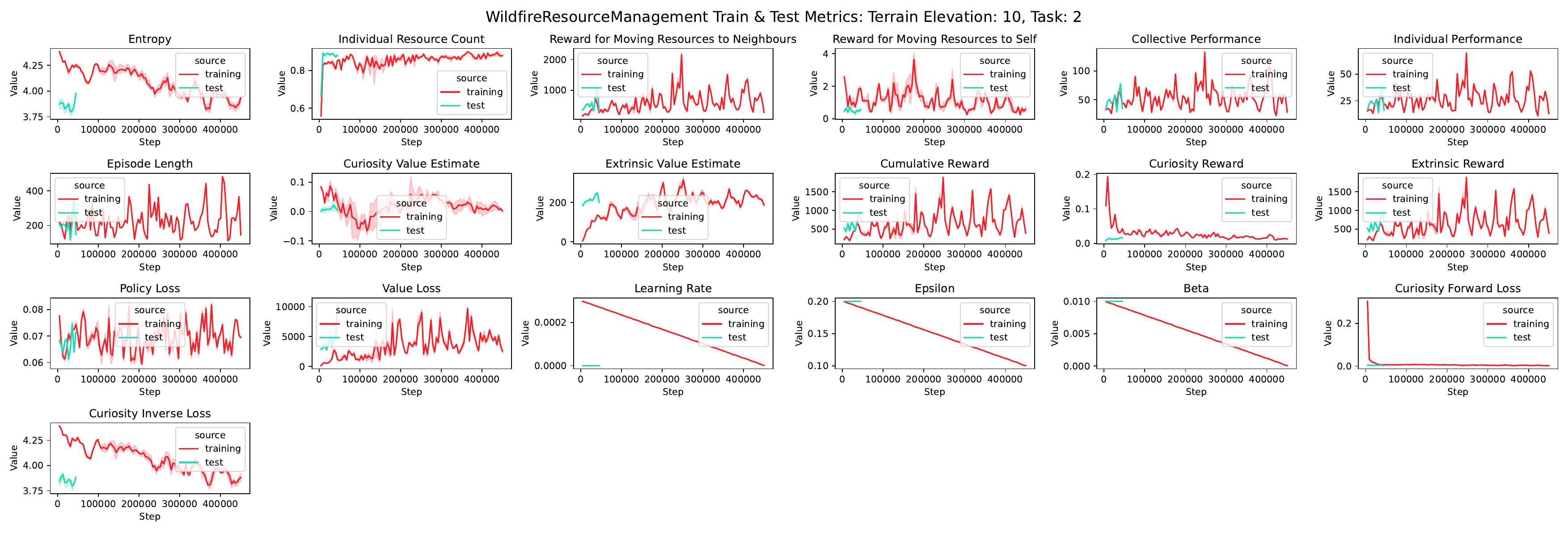}
\vspace{-0.6cm}
\caption{Wildfire Resource Management: Train \& Test Metrics: Terrain Elevation 10, Task 2.}
\end{figure}

\clearpage

\subsubsection{Wildfire Resource Management: Average Test Metric - Task VS Pattern}

\begin{figure}[h!]
\centering
\includegraphics[width=\linewidth]{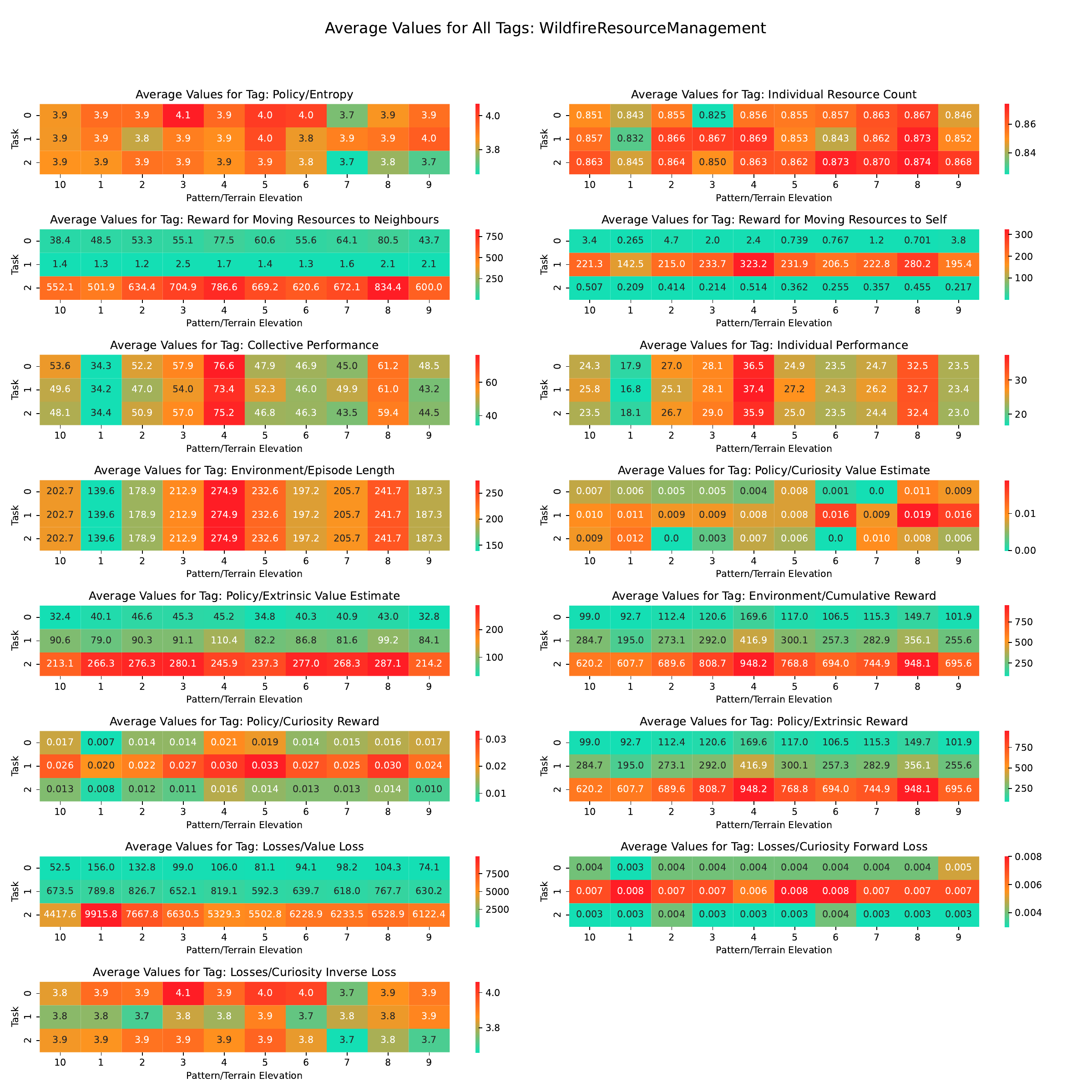}
\caption{Wildfire Resource Management: Average Train \& Test Metrics.}
\end{figure}

\clearpage

\subsubsection{Ocean Plastic Collection: Train \& Test Metrics}

\begin{figure}[h!]
\centering
\includegraphics[width=\linewidth]{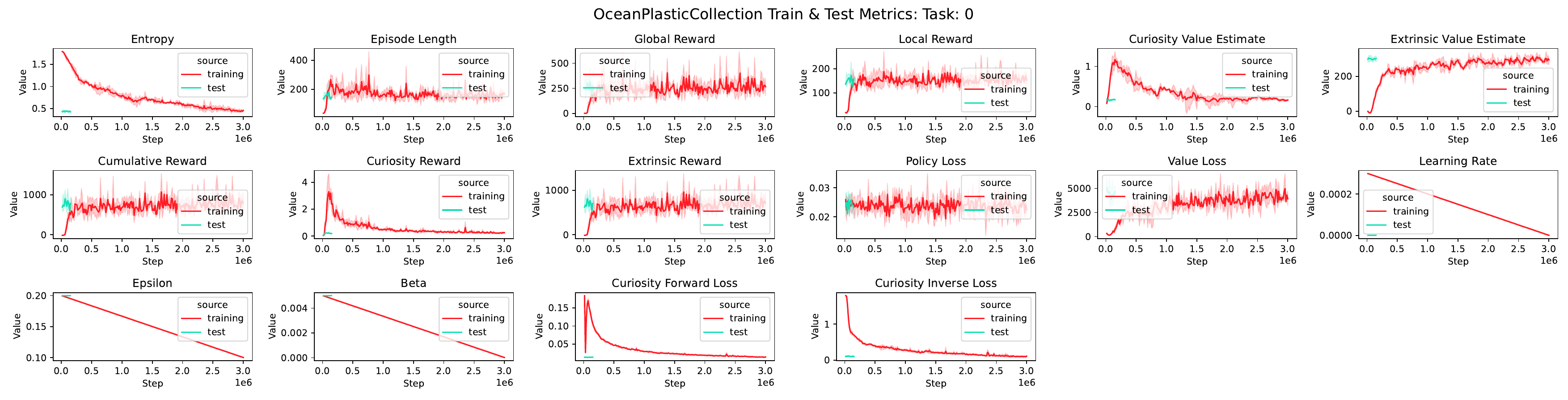}
\vspace{-0.6cm}
\caption{Ocean Plastic Collection: Train \& Test Metrics: Task 0.}
\end{figure}

\begin{figure}[h!]
\centering
\includegraphics[width=\linewidth]{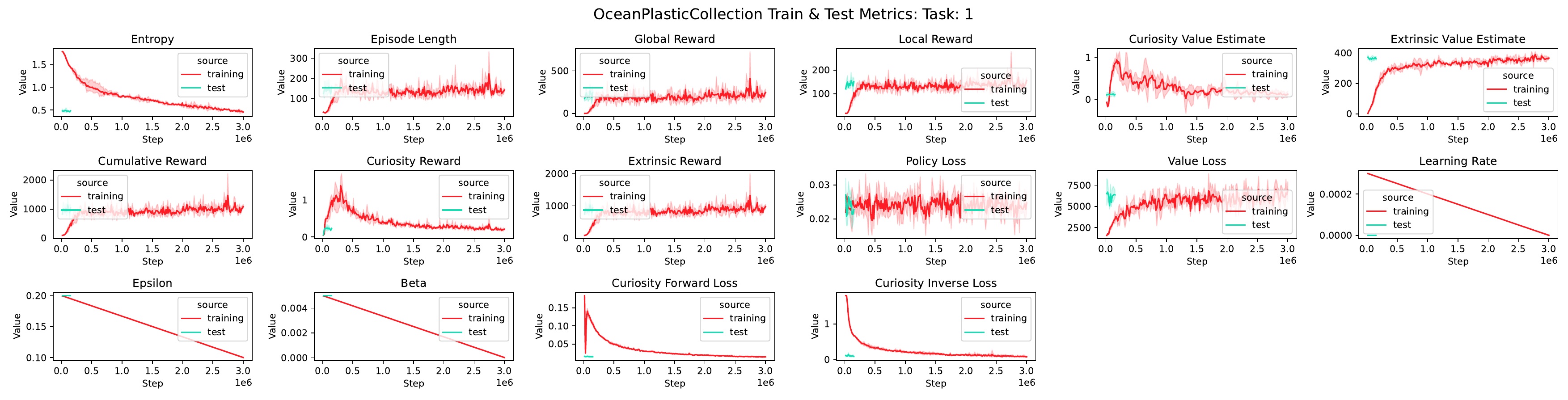}
\vspace{-0.6cm}
\caption{Ocean Plastic Collection: Train \& Test Metrics: Task 1.}
\end{figure}

\begin{figure}[h!]
\centering
\includegraphics[width=\linewidth]{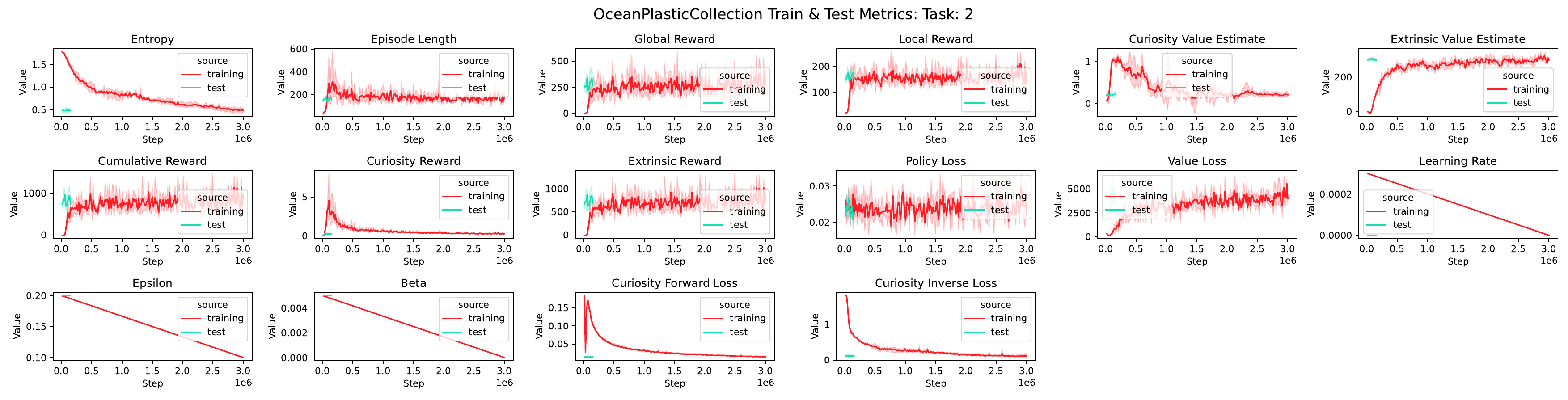}
\vspace{-0.6cm}
\caption{Ocean Plastic Collection: Train \& Test Metrics: Task 2.}
\end{figure}

\begin{figure}[h!]
\centering
\includegraphics[width=\linewidth]{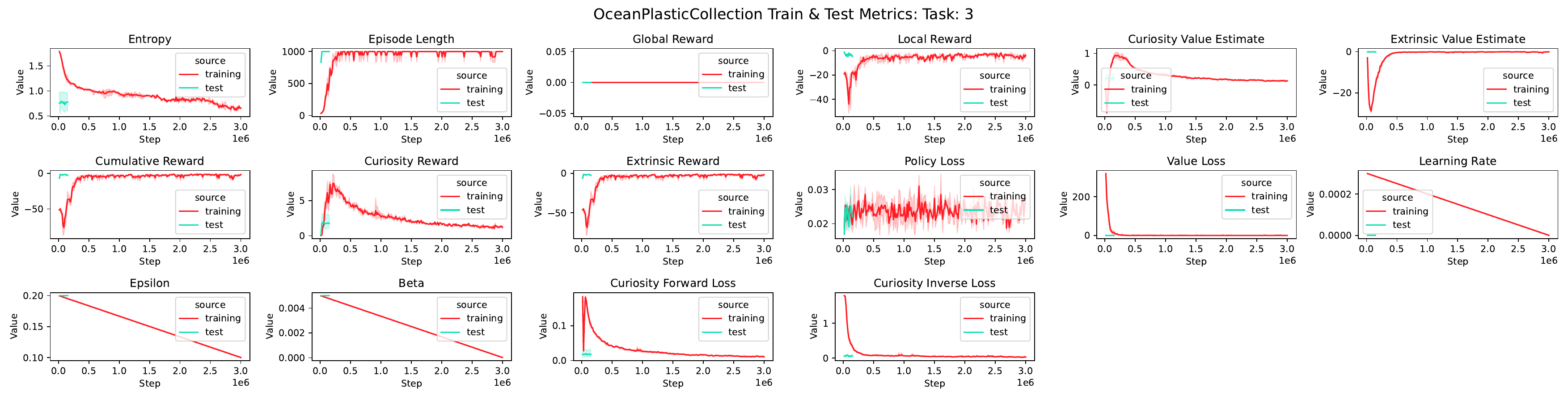}
\vspace{-0.6cm}
\caption{Ocean Plastic Collection: Train \& Test Metrics: Task 3.}
\end{figure}

\clearpage

\subsubsection{Ocean Plastic Collection: Average Test Metric - Task VS Pattern}
\begin{figure}[h!]
\centering
\includegraphics[width=\linewidth]{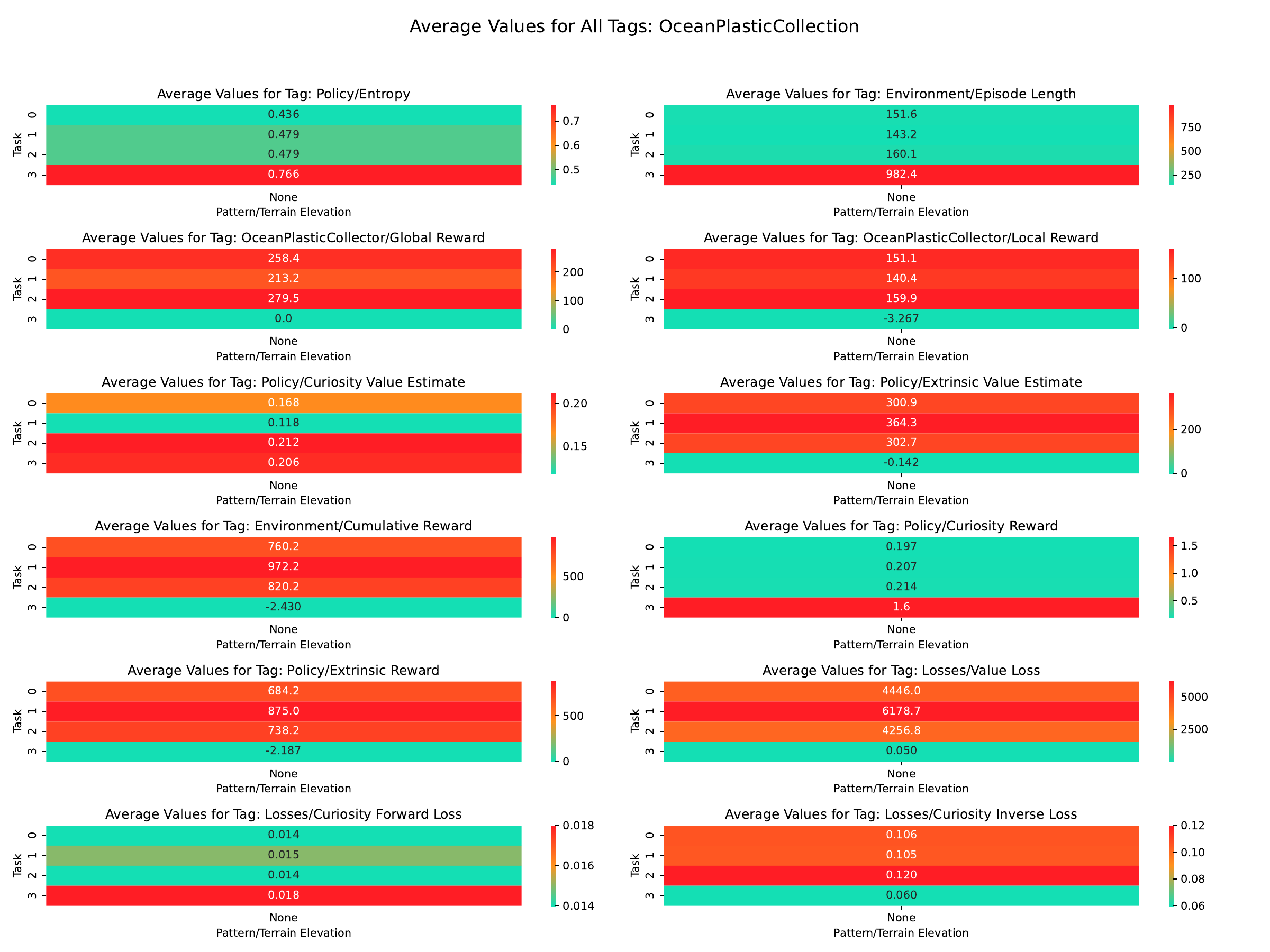}
\caption{Ocean Plastic Collection: Average Train \& Test Metrics.}
\end{figure}

\clearpage

\subsubsection{Drone-Based Reforestation: Train \& Test Metrics}
\begin{figure}[h!]
\centering
\includegraphics[width=\linewidth]{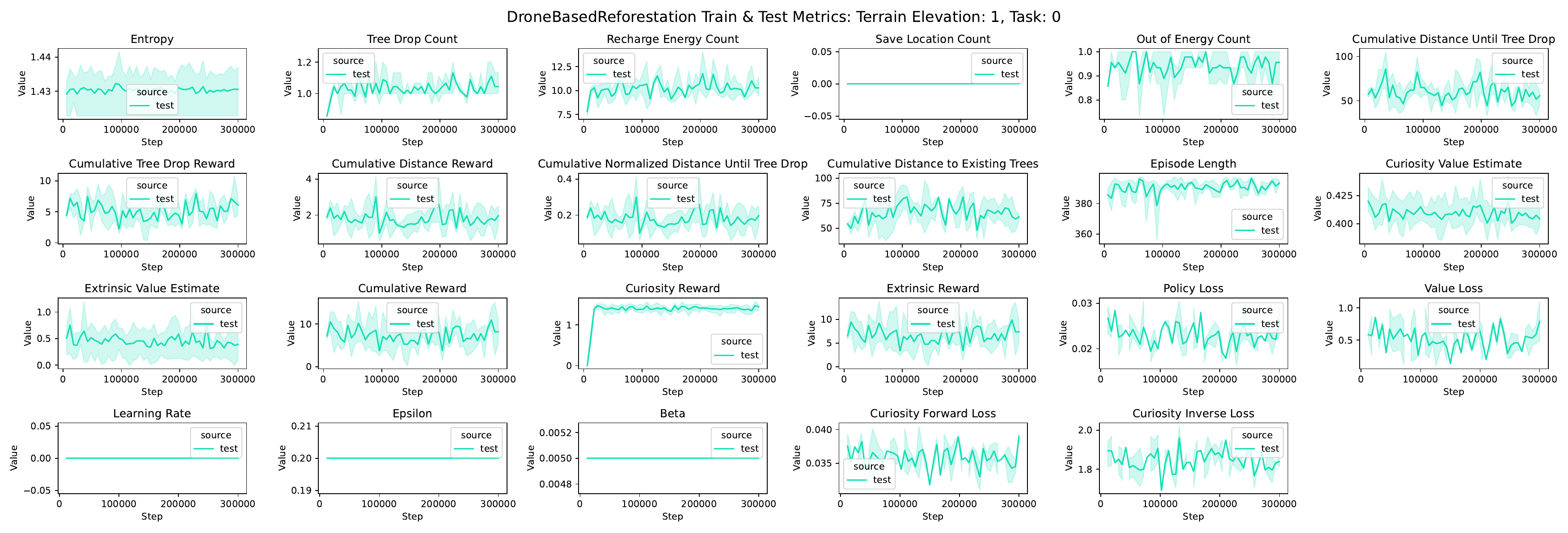}
\vspace{-0.6cm}
\caption{Drone-Based Reforestation: Train \& Test Metrics: Terrain Elevation 1, Task 0.}
\end{figure}

\begin{figure}[h!]
\centering
\includegraphics[width=\linewidth]{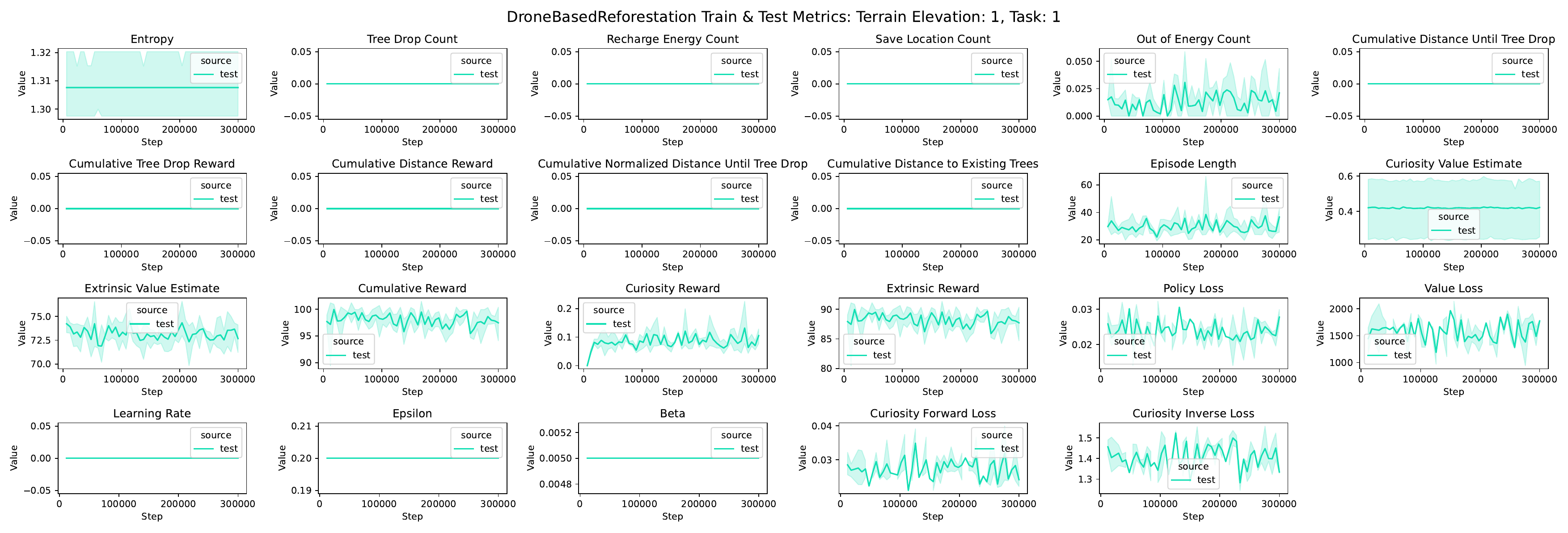}
\vspace{-0.6cm}
\caption{Drone-Based Reforestation: Train \& Test Metrics: Terrain Elevation 1, Task 1.}
\end{figure}

\begin{figure}[h!]
\centering
\includegraphics[width=\linewidth]{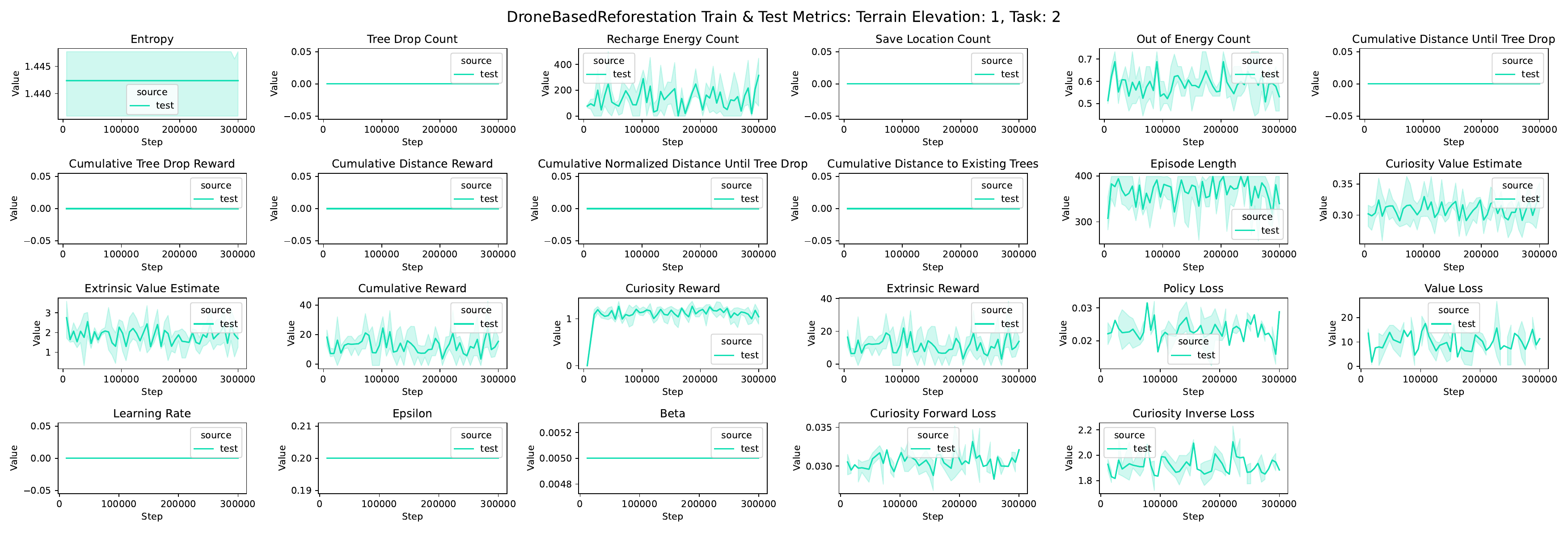}
\vspace{-0.6cm}
\caption{Drone-Based Reforestation: Train \& Test Metrics: Terrain Elevation 1, Task 2.}
\end{figure}

\clearpage

\begin{figure}[h!]
\centering
\includegraphics[width=\linewidth]{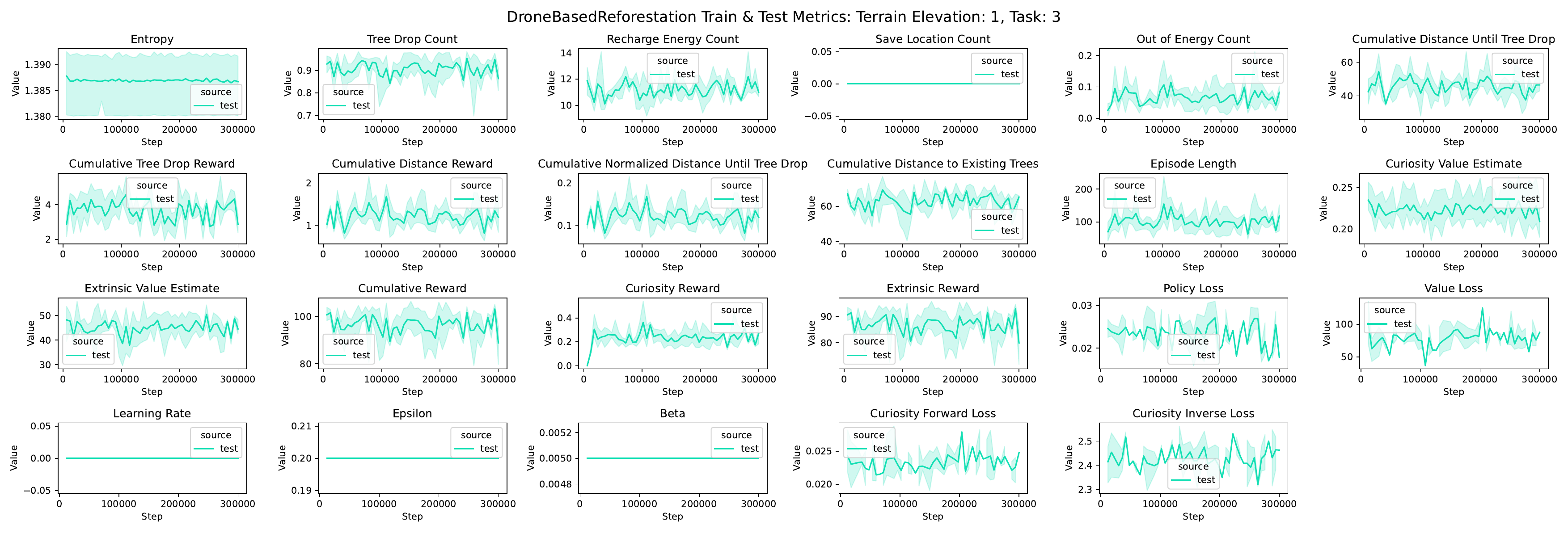}
\vspace{-0.6cm}
\caption{Drone-Based Reforestation: Train \& Test Metrics: Terrain Elevation 1, Task 3.}
\end{figure}

\begin{figure}[h!]
\centering
\includegraphics[width=\linewidth]{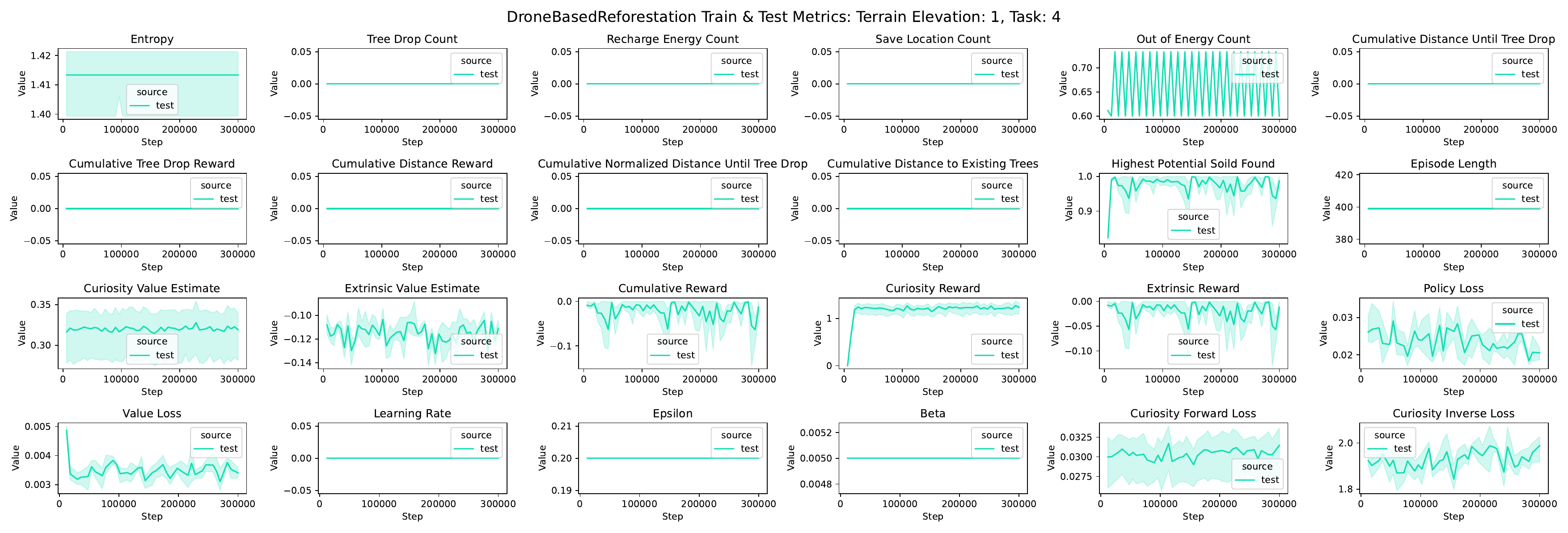}
\vspace{-0.6cm}
\caption{Drone-Based Reforestation: Train \& Test Metrics: Terrain Elevation 1, Task 4.}
\end{figure}

\begin{figure}[h!]
\centering
\includegraphics[width=\linewidth]{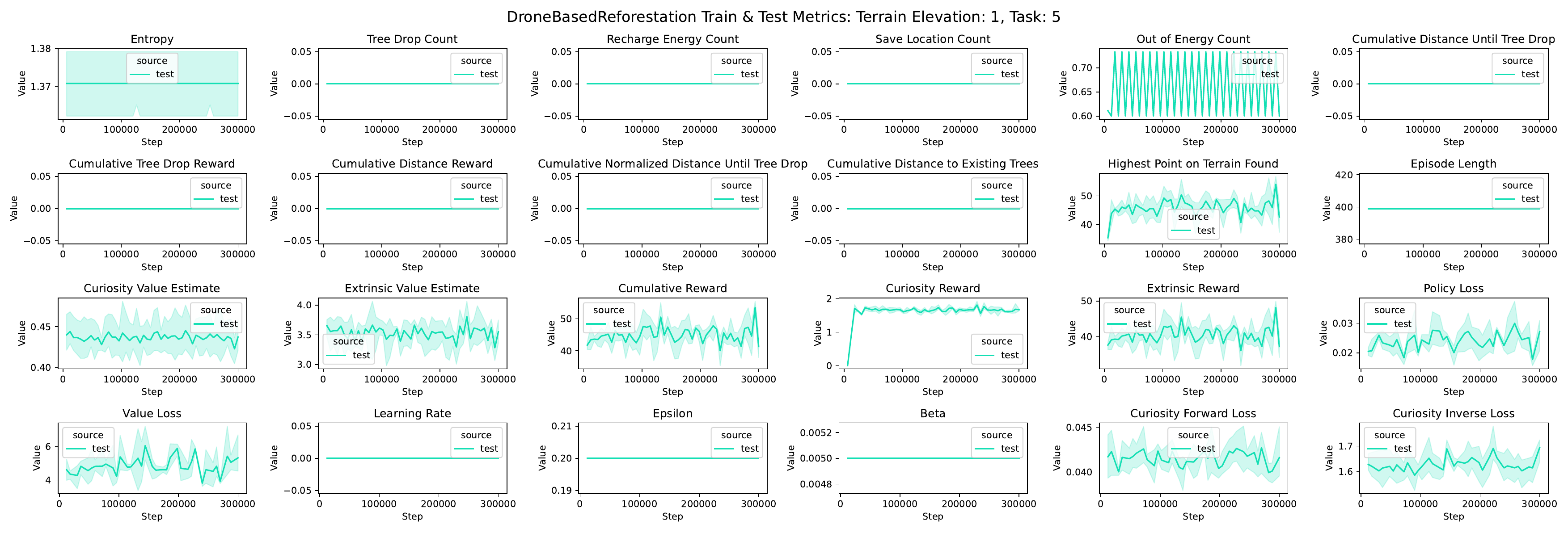}
\vspace{-0.6cm}
\caption{Drone-Based Reforestation: Train \& Test Metrics: Terrain Elevation 1, Task 5.}
\end{figure}

\clearpage

\begin{figure}[h!]
\centering
\includegraphics[width=\linewidth]{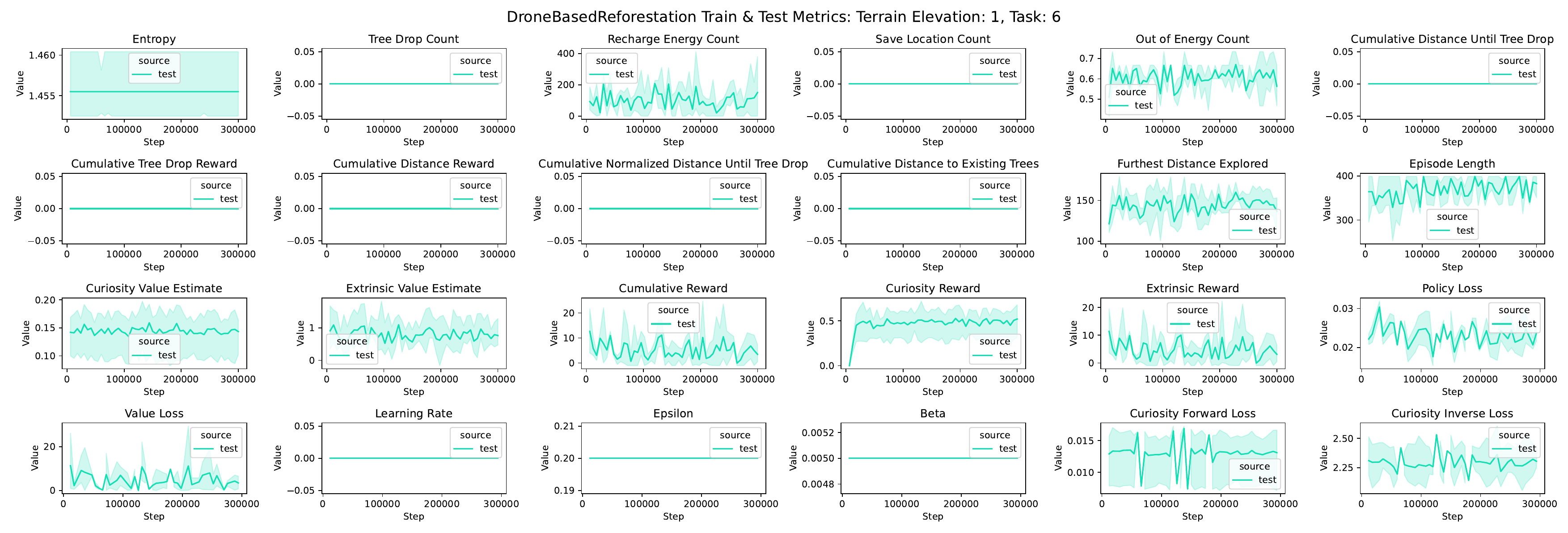}
\vspace{-0.6cm}
\caption{Drone-Based Reforestation: Train \& Test Metrics: Terrain Elevation 1, Task 6.}
\end{figure}

\begin{figure}[h!]
\centering
\includegraphics[width=\linewidth]{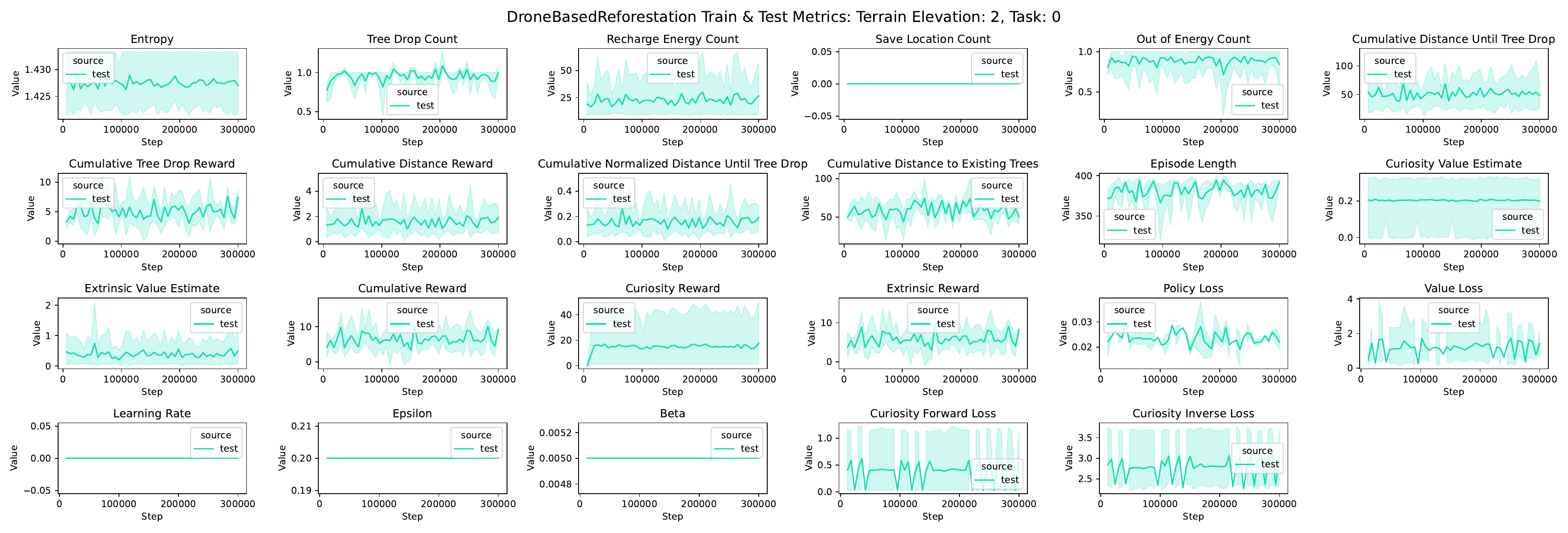}
\vspace{-0.6cm}
\caption{Drone-Based Reforestation: Train \& Test Metrics: Terrain Elevation 2, Task 0.}
\end{figure}

\begin{figure}[h!]
\centering
\includegraphics[width=\linewidth]{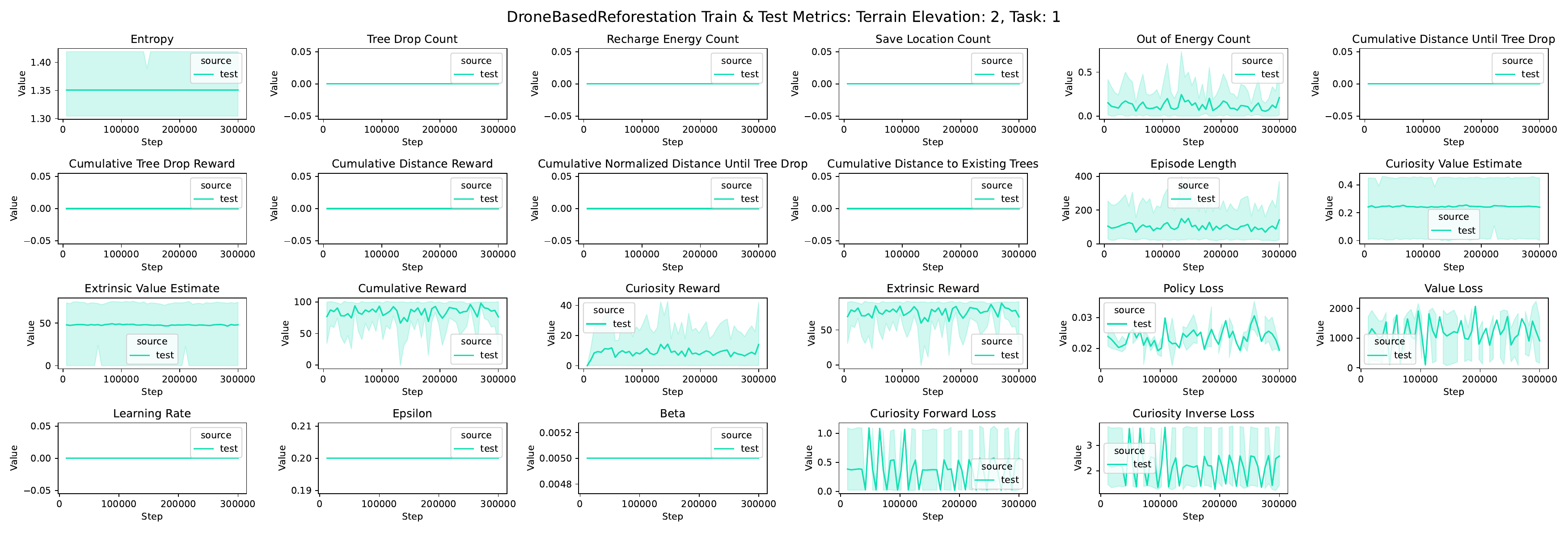}
\vspace{-0.6cm}
\caption{Drone-Based Reforestation: Train \& Test Metrics: Terrain Elevation 2, Task 1.}
\end{figure}

\clearpage

\begin{figure}[h!]
\centering
\includegraphics[width=\linewidth]{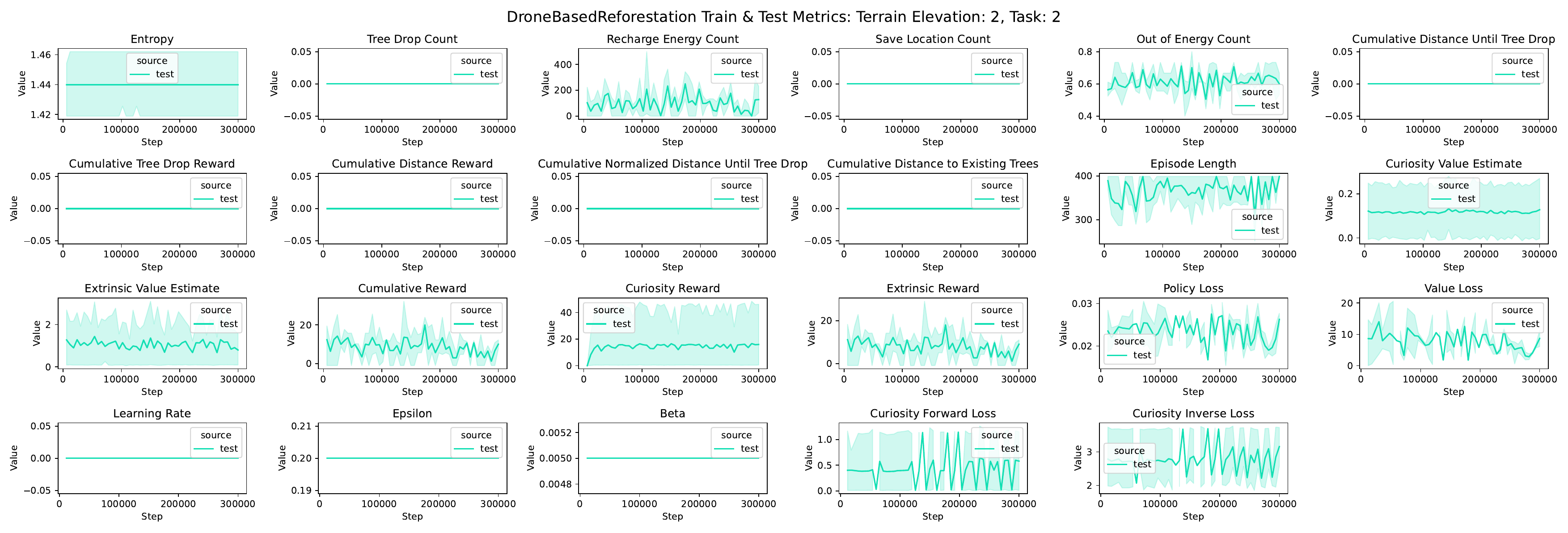}
\vspace{-0.6cm}
\caption{Drone-Based Reforestation: Train \& Test Metrics: Terrain Elevation 2, Task 2.}
\end{figure}

\begin{figure}[h!]
\centering
\includegraphics[width=\linewidth]{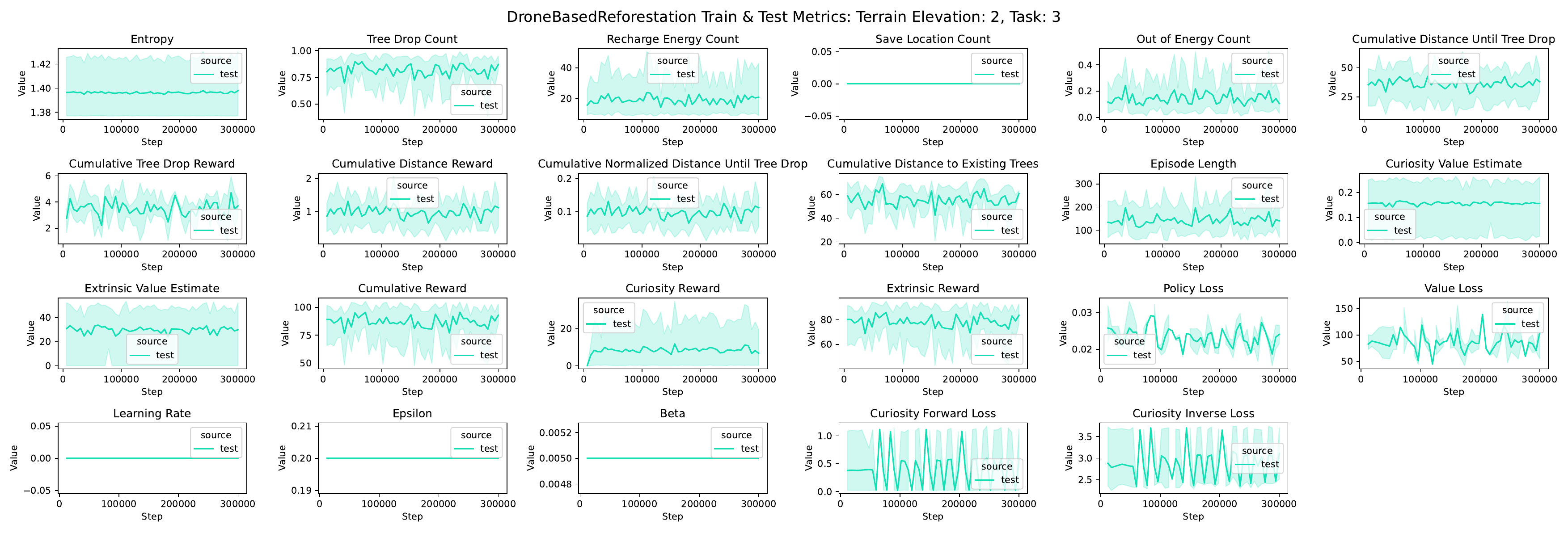}
\vspace{-0.6cm}
\caption{Drone-Based Reforestation: Train \& Test Metrics: Terrain Elevation 2, Task 3.}
\end{figure}

\begin{figure}[h!]
\centering
\includegraphics[width=\linewidth]{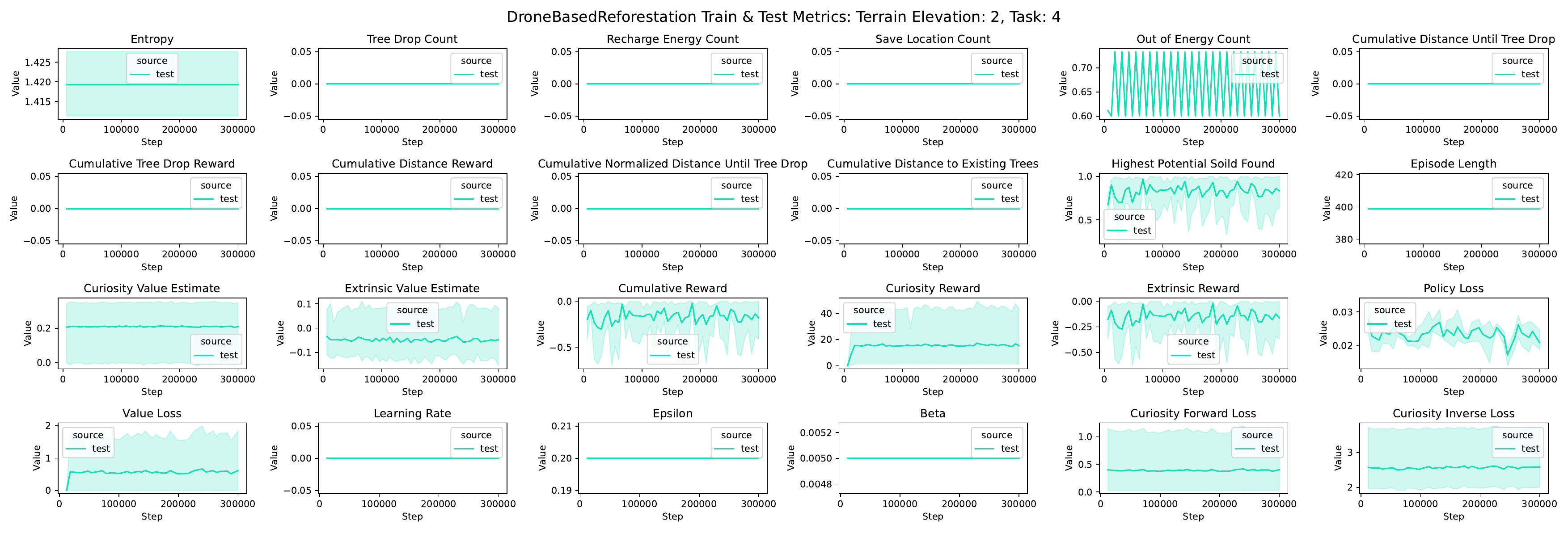}
\vspace{-0.6cm}
\caption{Drone-Based Reforestation: Train \& Test Metrics: Terrain Elevation 2, Task 4.}
\end{figure}

\clearpage

\begin{figure}[h!]
\centering
\includegraphics[width=\linewidth]{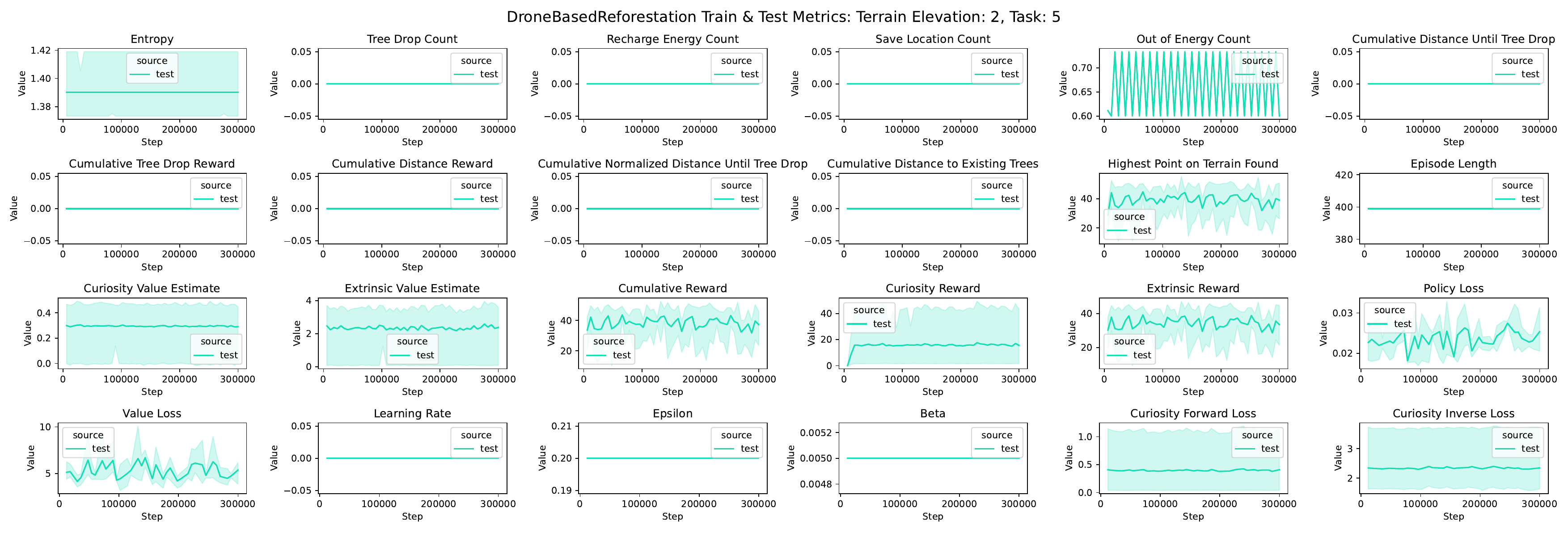}
\vspace{-0.6cm}
\caption{Drone-Based Reforestation: Train \& Test Metrics: Terrain Elevation 2, Task 5.}
\end{figure}

\begin{figure}[h!]
\centering
\includegraphics[width=\linewidth]{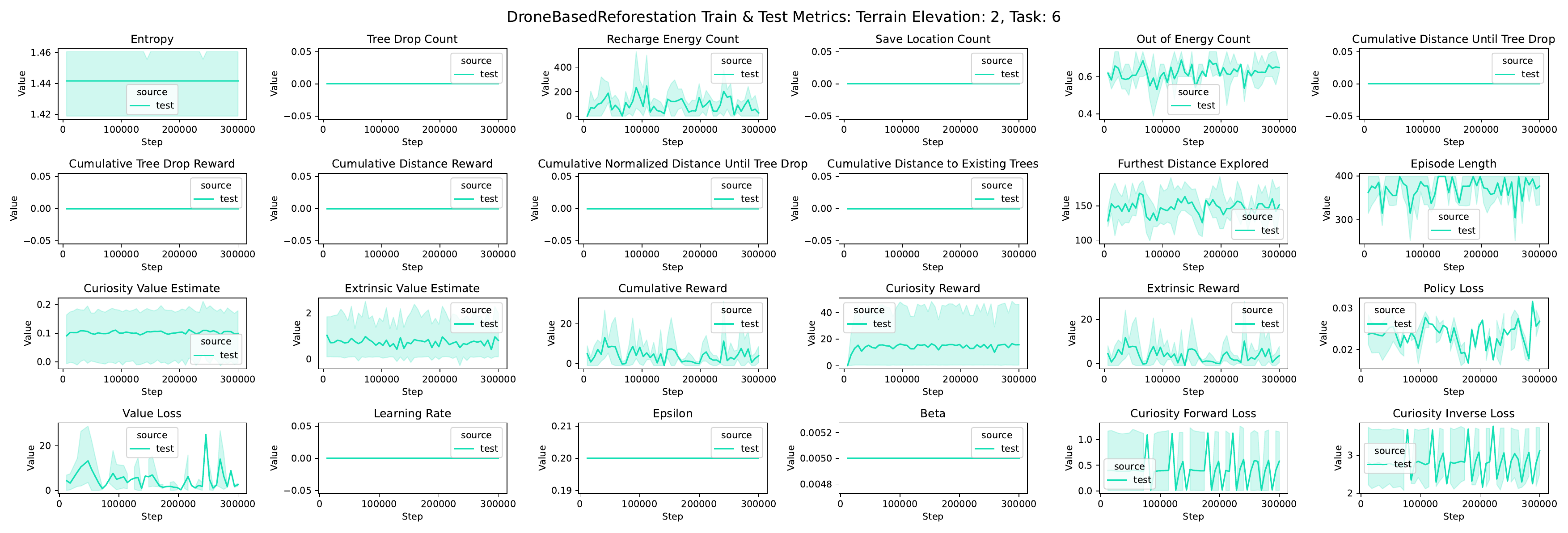}
\vspace{-0.6cm}
\caption{Drone-Based Reforestation: Train \& Test Metrics: Terrain Elevation 2, Task 6.}
\end{figure}

\begin{figure}[h!]
\centering
\includegraphics[width=\linewidth]{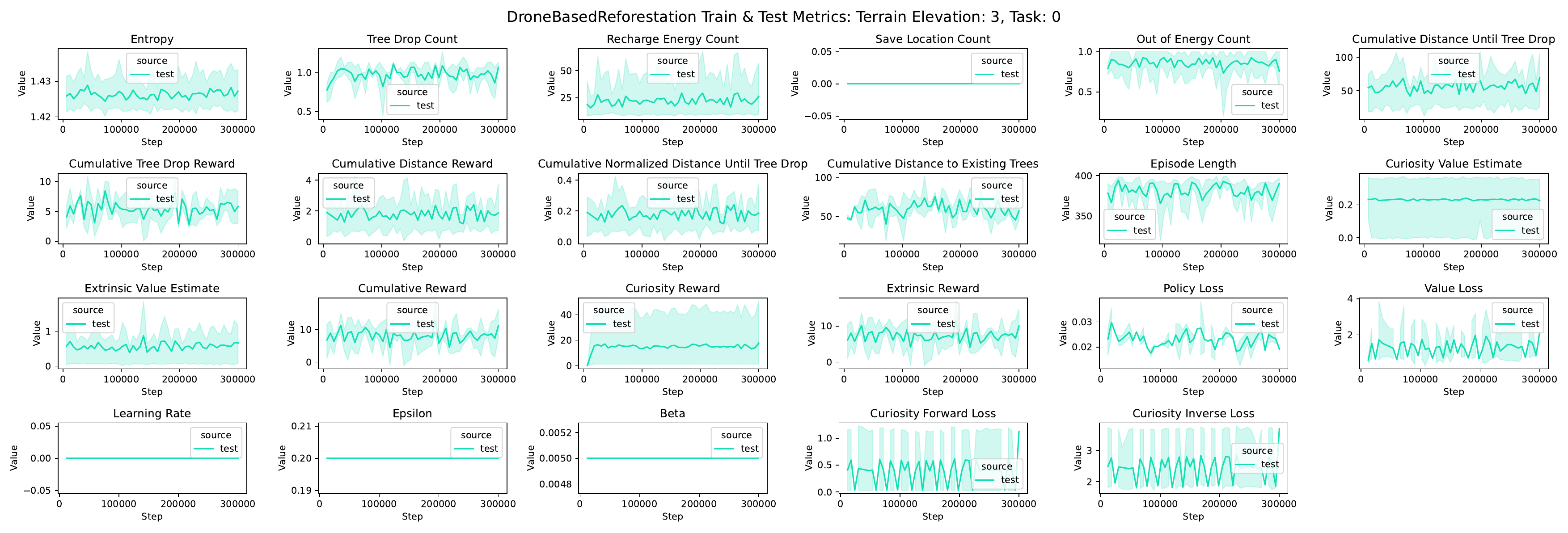}
\vspace{-0.6cm}
\caption{Drone-Based Reforestation: Train \& Test Metrics: Terrain Elevation 3, Task 0.}
\end{figure}

\clearpage

\begin{figure}[h!]
\centering
\includegraphics[width=\linewidth]{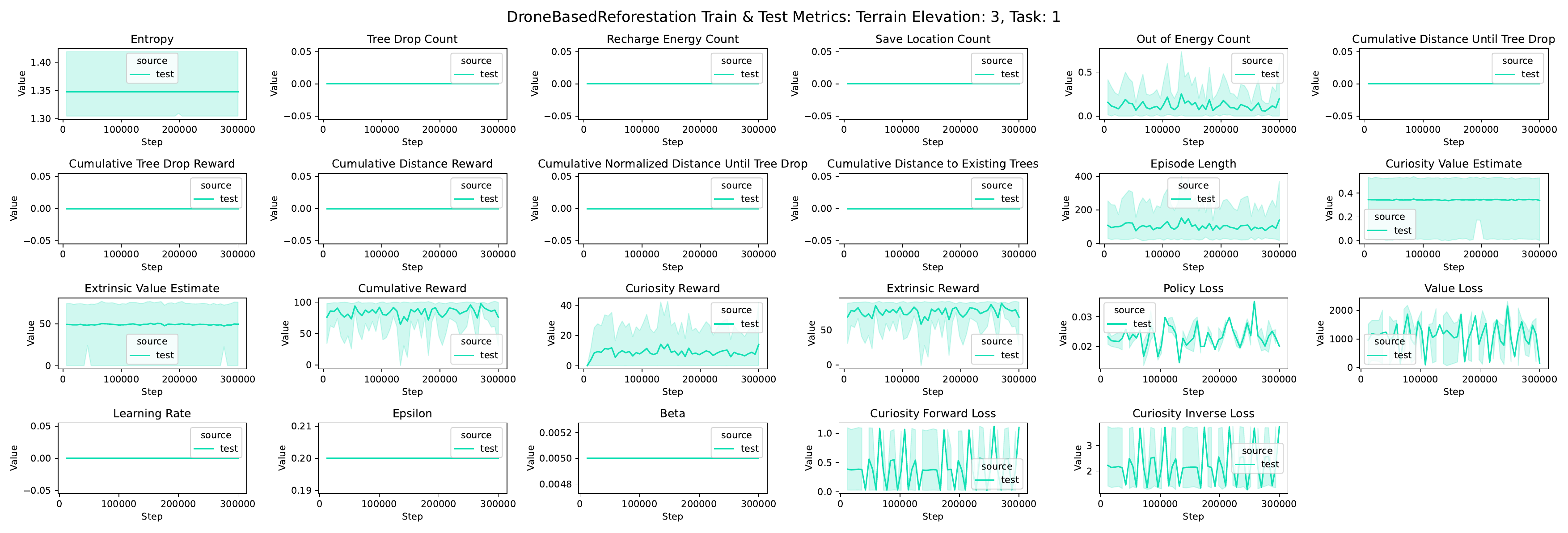}
\vspace{-0.6cm}
\caption{Drone-Based Reforestation: Train \& Test Metrics: Terrain Elevation 3, Task 1.}
\end{figure}

\begin{figure}[h!]
\centering
\includegraphics[width=\linewidth]{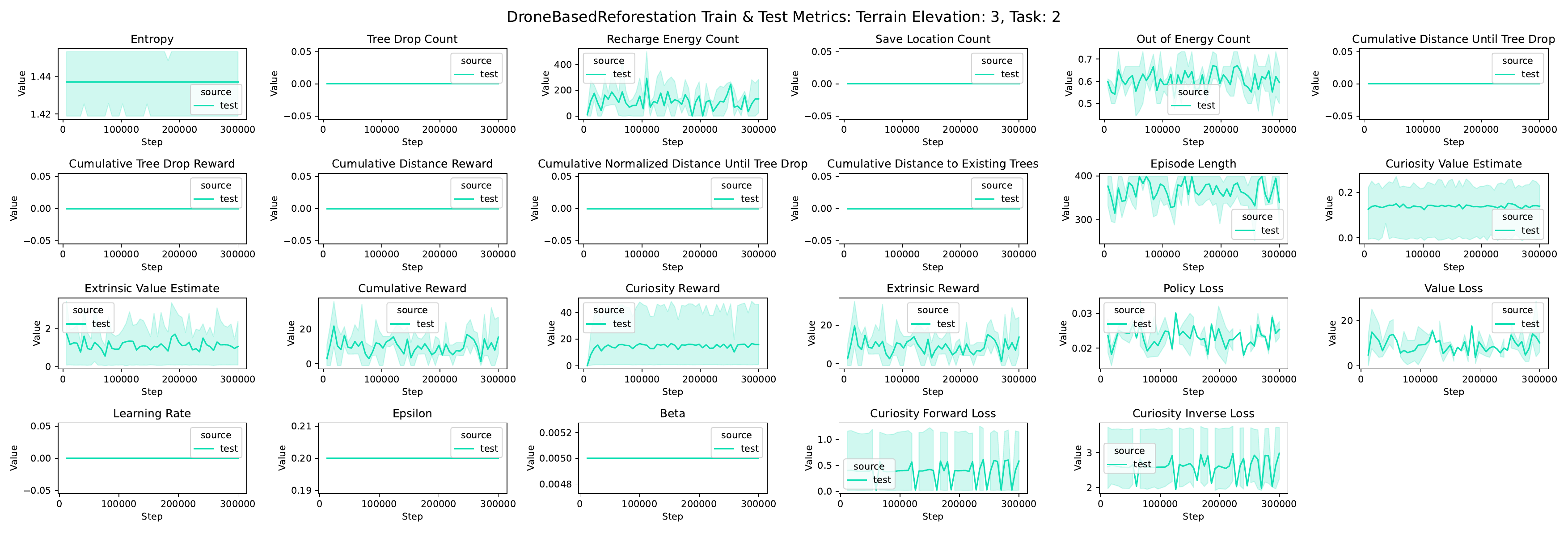}
\vspace{-0.6cm}
\caption{Drone-Based Reforestation: Train \& Test Metrics: Terrain Elevation 3, Task 2.}
\end{figure}

\begin{figure}[h!]
\centering
\includegraphics[width=\linewidth]{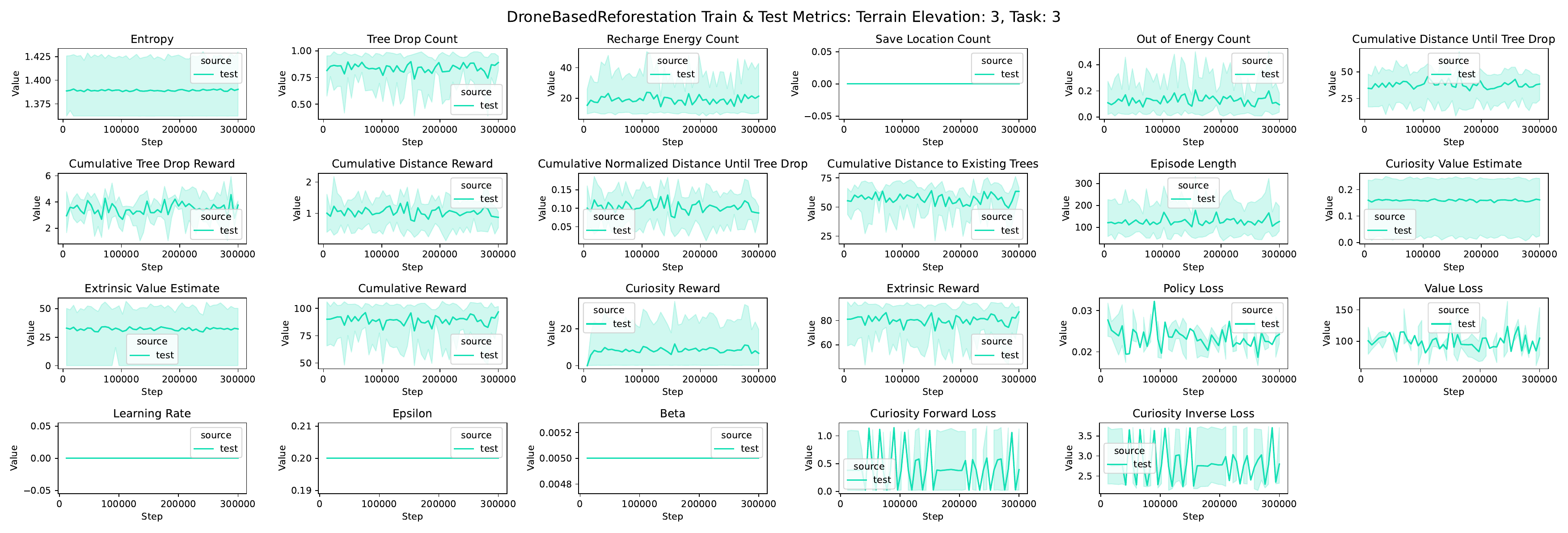}
\vspace{-0.6cm}
\caption{Drone-Based Reforestation: Train \& Test Metrics: Terrain Elevation 3, Task 3.}
\end{figure}

\clearpage

\begin{figure}[h!]
\centering
\includegraphics[width=\linewidth]{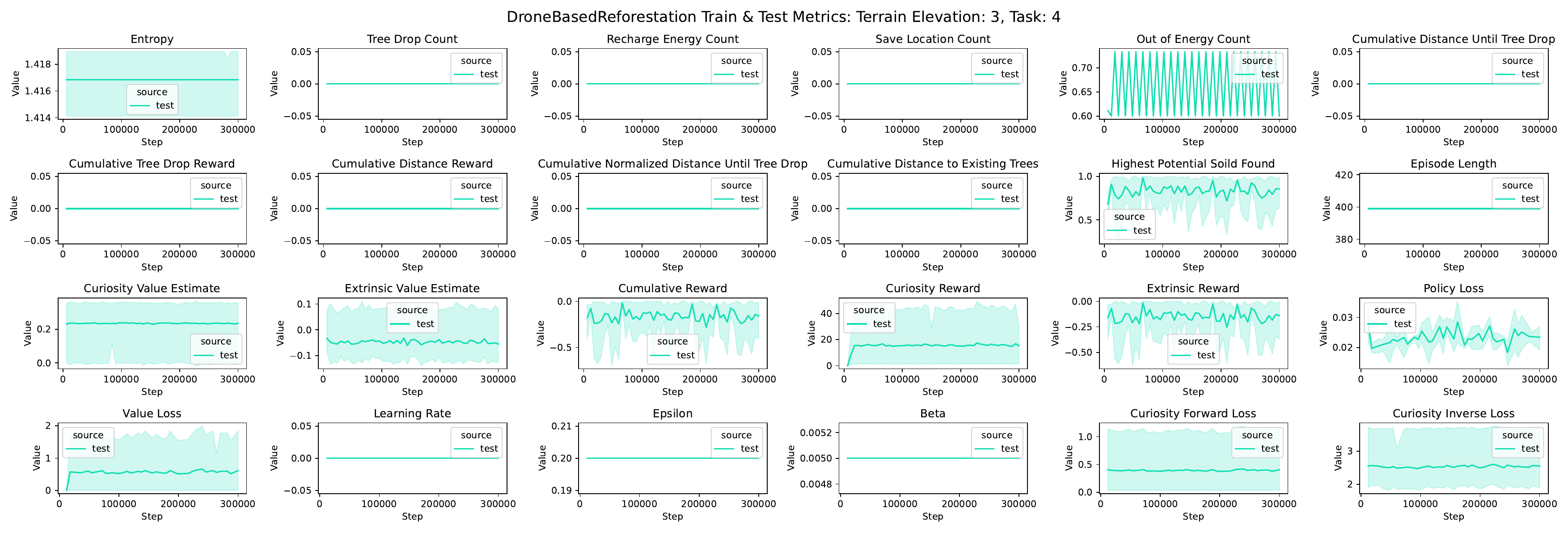}
\vspace{-0.6cm}
\caption{Drone-Based Reforestation: Train \& Test Metrics: Terrain Elevation 3, Task 4.}
\end{figure}

\begin{figure}[h!]
\centering
\includegraphics[width=\linewidth]{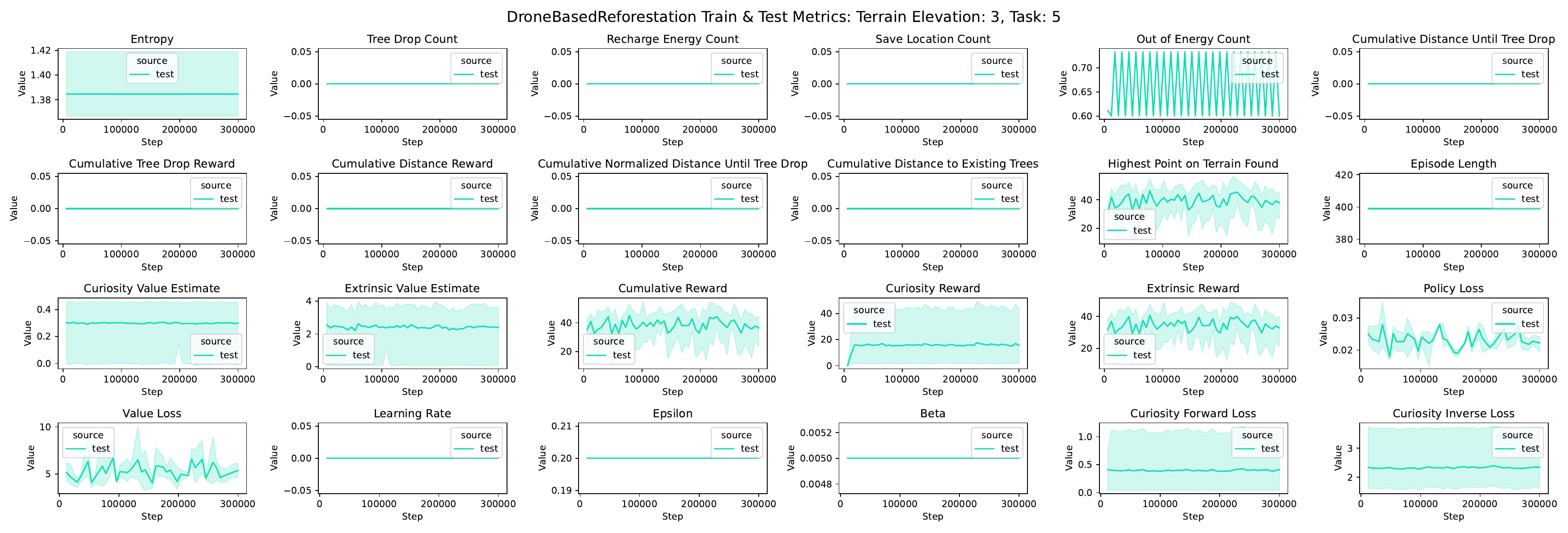}
\vspace{-0.6cm}
\caption{Drone-Based Reforestation: Train \& Test Metrics: Terrain Elevation 3, Task 5.}
\end{figure}

\begin{figure}[h!]
\centering
\includegraphics[width=\linewidth]{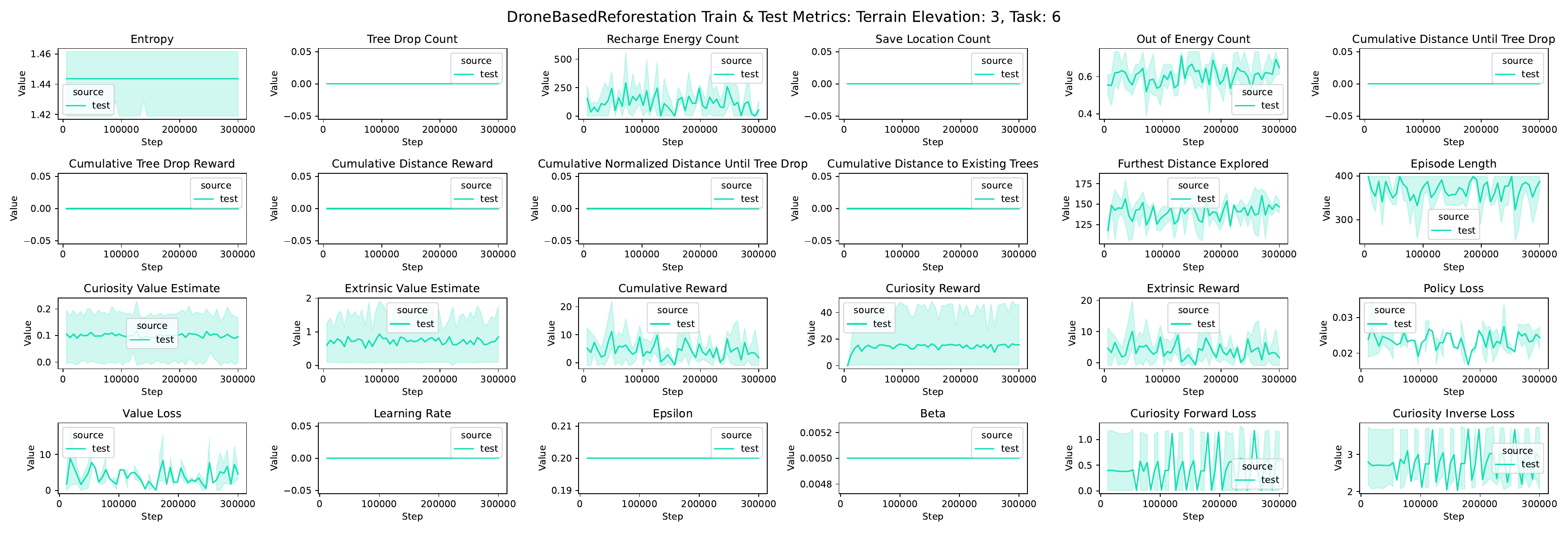}
\vspace{-0.6cm}
\caption{Drone-Based Reforestation: Train \& Test Metrics: Terrain Elevation 3, Task 6.}
\end{figure}

\clearpage

\begin{figure}[h!]
\centering
\includegraphics[width=\linewidth]{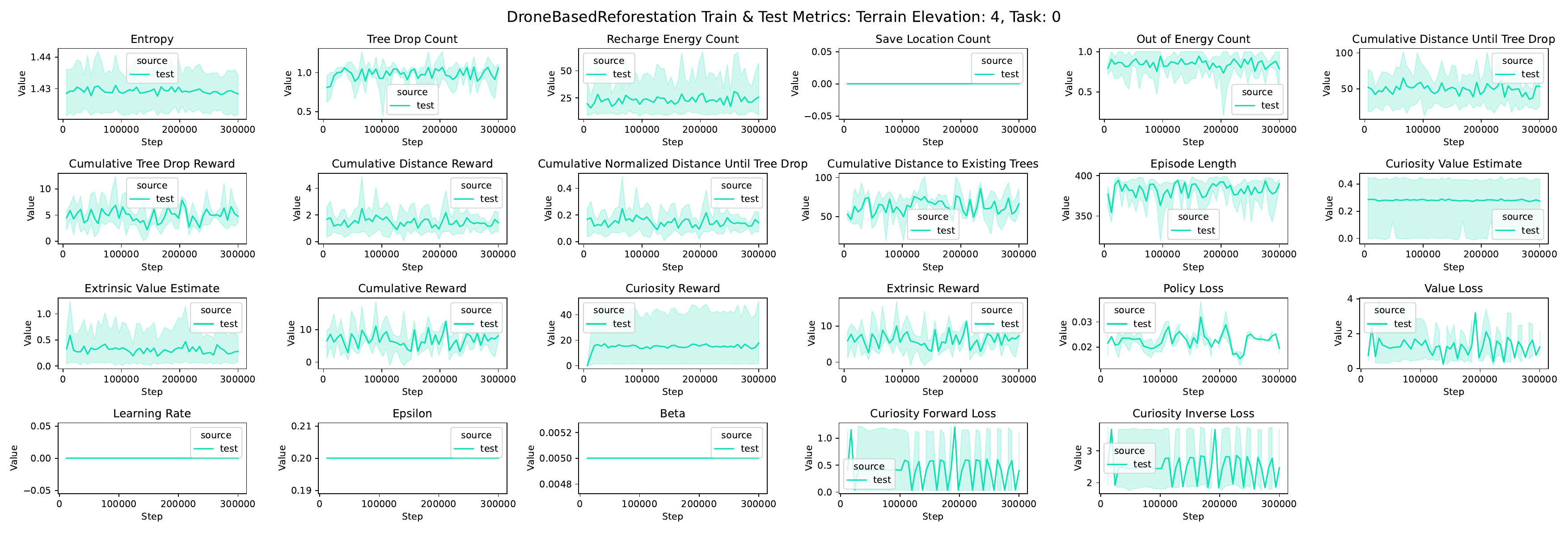}
\vspace{-0.6cm}
\caption{Drone-Based Reforestation: Train \& Test Metrics: Terrain Elevation 4, Task 0.}
\end{figure}

\begin{figure}[h!]
\centering
\includegraphics[width=\linewidth]{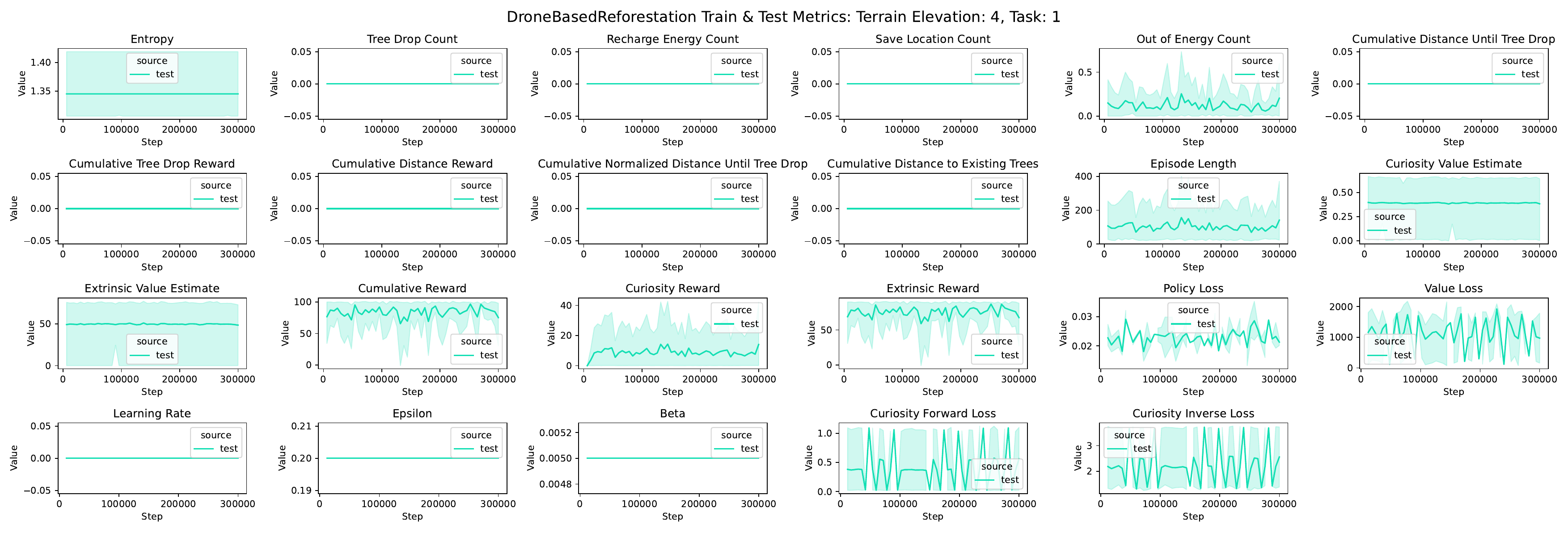}
\vspace{-0.6cm}
\caption{Drone-Based Reforestation: Train \& Test Metrics: Terrain Elevation 4, Task 1.}
\end{figure}

\begin{figure}[h!]
\centering
\includegraphics[width=\linewidth]{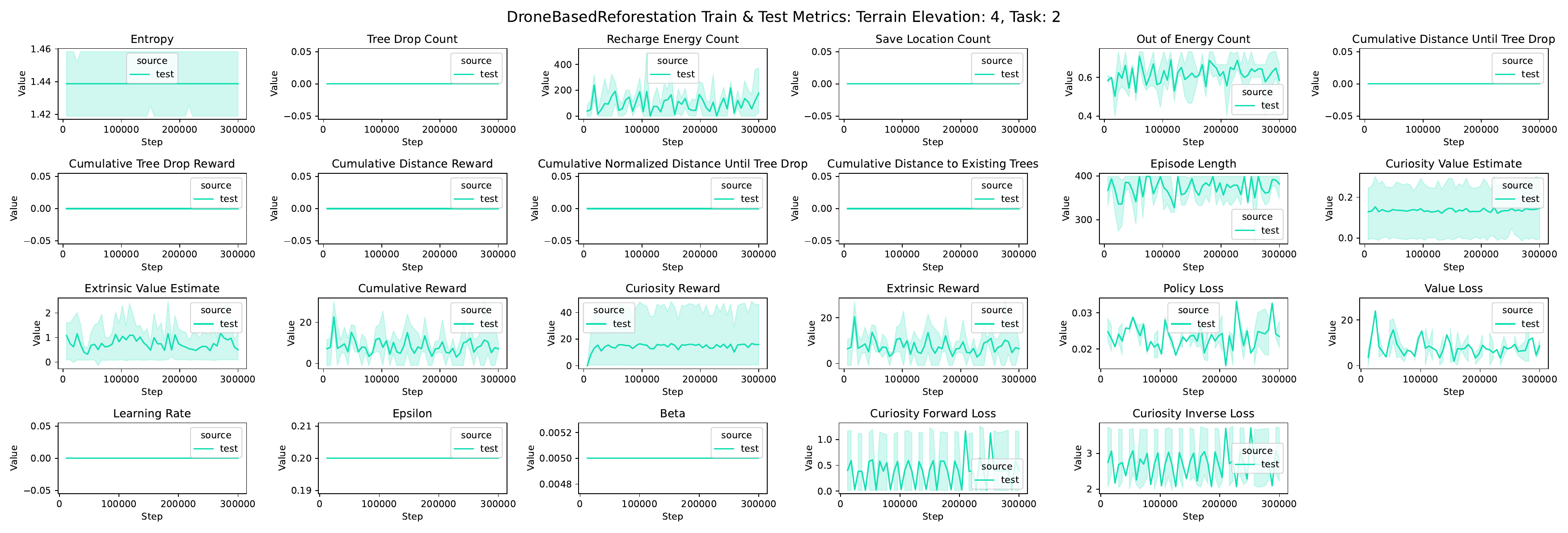}
\vspace{-0.6cm}
\caption{Drone-Based Reforestation: Train \& Test Metrics: Terrain Elevation 4, Task 2.}
\end{figure}

\clearpage

\begin{figure}[h!]
\centering
\includegraphics[width=\linewidth]{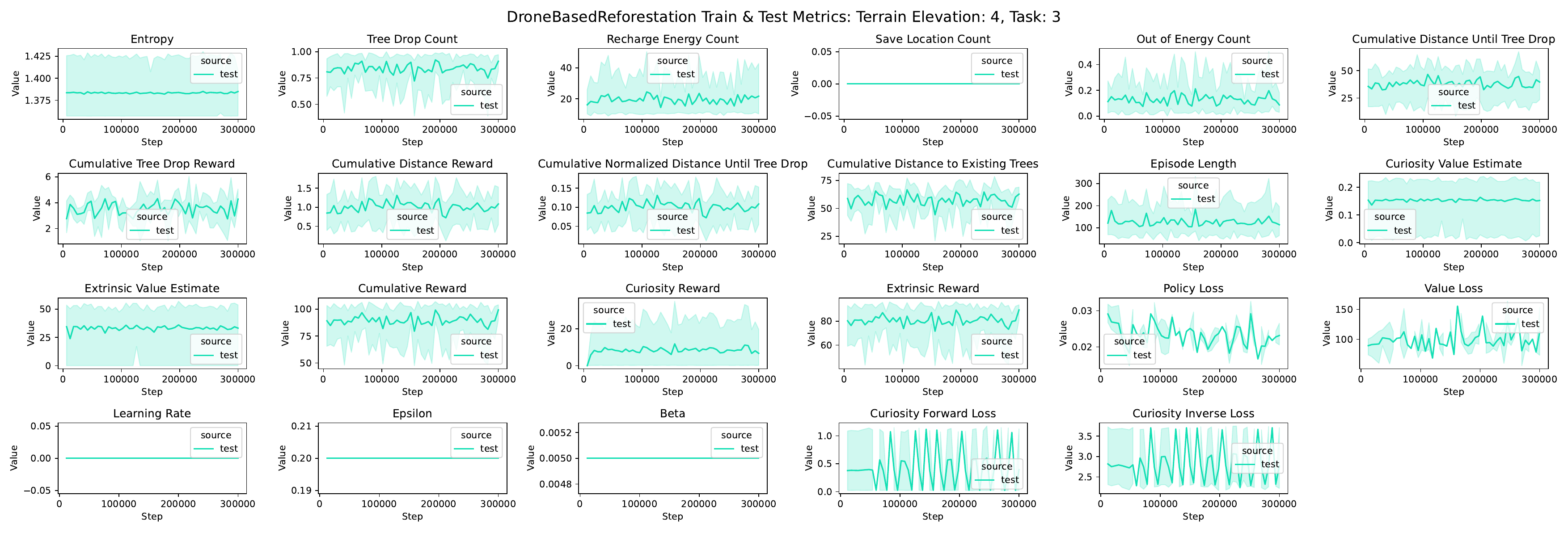}
\vspace{-0.6cm}
\caption{Drone-Based Reforestation: Train \& Test Metrics: Terrain Elevation 4, Task 3.}
\end{figure}

\begin{figure}[h!]
\centering
\includegraphics[width=\linewidth]{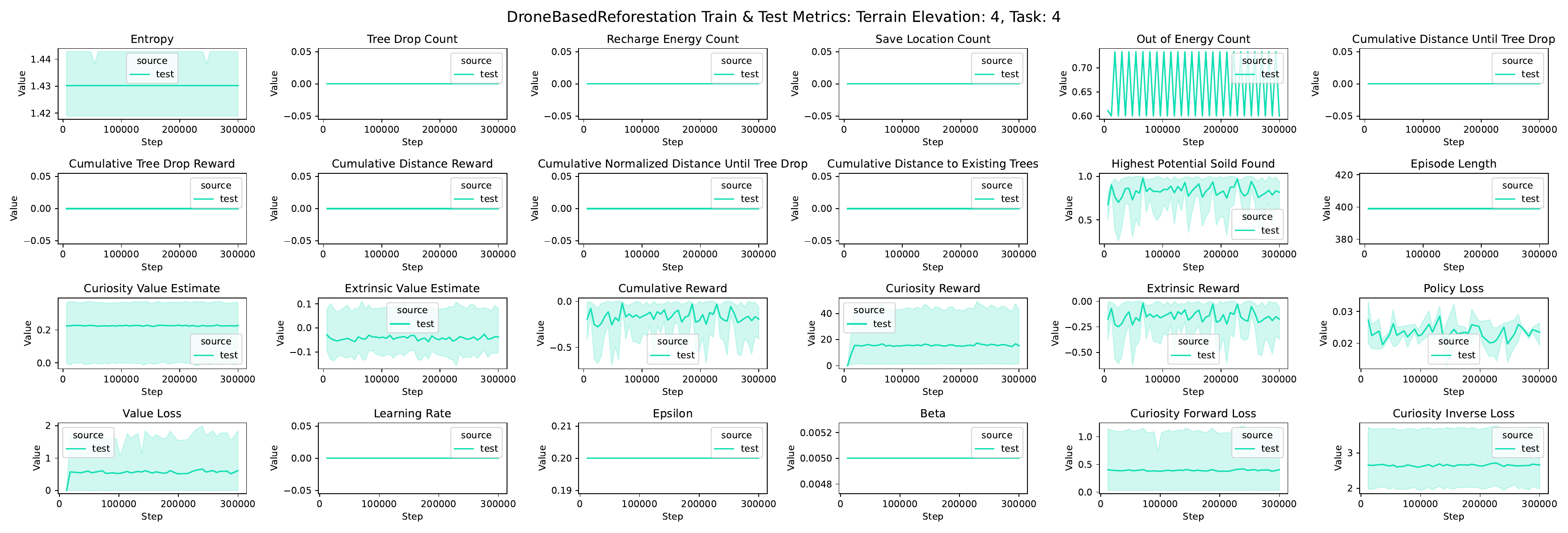}
\vspace{-0.6cm}
\caption{Drone-Based Reforestation: Train \& Test Metrics: Terrain Elevation 4, Task 4.}
\end{figure}

\begin{figure}[h!]
\centering
\includegraphics[width=\linewidth]{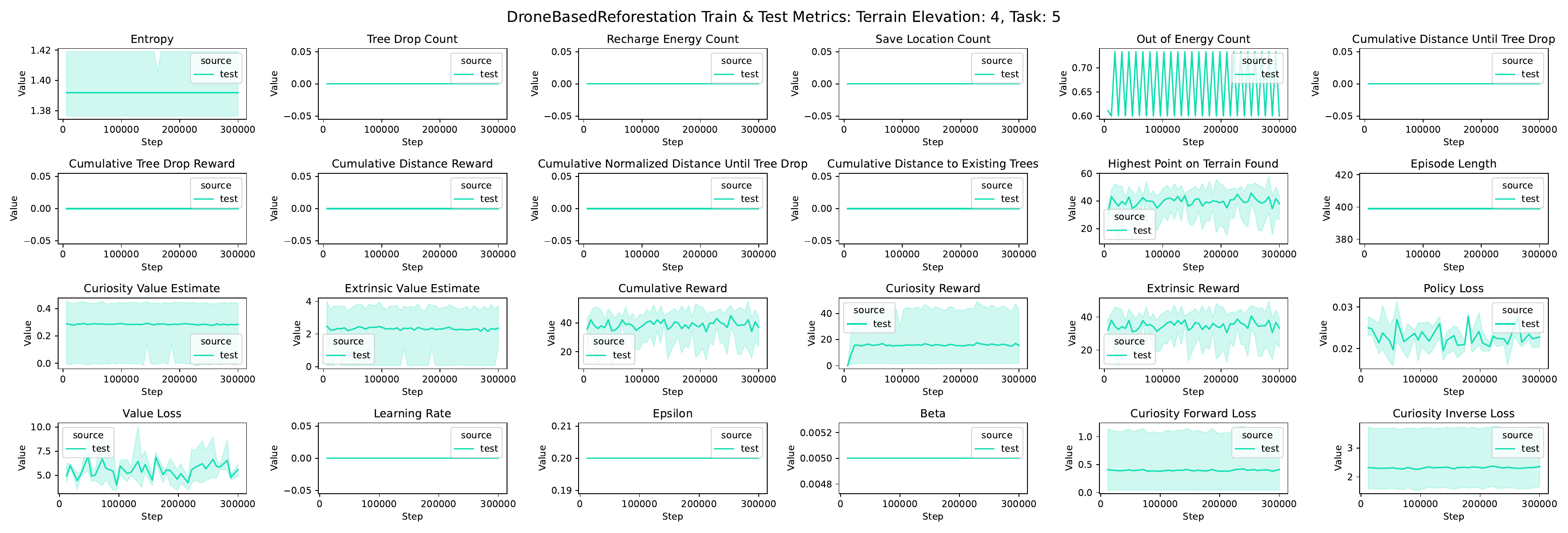}
\vspace{-0.6cm}
\caption{Drone-Based Reforestation: Train \& Test Metrics: Terrain Elevation 4, Task 5.}
\end{figure}

\clearpage

\begin{figure}[h!]
\centering
\includegraphics[width=\linewidth]{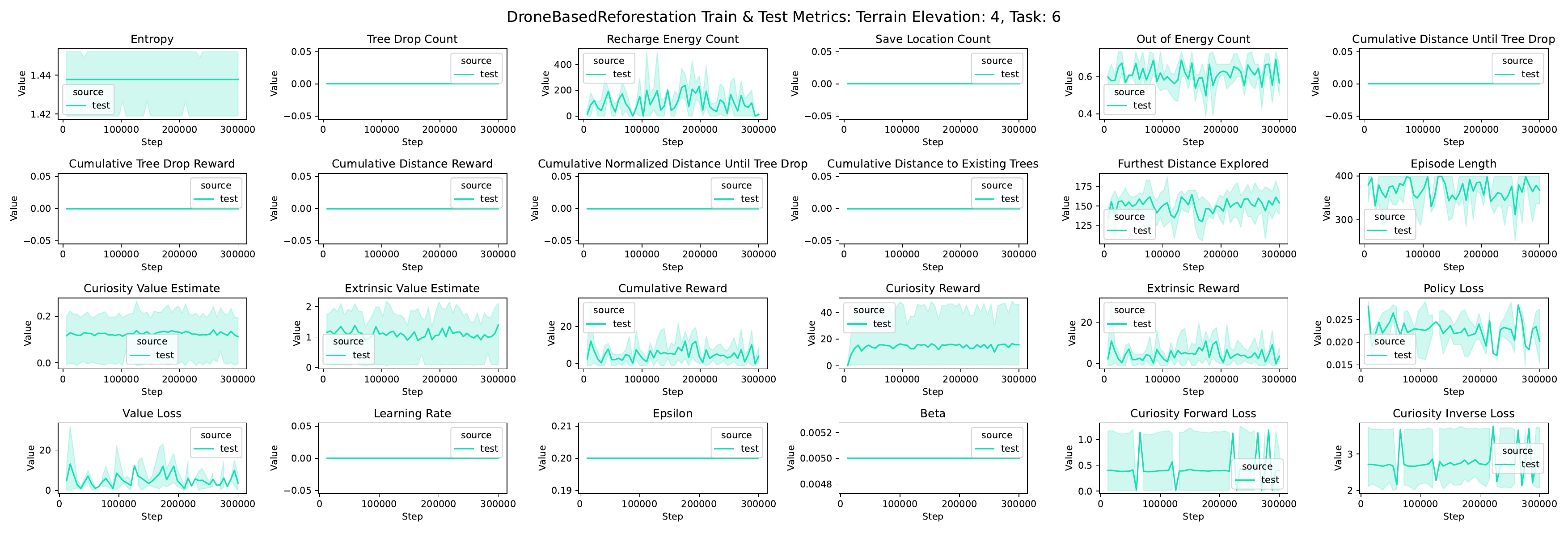}
\vspace{-0.6cm}
\caption{Drone-Based Reforestation: Train \& Test Metrics: Terrain Elevation 4, Task 6.}
\end{figure}

\begin{figure}[h!]
\centering
\includegraphics[width=\linewidth]{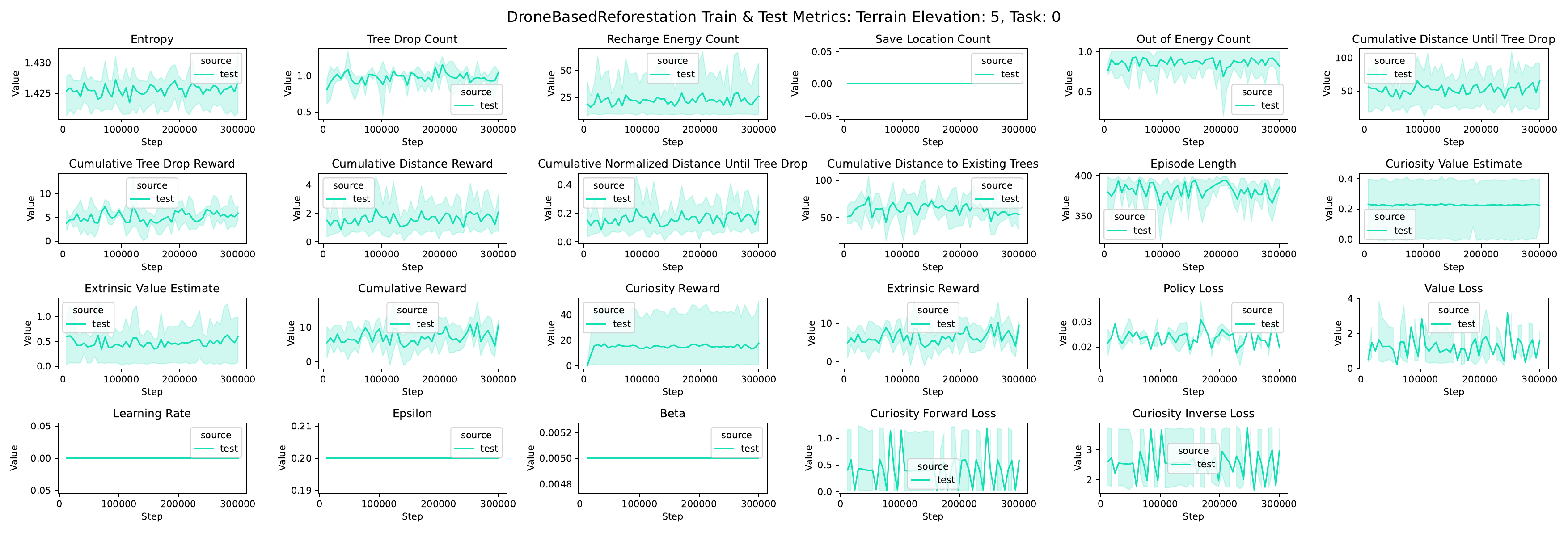}
\vspace{-0.6cm}
\caption{Drone-Based Reforestation: Train \& Test Metrics: Terrain Elevation 5, Task 0.}
\end{figure}

\begin{figure}[h!]
\centering
\includegraphics[width=\linewidth]{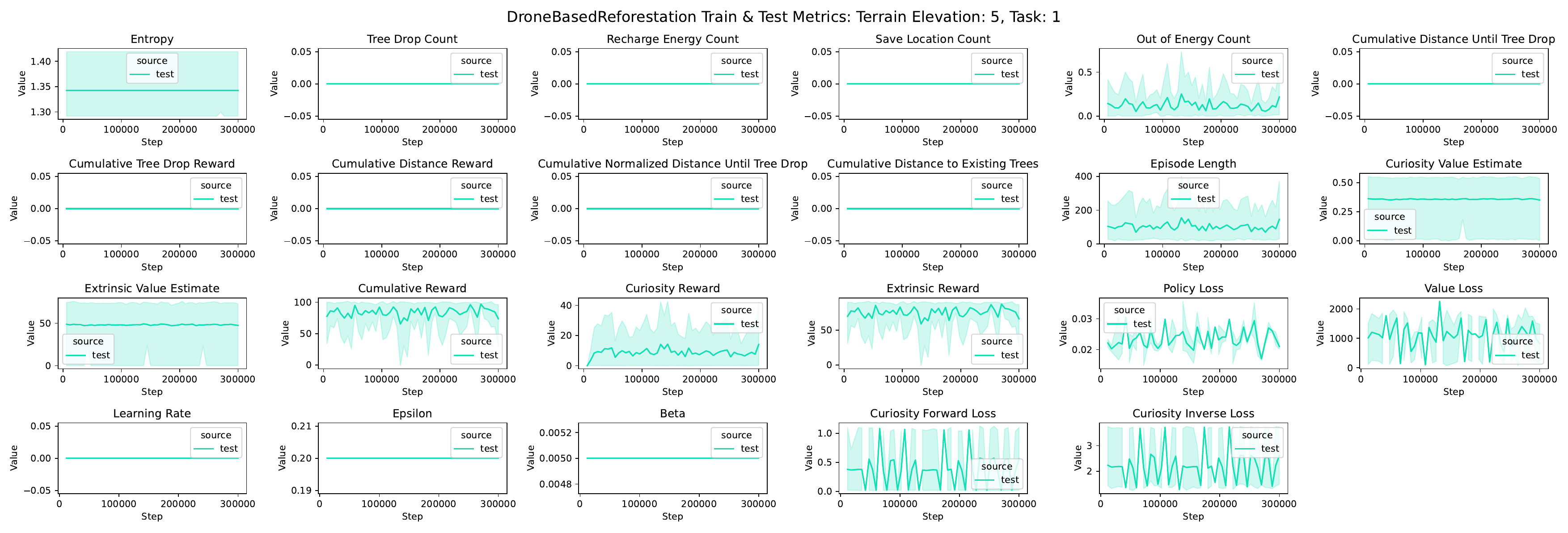}
\vspace{-0.6cm}
\caption{Drone-Based Reforestation: Train \& Test Metrics: Terrain Elevation 5, Task 1.}
\end{figure}

\clearpage

\begin{figure}[h!]
\centering
\includegraphics[width=\linewidth]{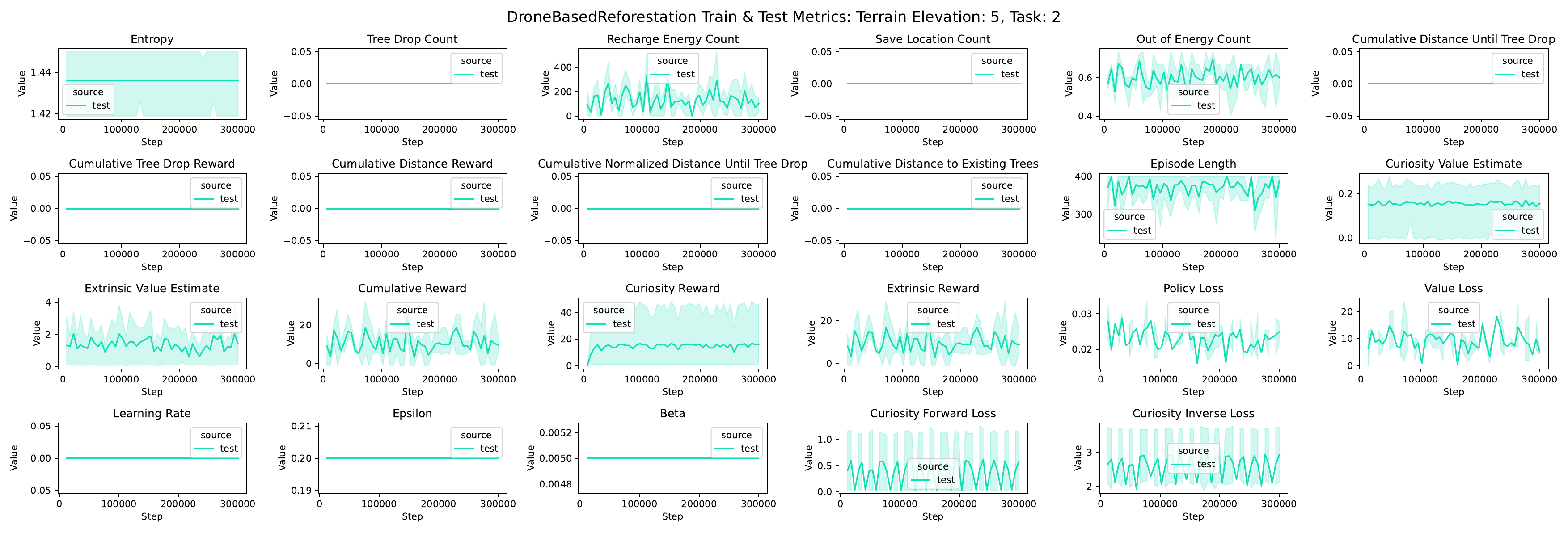}
\vspace{-0.6cm}
\caption{Drone-Based Reforestation: Train \& Test Metrics: Terrain Elevation 5, Task 2.}
\end{figure}

\begin{figure}[h!]
\centering
\includegraphics[width=\linewidth]{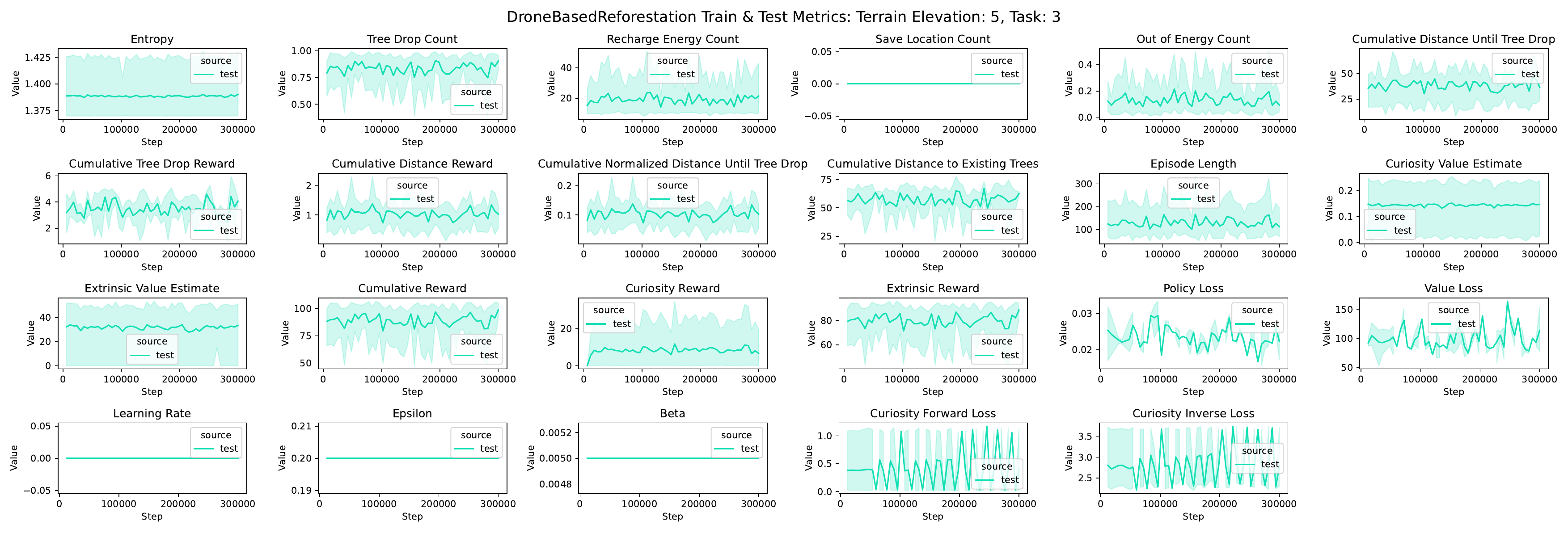}
\vspace{-0.6cm}
\caption{Drone-Based Reforestation: Train \& Test Metrics: Terrain Elevation 5, Task 3.}
\end{figure}

\begin{figure}[h!]
\centering
\includegraphics[width=\linewidth]{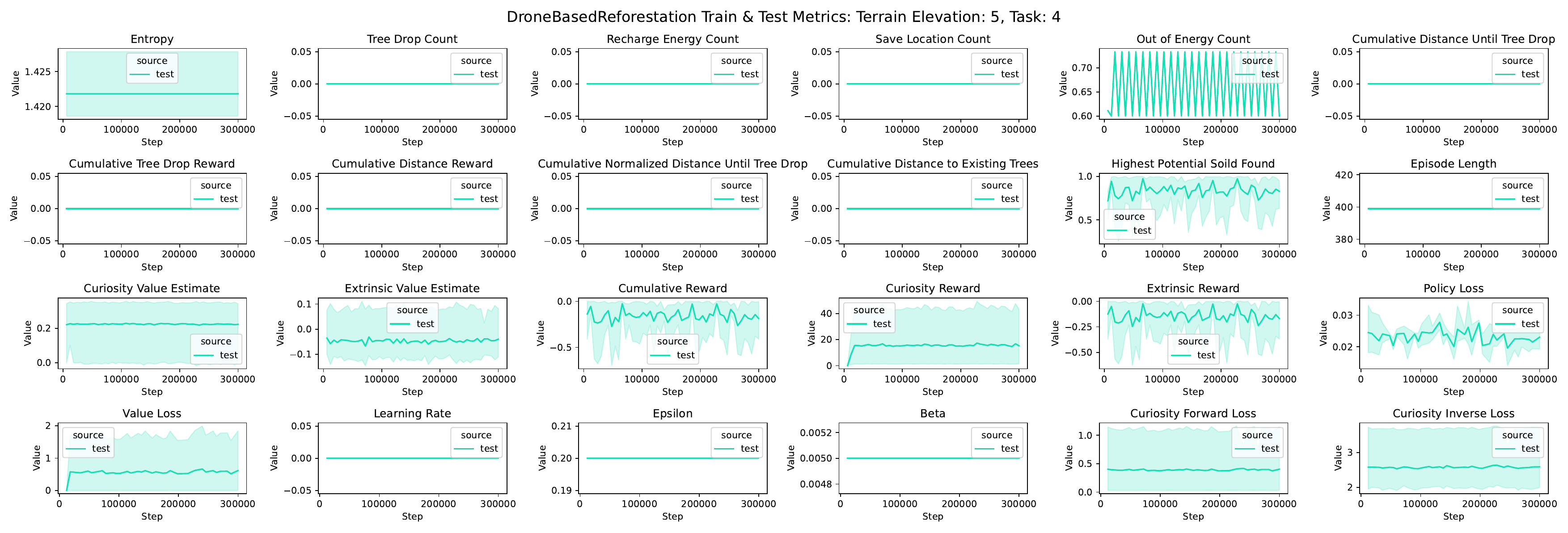}
\vspace{-0.6cm}
\caption{Drone-Based Reforestation: Train \& Test Metrics: Terrain Elevation 5, Task 4.}
\end{figure}

\clearpage

\begin{figure}[h!]
\centering
\includegraphics[width=\linewidth]{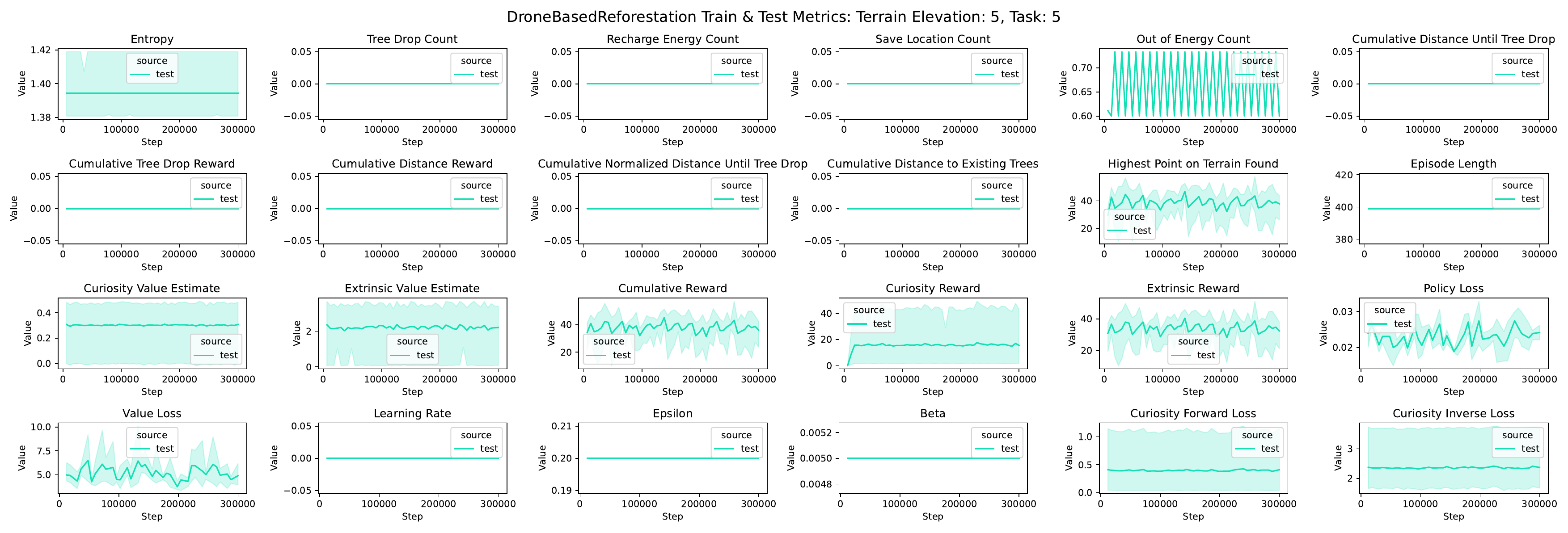}
\vspace{-0.6cm}
\caption{Drone-Based Reforestation: Train \& Test Metrics: Terrain Elevation 5, Task 5.}
\end{figure}

\begin{figure}[h!]
\centering
\includegraphics[width=\linewidth]{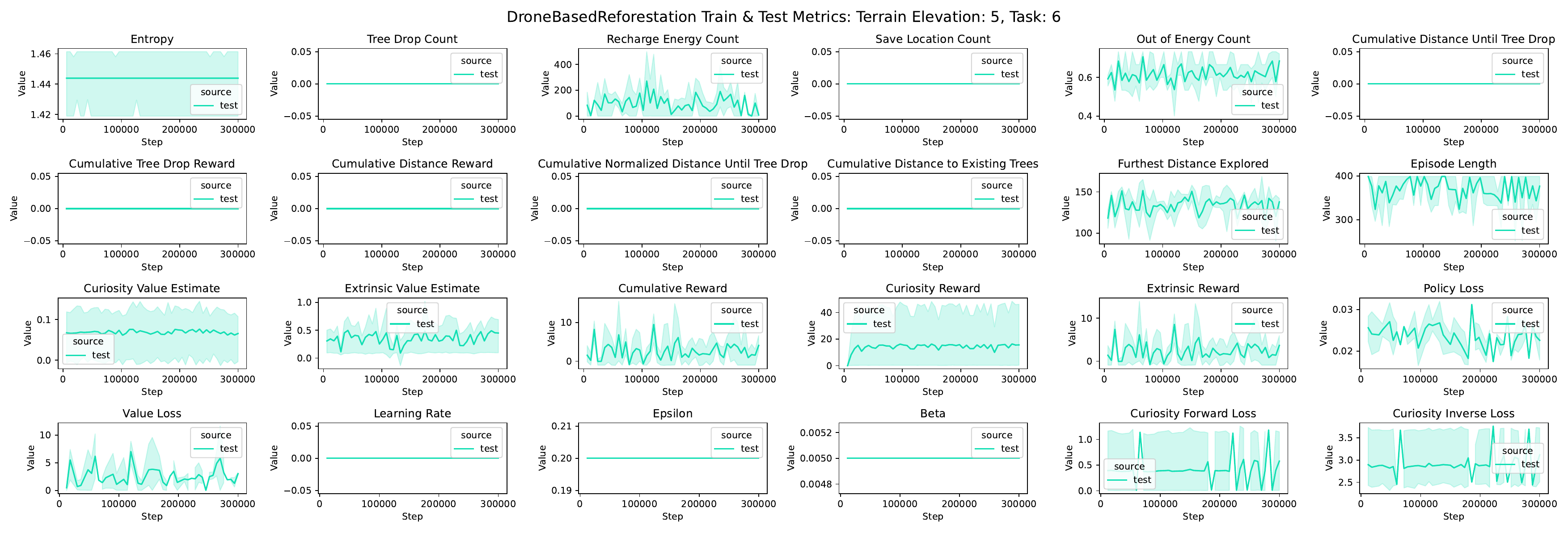}
\vspace{-0.6cm}
\caption{Drone-Based Reforestation: Train \& Test Metrics: Terrain Elevation 5, Task 6.}
\end{figure}

\begin{figure}[h!]
\centering
\includegraphics[width=\linewidth]{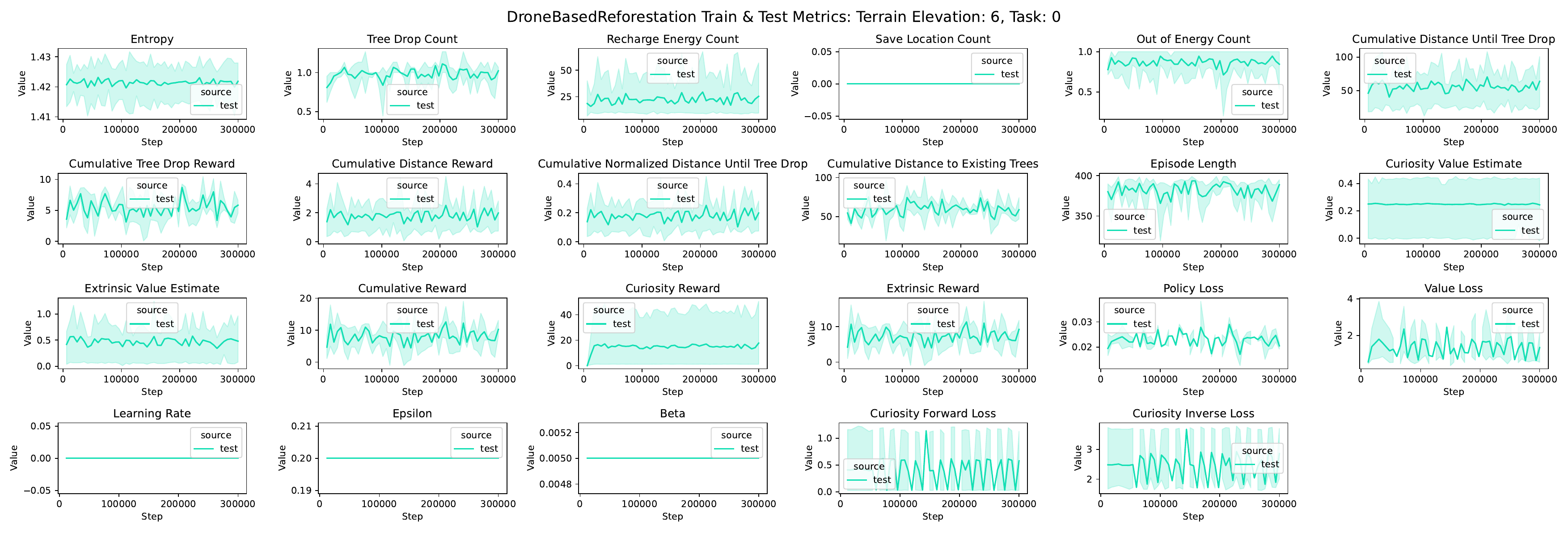}
\vspace{-0.6cm}
\caption{Drone-Based Reforestation: Train \& Test Metrics: Terrain Elevation 6, Task 0.}
\end{figure}

\clearpage

\begin{figure}[h!]
\centering
\includegraphics[width=\linewidth]{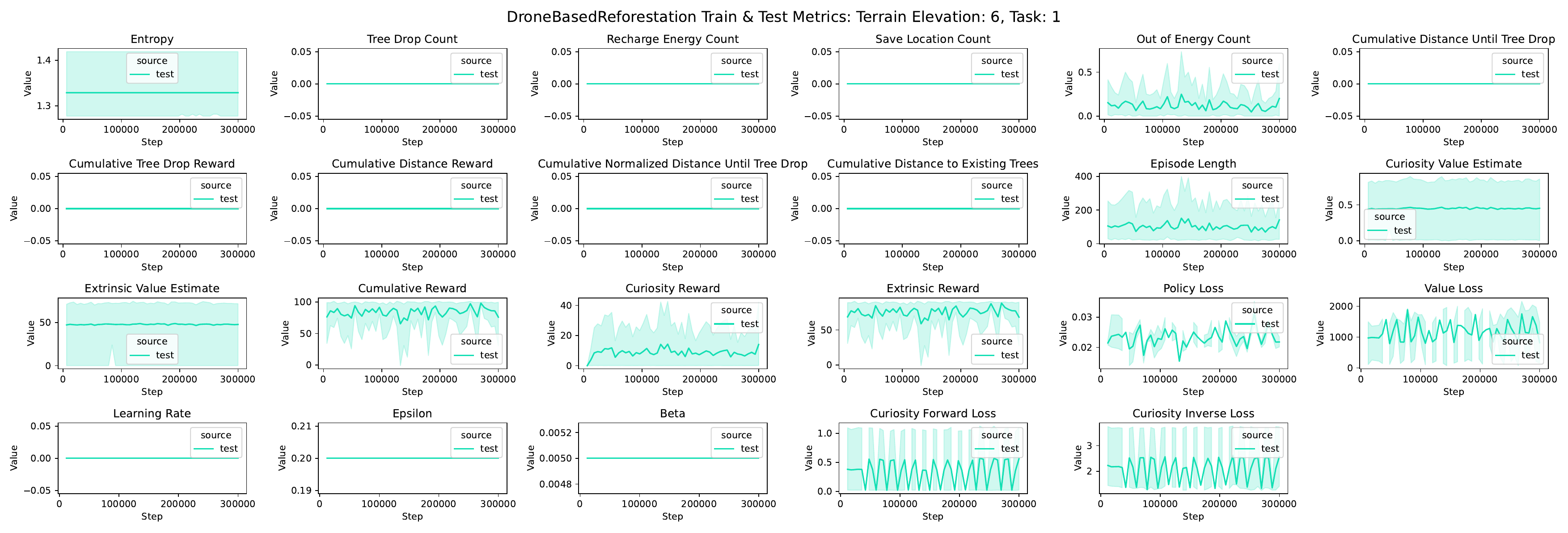}
\vspace{-0.6cm}
\caption{Drone-Based Reforestation: Train \& Test Metrics: Terrain Elevation 6, Task 1.}
\end{figure}

\begin{figure}[h!]
\centering
\includegraphics[width=\linewidth]{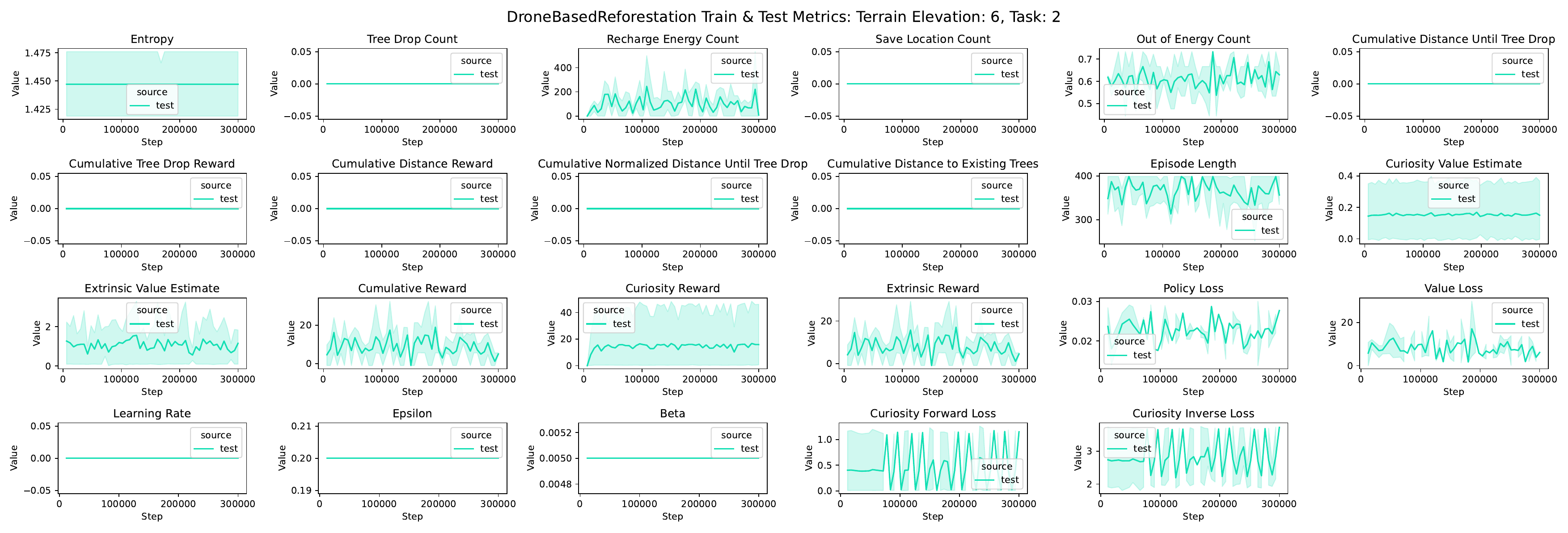}
\vspace{-0.6cm}
\caption{Drone-Based Reforestation: Train \& Test Metrics: Terrain Elevation 6, Task 2.}
\end{figure}

\begin{figure}[h!]
\centering
\includegraphics[width=\linewidth]{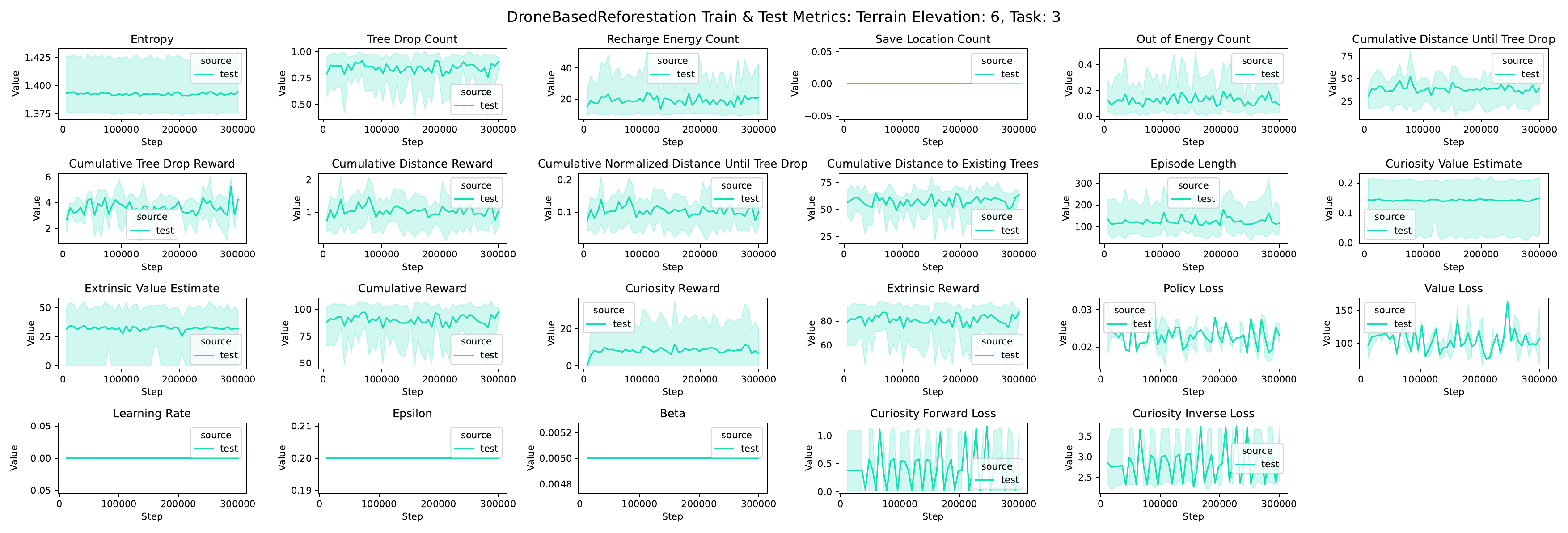}
\vspace{-0.6cm}
\caption{Drone-Based Reforestation: Train \& Test Metrics: Terrain Elevation 6, Task 3.}
\end{figure}

\clearpage

\begin{figure}[h!]
\centering
\includegraphics[width=\linewidth]{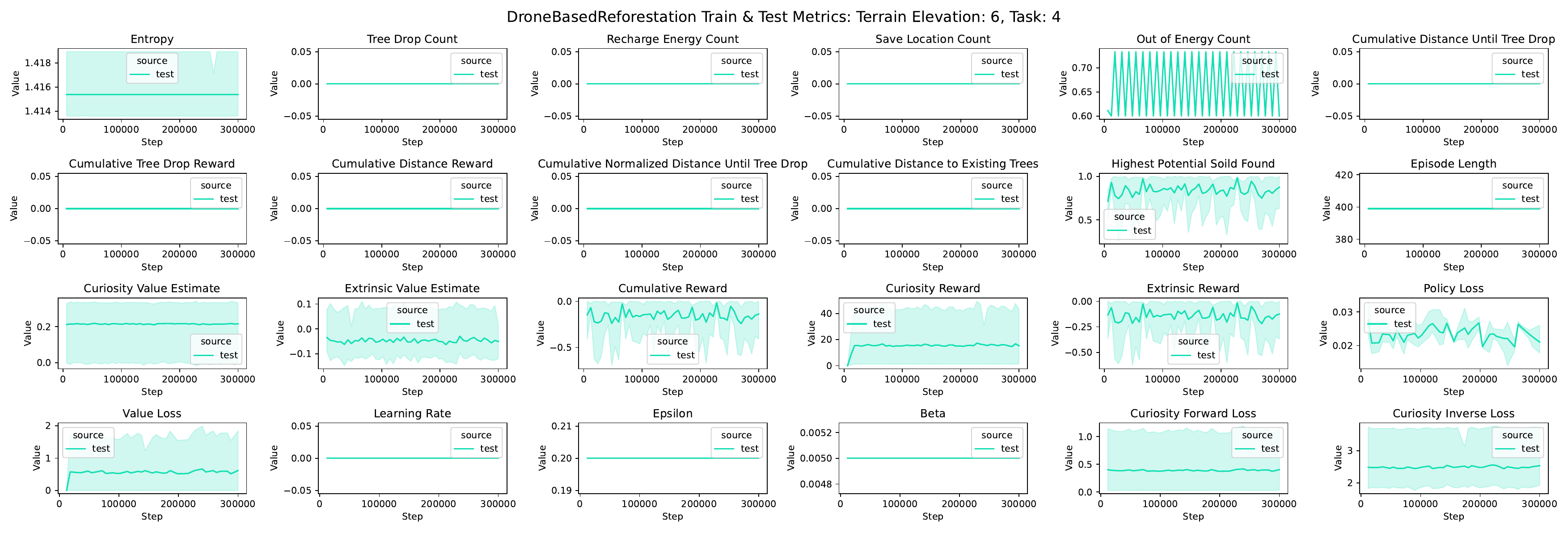}
\vspace{-0.6cm}
\caption{Drone-Based Reforestation: Train \& Test Metrics: Terrain Elevation 6, Task 4.}
\end{figure}

\begin{figure}[h!]
\centering
\includegraphics[width=\linewidth]{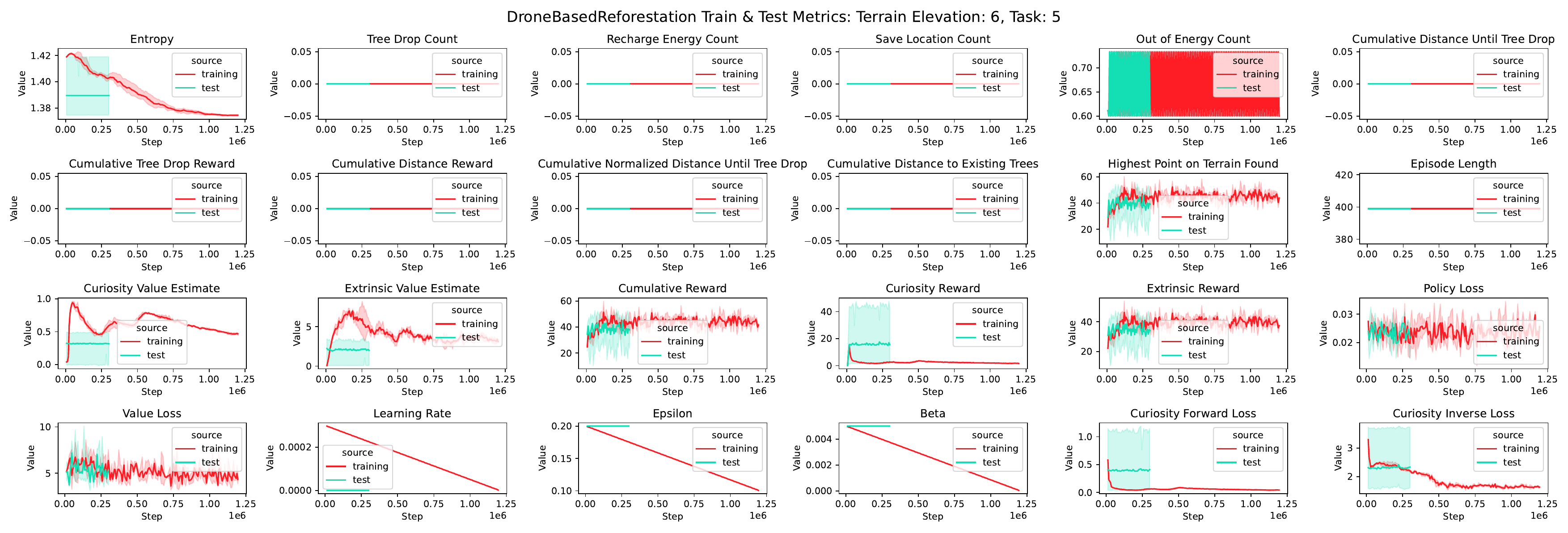}
\vspace{-0.6cm}
\caption{Drone-Based Reforestation: Train \& Test Metrics: Terrain Elevation 6, Task 5.}
\end{figure}

\begin{figure}[h!]
\centering
\includegraphics[width=\linewidth]{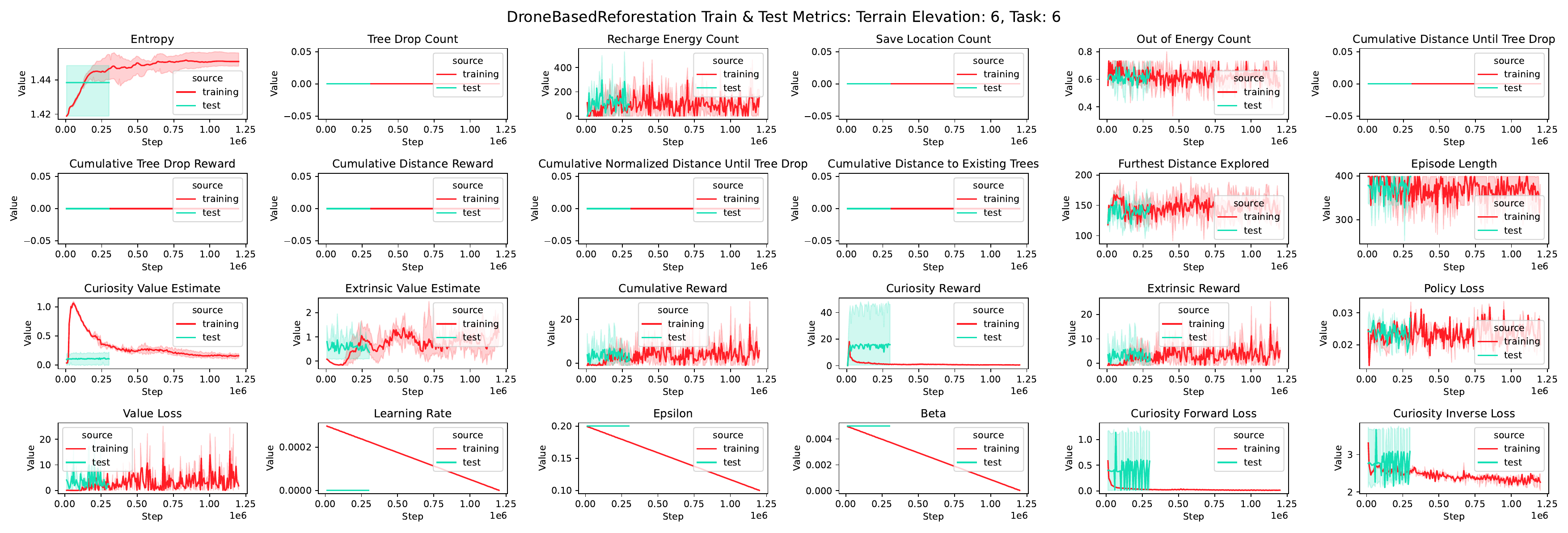}
\vspace{-0.6cm}
\caption{Drone-Based Reforestation: Train \& Test Metrics: Terrain Elevation 6, Task 6.}
\end{figure}

\clearpage

\begin{figure}[h!]
\centering
\includegraphics[width=\linewidth]{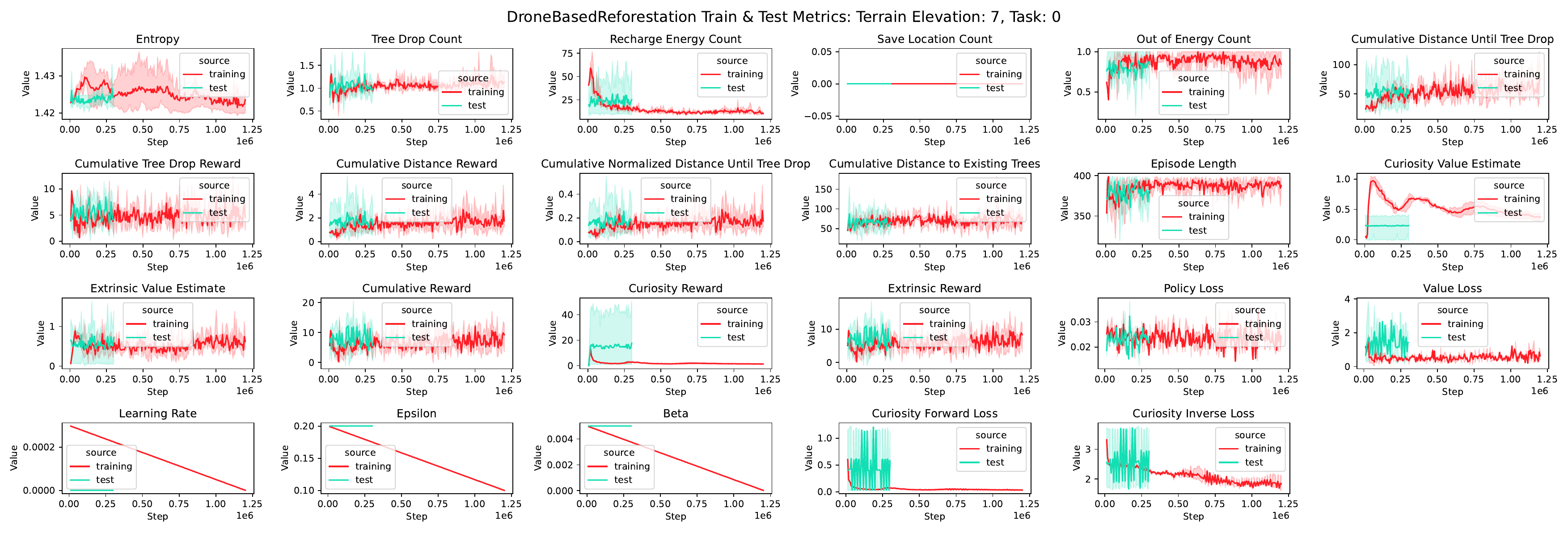}
\vspace{-0.6cm}
\caption{Drone-Based Reforestation: Train \& Test Metrics: Terrain Elevation 7, Task 0.}
\end{figure}

\begin{figure}[h!]
\centering
\includegraphics[width=\linewidth]{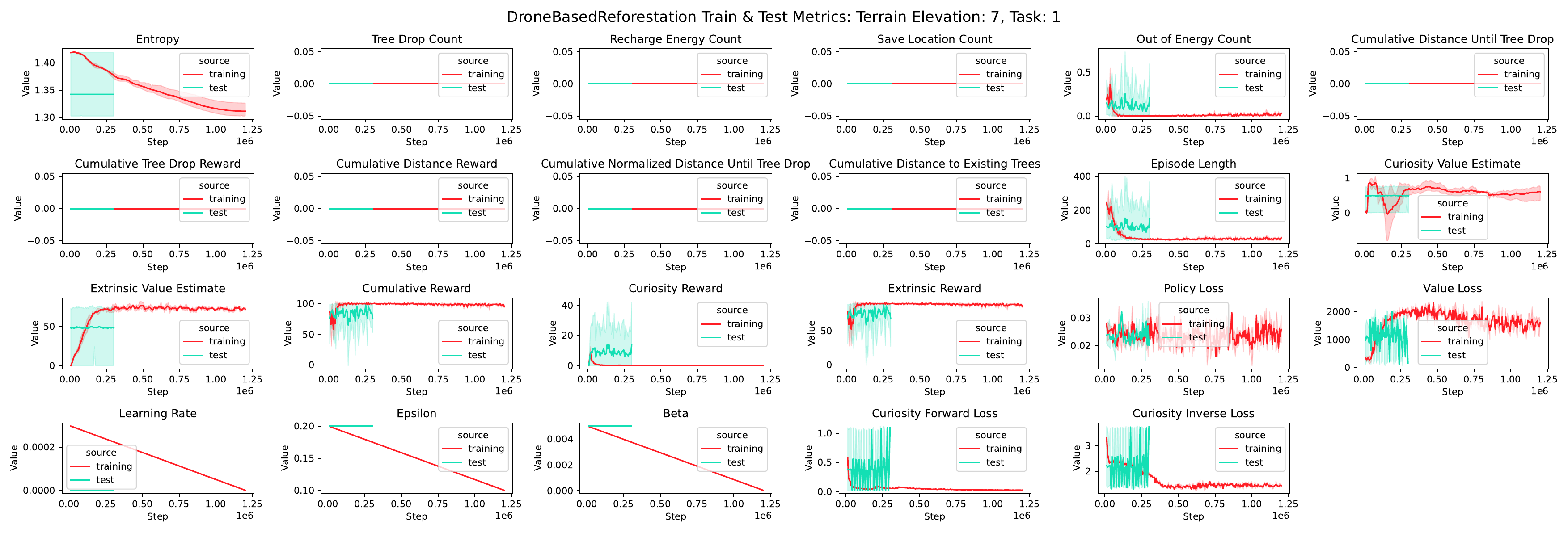}
\vspace{-0.6cm}
\caption{Drone-Based Reforestation: Train \& Test Metrics: Terrain Elevation 7, Task 1.}
\end{figure}

\begin{figure}[h!]
\centering
\includegraphics[width=\linewidth]{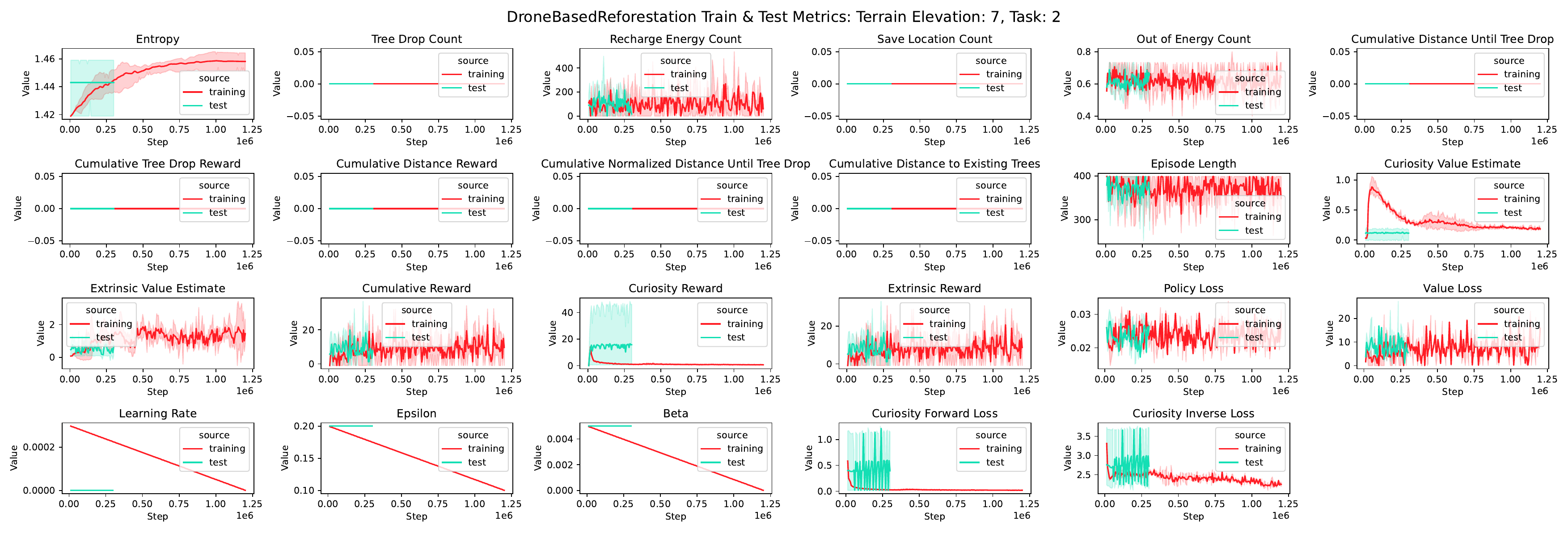}
\vspace{-0.6cm}
\caption{Drone-Based Reforestation: Train \& Test Metrics: Terrain Elevation 7, Task 2.}
\end{figure}

\clearpage

\begin{figure}[h!]
\centering
\includegraphics[width=\linewidth]{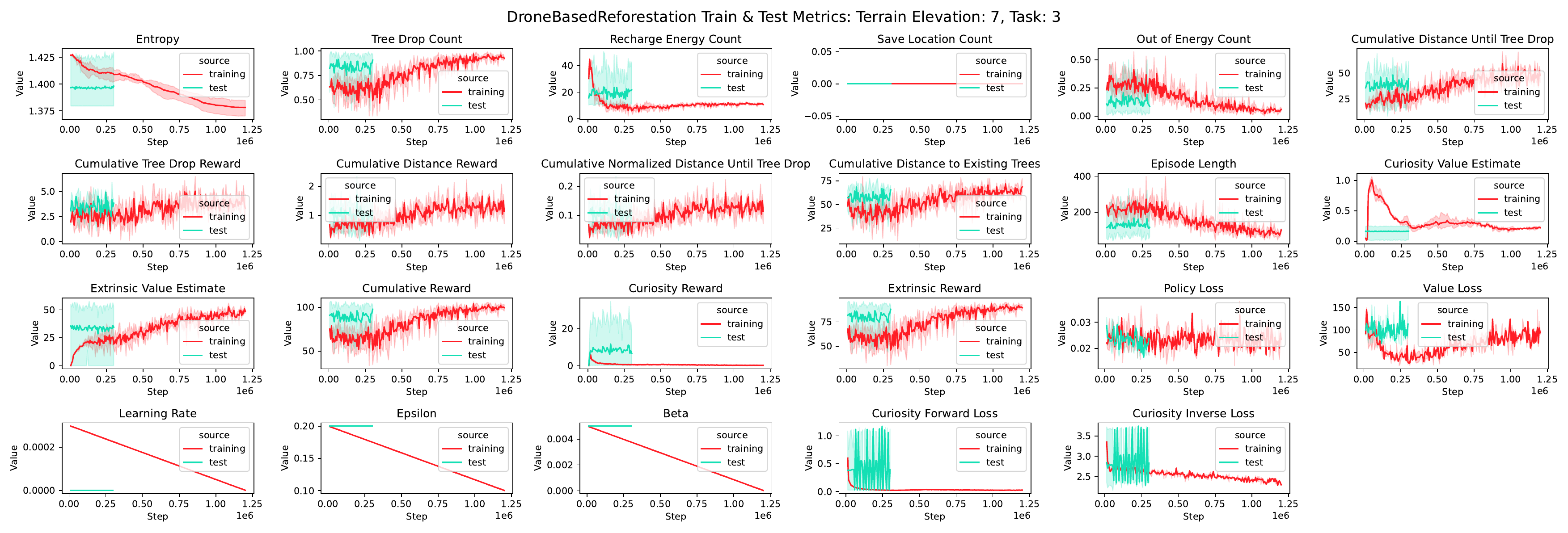}
\vspace{-0.6cm}
\caption{Drone-Based Reforestation: Train \& Test Metrics: Terrain Elevation 7, Task 3.}
\end{figure}

\begin{figure}[h!]
\centering
\includegraphics[width=\linewidth]{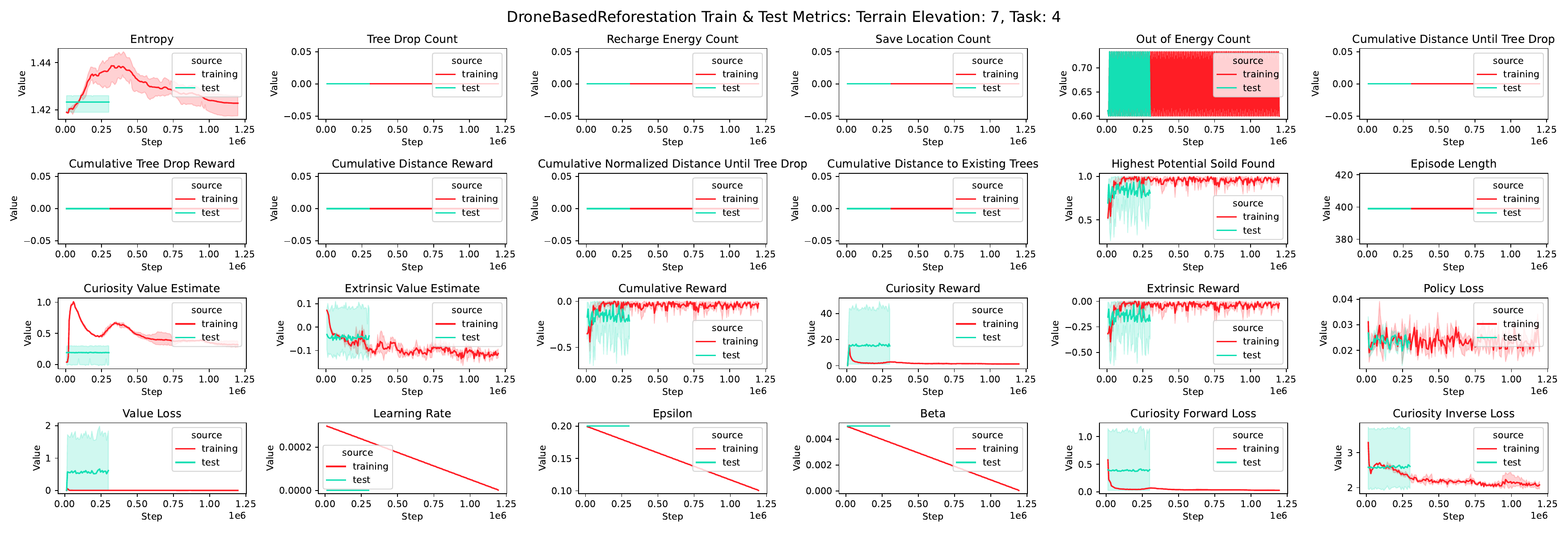}
\vspace{-0.6cm}
\caption{Drone-Based Reforestation: Train \& Test Metrics: Terrain Elevation 7, Task 4.}
\end{figure}

\begin{figure}[h!]
\centering
\includegraphics[width=\linewidth]{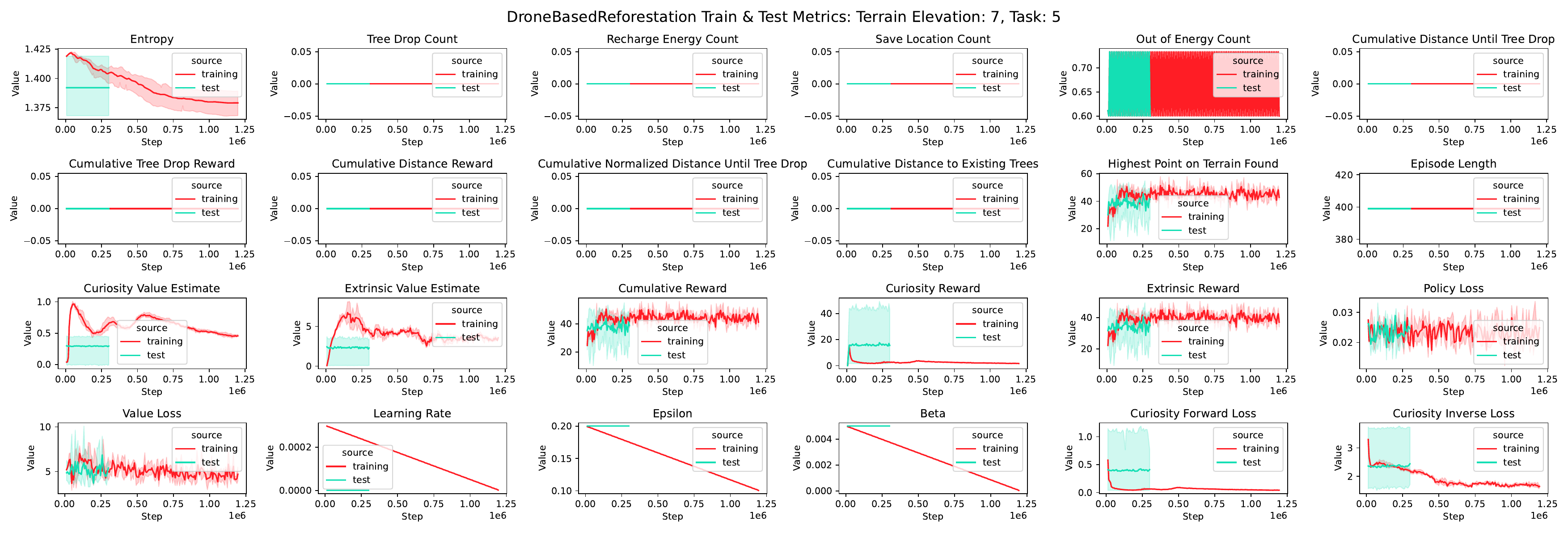}
\vspace{-0.6cm}
\caption{Drone-Based Reforestation: Train \& Test Metrics: Terrain Elevation 7, Task 5.}
\end{figure}

\clearpage

\begin{figure}[h!]
\centering
\includegraphics[width=\linewidth]{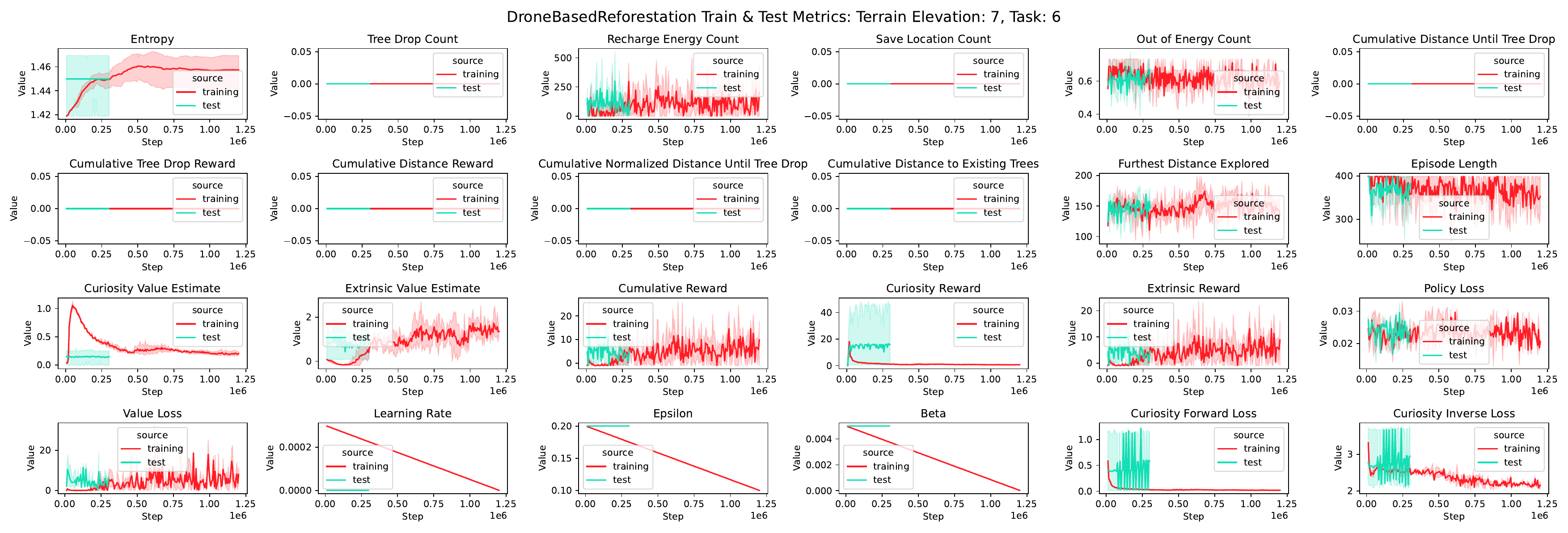}
\vspace{-0.6cm}
\caption{Drone-Based Reforestation: Train \& Test Metrics: Terrain Elevation 7, Task 6.}
\end{figure}

\begin{figure}[h!]
\centering
\includegraphics[width=\linewidth]{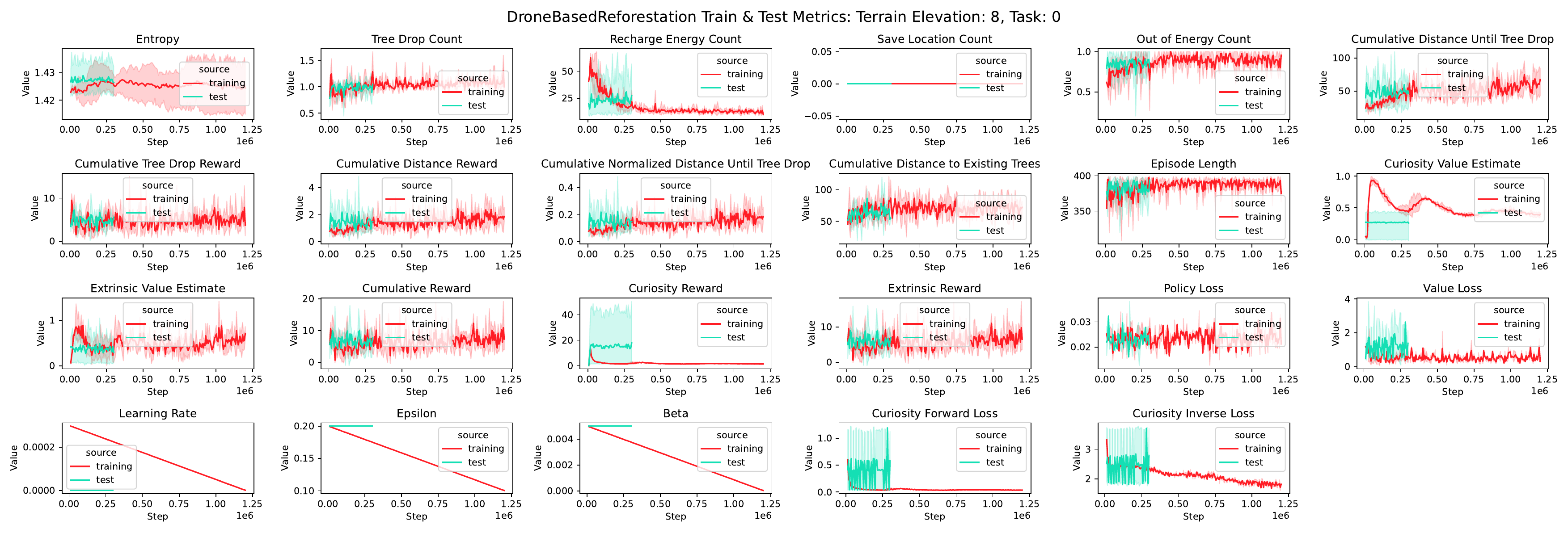}
\vspace{-0.6cm}
\caption{Drone-Based Reforestation: Train \& Test Metrics: Terrain Elevation 8, Task 0.}
\end{figure}

\begin{figure}[h!]
\centering
\includegraphics[width=\linewidth]{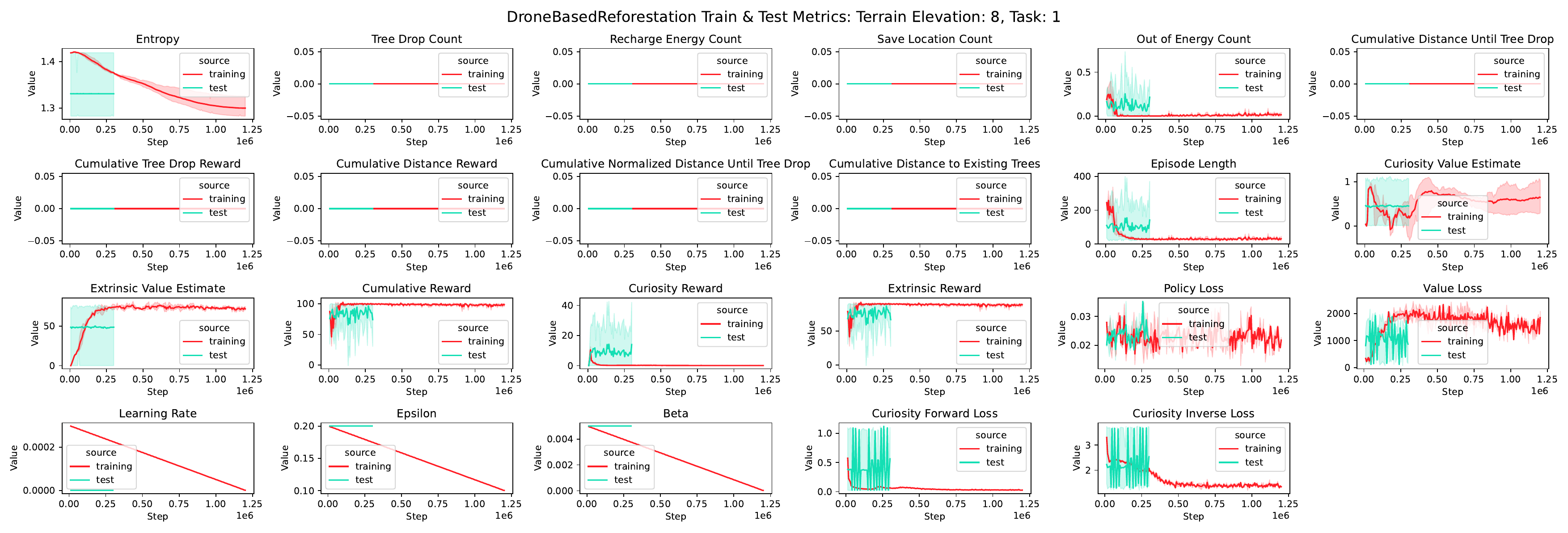}
\vspace{-0.6cm}
\caption{Drone-Based Reforestation: Train \& Test Metrics: Terrain Elevation 8, Task 1.}
\end{figure}

\clearpage

\begin{figure}[h!]
\centering
\includegraphics[width=\linewidth]{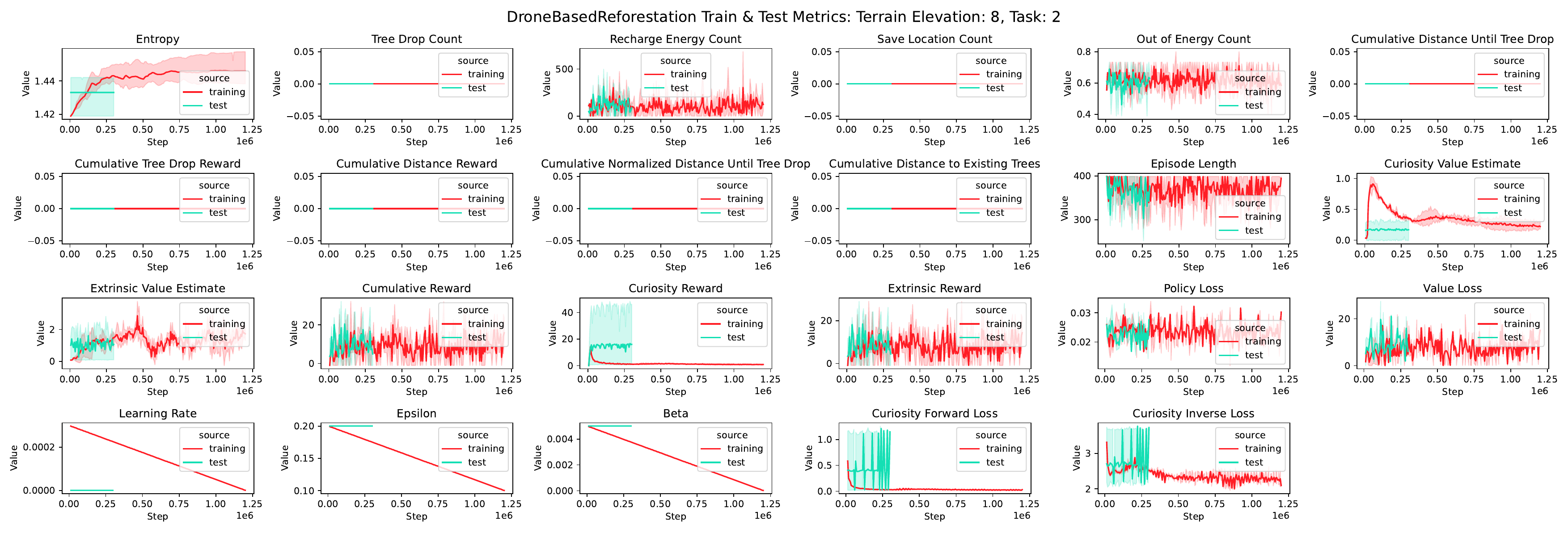}
\vspace{-0.6cm}
\caption{Drone-Based Reforestation: Train \& Test Metrics: Terrain Elevation 8, Task 2.}
\end{figure}

\begin{figure}[h!]
\centering
\includegraphics[width=\linewidth]{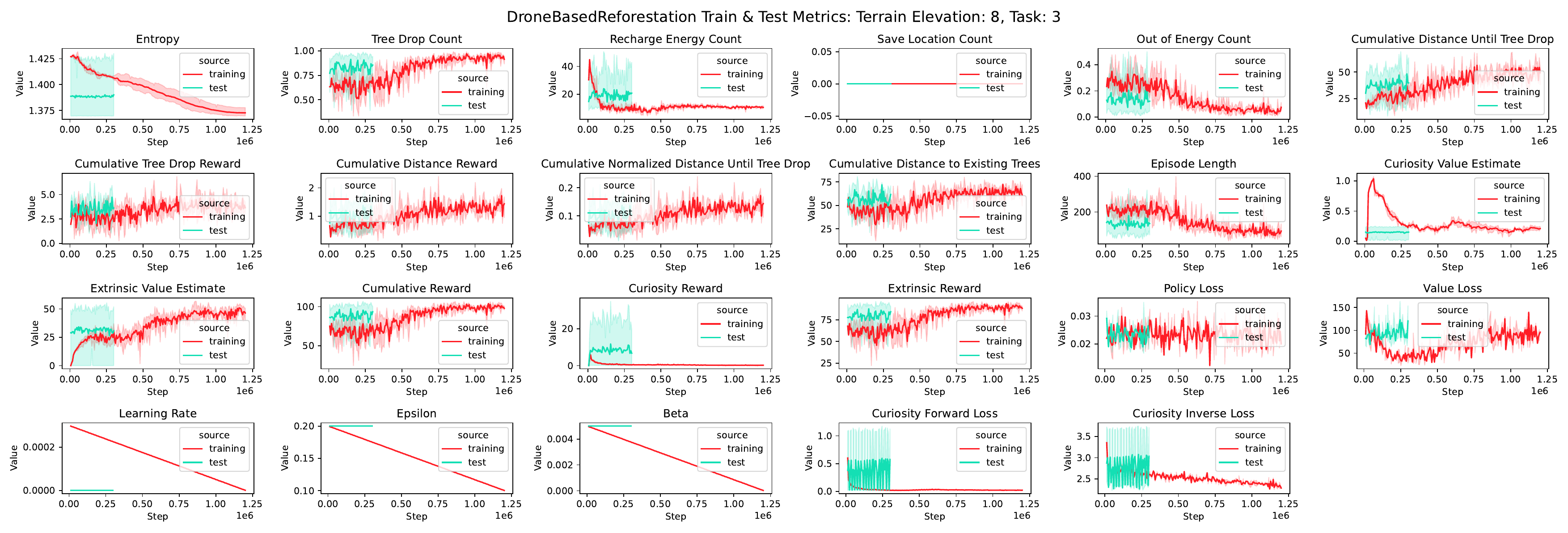}
\vspace{-0.6cm}
\caption{Drone-Based Reforestation: Train \& Test Metrics: Terrain Elevation 8, Task 3.}
\end{figure}

\begin{figure}[h!]
\centering
\includegraphics[width=\linewidth]{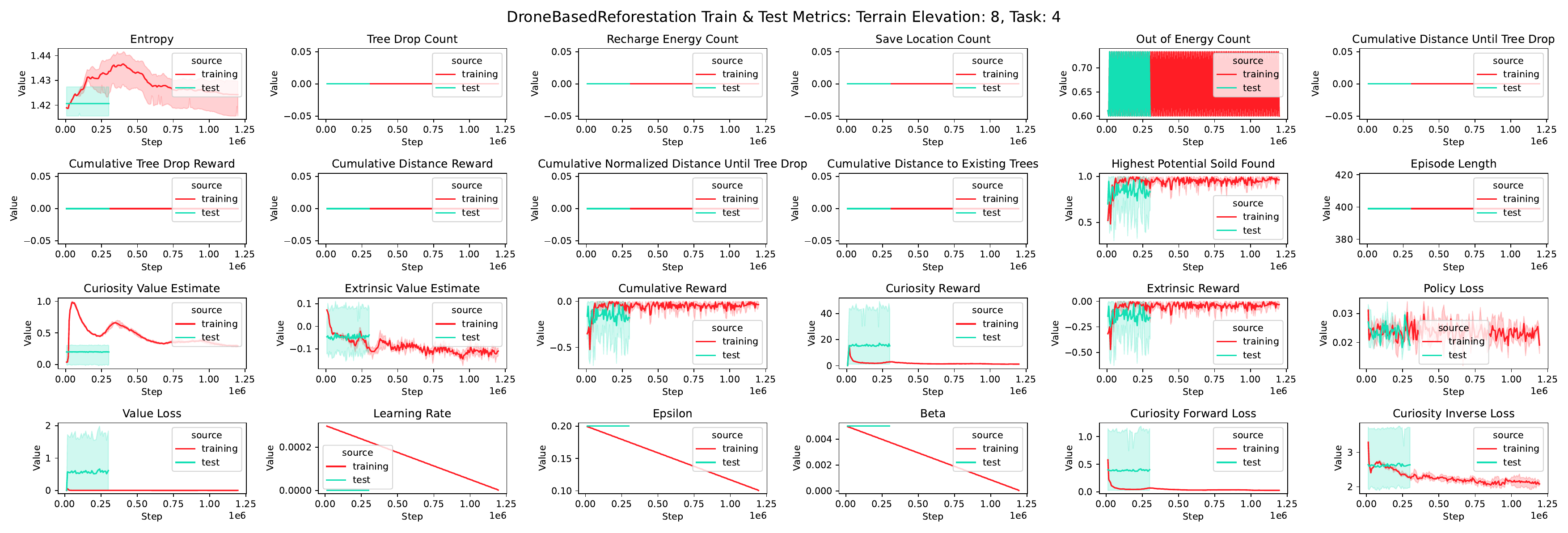}
\vspace{-0.6cm}
\caption{Drone-Based Reforestation: Train \& Test Metrics: Terrain Elevation 8, Task 4.}
\end{figure}

\clearpage

\begin{figure}[h!]
\centering
\includegraphics[width=\linewidth]{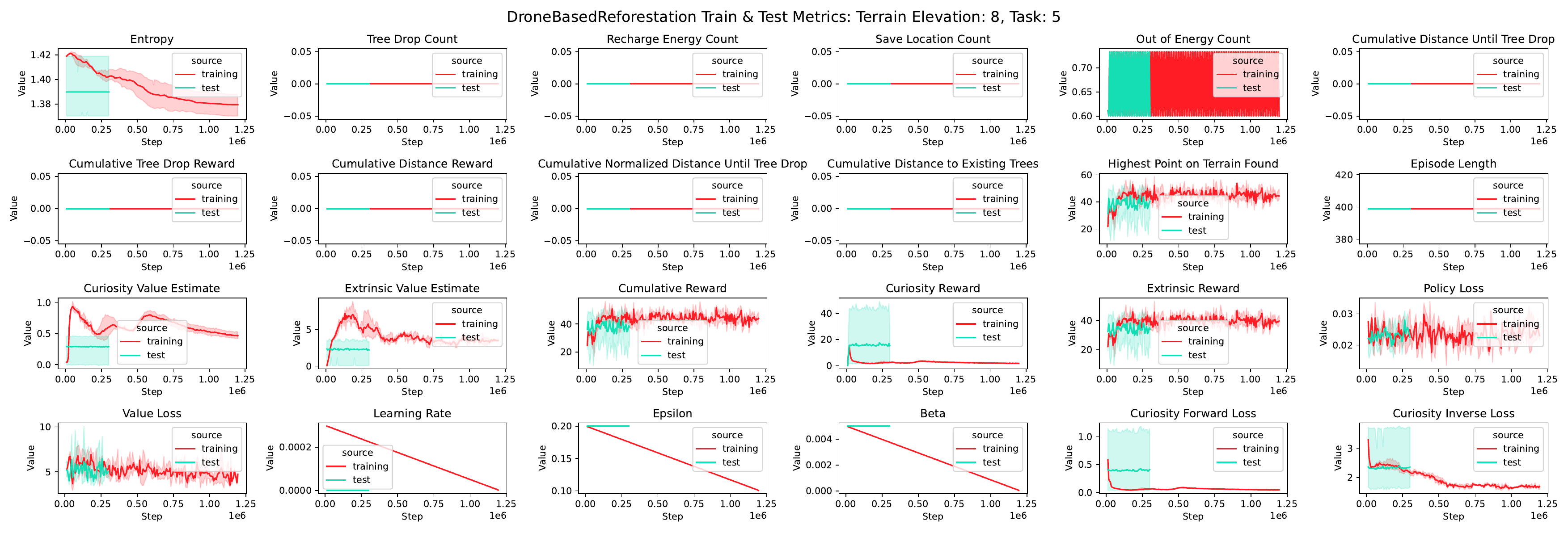}
\vspace{-0.6cm}
\caption{Drone-Based Reforestation: Train \& Test Metrics: Terrain Elevation 8, Task 5.}
\end{figure}

\begin{figure}[h!]
\centering
\includegraphics[width=\linewidth]{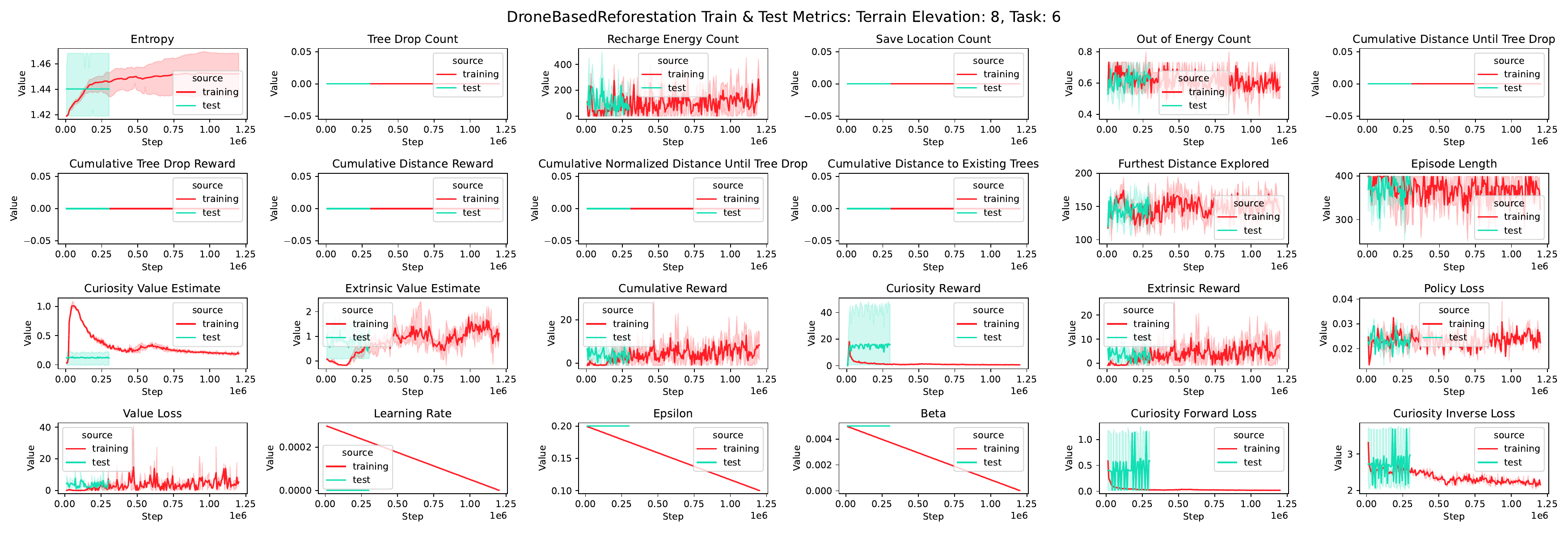}
\vspace{-0.6cm}
\caption{Drone-Based Reforestation: Train \& Test Metrics: Terrain Elevation 8, Task 6.}
\end{figure}

\begin{figure}[h!]
\centering
\includegraphics[width=\linewidth]{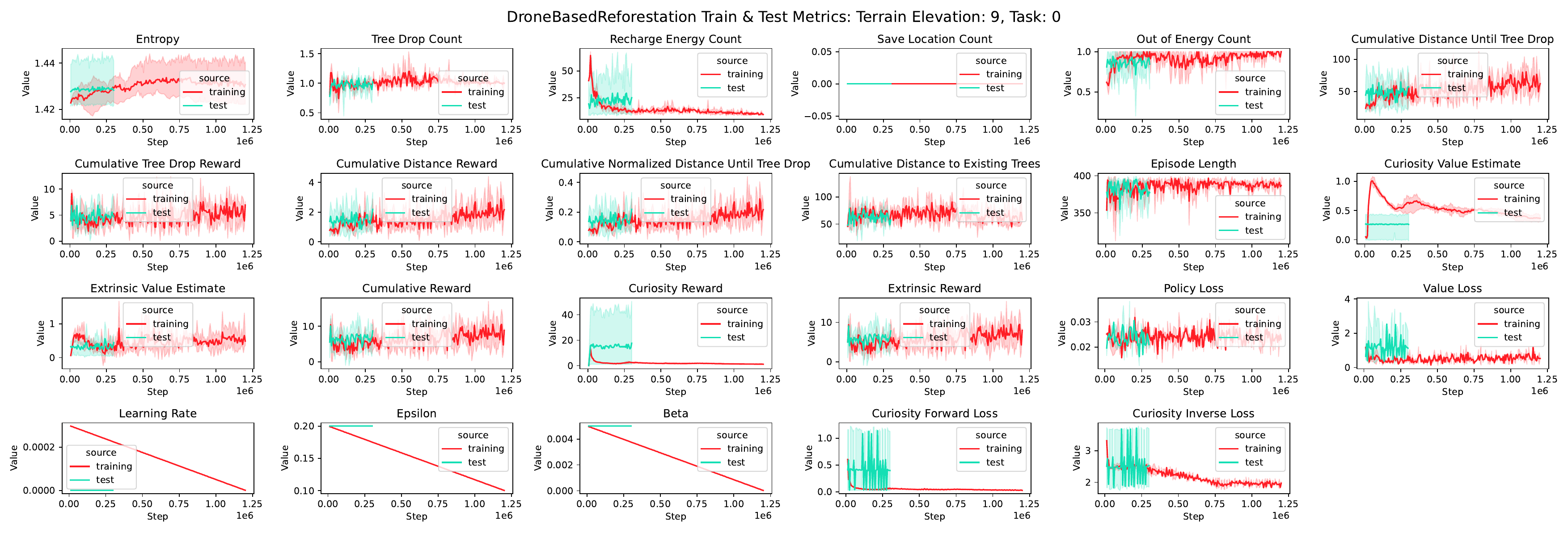}
\vspace{-0.6cm}
\caption{Drone-Based Reforestation: Train \& Test Metrics: Terrain Elevation 9, Task 0.}
\end{figure}

\clearpage

\begin{figure}[h!]
\centering
\includegraphics[width=\linewidth]{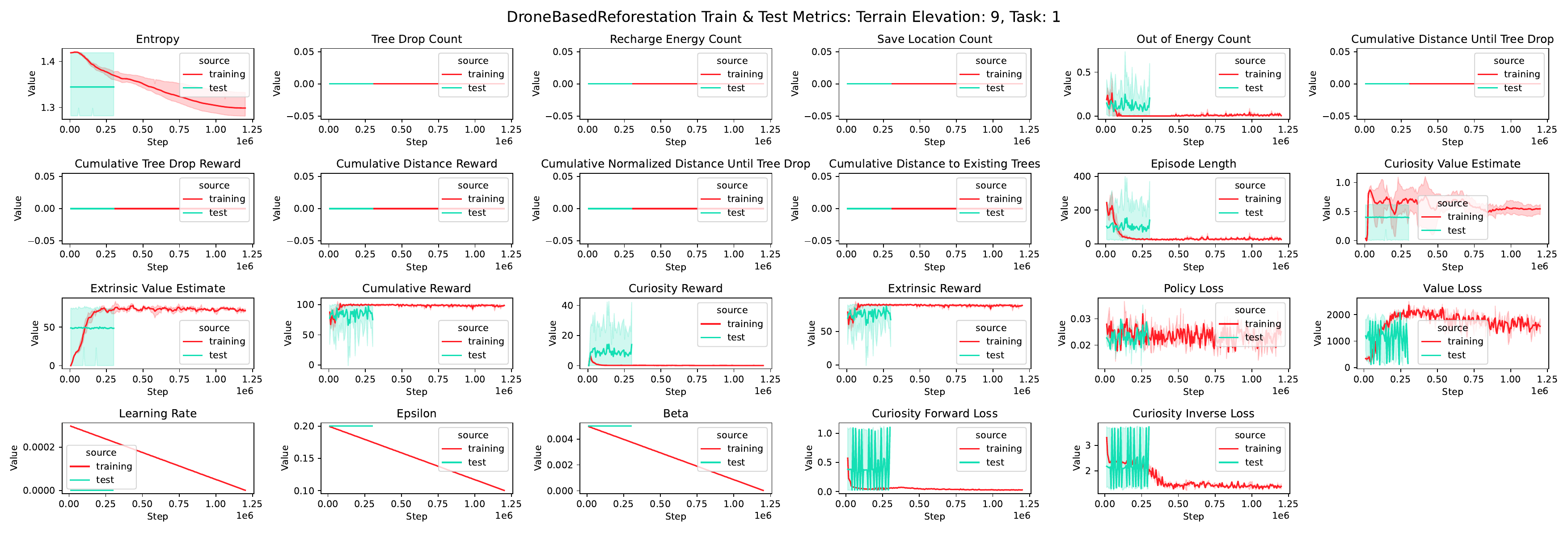}
\vspace{-0.6cm}
\caption{Drone-Based Reforestation: Train \& Test Metrics: Terrain Elevation 9, Task 1.}
\end{figure}

\begin{figure}[h!]
\centering
\includegraphics[width=\linewidth]{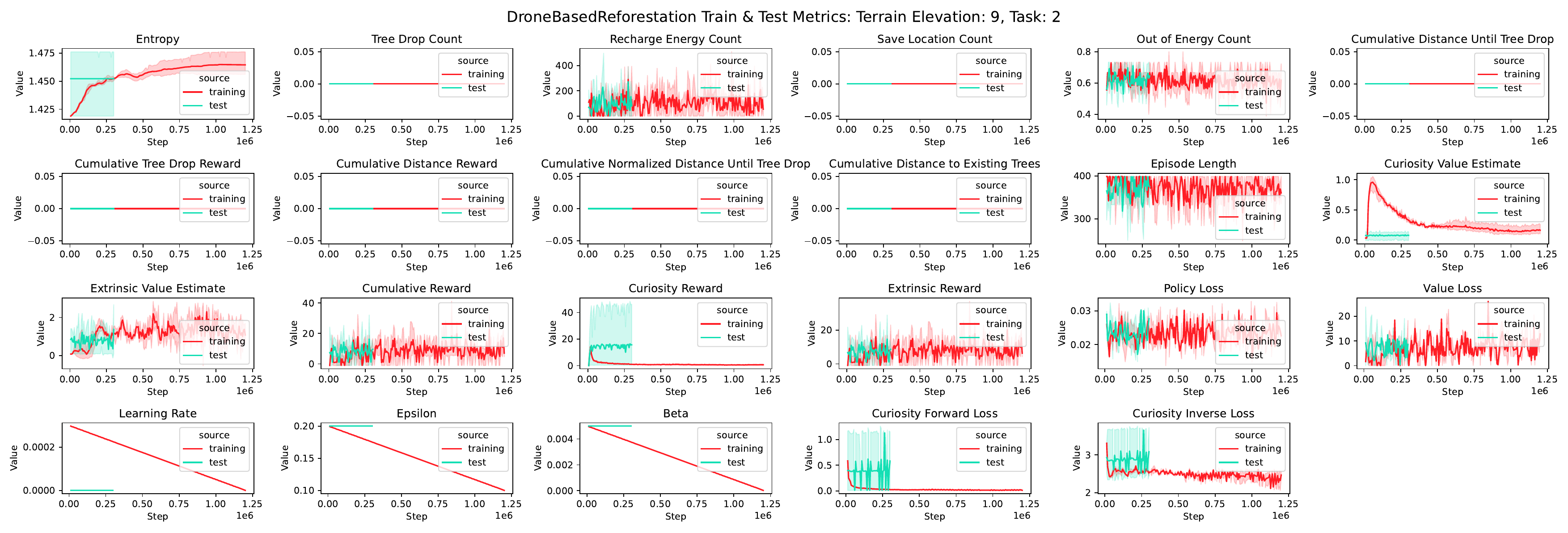}
\vspace{-0.6cm}
\caption{Drone-Based Reforestation: Train \& Test Metrics: Terrain Elevation 9, Task 2.}
\end{figure}

\begin{figure}[h!]
\centering
\includegraphics[width=\linewidth]{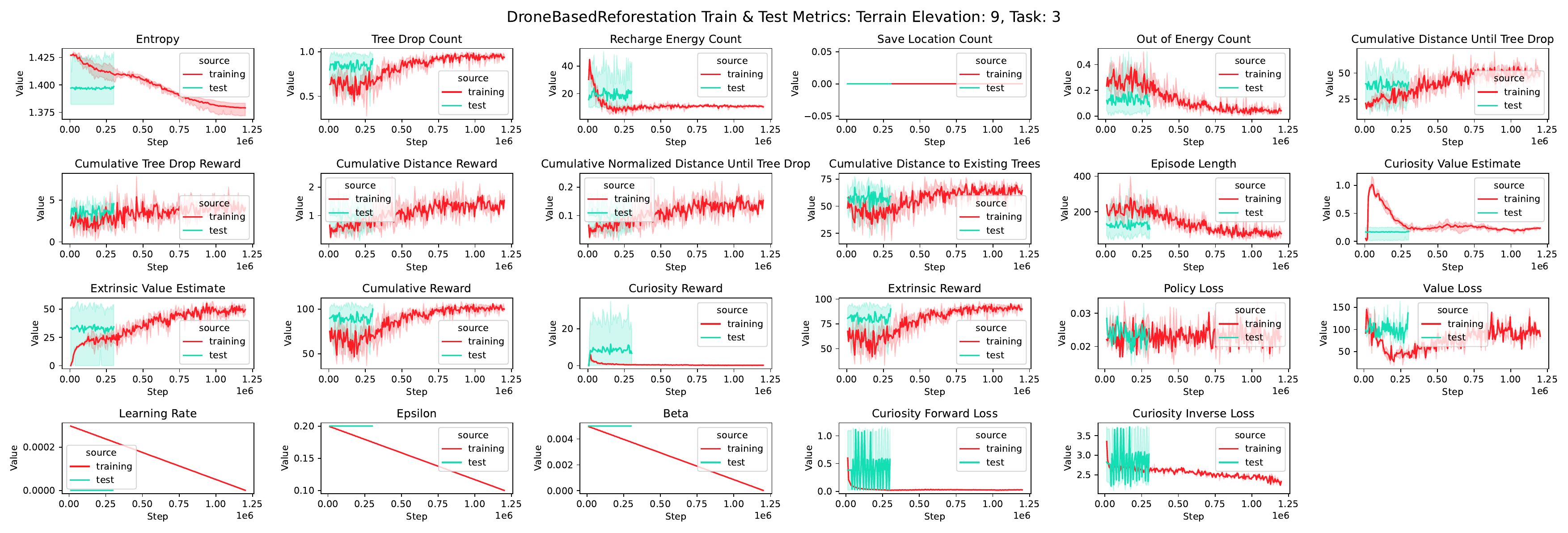}
\vspace{-0.6cm}
\caption{Drone-Based Reforestation: Train \& Test Metrics: Terrain Elevation 9, Task 3.}
\end{figure}

\clearpage

\begin{figure}[h!]
\centering
\includegraphics[width=\linewidth]{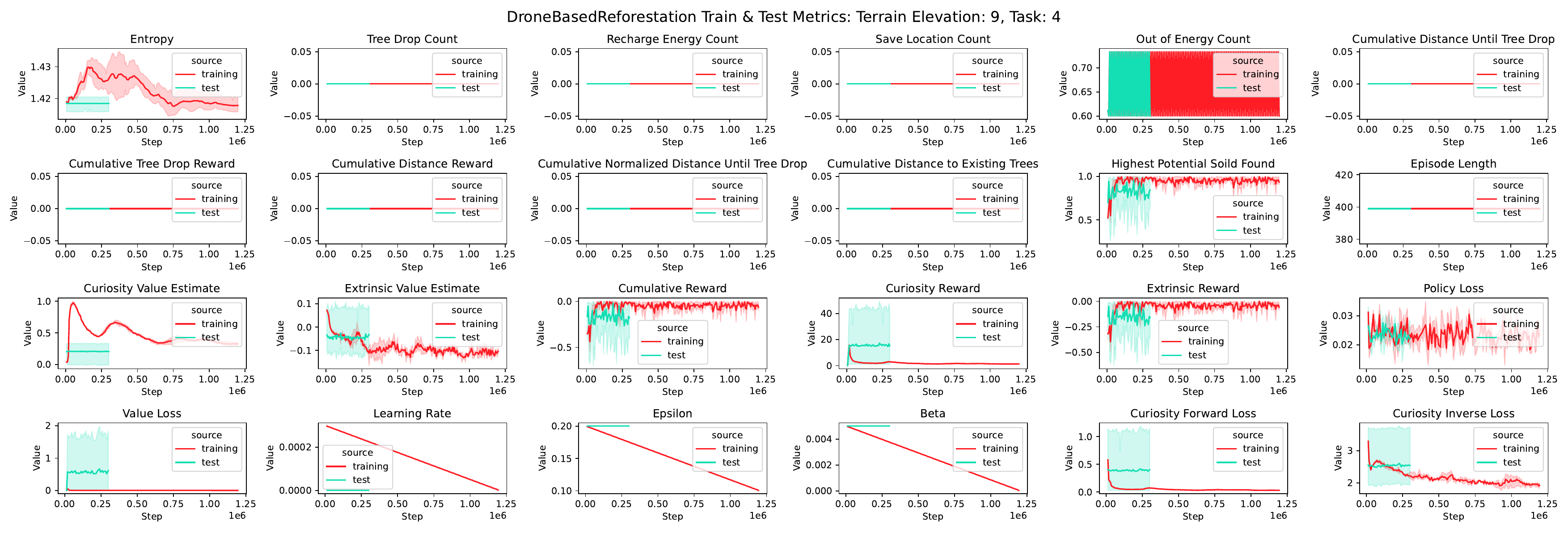}
\vspace{-0.6cm}
\caption{Drone-Based Reforestation: Train \& Test Metrics: Terrain Elevation 9, Task 4.}
\end{figure}

\begin{figure}[h!]
\centering
\includegraphics[width=\linewidth]{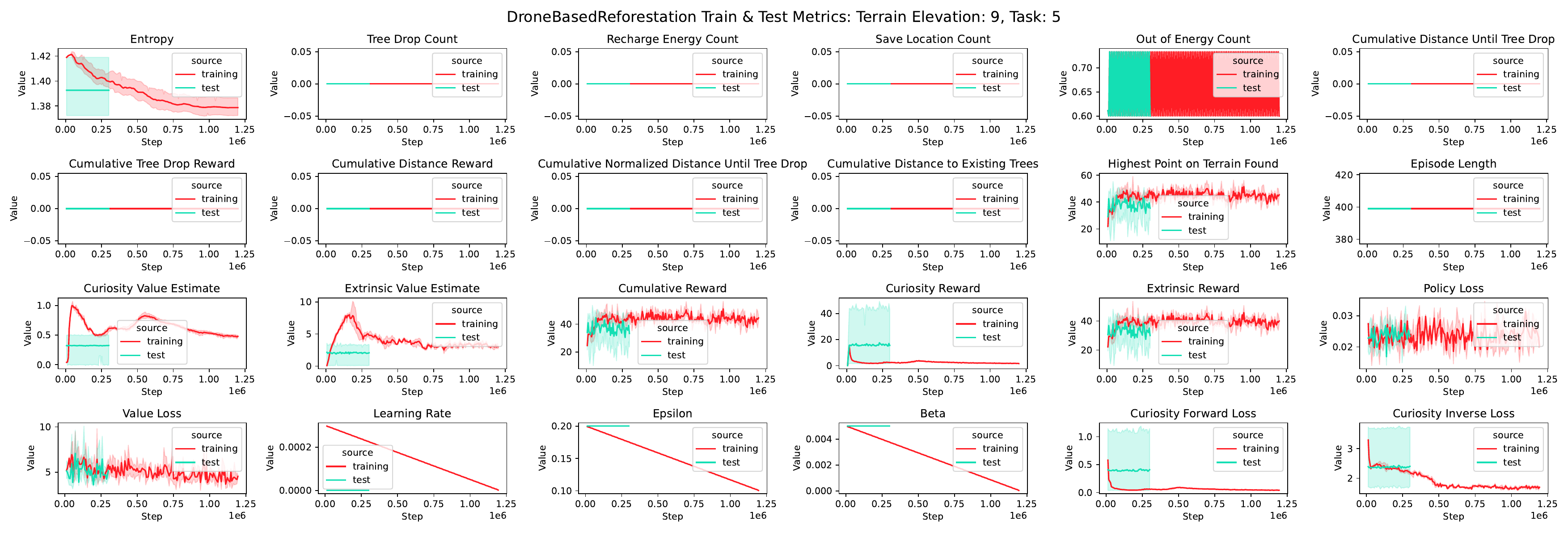}
\vspace{-0.6cm}
\caption{Drone-Based Reforestation: Train \& Test Metrics: Terrain Elevation 9, Task 5.}
\end{figure}

\begin{figure}[h!]
\centering
\includegraphics[width=\linewidth]{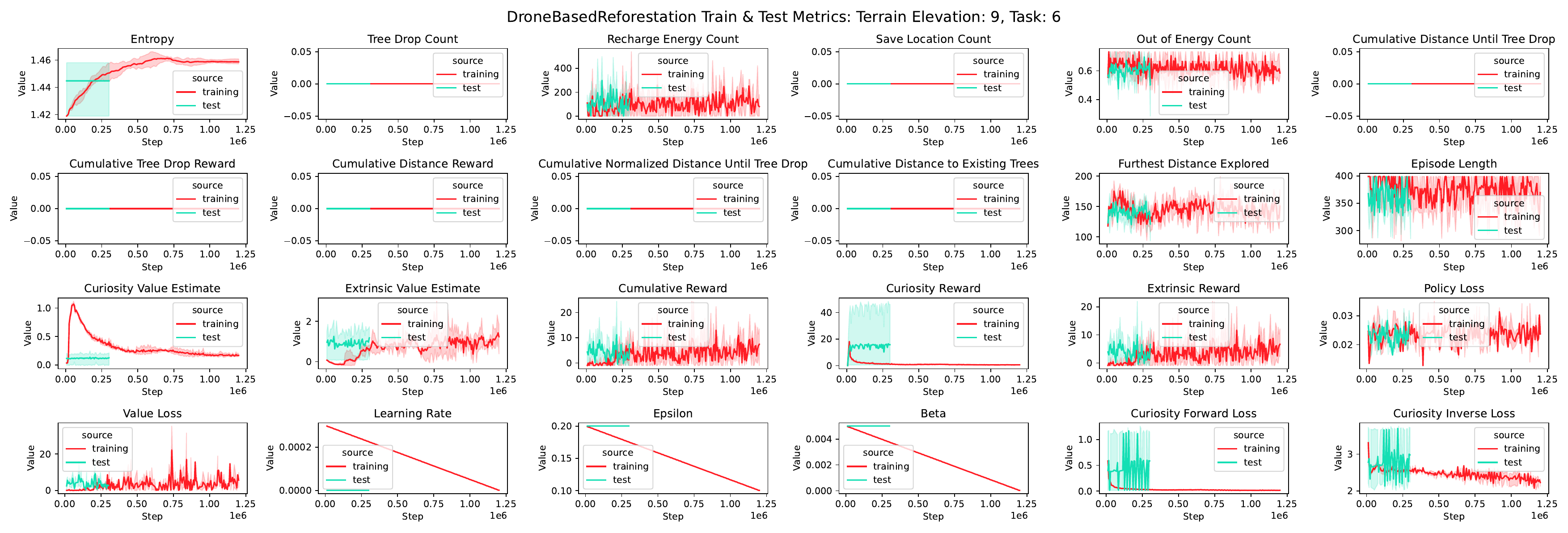}
\vspace{-0.6cm}
\caption{Drone-Based Reforestation: Train \& Test Metrics: Terrain Elevation 9, Task 6.}
\end{figure}

\clearpage

\begin{figure}[h!]
\centering
\includegraphics[width=\linewidth]{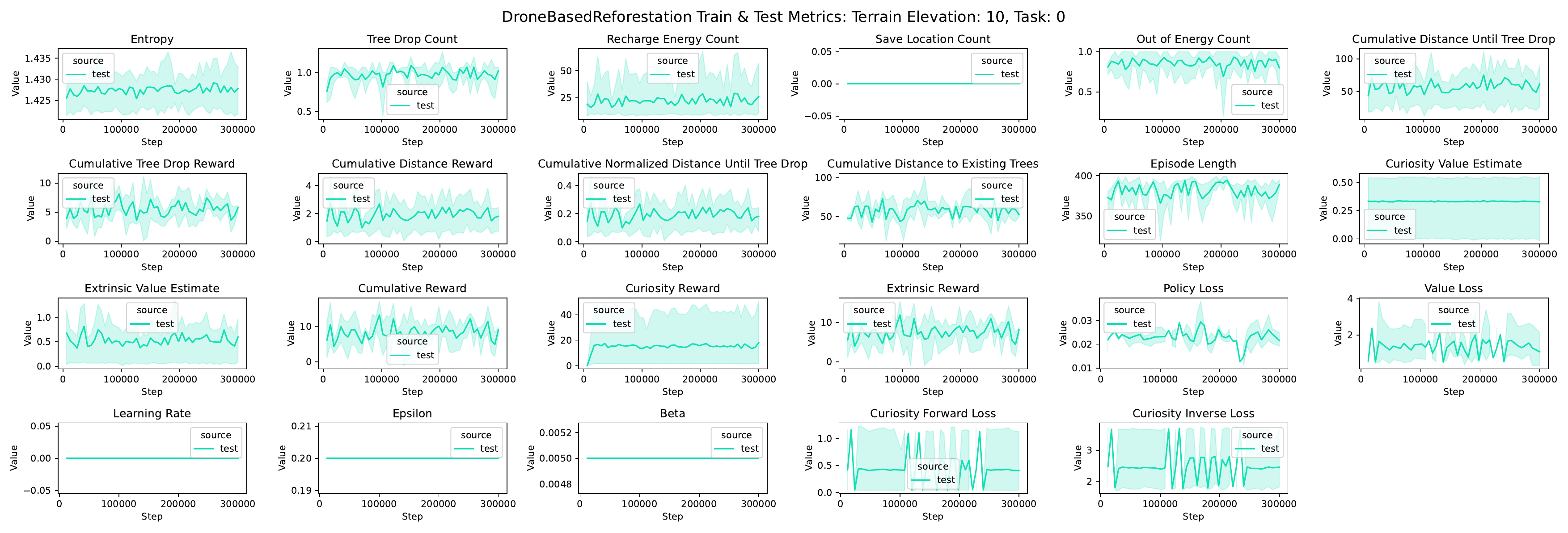}
\vspace{-0.6cm}
\caption{Drone-Based Reforestation: Train \& Test Metrics: Terrain Elevation 10, Task 0.}
\end{figure}

\begin{figure}[h!]
\centering
\includegraphics[width=\linewidth]{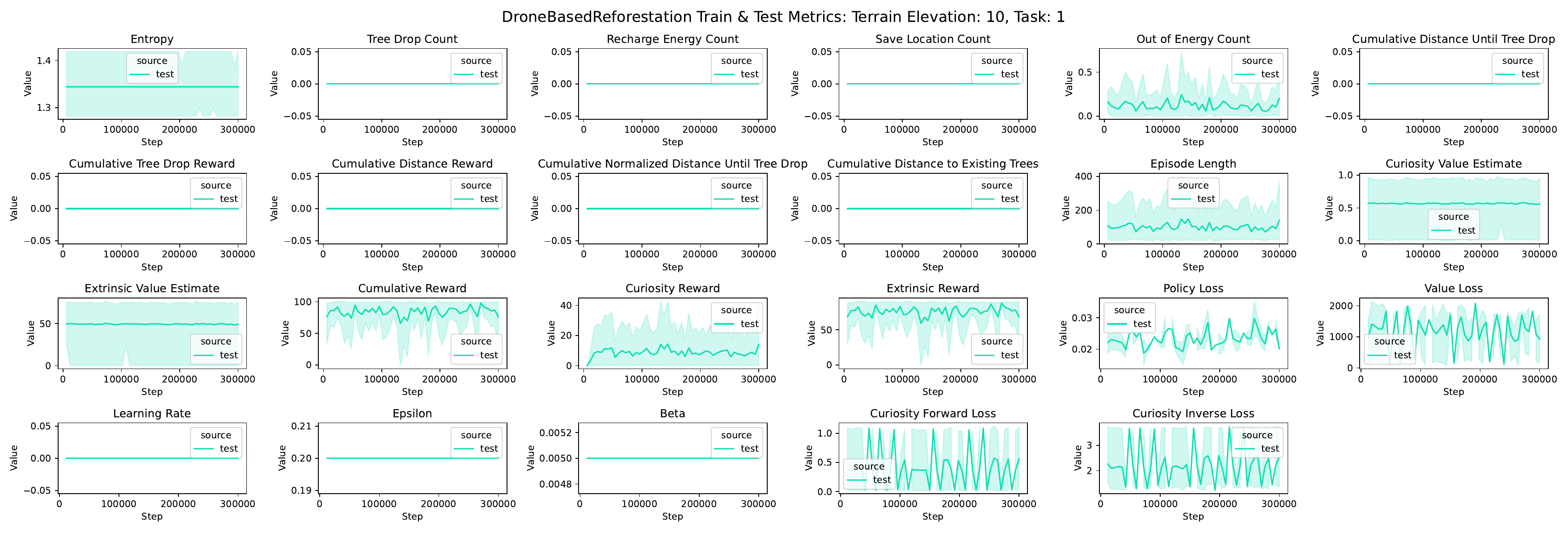}
\vspace{-0.6cm}
\caption{Drone-Based Reforestation: Train \& Test Metrics: Terrain Elevation 10, Task 1.}
\end{figure}

\begin{figure}[h!]
\centering
\includegraphics[width=\linewidth]{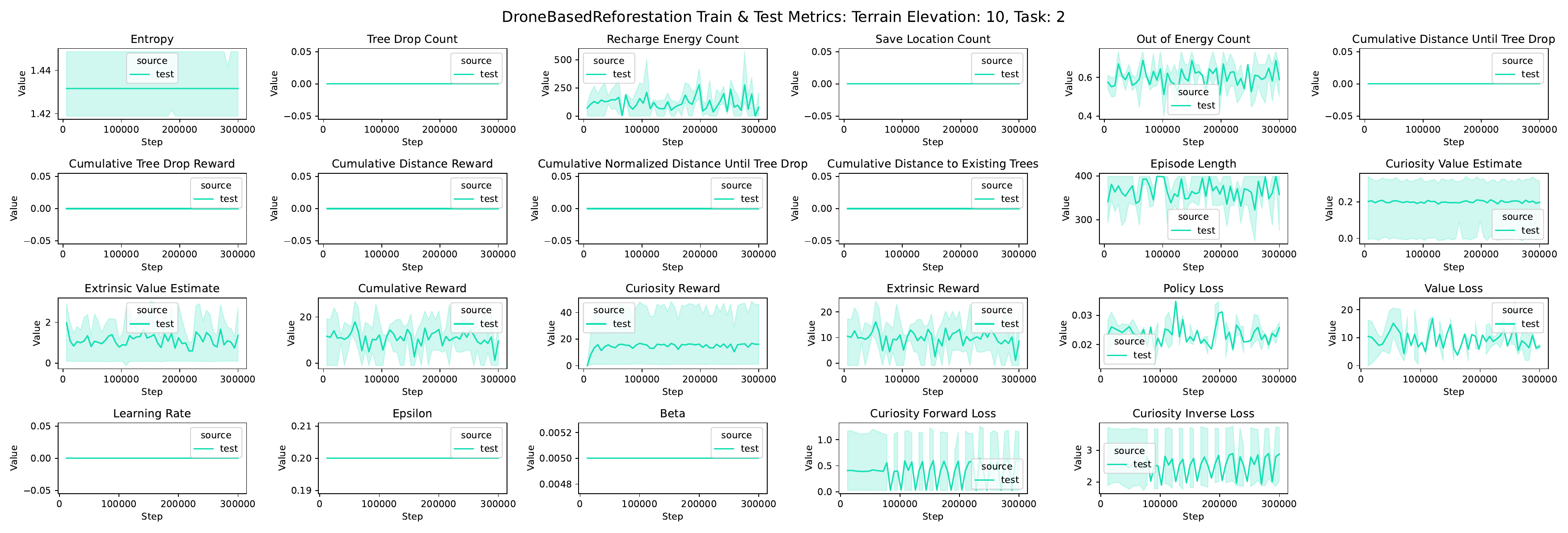}
\vspace{-0.6cm}
\caption{Drone-Based Reforestation: Train \& Test Metrics: Terrain Elevation 10, Task 2.}
\end{figure}

\clearpage

\begin{figure}[h!]
\centering
\includegraphics[width=\linewidth]{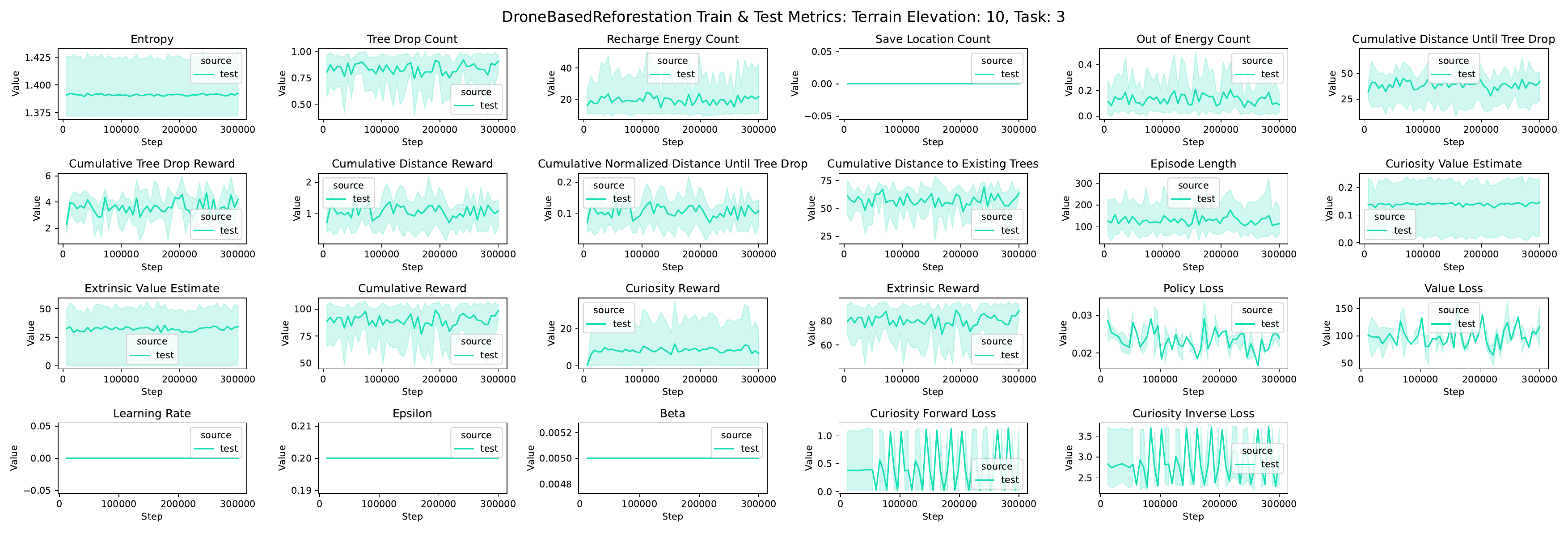}
\vspace{-0.6cm}
\caption{Drone-Based Reforestation: Train \& Test Metrics: Terrain Elevation 10, Task 3.}
\end{figure}

\begin{figure}[h!]
\centering
\includegraphics[width=\linewidth]{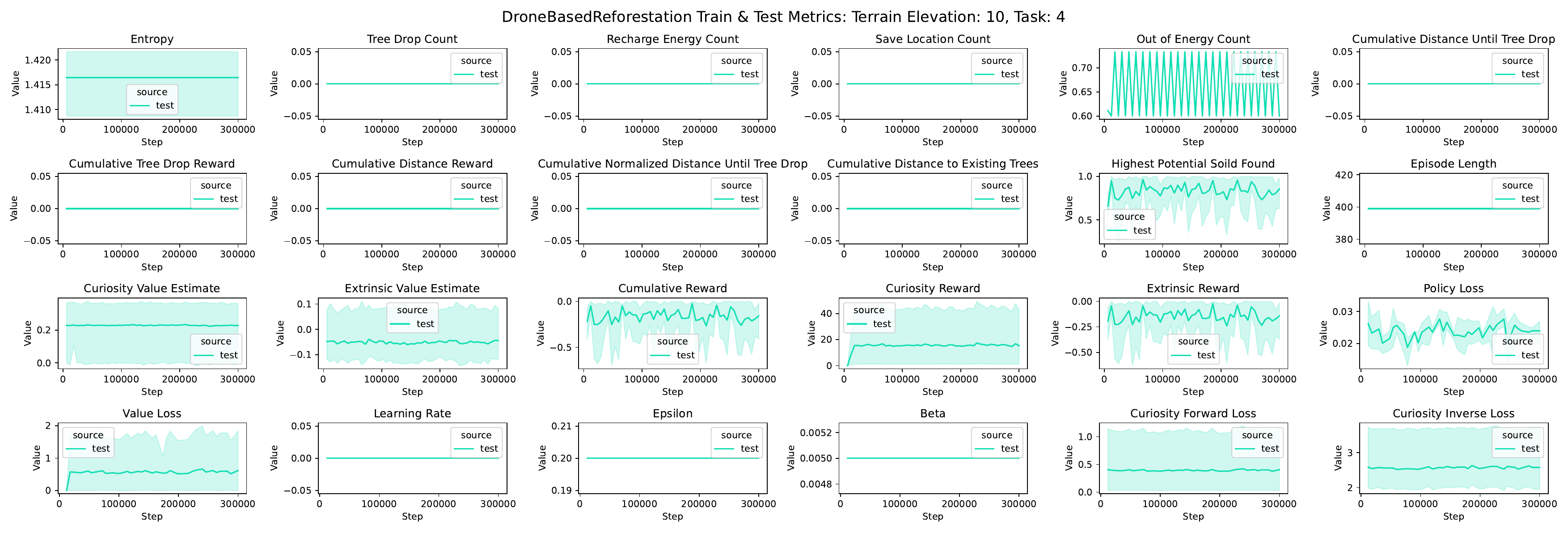}
\vspace{-0.6cm}
\caption{Drone-Based Reforestation: Train \& Test Metrics: Terrain Elevation 10, Task 4.}
\end{figure}

\begin{figure}[h!]
\centering
\includegraphics[width=\linewidth]{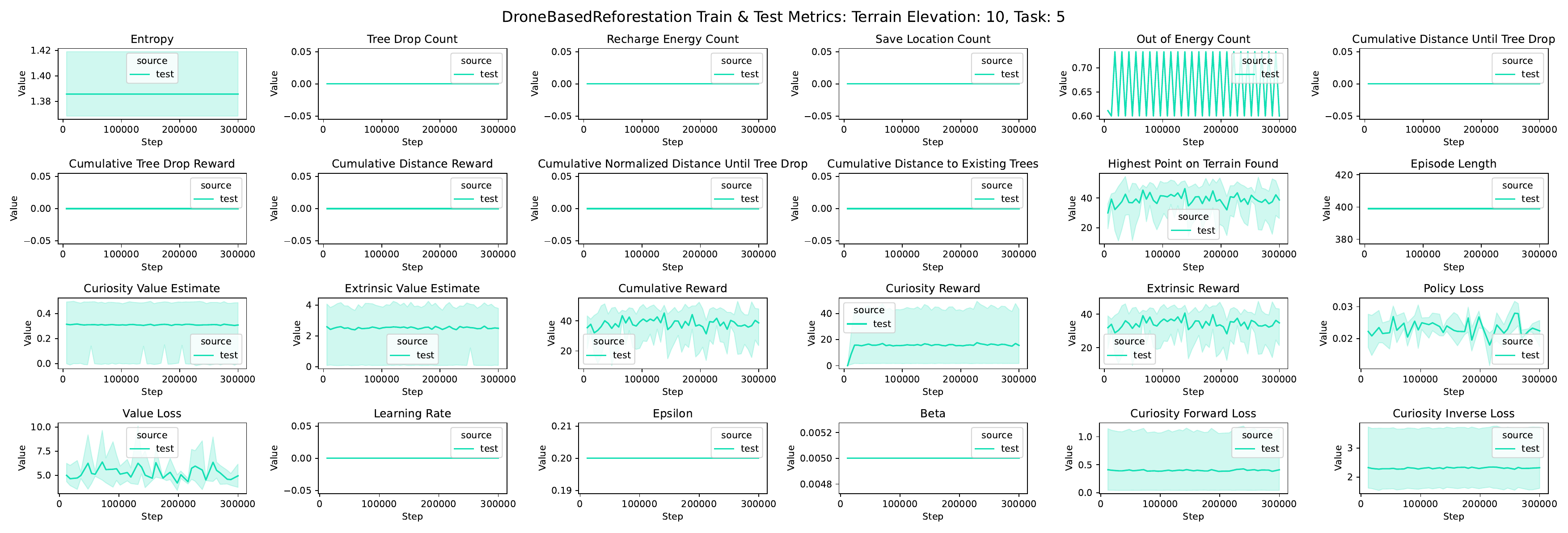}
\vspace{-0.6cm}
\caption{Drone-Based Reforestation: Train \& Test Metrics: Terrain Elevation 10, Task 5.}
\end{figure}

\clearpage

\begin{figure}[h!]
\centering
\includegraphics[width=\linewidth]{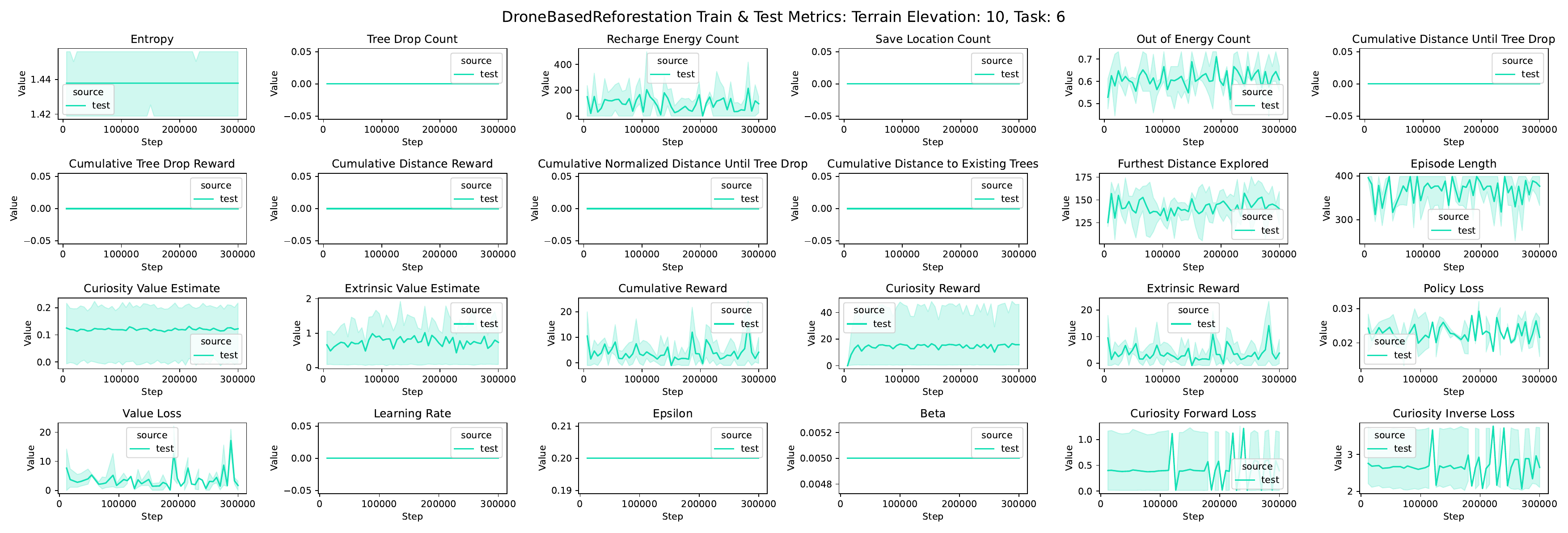}
\vspace{-0.6cm}
\caption{Drone-Based Reforestation: Train \& Test Metrics: Terrain Elevation 10, Task 6.}
\end{figure}

\clearpage

\subsubsection{Drone-Based Reforestation: Average Test Metric - Task VS Pattern}

\begin{figure}[h!]
\centering
\includegraphics[width=\linewidth]{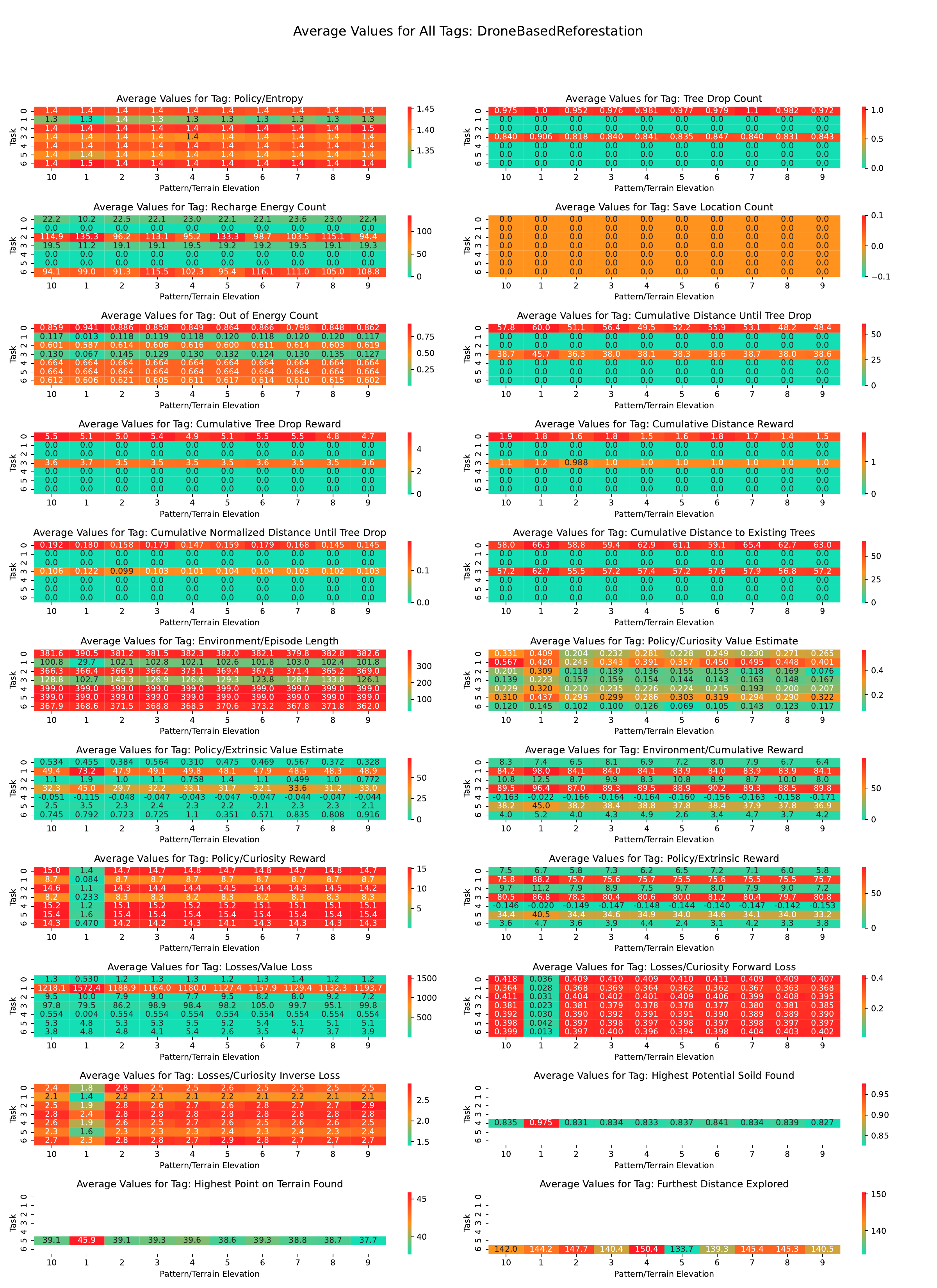}
\caption{Drone-Based Reforestation: Average Train \& Test Metrics.}
\end{figure}

\clearpage

\subsubsection{Aerial Wildfire Suppression: Train \& Test Metrics}
\begin{figure}[h!]
\centering
\includegraphics[width=\linewidth]{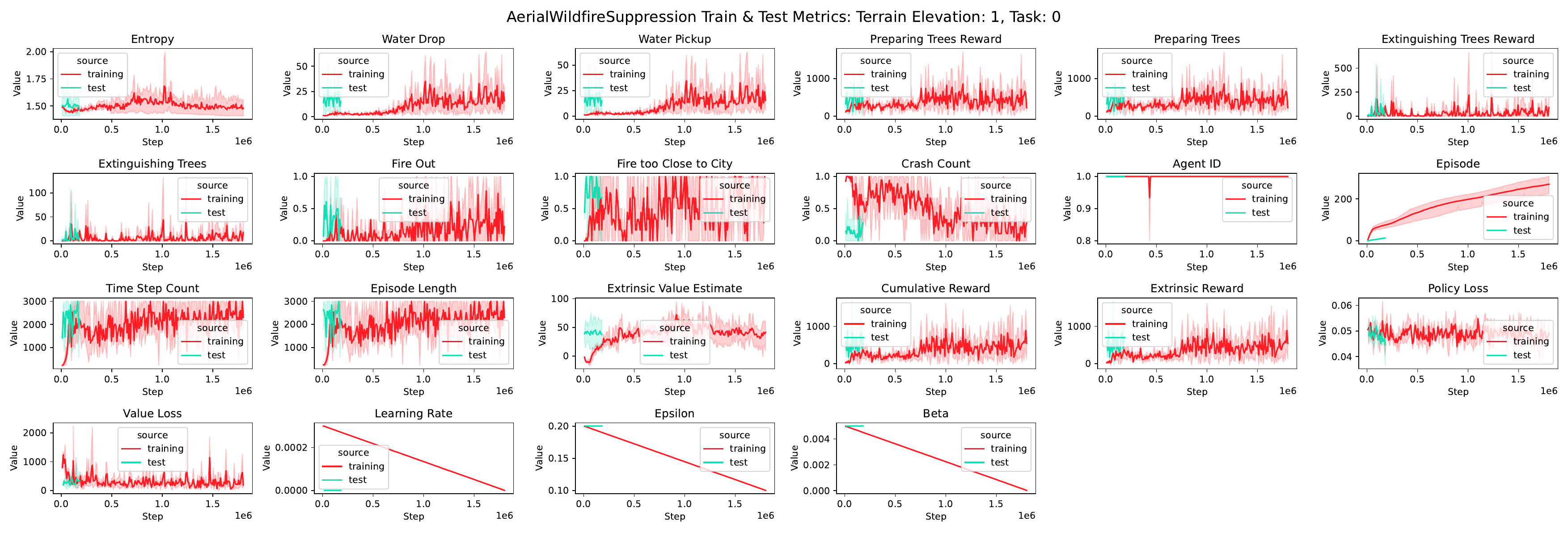}
\vspace{-0.6cm}
\caption{Aerial Wildfire Suppression: Train \& Test Metrics: Terrain Elevation 1, Task 0.}
\end{figure}

\begin{figure}[h!]
\centering
\includegraphics[width=\linewidth]{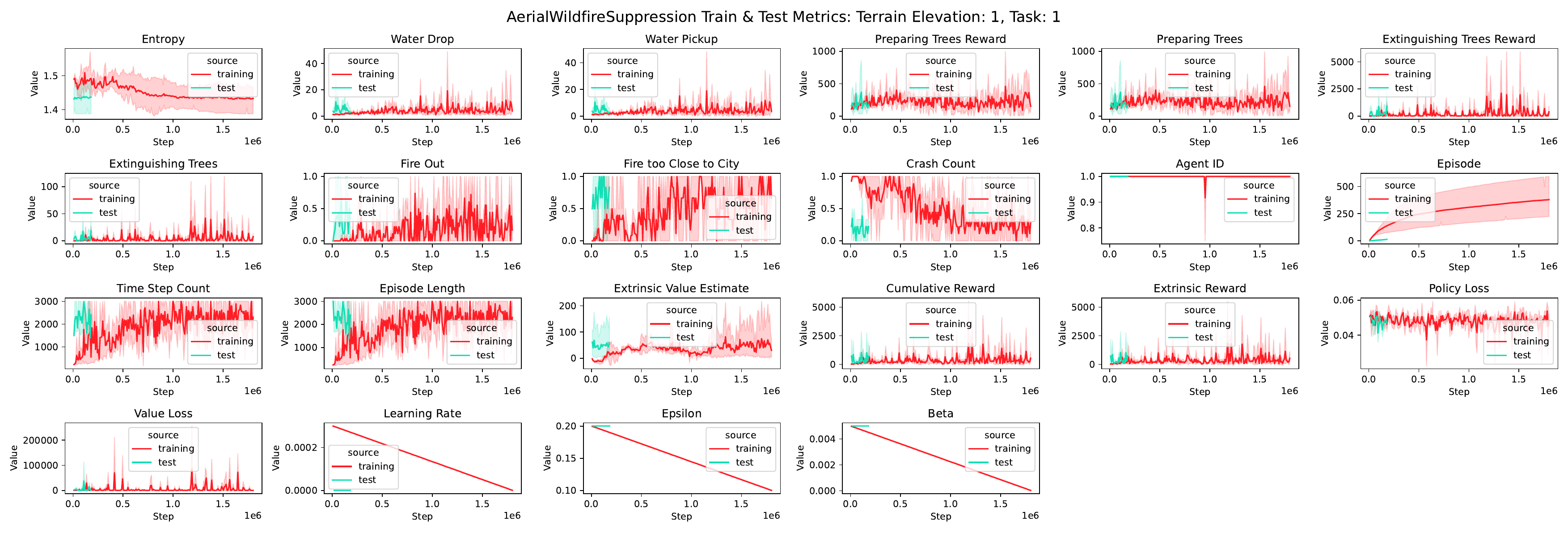}
\vspace{-0.6cm}
\caption{Aerial Wildfire Suppression: Train \& Test Metrics: Terrain Elevation 1, Task 1.}
\end{figure}

\begin{figure}[h!]
\centering
\includegraphics[width=\linewidth]{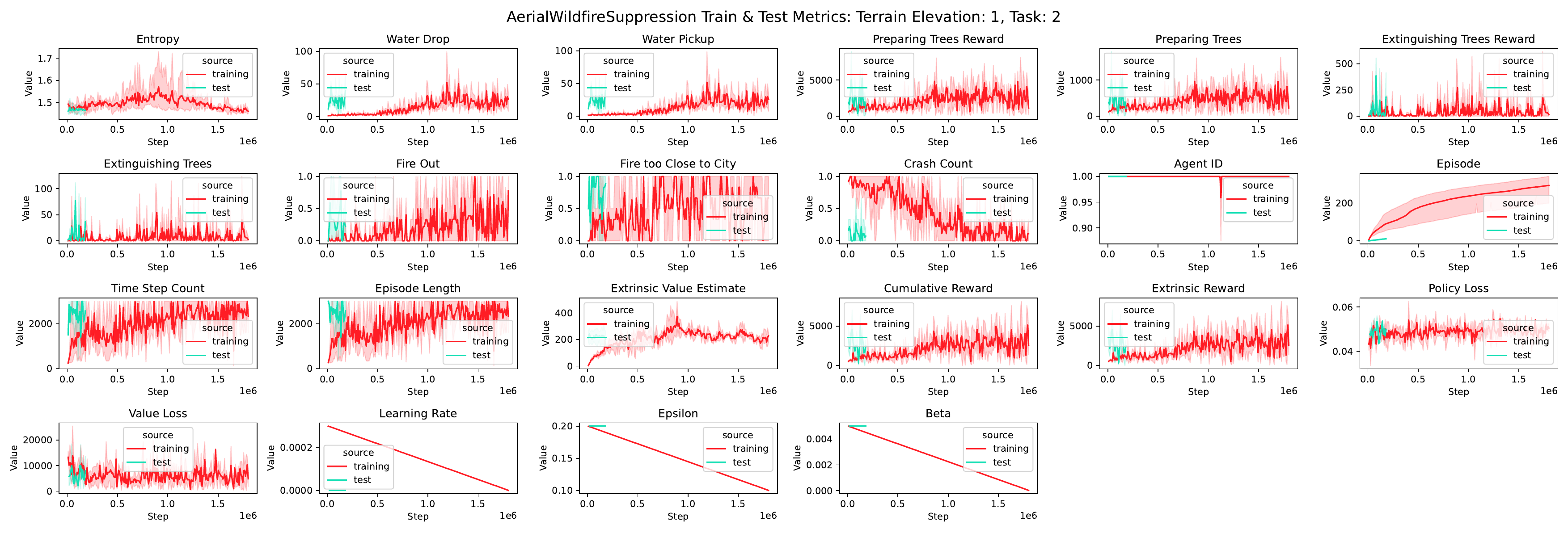}
\vspace{-0.6cm}
\caption{Aerial Wildfire Suppression: Train \& Test Metrics: Terrain Elevation 1, Task 2.}
\end{figure}

\clearpage

\begin{figure}[h!]
\centering
\includegraphics[width=\linewidth]{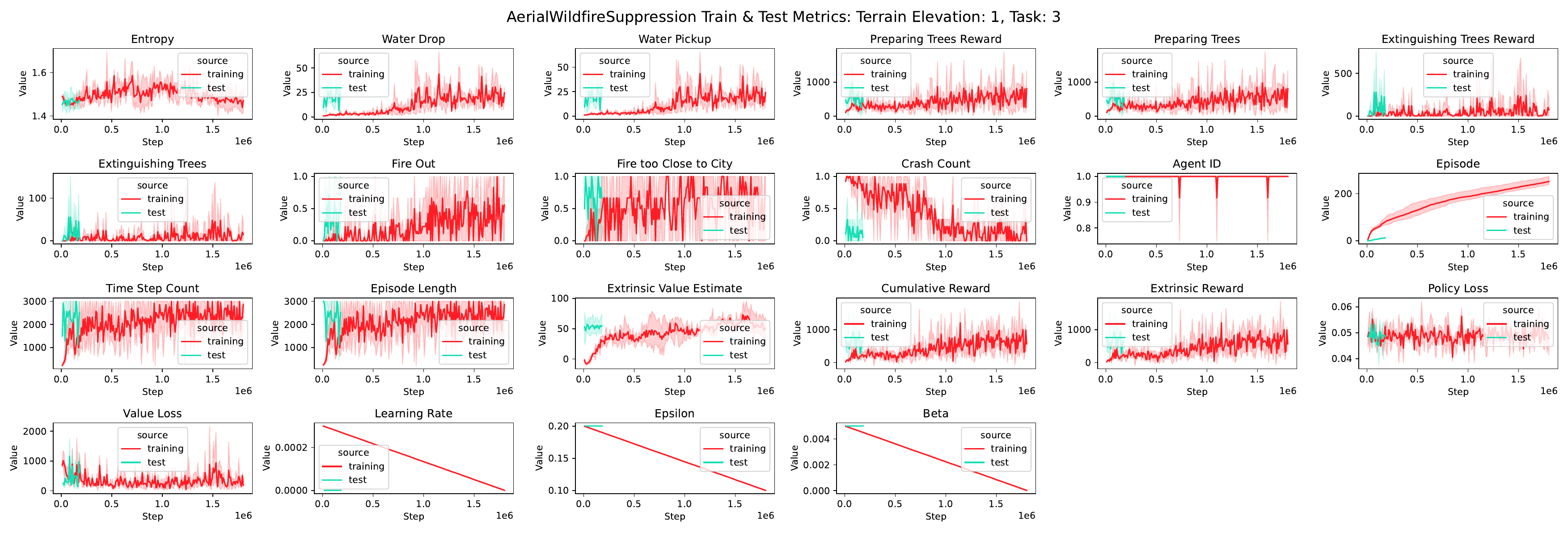}
\vspace{-0.6cm}
\caption{Aerial Wildfire Suppression: Train \& Test Metrics: Terrain Elevation 1, Task 3.}
\end{figure}

\begin{figure}[h!]
\centering
\includegraphics[width=\linewidth]{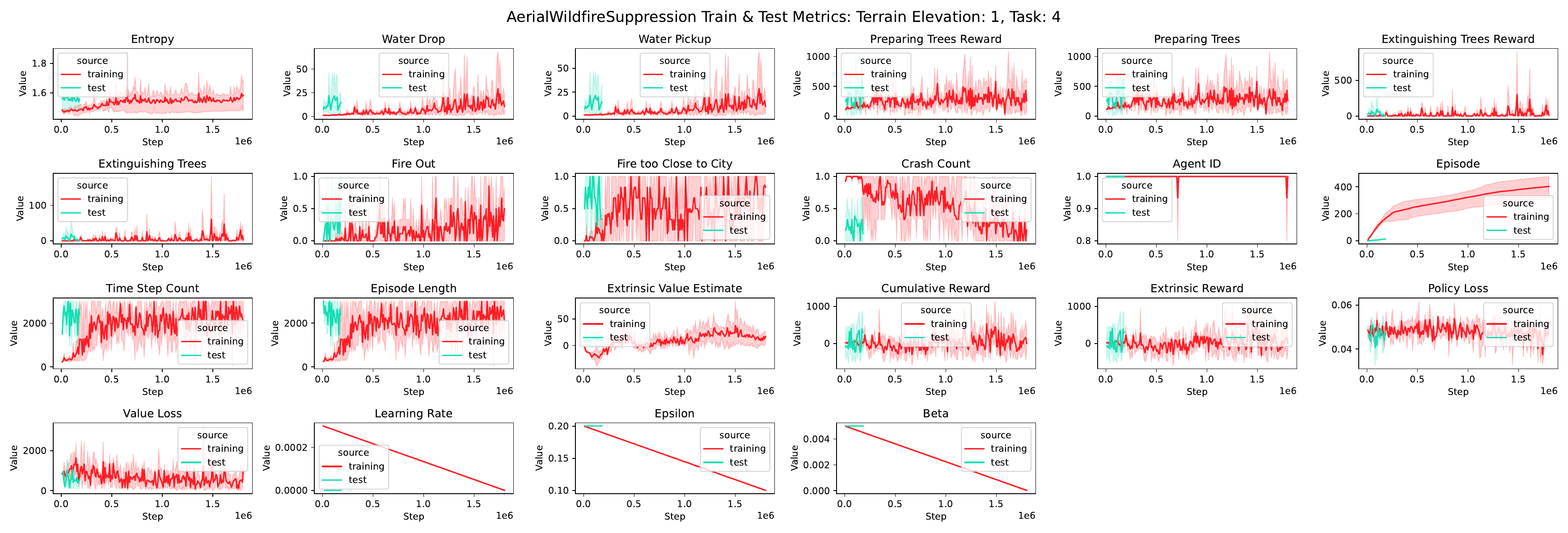}
\vspace{-0.6cm}
\caption{Aerial Wildfire Suppression: Train \& Test Metrics: Terrain Elevation 1, Task 4.}
\end{figure}

\begin{figure}[h!]
\centering
\includegraphics[width=\linewidth]{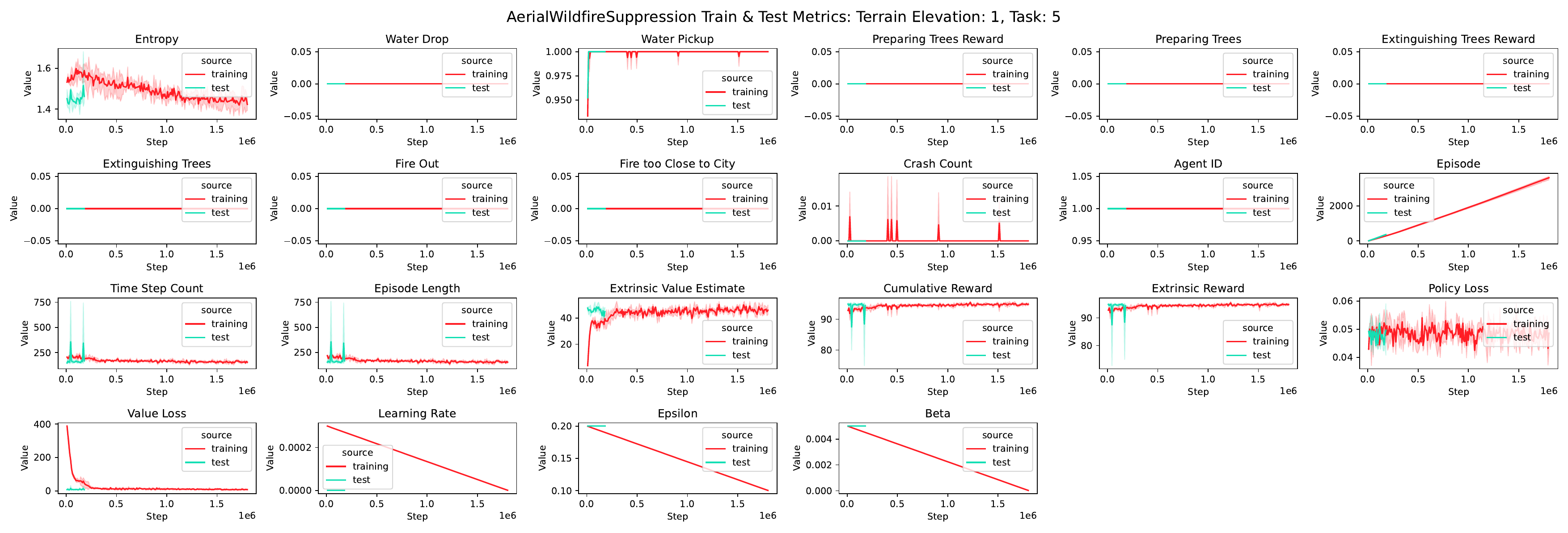}
\vspace{-0.6cm}
\caption{Aerial Wildfire Suppression: Train \& Test Metrics: Terrain Elevation 1, Task 5.}
\end{figure}

\clearpage

\begin{figure}[h!]
\centering
\includegraphics[width=\linewidth]{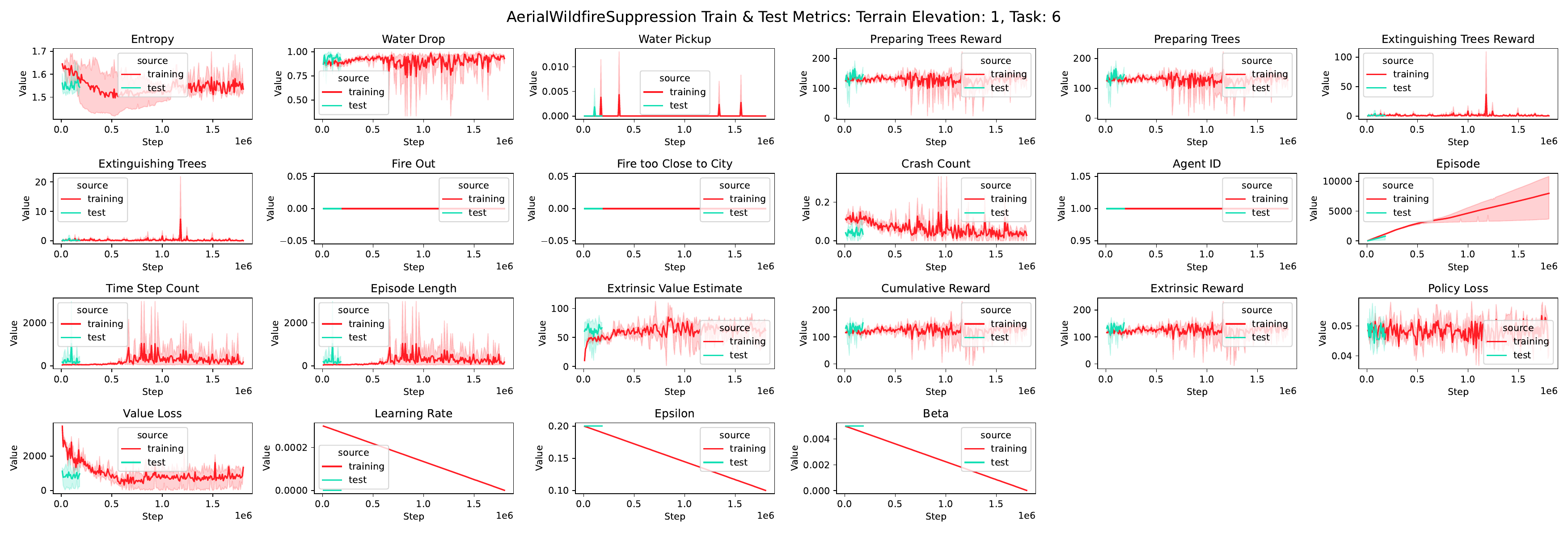}
\vspace{-0.6cm}
\caption{Aerial Wildfire Suppression: Train \& Test Metrics: Terrain Elevation 1, Task 6.}
\end{figure}

\begin{figure}[h!]
\centering
\includegraphics[width=\linewidth]{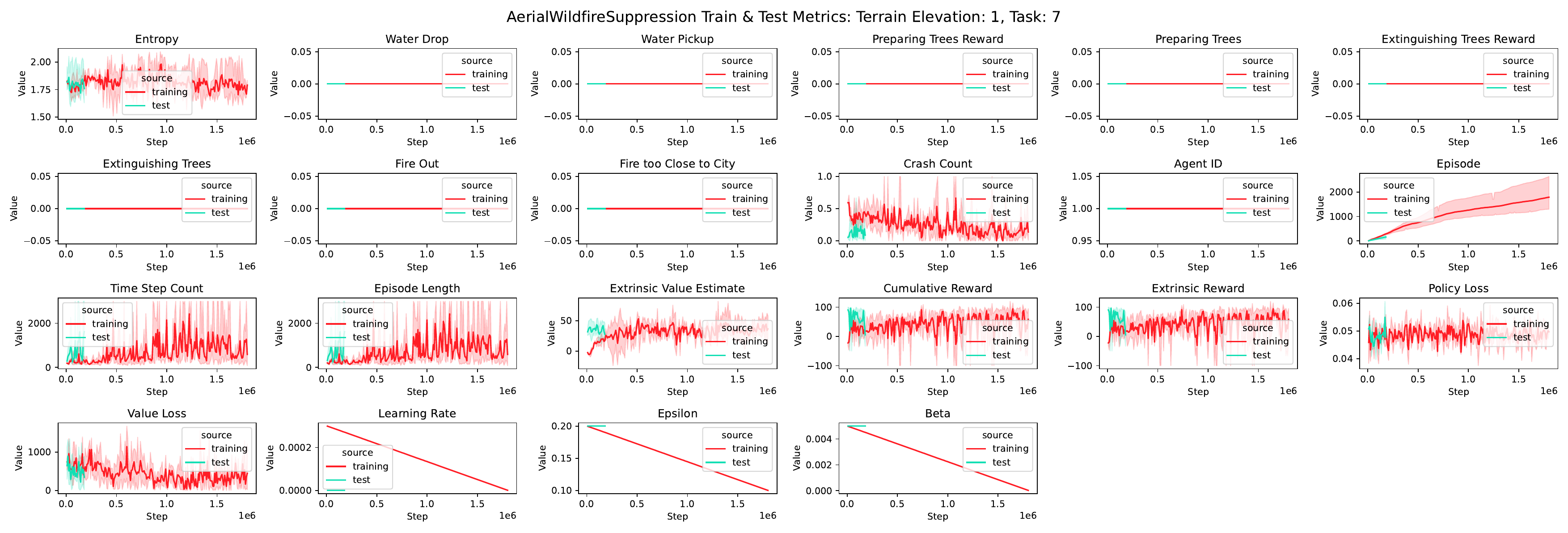}
\vspace{-0.6cm}
\caption{Aerial Wildfire Suppression: Train \& Test Metrics: Terrain Elevation 1, Task 7.}
\end{figure}

\begin{figure}[h!]
\centering
\includegraphics[width=\linewidth]{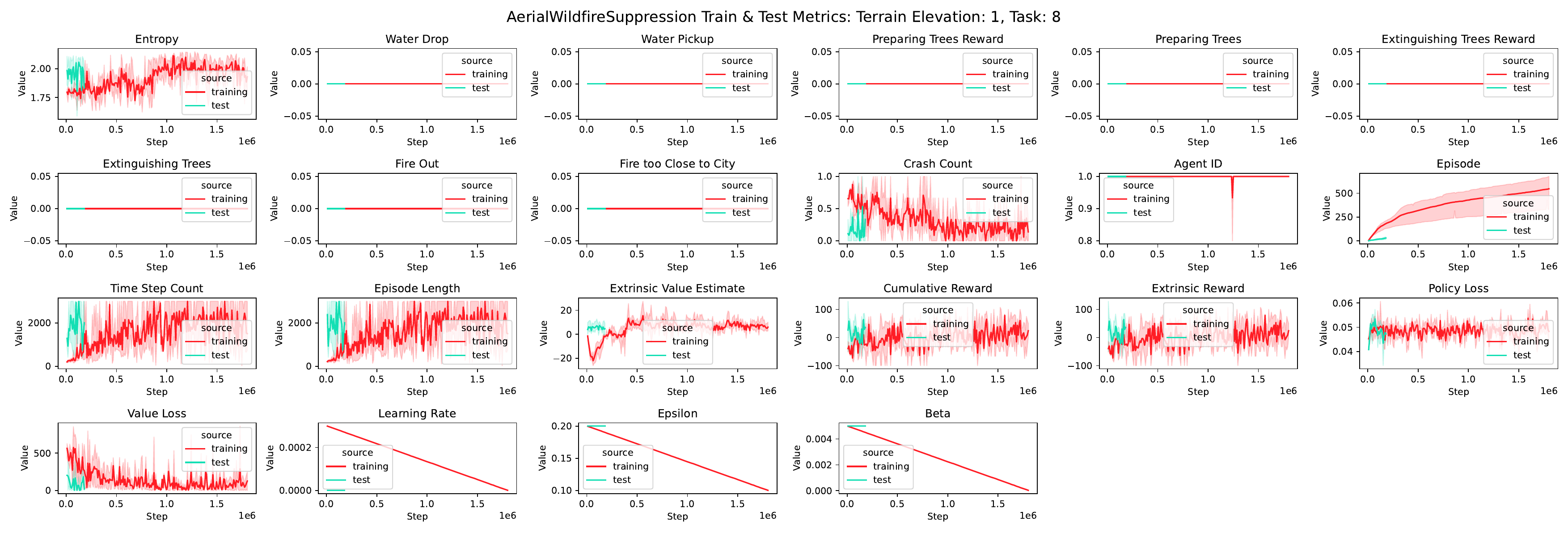}
\vspace{-0.6cm}
\caption{Aerial Wildfire Suppression: Train \& Test Metrics: Terrain Elevation 1, Task 8.}
\end{figure}

\clearpage

\begin{figure}[h!]
\centering
\includegraphics[width=\linewidth]{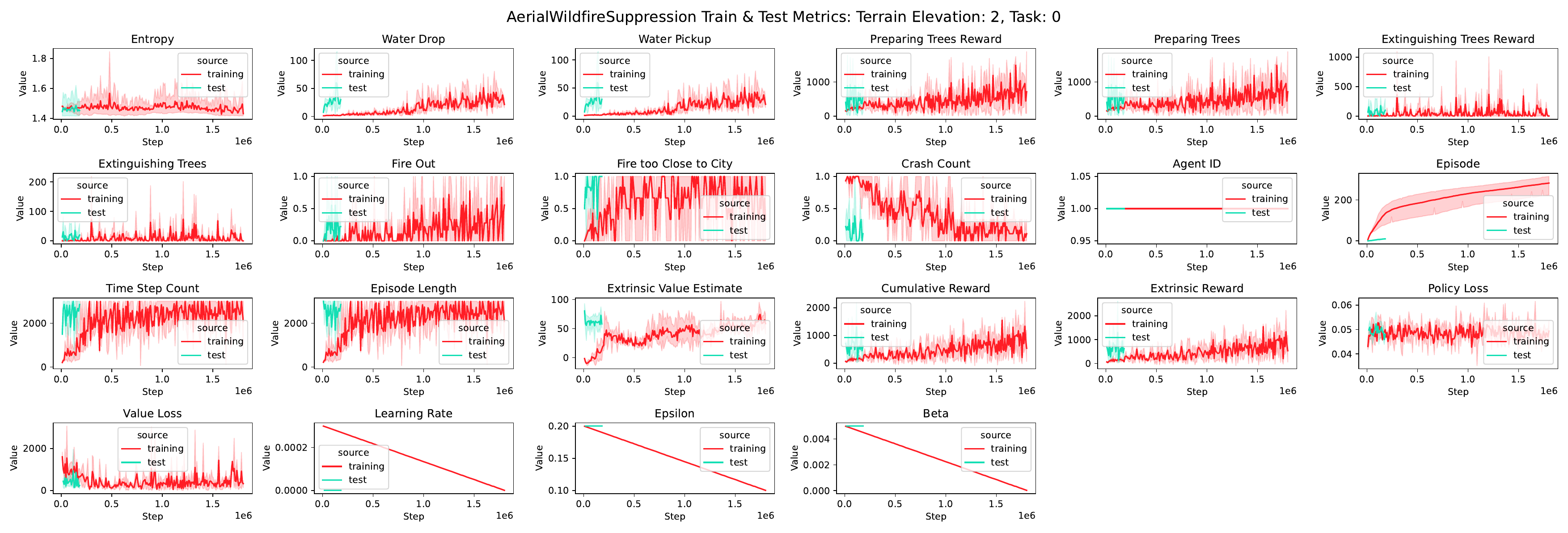}
\vspace{-0.6cm}
\caption{Aerial Wildfire Suppression: Train \& Test Metrics: Terrain Elevation 2, Task 0.}
\end{figure}

\begin{figure}[h!]
\centering
\includegraphics[width=\linewidth]{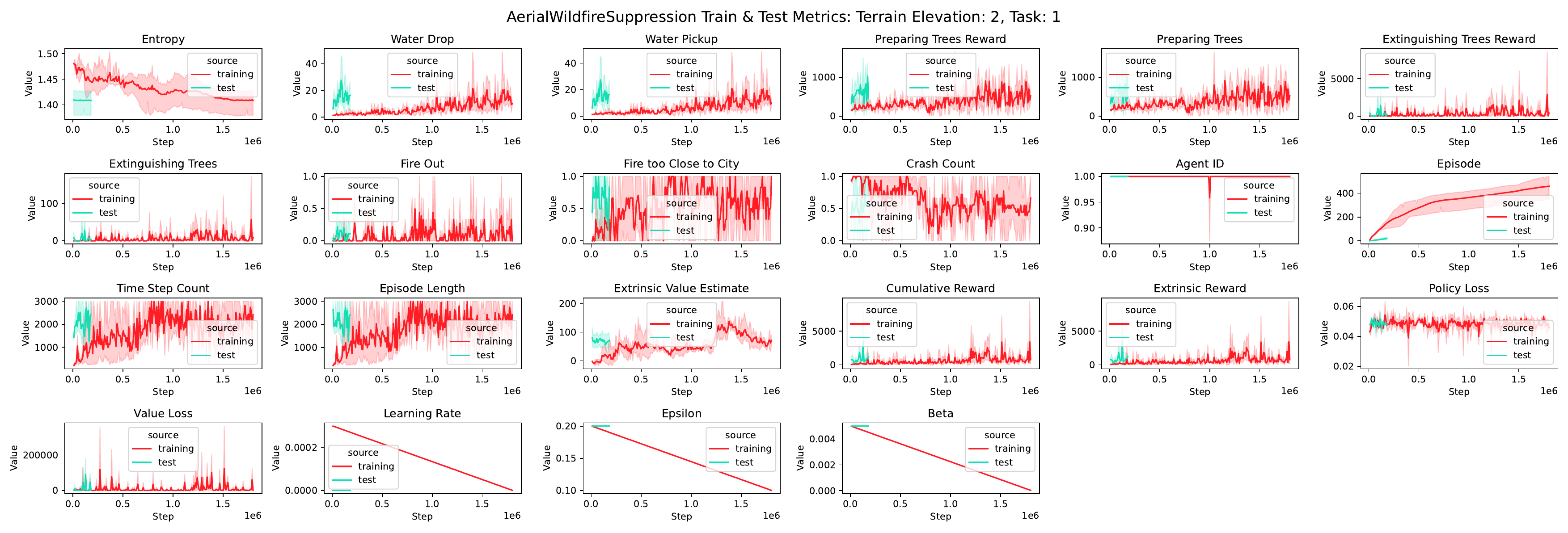}
\vspace{-0.6cm}
\caption{Aerial Wildfire Suppression: Train \& Test Metrics: Terrain Elevation 2, Task 1.}
\end{figure}

\begin{figure}[h!]
\centering
\includegraphics[width=\linewidth]{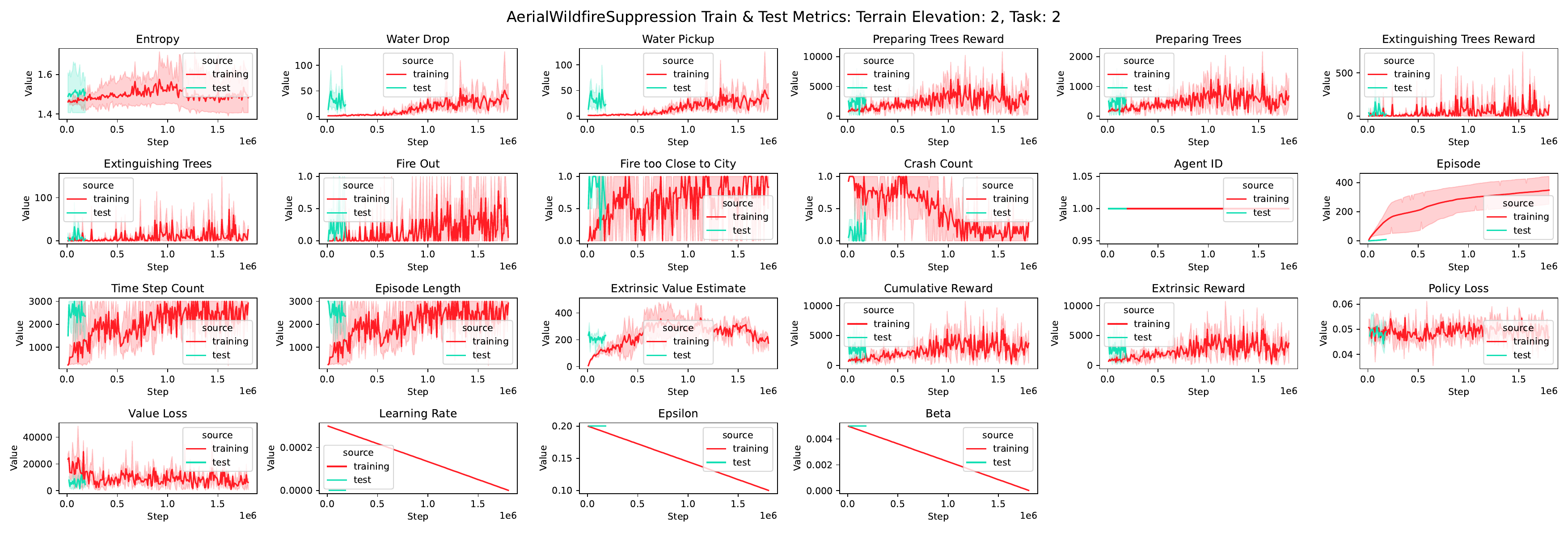}
\vspace{-0.6cm}
\caption{Aerial Wildfire Suppression: Train \& Test Metrics: Terrain Elevation 2, Task 2.}
\end{figure}

\clearpage

\begin{figure}[h!]
\centering
\includegraphics[width=\linewidth]{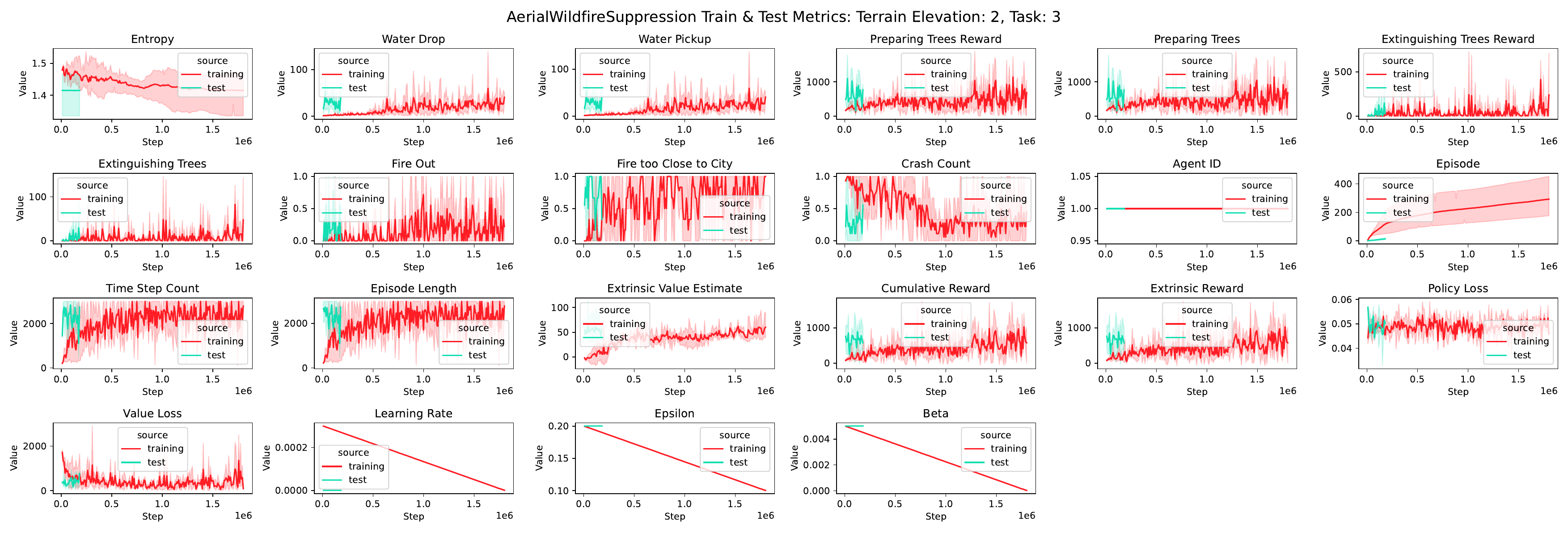}
\vspace{-0.6cm}
\caption{Aerial Wildfire Suppression: Train \& Test Metrics: Terrain Elevation 2, Task 3.}
\end{figure}

\begin{figure}[h!]
\centering
\includegraphics[width=\linewidth]{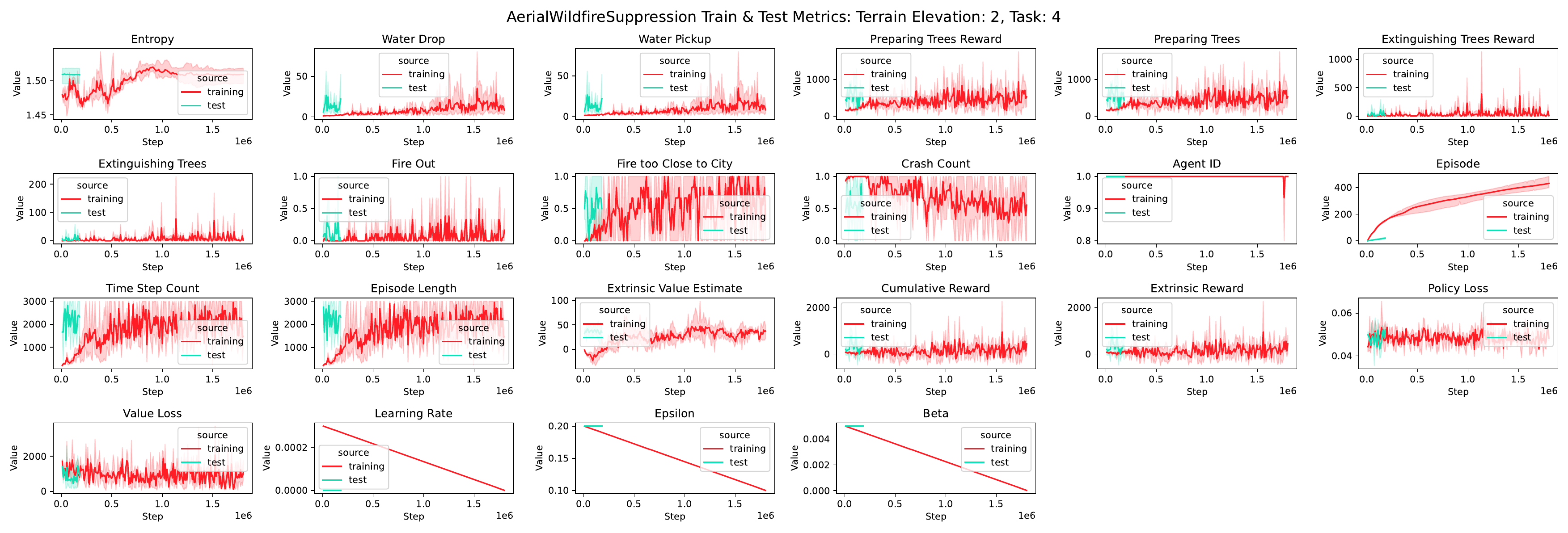}
\vspace{-0.6cm}
\caption{Aerial Wildfire Suppression: Train \& Test Metrics: Terrain Elevation 2, Task 4.}
\end{figure}

\begin{figure}[h!]
\centering
\includegraphics[width=\linewidth]{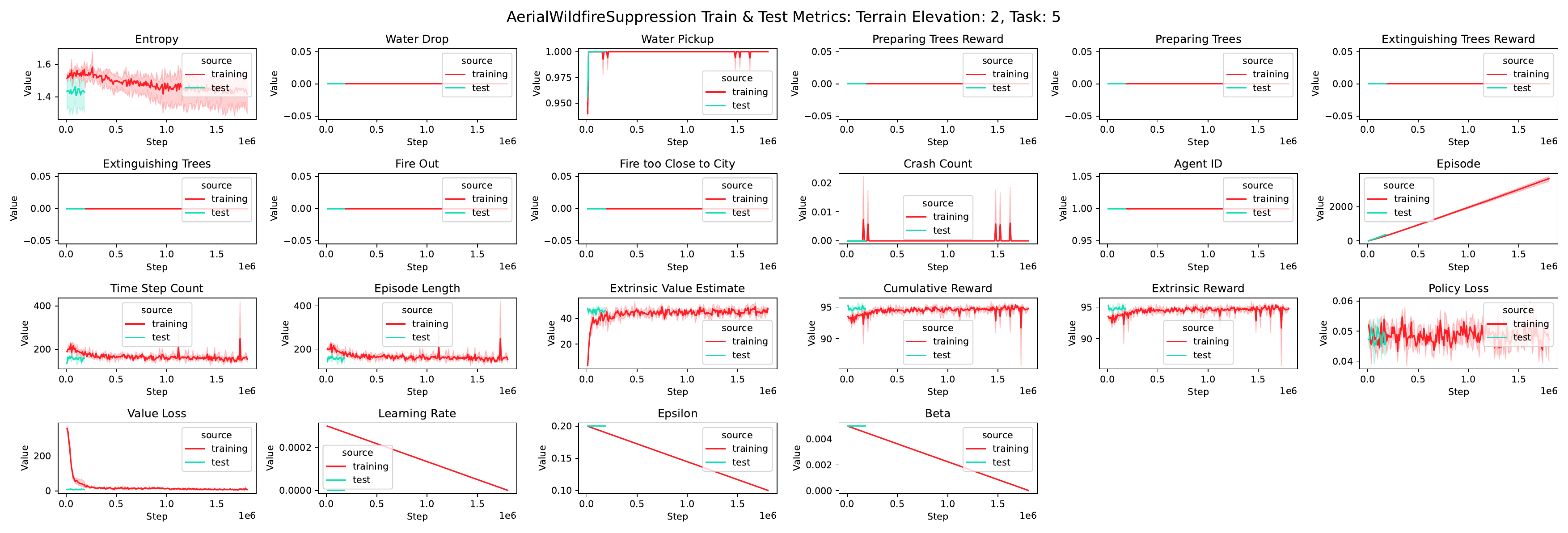}
\vspace{-0.6cm}
\caption{Aerial Wildfire Suppression: Train \& Test Metrics: Terrain Elevation 2, Task 5.}
\end{figure}

\clearpage

\begin{figure}[h!]
\centering
\includegraphics[width=\linewidth]{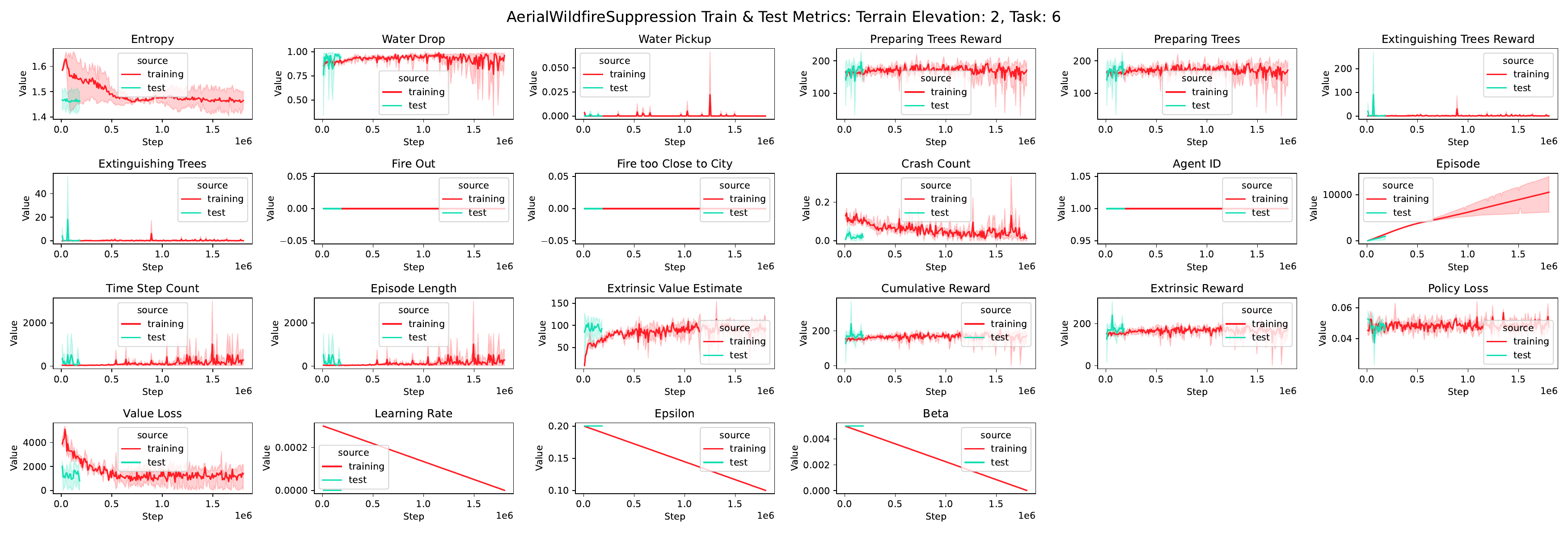}
\vspace{-0.6cm}
\caption{Aerial Wildfire Suppression: Train \& Test Metrics: Terrain Elevation 2, Task 6.}
\end{figure}

\begin{figure}[h!]
\centering
\includegraphics[width=\linewidth]{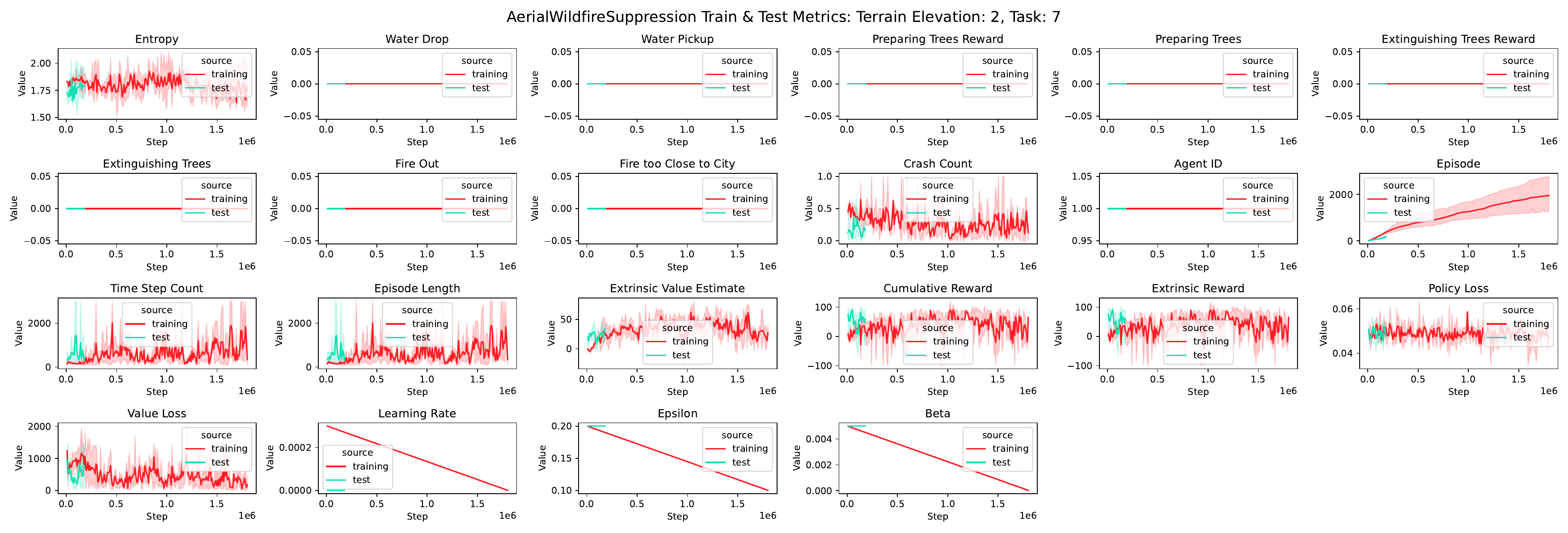}
\vspace{-0.6cm}
\caption{Aerial Wildfire Suppression: Train \& Test Metrics: Terrain Elevation 2, Task 7.}
\end{figure}

\begin{figure}[h!]
\centering
\includegraphics[width=\linewidth]{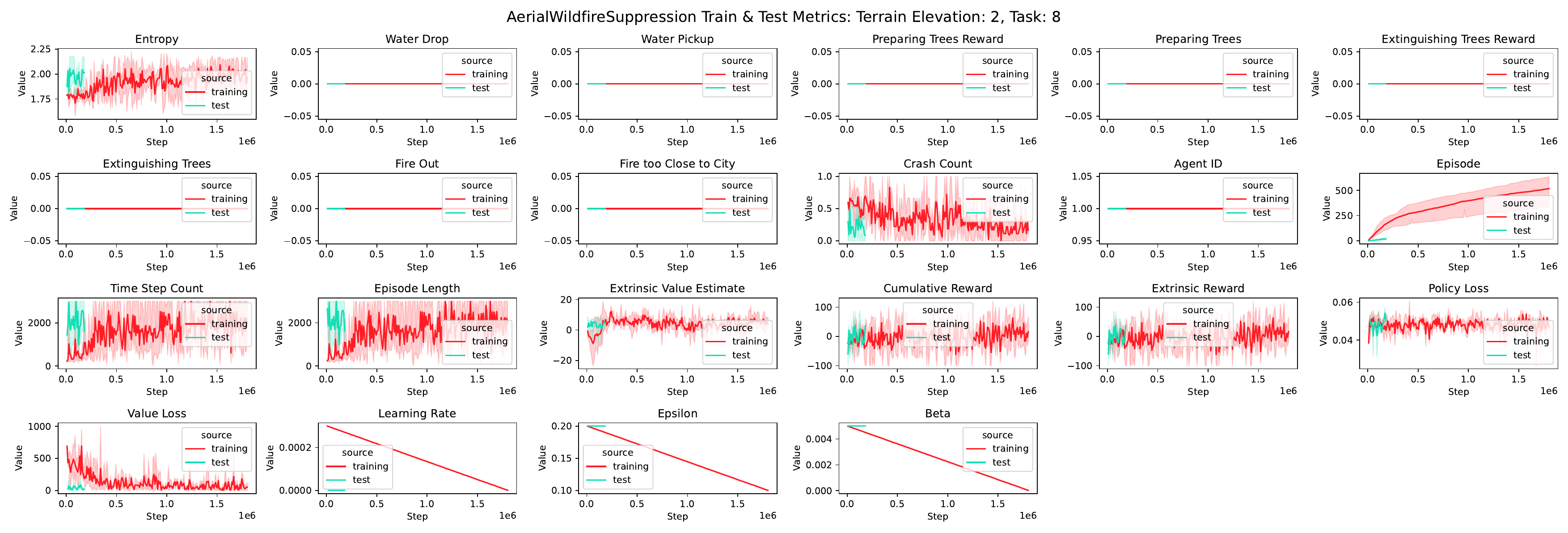}
\vspace{-0.6cm}
\caption{Aerial Wildfire Suppression: Train \& Test Metrics: Terrain Elevation 2, Task 8.}
\end{figure}

\clearpage

\begin{figure}[h!]
\centering
\includegraphics[width=\linewidth]{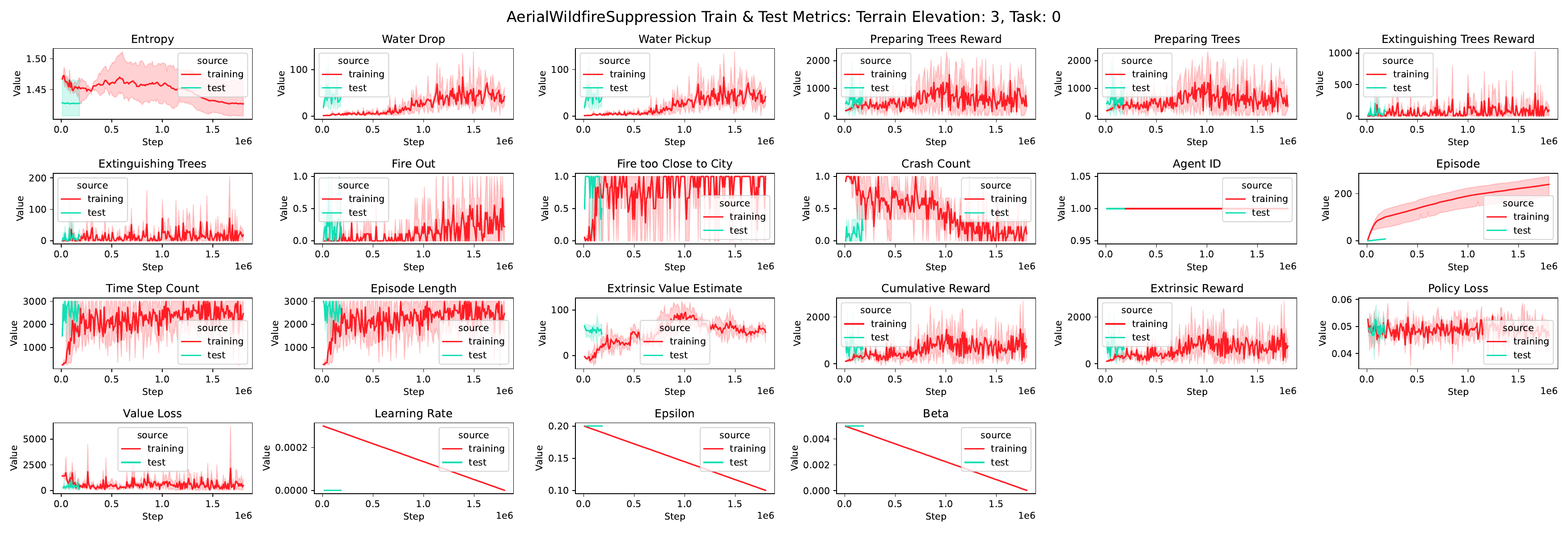}
\vspace{-0.6cm}
\caption{Aerial Wildfire Suppression: Train \& Test Metrics: Terrain Elevation 3, Task 0.}
\end{figure}

\begin{figure}[h!]
\centering
\includegraphics[width=\linewidth]{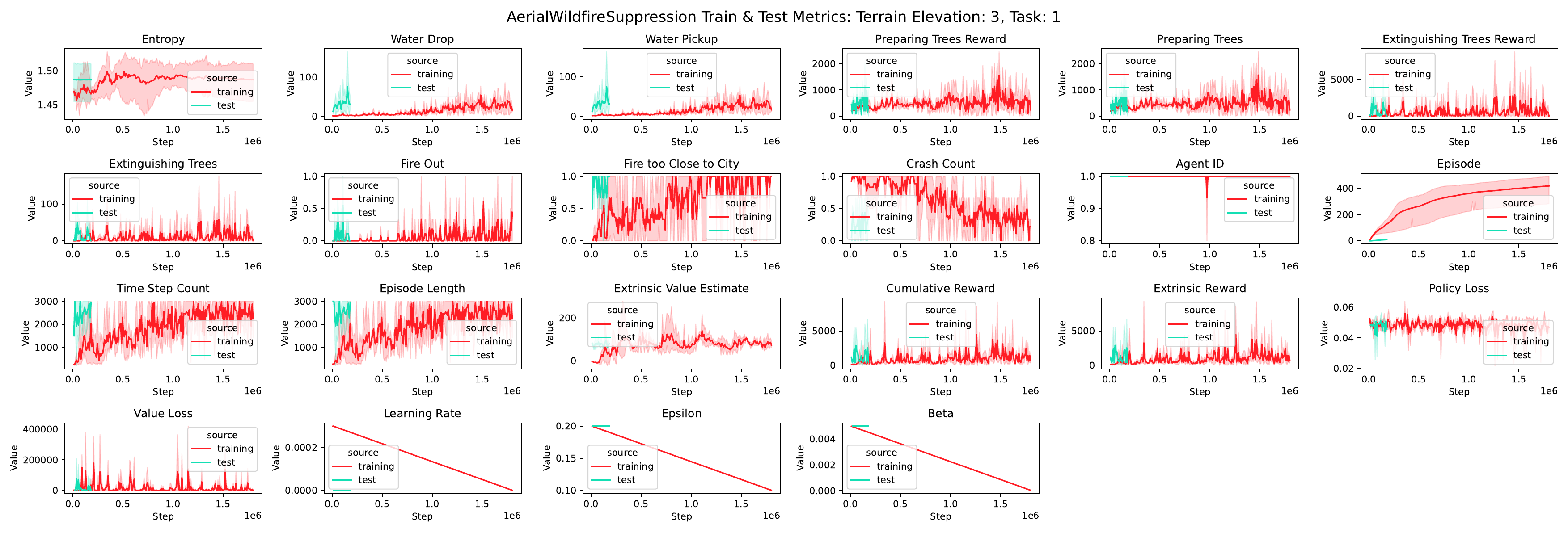}
\vspace{-0.6cm}
\caption{Aerial Wildfire Suppression: Train \& Test Metrics: Terrain Elevation 3, Task 1.}
\end{figure}

\begin{figure}[h!]
\centering
\includegraphics[width=\linewidth]{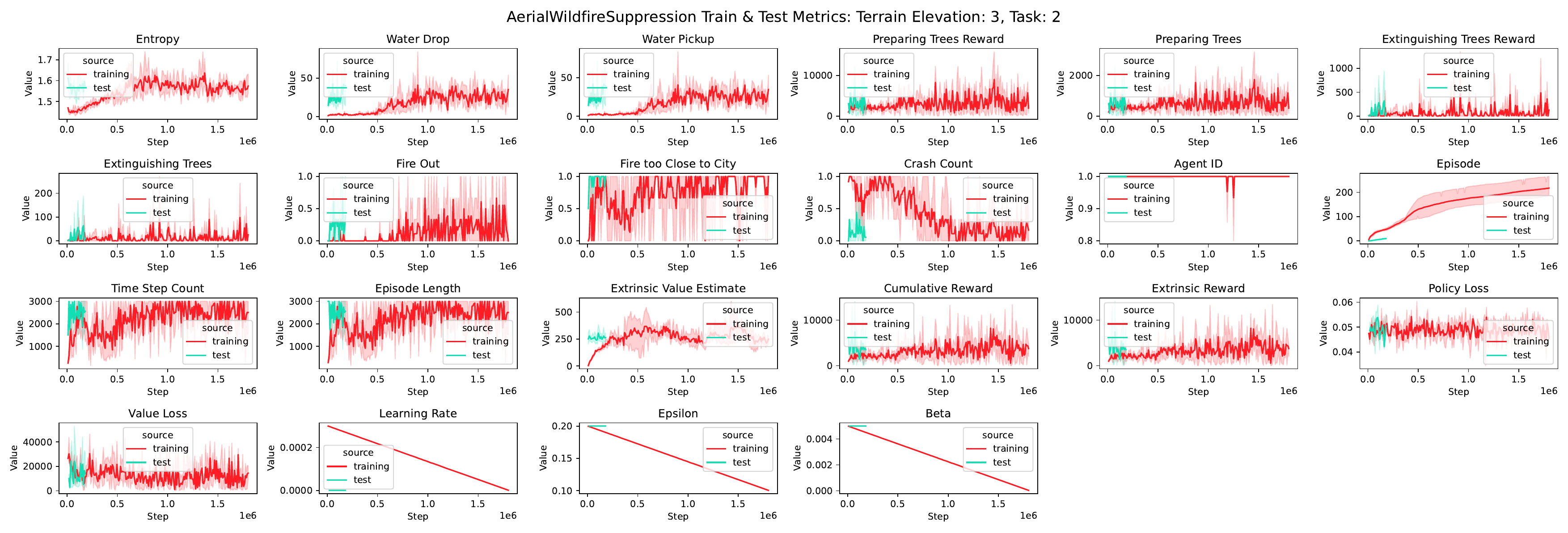}
\vspace{-0.6cm}
\caption{Aerial Wildfire Suppression: Train \& Test Metrics: Terrain Elevation 3, Task 2.}
\end{figure}

\clearpage

\begin{figure}[h!]
\centering
\includegraphics[width=\linewidth]{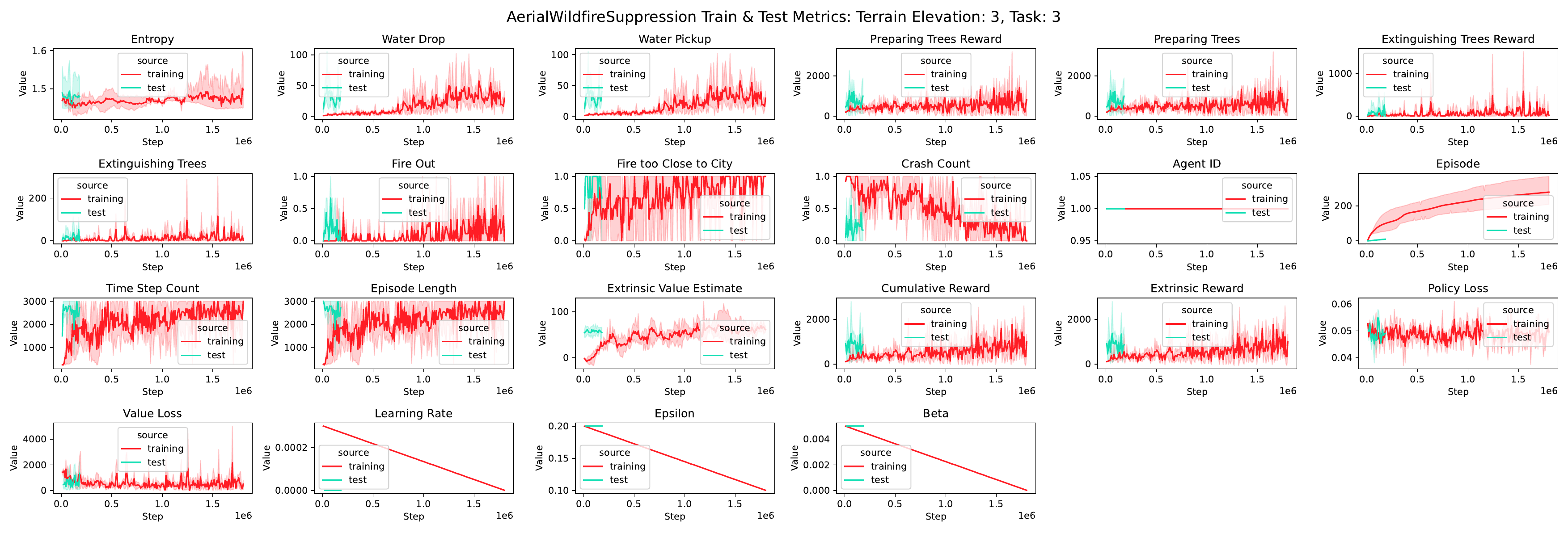}
\vspace{-0.6cm}
\caption{Aerial Wildfire Suppression: Train \& Test Metrics: Terrain Elevation 3, Task 3.}
\end{figure}

\begin{figure}[h!]
\centering
\includegraphics[width=\linewidth]{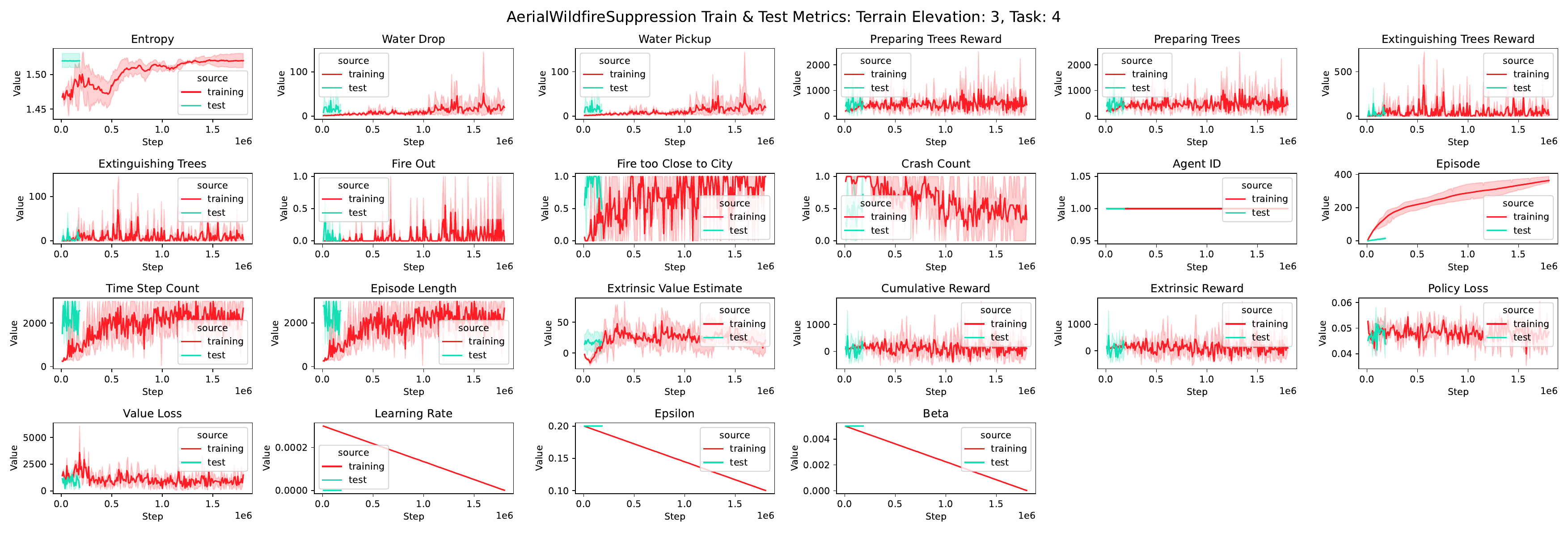}
\vspace{-0.6cm}
\caption{Aerial Wildfire Suppression: Train \& Test Metrics: Terrain Elevation 3, Task 4.}
\end{figure}

\begin{figure}[h!]
\centering
\includegraphics[width=\linewidth]{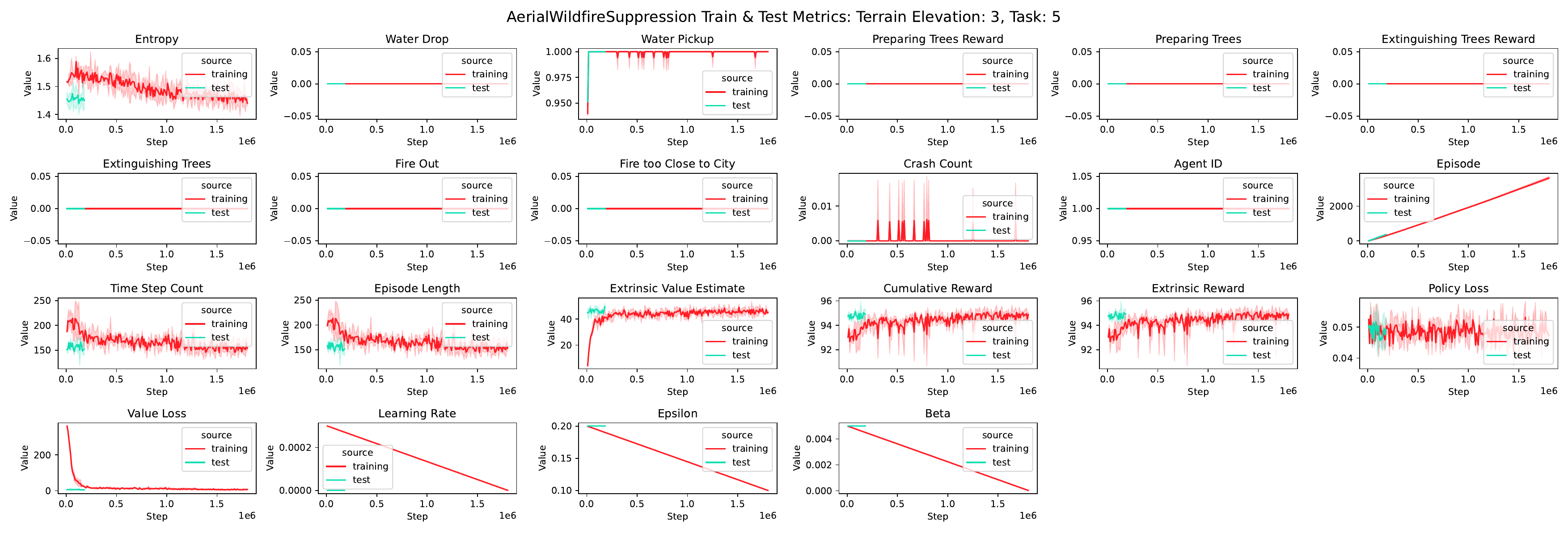}
\vspace{-0.6cm}
\caption{Aerial Wildfire Suppression: Train \& Test Metrics: Terrain Elevation 3, Task 5.}
\end{figure}

\clearpage

\begin{figure}[h!]
\centering
\includegraphics[width=\linewidth]{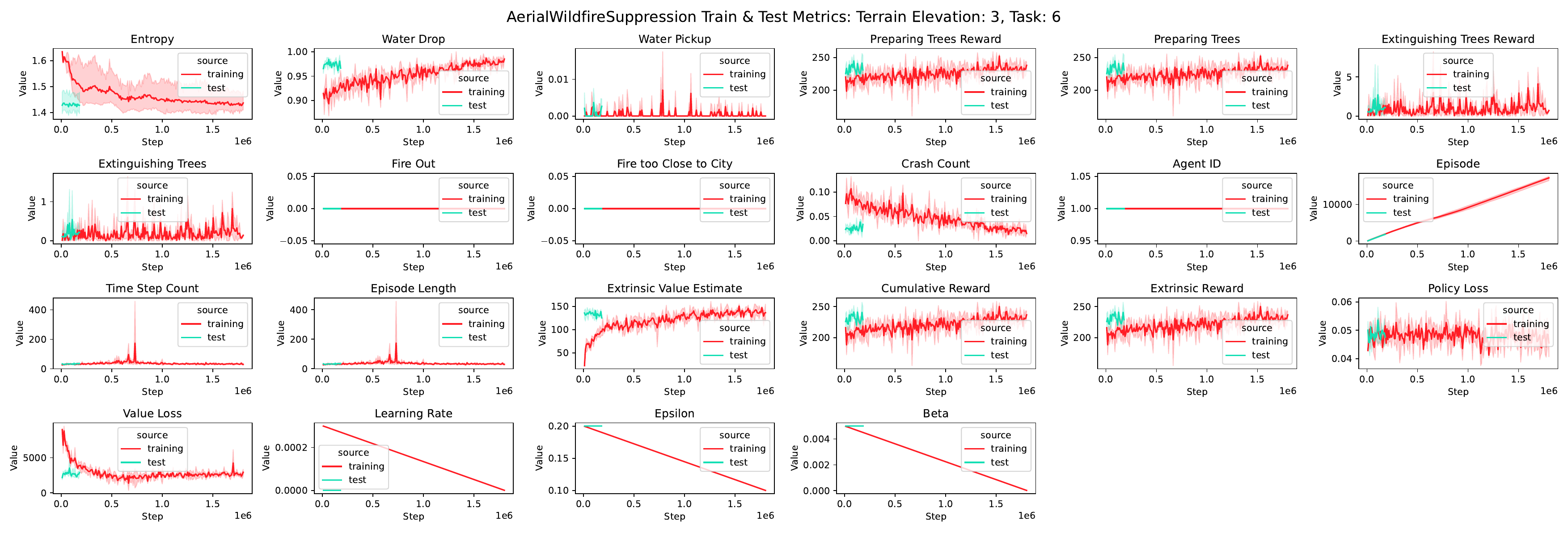}
\vspace{-0.6cm}
\caption{Aerial Wildfire Suppression: Train \& Test Metrics: Terrain Elevation 3, Task 6.}
\end{figure}

\begin{figure}[h!]
\centering
\includegraphics[width=\linewidth]{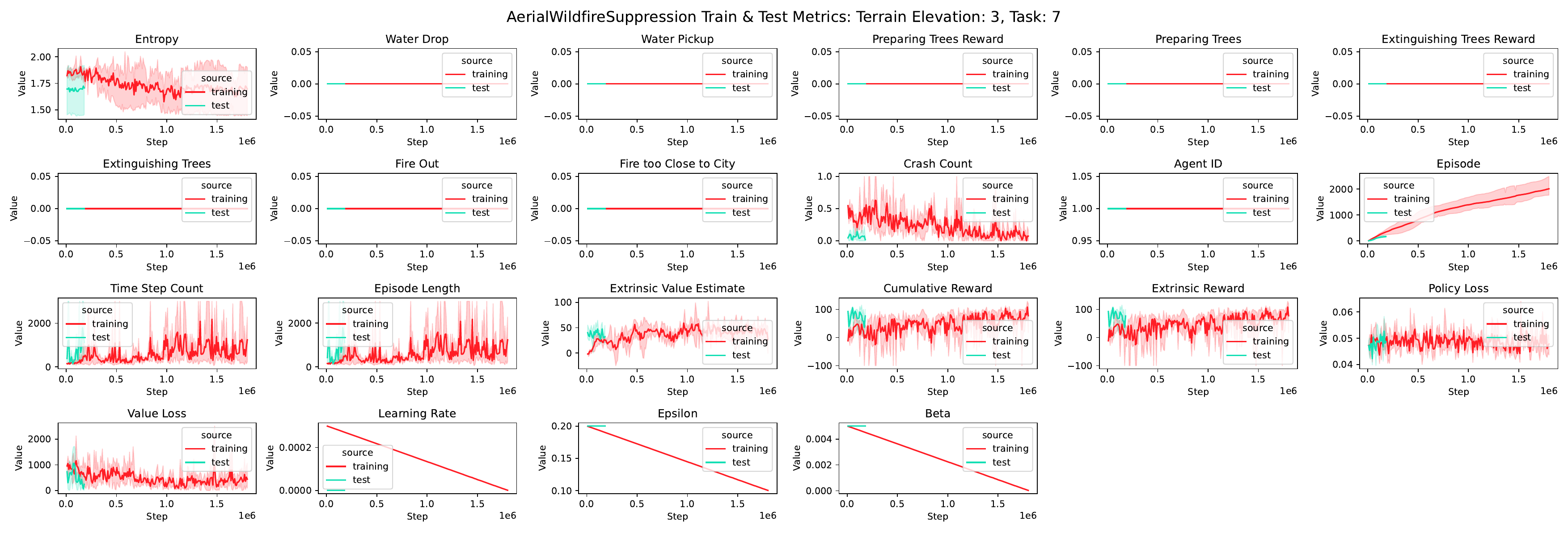}
\vspace{-0.6cm}
\caption{Aerial Wildfire Suppression: Train \& Test Metrics: Terrain Elevation 3, Task 7.}
\end{figure}

\begin{figure}[h!]
\centering
\includegraphics[width=\linewidth]{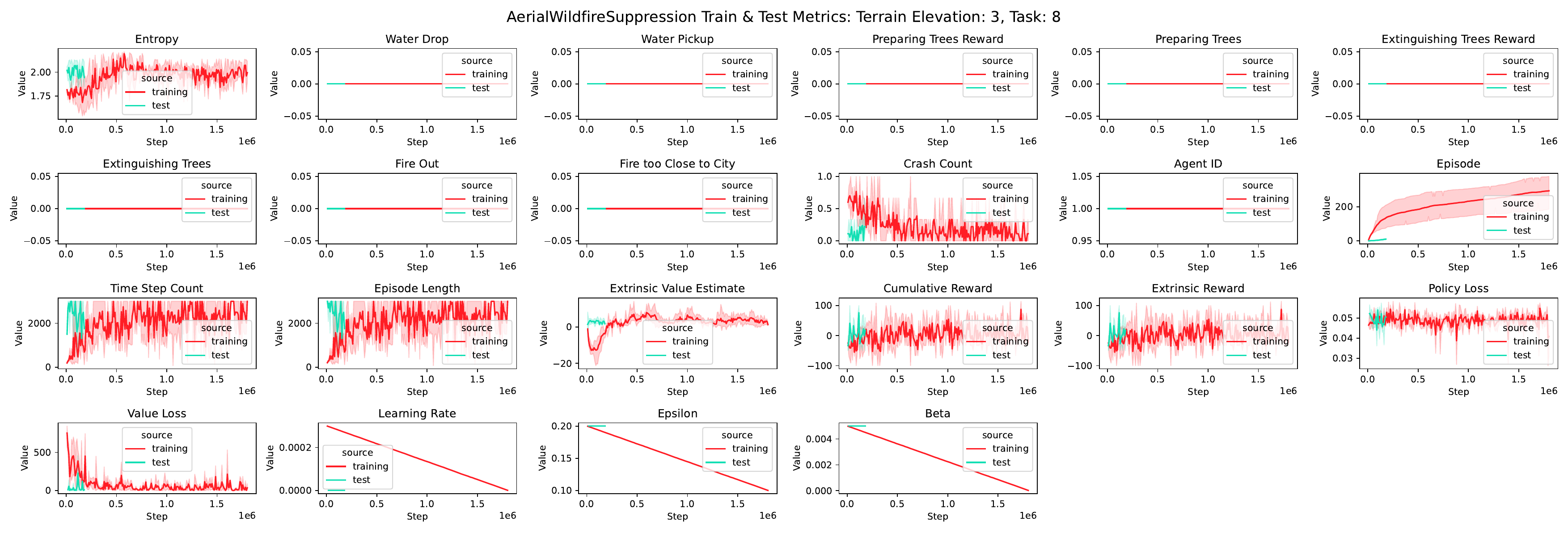}
\vspace{-0.6cm}
\caption{Aerial Wildfire Suppression: Train \& Test Metrics: Terrain Elevation 3, Task 8.}
\end{figure}

\clearpage

\begin{figure}[h!]
\centering
\includegraphics[width=\linewidth]{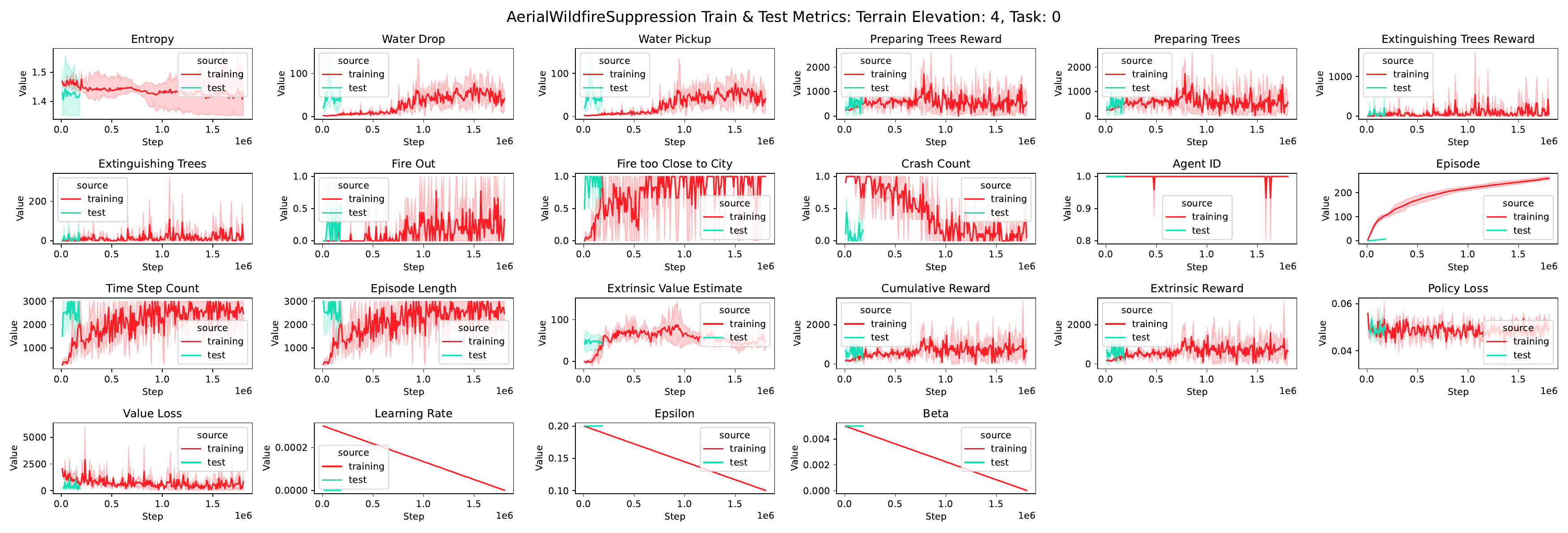}
\vspace{-0.6cm}
\caption{Aerial Wildfire Suppression: Train \& Test Metrics: Terrain Elevation 4, Task 0.}
\end{figure}

\begin{figure}[h!]
\centering
\includegraphics[width=\linewidth]{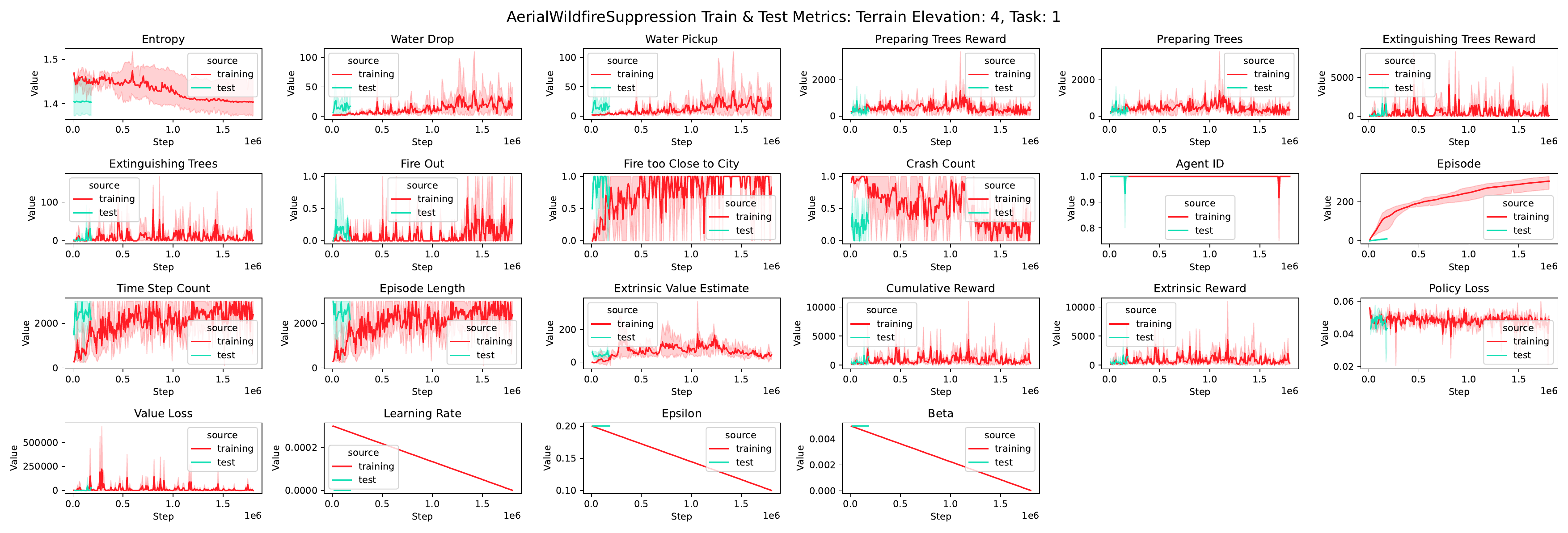}
\vspace{-0.6cm}
\caption{Aerial Wildfire Suppression: Train \& Test Metrics: Terrain Elevation 4, Task 1.}
\end{figure}

\begin{figure}[h!]
\centering
\includegraphics[width=\linewidth]{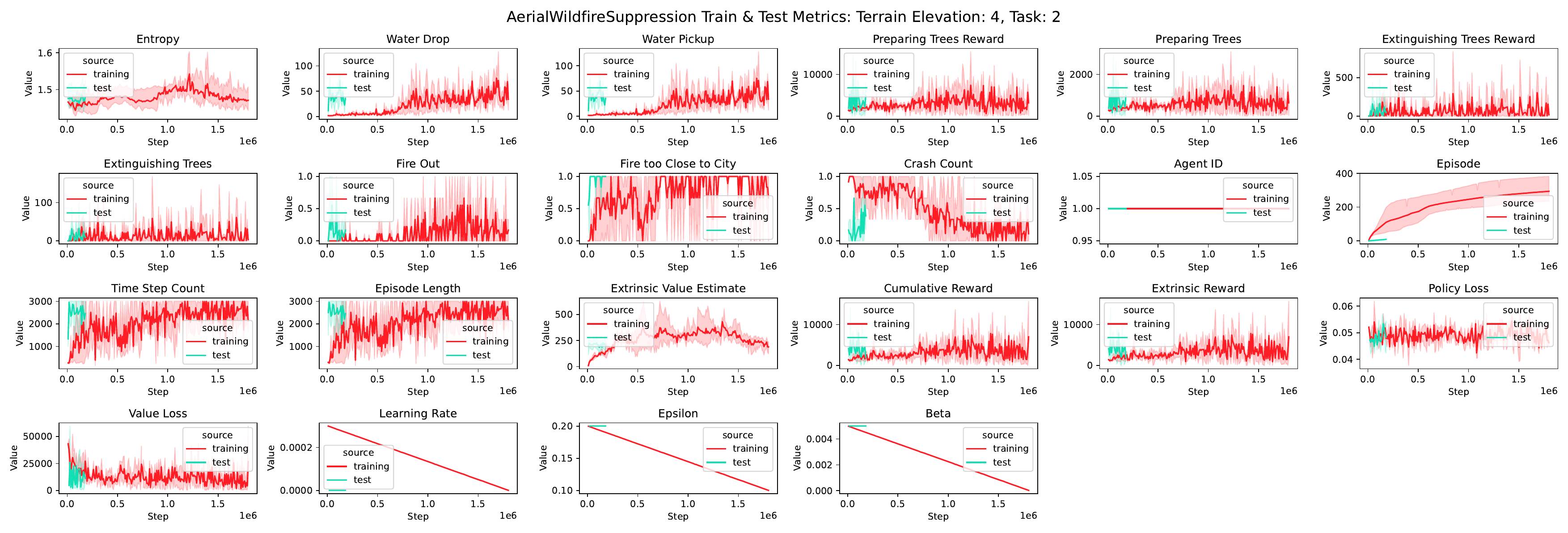}
\vspace{-0.6cm}
\caption{Aerial Wildfire Suppression: Train \& Test Metrics: Terrain Elevation 4, Task 2.}
\end{figure}

\clearpage

\begin{figure}[h!]
\centering
\includegraphics[width=\linewidth]{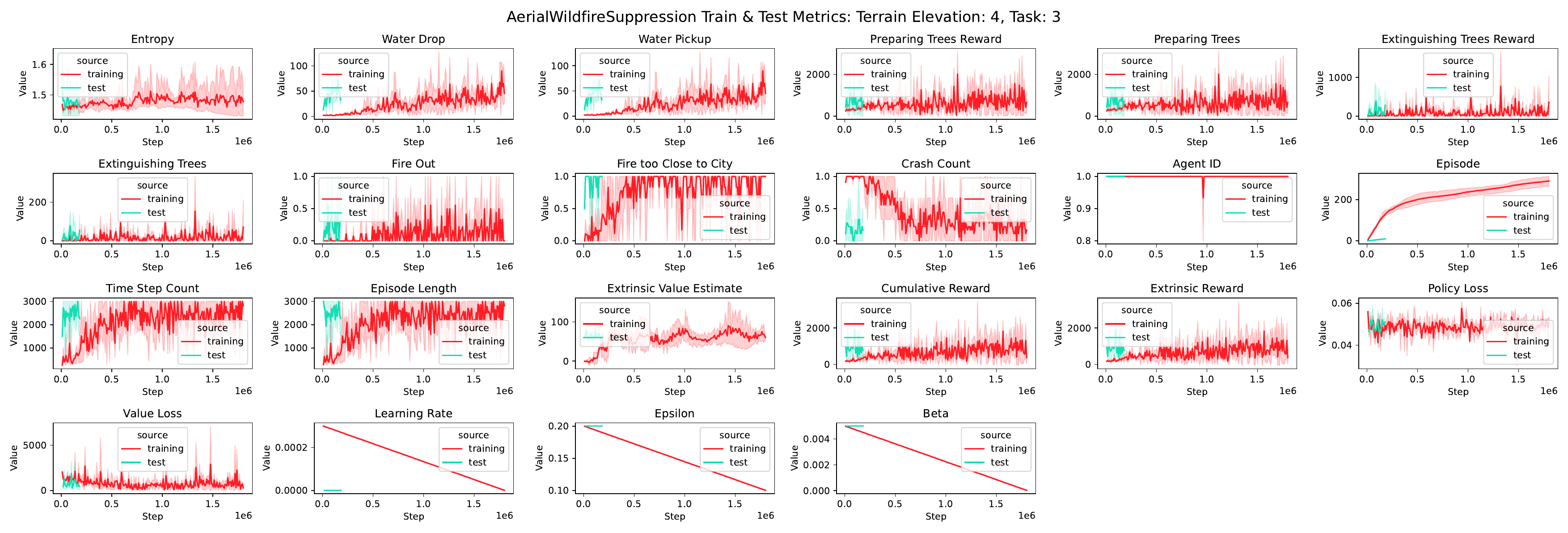}
\vspace{-0.6cm}
\caption{Aerial Wildfire Suppression: Train \& Test Metrics: Terrain Elevation 4, Task 3.}
\end{figure}

\begin{figure}[h!]
\centering
\includegraphics[width=\linewidth]{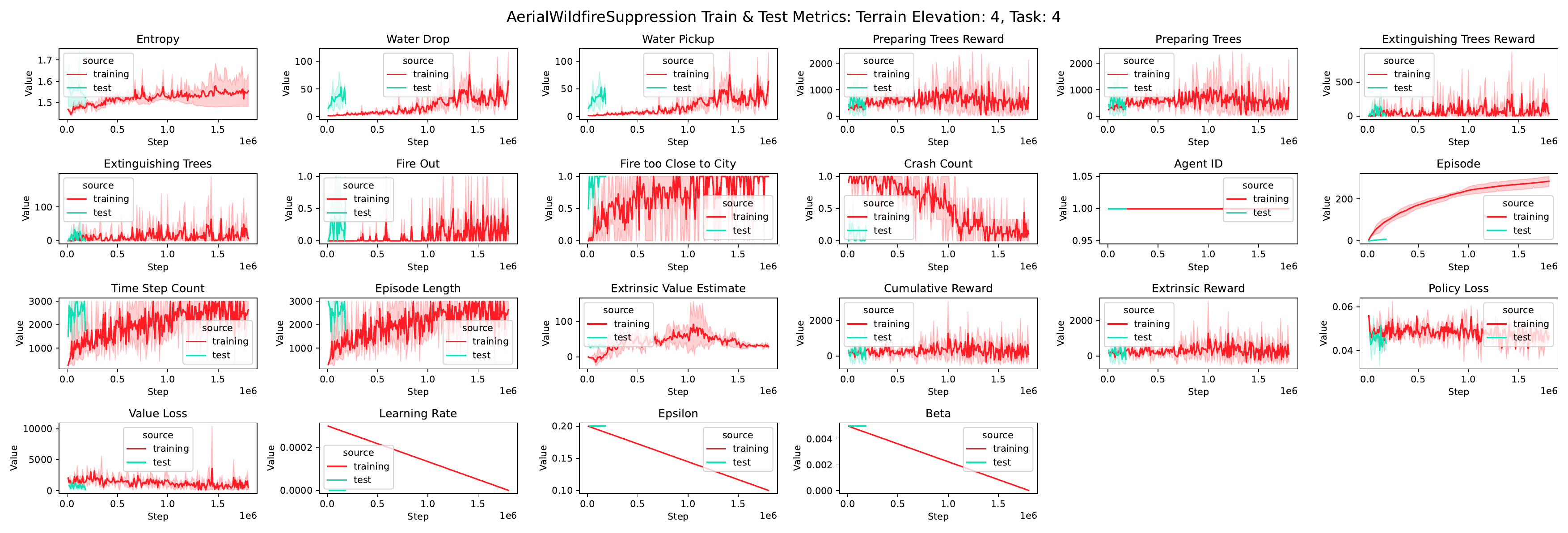}
\vspace{-0.6cm}
\caption{Aerial Wildfire Suppression: Train \& Test Metrics: Terrain Elevation 4, Task 4.}
\end{figure}

\begin{figure}[h!]
\centering
\includegraphics[width=\linewidth]{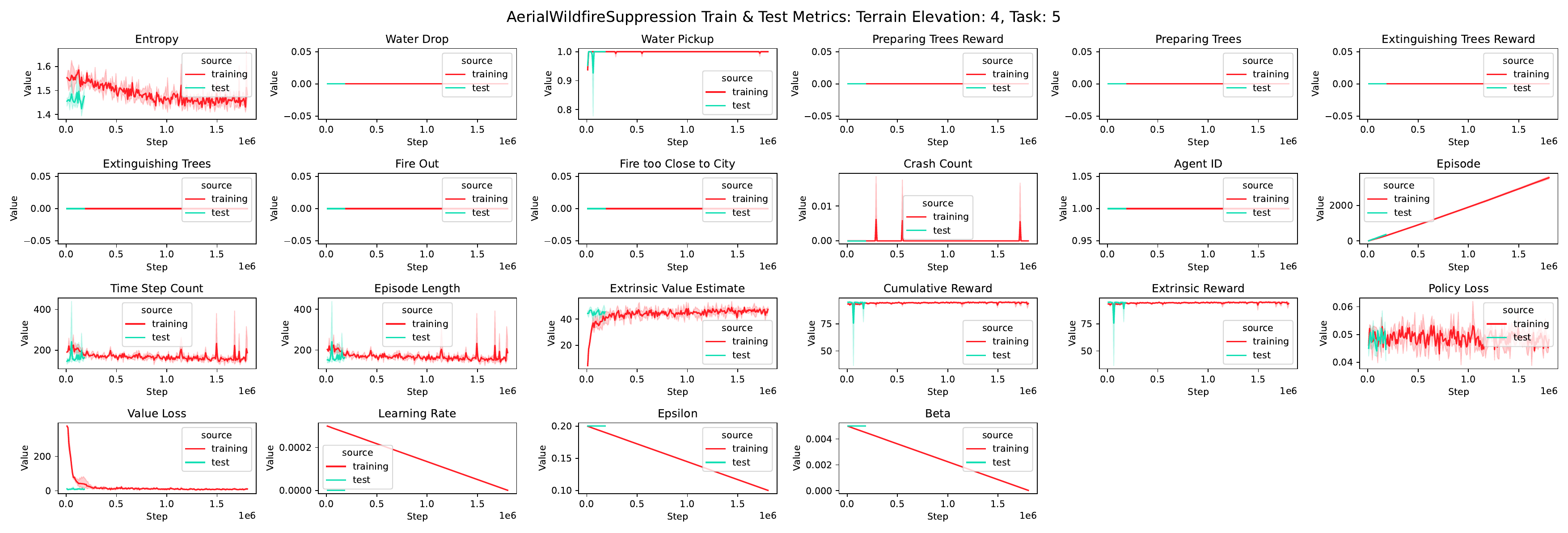}
\vspace{-0.6cm}
\caption{Aerial Wildfire Suppression: Train \& Test Metrics: Terrain Elevation 4, Task 5.}
\end{figure}

\clearpage

\begin{figure}[h!]
\centering
\includegraphics[width=\linewidth]{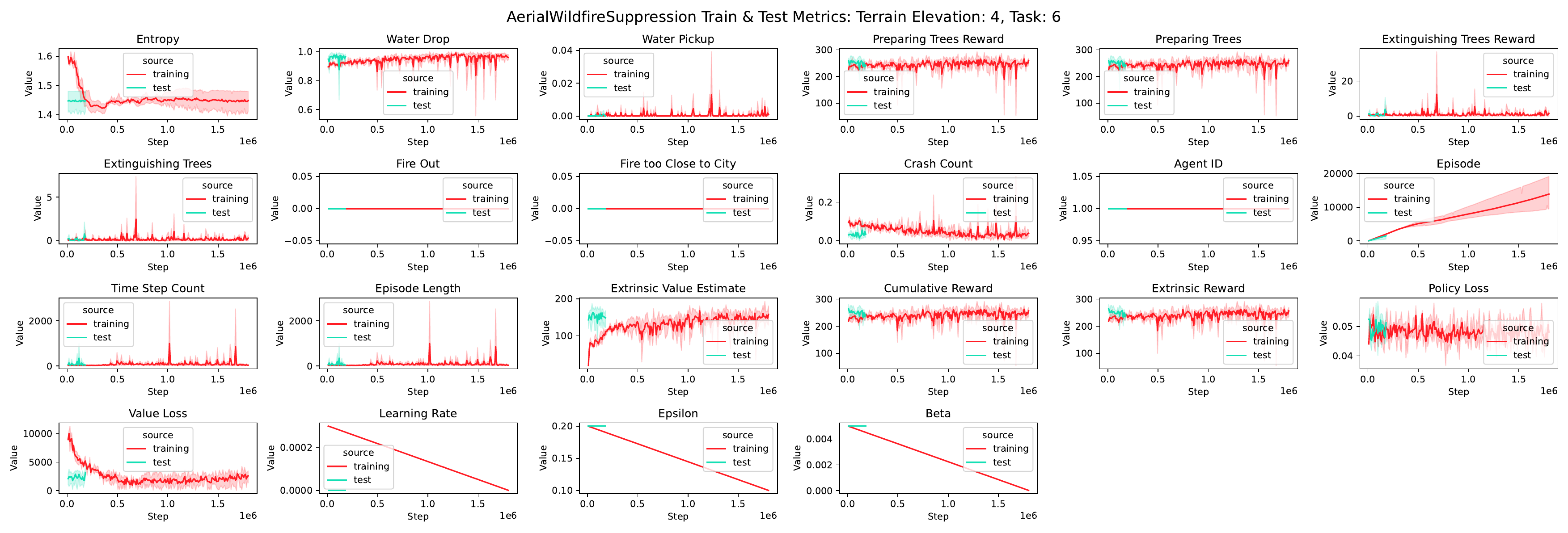}
\vspace{-0.6cm}
\caption{Aerial Wildfire Suppression: Train \& Test Metrics: Terrain Elevation 4, Task 6.}
\end{figure}

\begin{figure}[h!]
\centering
\includegraphics[width=\linewidth]{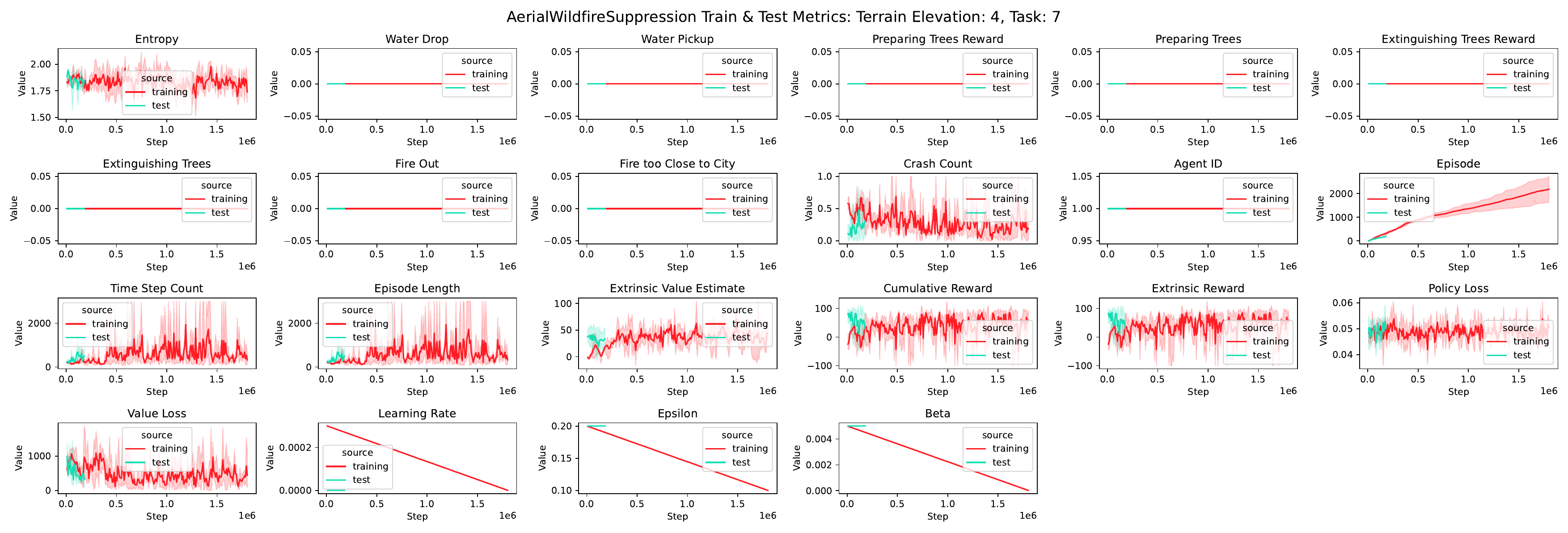}
\vspace{-0.6cm}
\caption{Aerial Wildfire Suppression: Train \& Test Metrics: Terrain Elevation 4, Task 7.}
\end{figure}

\begin{figure}[h!]
\centering
\includegraphics[width=\linewidth]{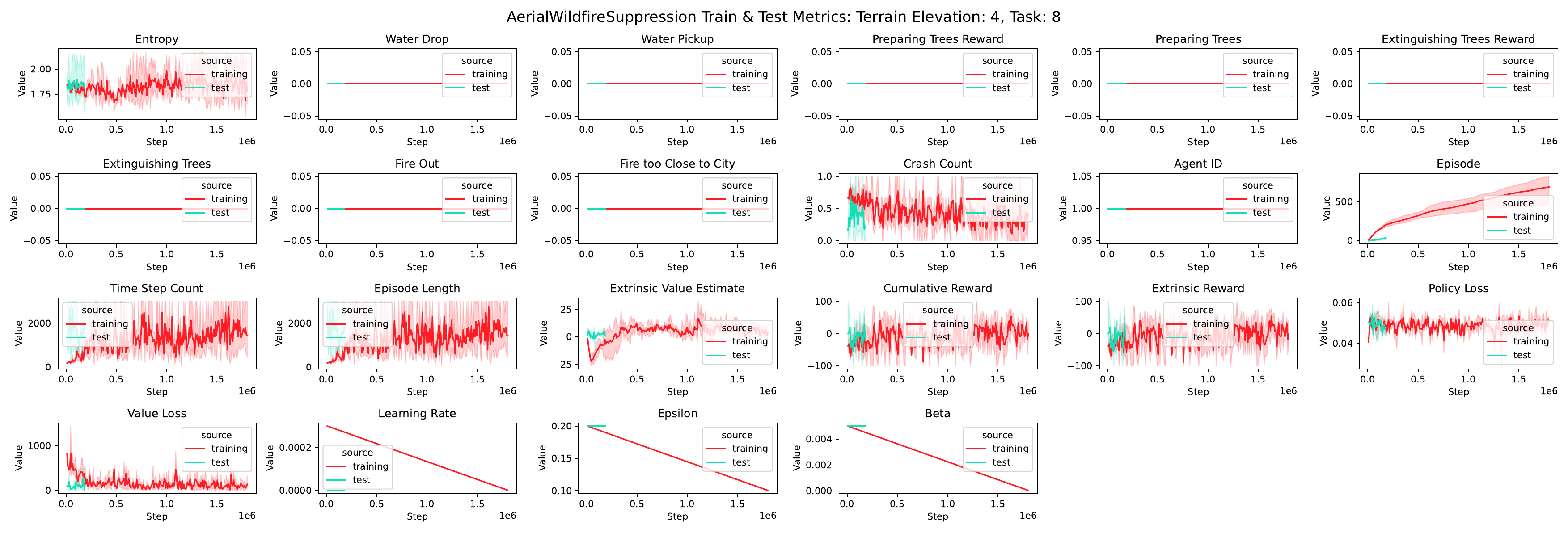}
\vspace{-0.6cm}
\caption{Aerial Wildfire Suppression: Train \& Test Metrics: Terrain Elevation 4, Task 8.}
\end{figure}

\clearpage

\begin{figure}[h!]
\centering
\includegraphics[width=\linewidth]{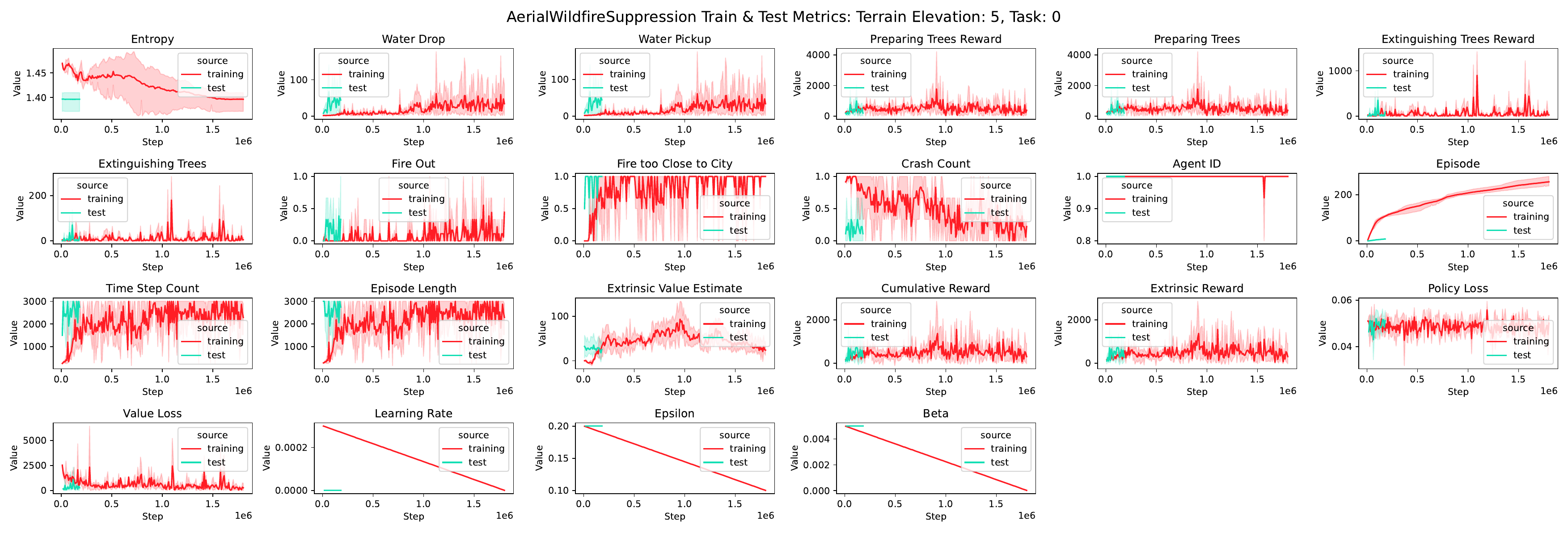}
\vspace{-0.6cm}
\caption{Aerial Wildfire Suppression: Train \& Test Metrics: Terrain Elevation 5, Task 0.}
\end{figure}

\begin{figure}[h!]
\centering
\includegraphics[width=\linewidth]{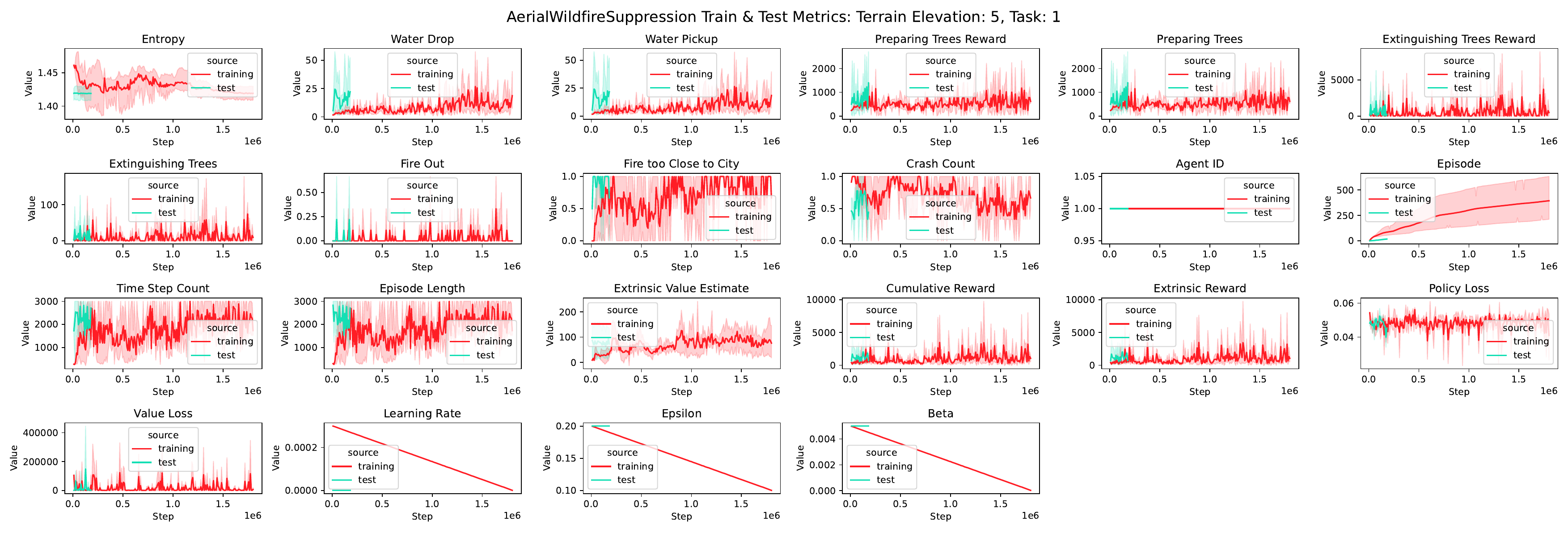}
\vspace{-0.6cm}
\caption{Aerial Wildfire Suppression: Train \& Test Metrics: Terrain Elevation 5, Task 1.}
\end{figure}

\begin{figure}[h!]
\centering
\includegraphics[width=\linewidth]{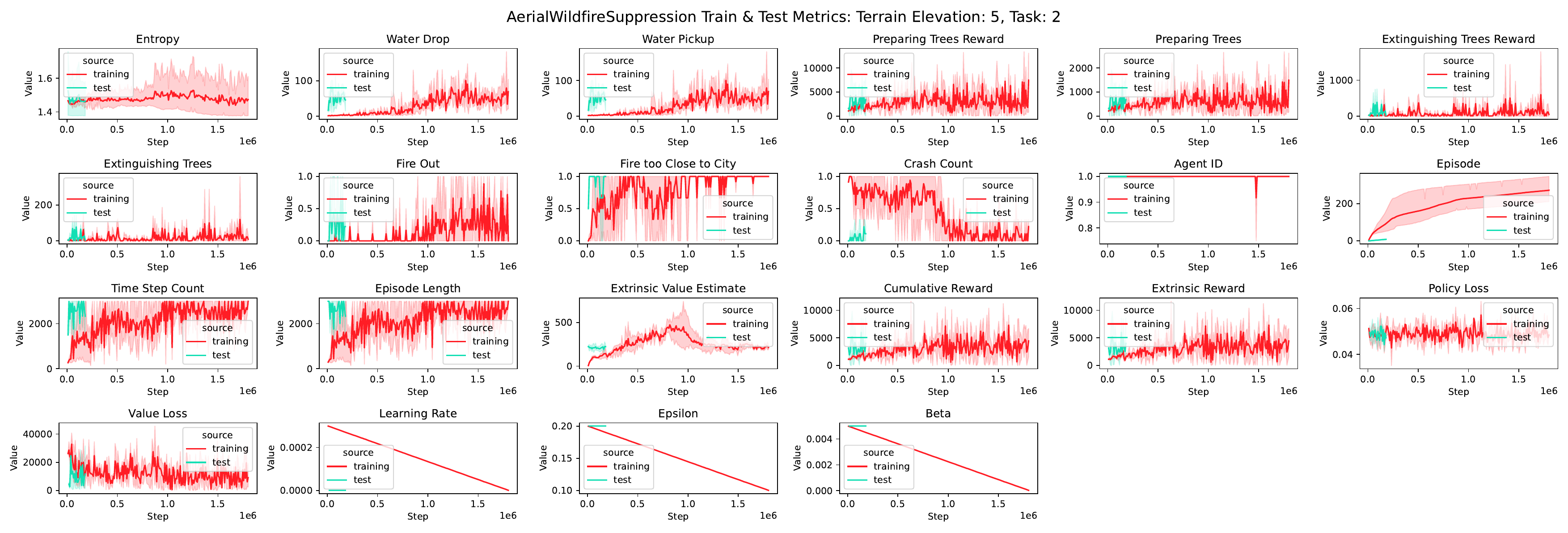}
\vspace{-0.6cm}
\caption{Aerial Wildfire Suppression: Train \& Test Metrics: Terrain Elevation 5, Task 2.}
\end{figure}

\clearpage

\begin{figure}[h!]
\centering
\includegraphics[width=\linewidth]{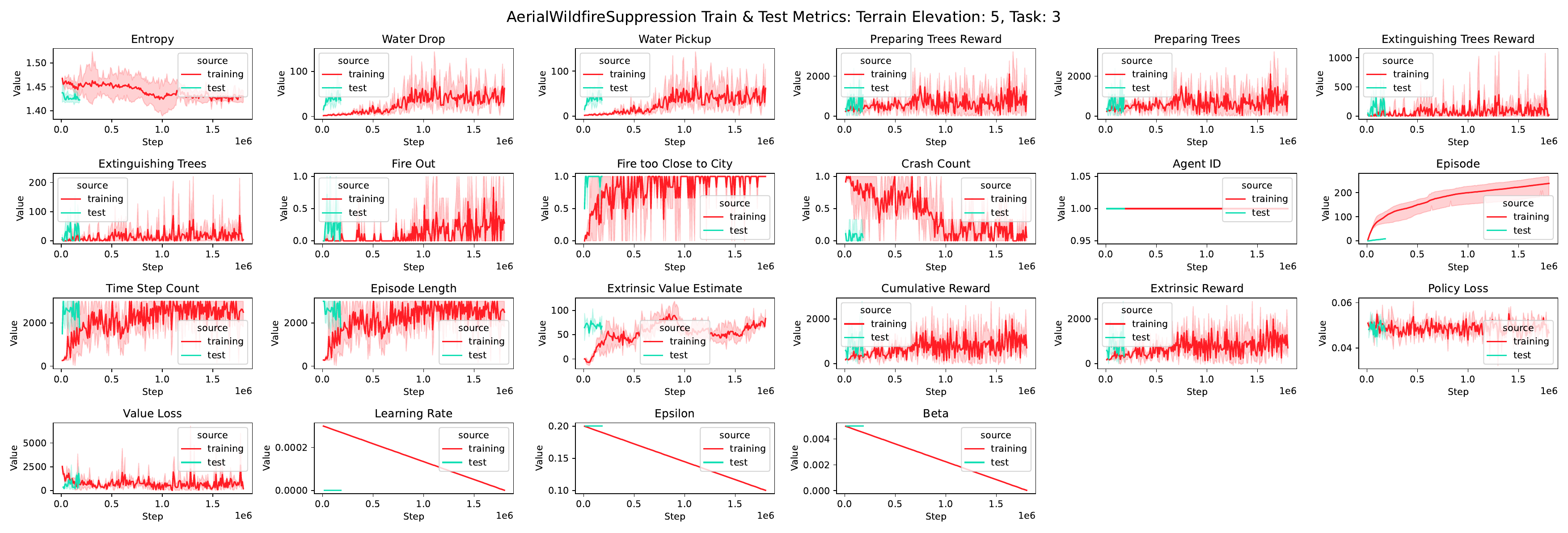}
\vspace{-0.6cm}
\caption{Aerial Wildfire Suppression: Train \& Test Metrics: Terrain Elevation 5, Task 3.}
\end{figure}

\begin{figure}[h!]
\centering
\includegraphics[width=\linewidth]{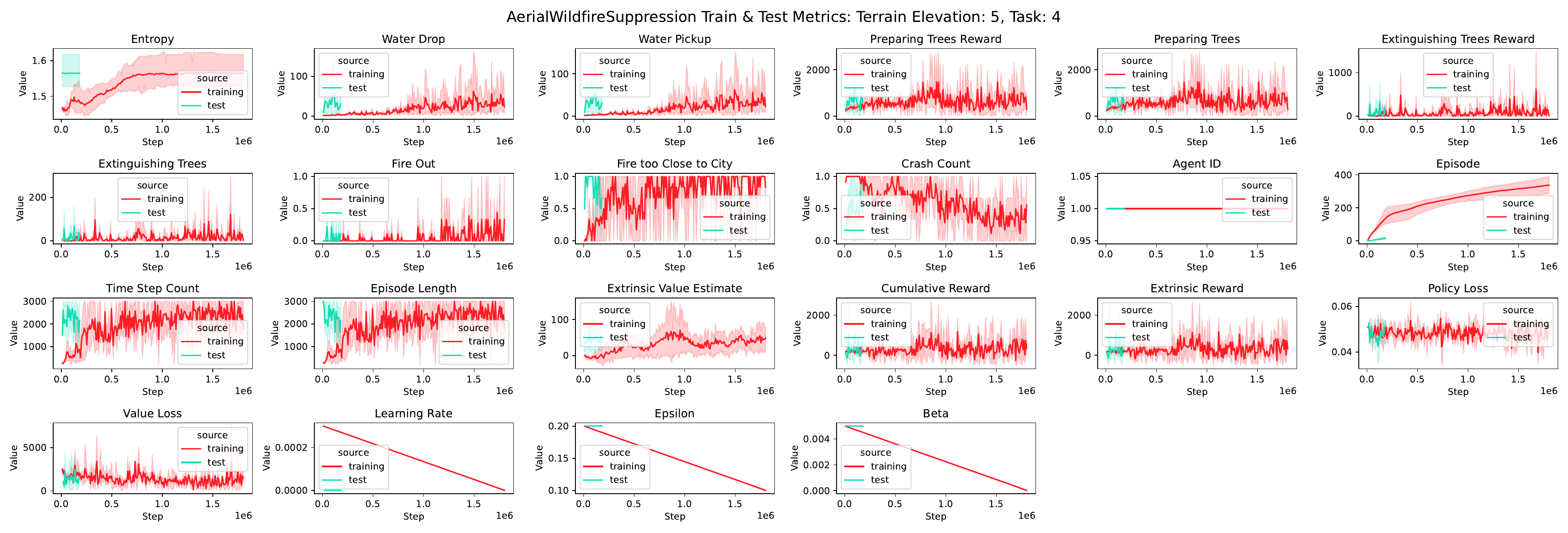}
\vspace{-0.6cm}
\caption{Aerial Wildfire Suppression: Train \& Test Metrics: Terrain Elevation 5, Task 4.}
\end{figure}

\begin{figure}[h!]
\centering
\includegraphics[width=\linewidth]{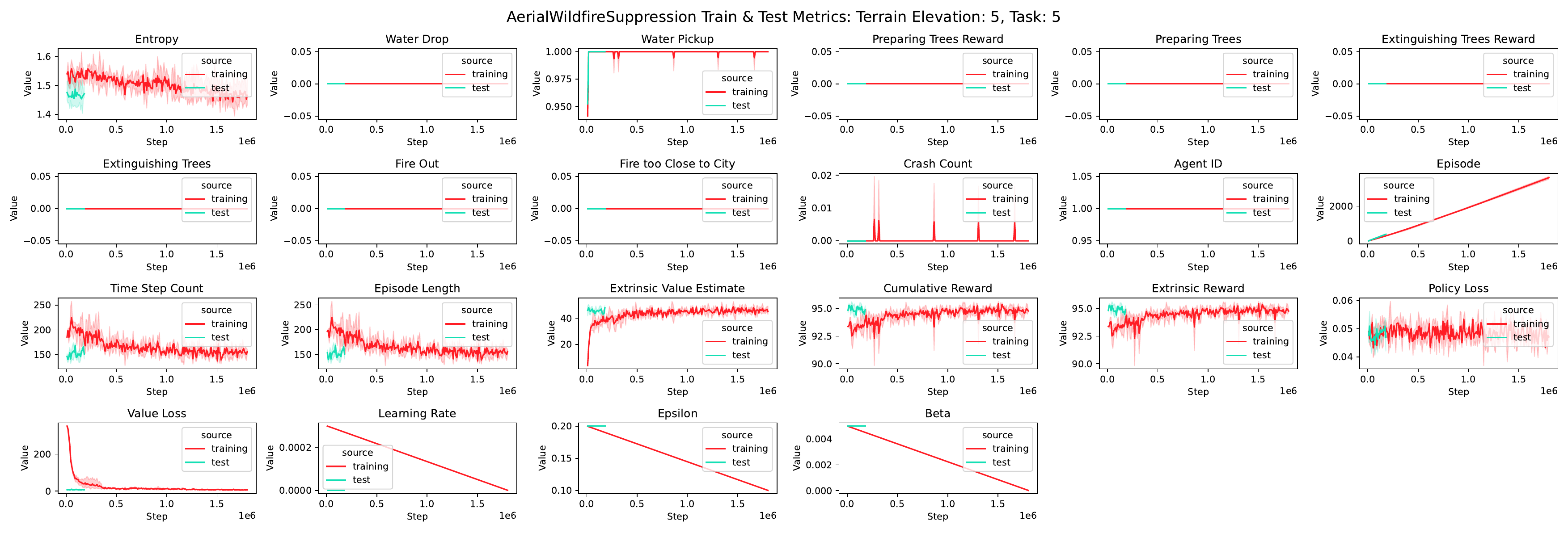}
\vspace{-0.6cm}
\caption{Aerial Wildfire Suppression: Train \& Test Metrics: Terrain Elevation 5, Task 5.}
\end{figure}

\clearpage

\begin{figure}[h!]
\centering
\includegraphics[width=\linewidth]{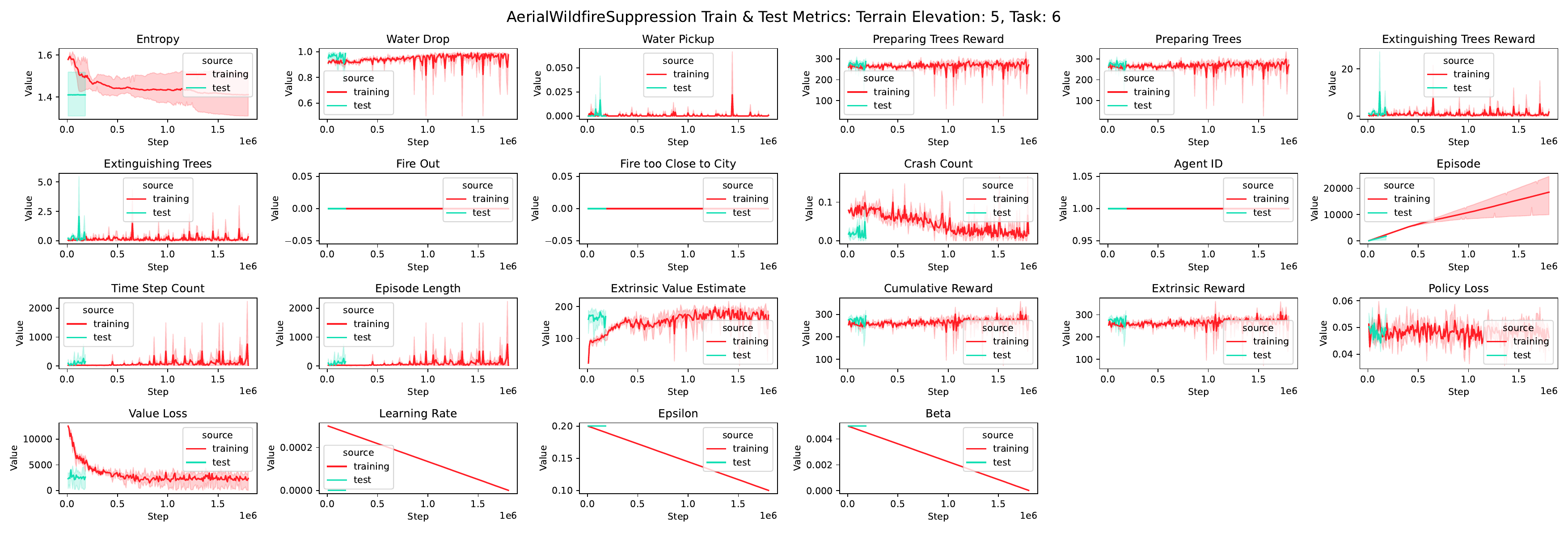}
\vspace{-0.6cm}
\caption{Aerial Wildfire Suppression: Train \& Test Metrics: Terrain Elevation 5, Task 6.}
\end{figure}

\begin{figure}[h!]
\centering
\includegraphics[width=\linewidth]{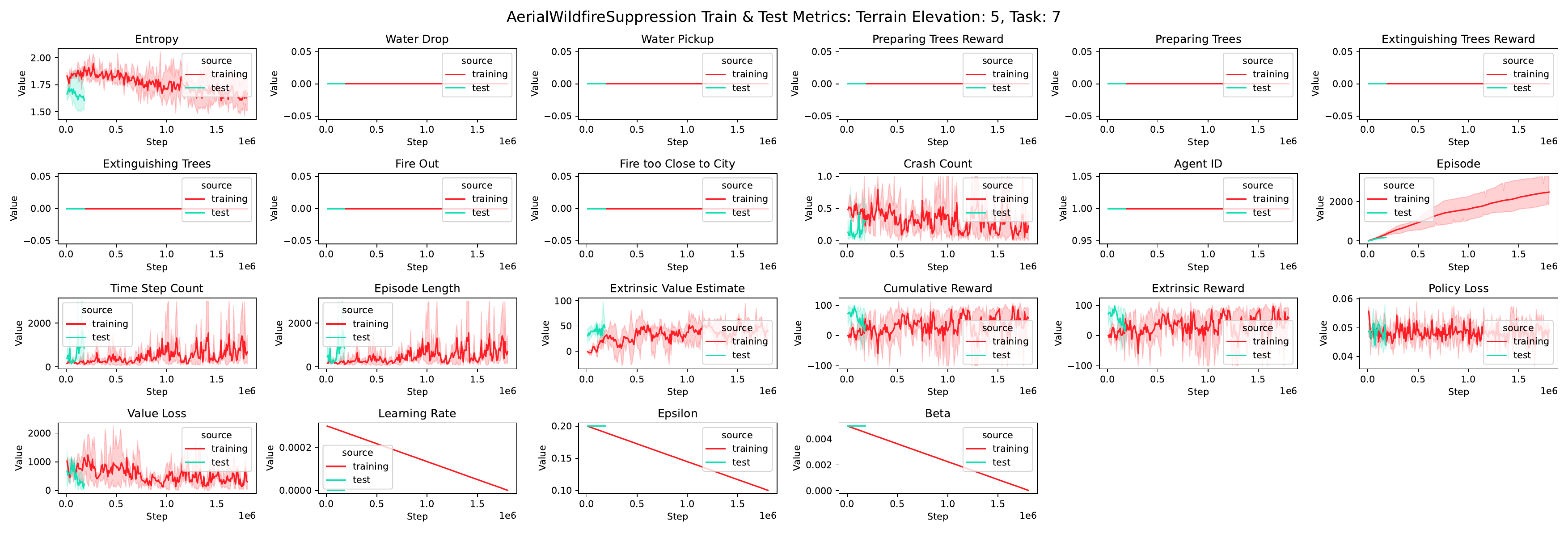}
\vspace{-0.6cm}
\caption{Aerial Wildfire Suppression: Train \& Test Metrics: Terrain Elevation 5, Task 7.}
\end{figure}

\begin{figure}[h!]
\centering
\includegraphics[width=\linewidth]{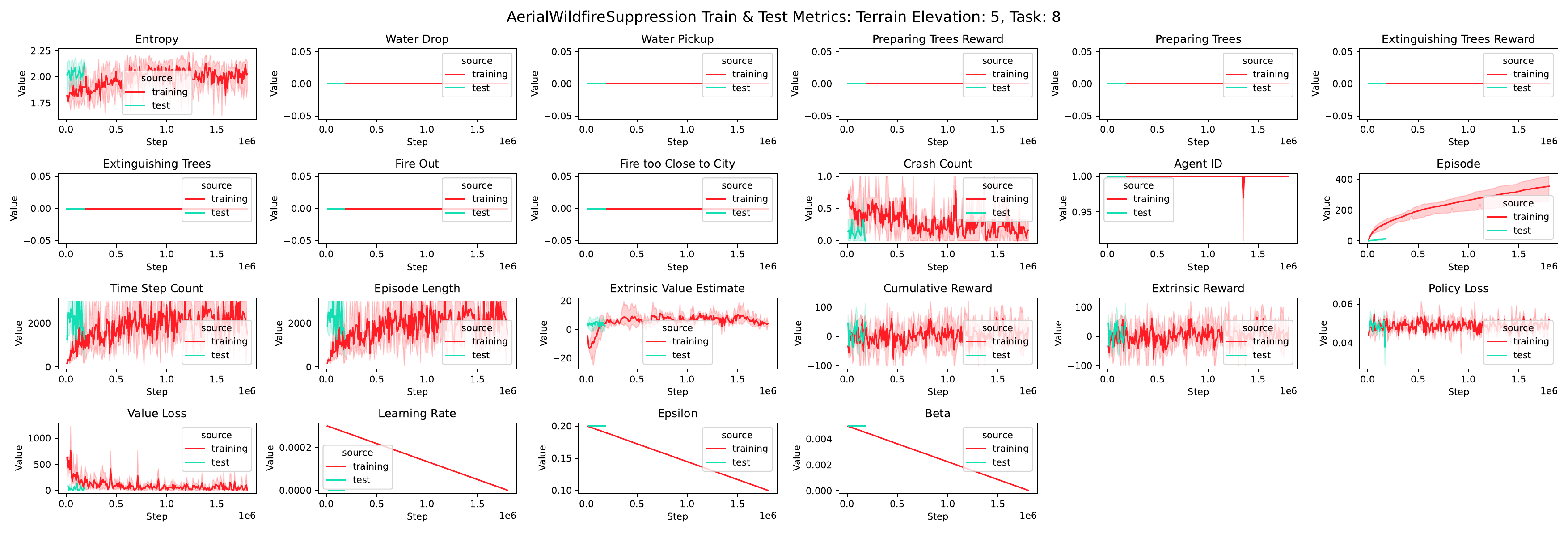}
\vspace{-0.6cm}
\caption{Aerial Wildfire Suppression: Train \& Test Metrics: Terrain Elevation 5, Task 8.}
\end{figure}

\clearpage

\begin{figure}[h!]
\centering
\includegraphics[width=\linewidth]{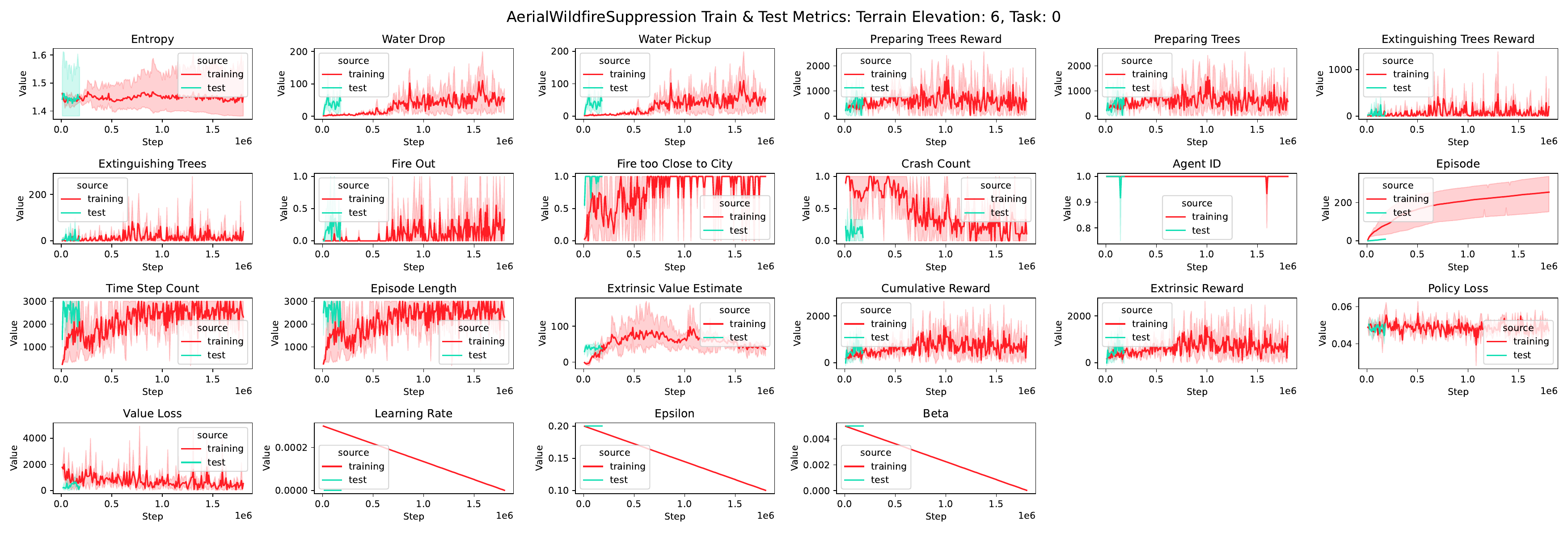}
\vspace{-0.6cm}
\caption{Aerial Wildfire Suppression: Train \& Test Metrics: Terrain Elevation 6, Task 0.}
\end{figure}

\begin{figure}[h!]
\centering
\includegraphics[width=\linewidth]{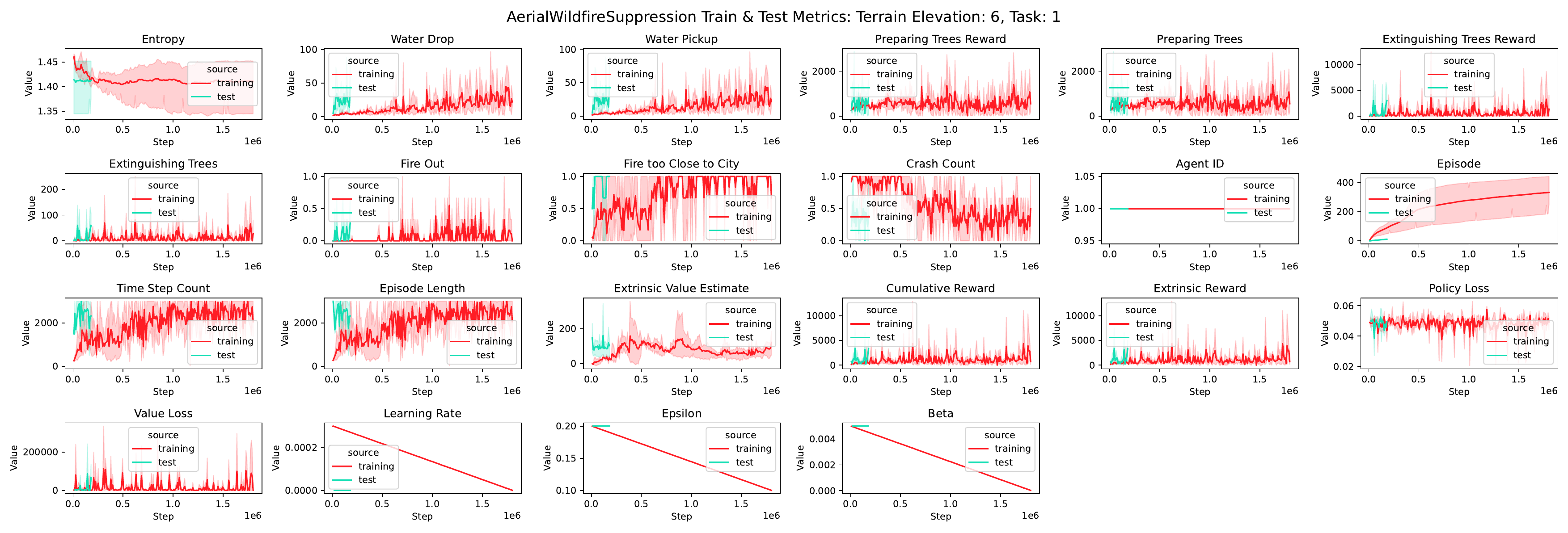}
\vspace{-0.6cm}
\caption{Aerial Wildfire Suppression: Train \& Test Metrics: Terrain Elevation 6, Task 1.}
\end{figure}

\begin{figure}[h!]
\centering
\includegraphics[width=\linewidth]{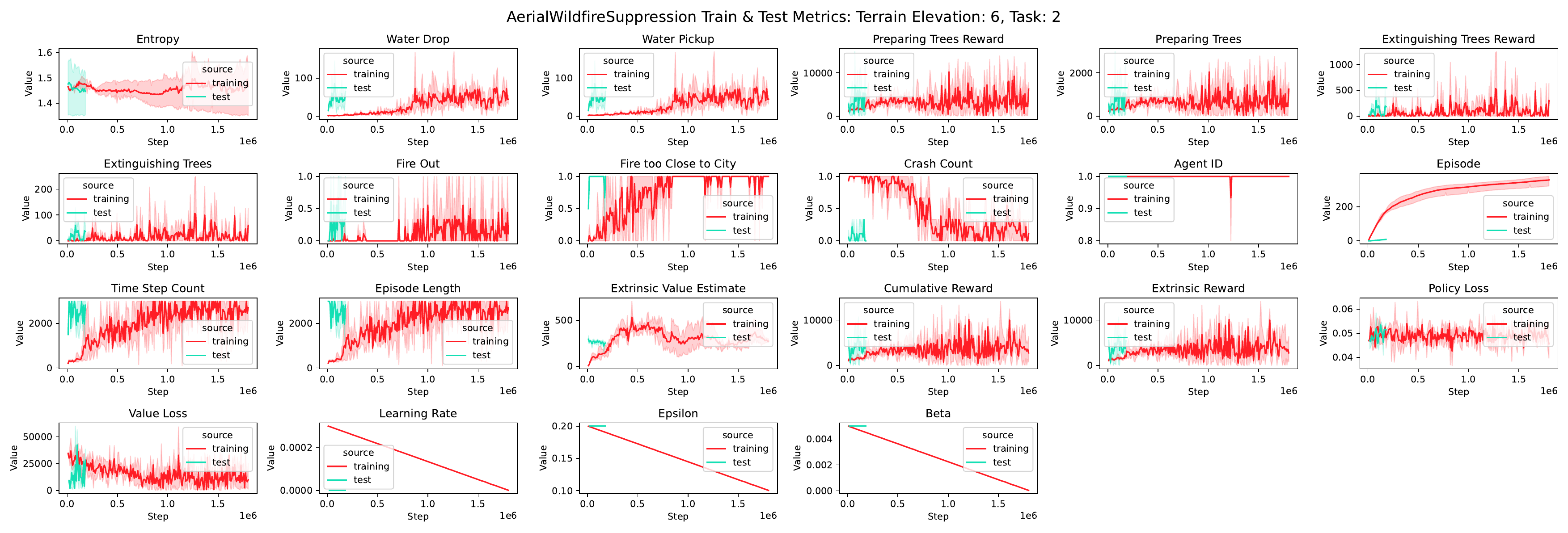}
\vspace{-0.6cm}
\caption{Aerial Wildfire Suppression: Train \& Test Metrics: Terrain Elevation 6, Task 2.}
\end{figure}

\clearpage

\begin{figure}[h!]
\centering
\includegraphics[width=\linewidth]{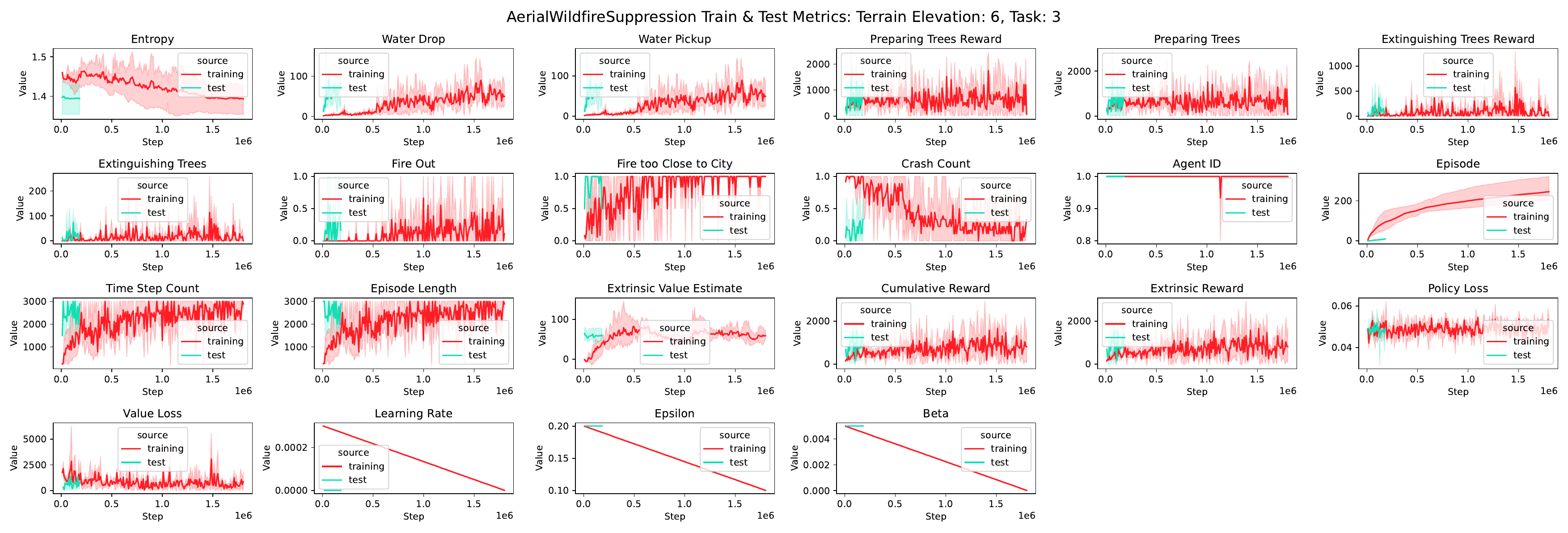}
\vspace{-0.6cm}
\caption{Aerial Wildfire Suppression: Train \& Test Metrics: Terrain Elevation 6, Task 3.}
\end{figure}

\begin{figure}[h!]
\centering
\includegraphics[width=\linewidth]{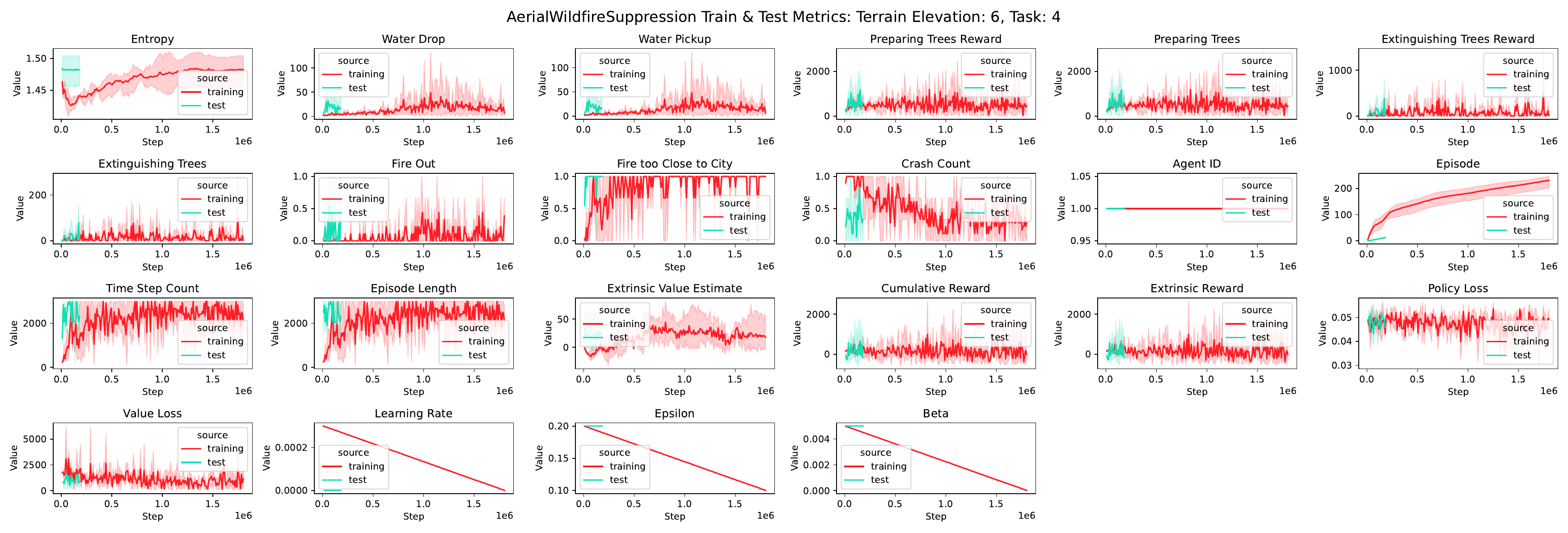}
\vspace{-0.6cm}
\caption{Aerial Wildfire Suppression: Train \& Test Metrics: Terrain Elevation 6, Task 4.}
\end{figure}

\begin{figure}[h!]
\centering
\includegraphics[width=\linewidth]{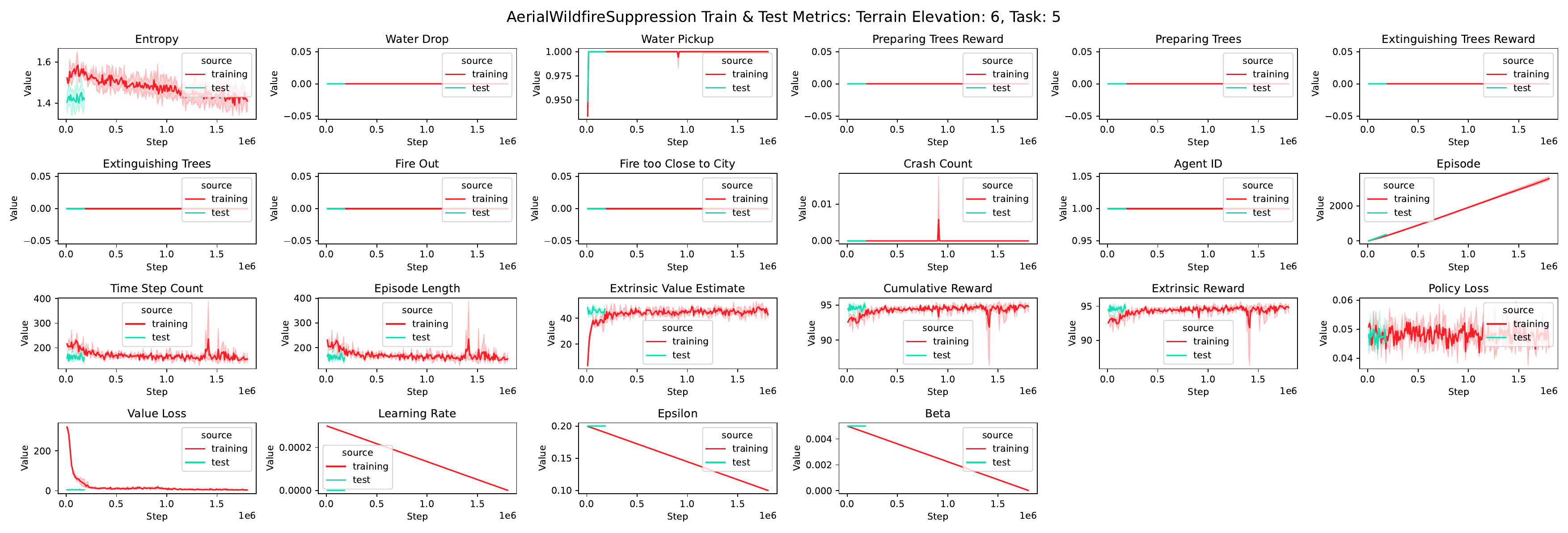}
\vspace{-0.6cm}
\caption{Aerial Wildfire Suppression: Train \& Test Metrics: Terrain Elevation 6, Task 5.}
\end{figure}

\clearpage

\begin{figure}[h!]
\centering
\includegraphics[width=\linewidth]{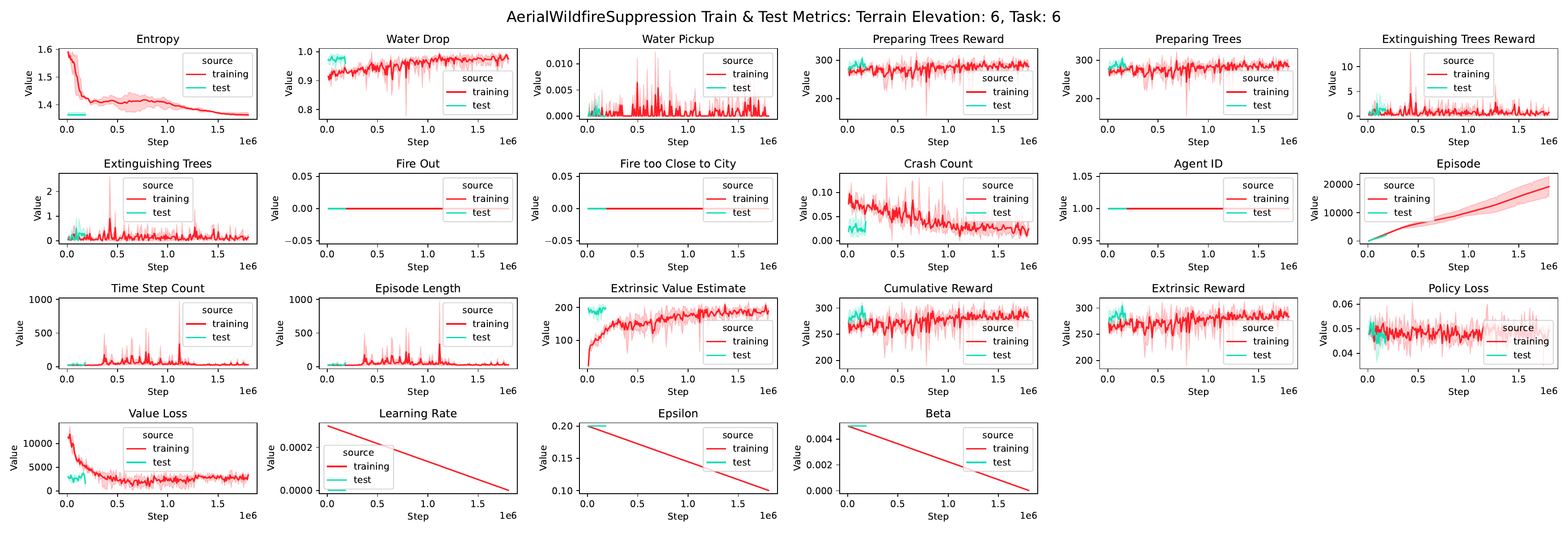}
\vspace{-0.6cm}
\caption{Aerial Wildfire Suppression: Train \& Test Metrics: Terrain Elevation 6, Task 6.}
\end{figure}

\begin{figure}[h!]
\centering
\includegraphics[width=\linewidth]{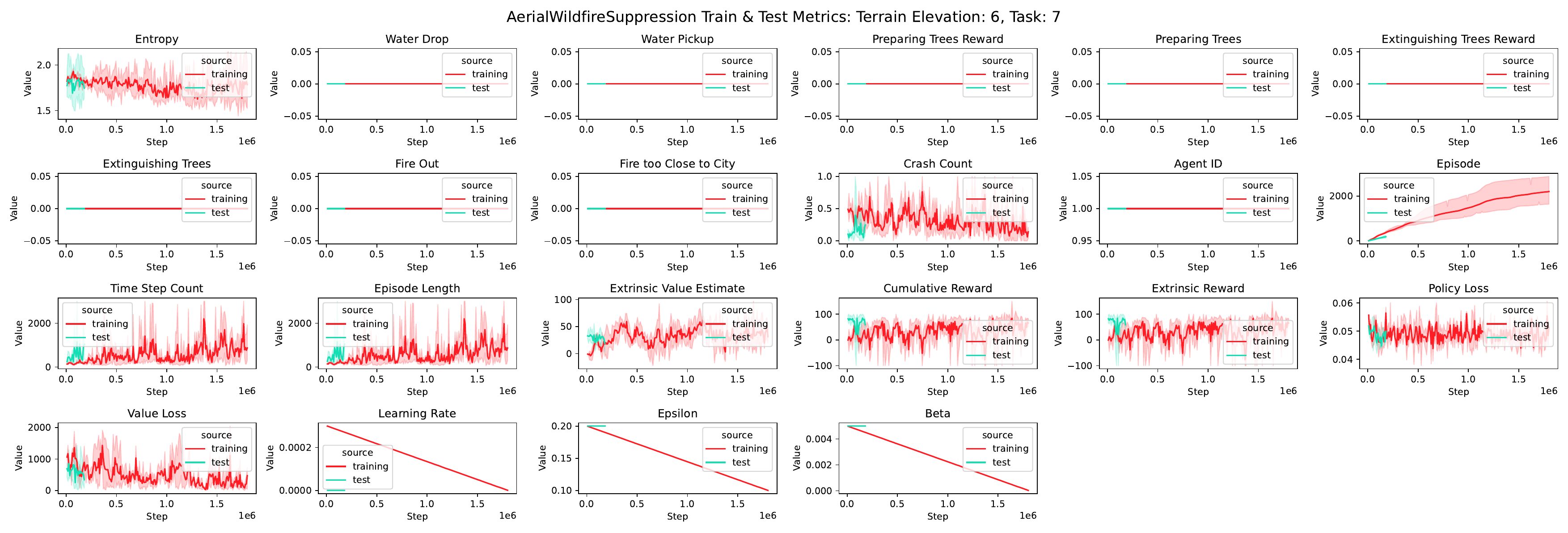}
\vspace{-0.6cm}
\caption{Aerial Wildfire Suppression: Train \& Test Metrics: Terrain Elevation 6, Task 7.}
\end{figure}

\begin{figure}[h!]
\centering
\includegraphics[width=\linewidth]{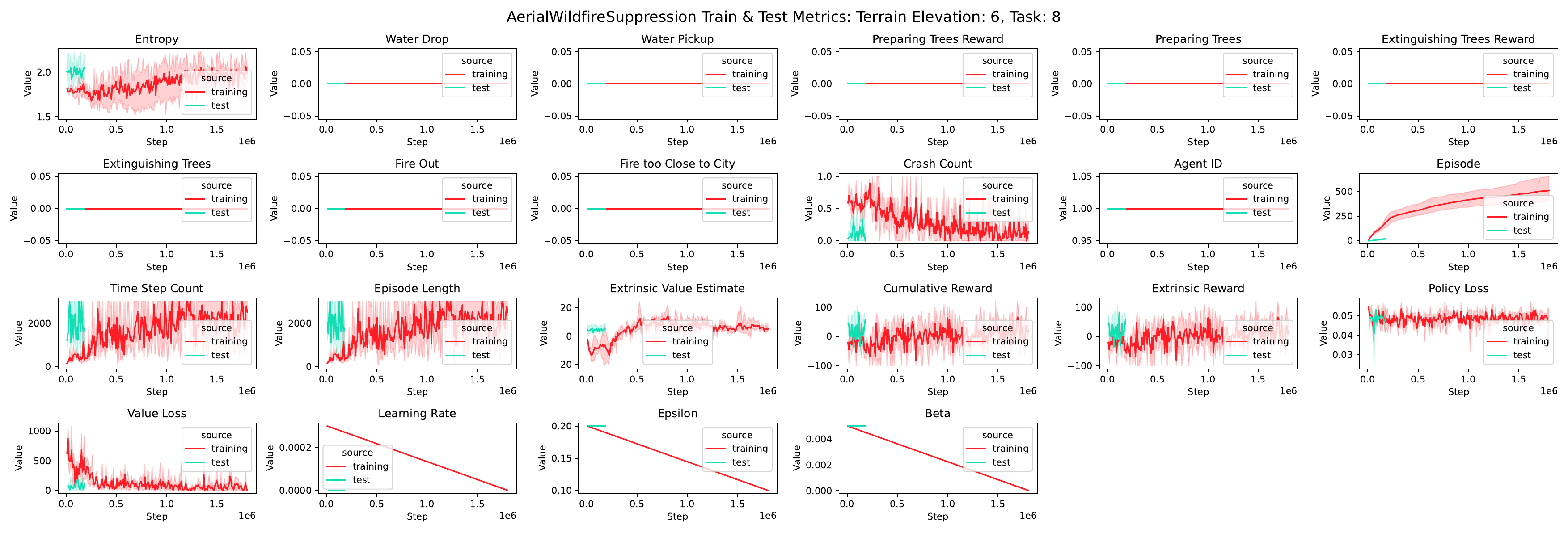}
\vspace{-0.6cm}
\caption{Aerial Wildfire Suppression: Train \& Test Metrics: Terrain Elevation 6, Task 8.}
\end{figure}

\clearpage

\begin{figure}[h!]
\centering
\includegraphics[width=\linewidth]{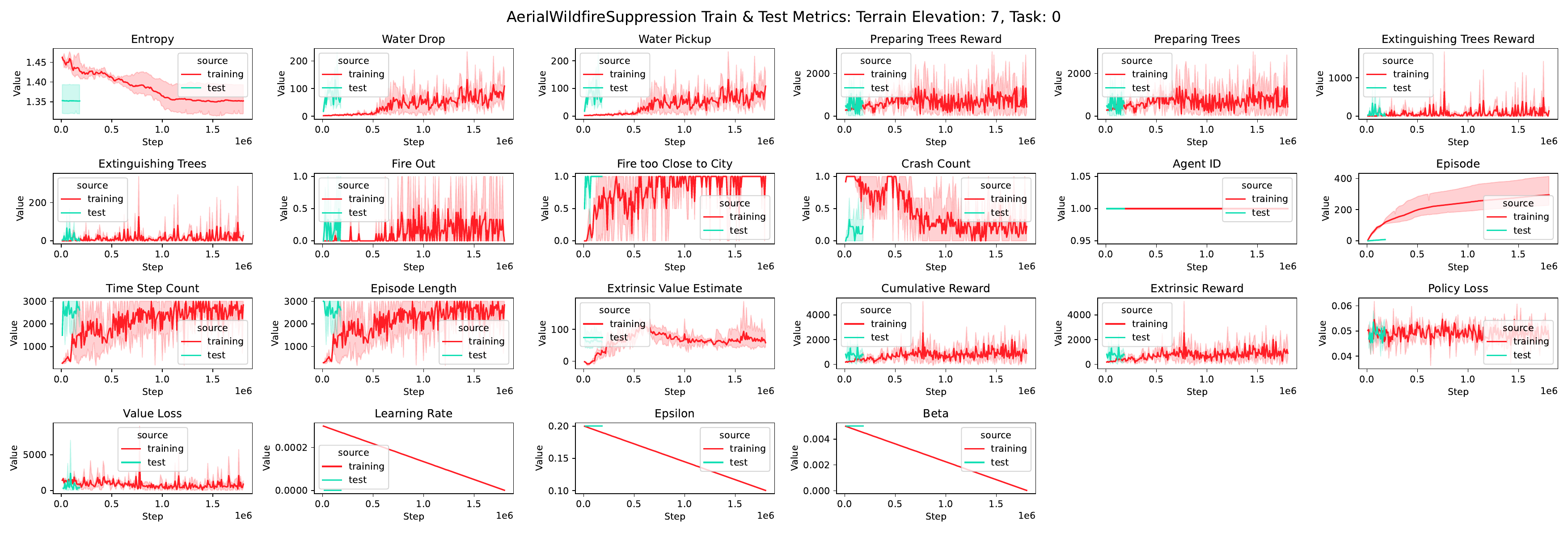}
\vspace{-0.6cm}
\caption{Aerial Wildfire Suppression: Train \& Test Metrics: Terrain Elevation 7, Task 0.}
\end{figure}

\begin{figure}[h!]
\centering
\includegraphics[width=\linewidth]{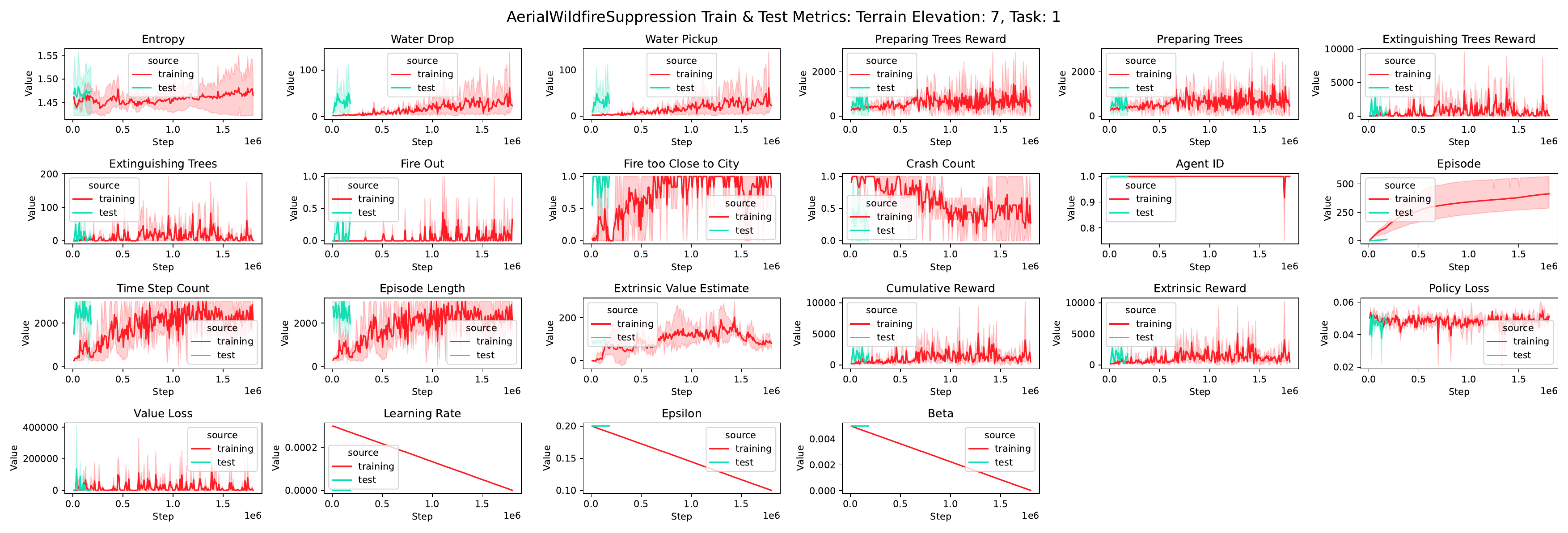}
\vspace{-0.6cm}
\caption{Aerial Wildfire Suppression: Train \& Test Metrics: Terrain Elevation 7, Task 1.}
\end{figure}

\begin{figure}[h!]
\centering
\includegraphics[width=\linewidth]{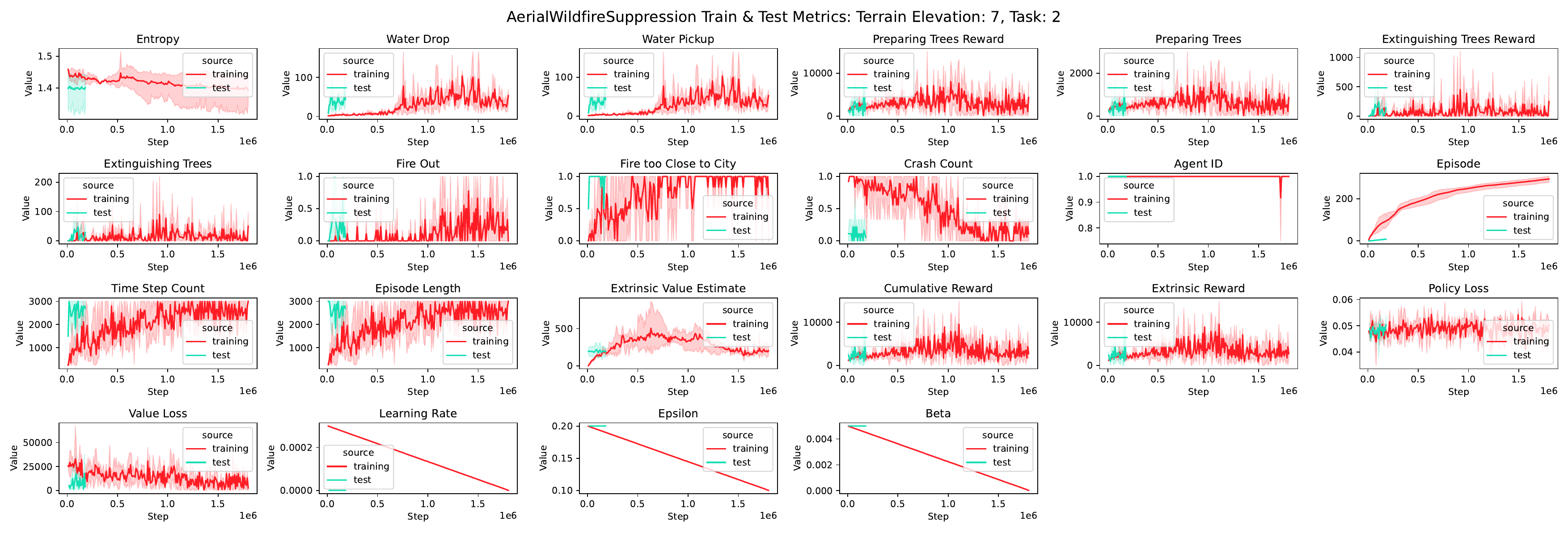}
\vspace{-0.6cm}
\caption{Aerial Wildfire Suppression: Train \& Test Metrics: Terrain Elevation 7, Task 2.}
\end{figure}

\clearpage

\begin{figure}[h!]
\centering
\includegraphics[width=\linewidth]{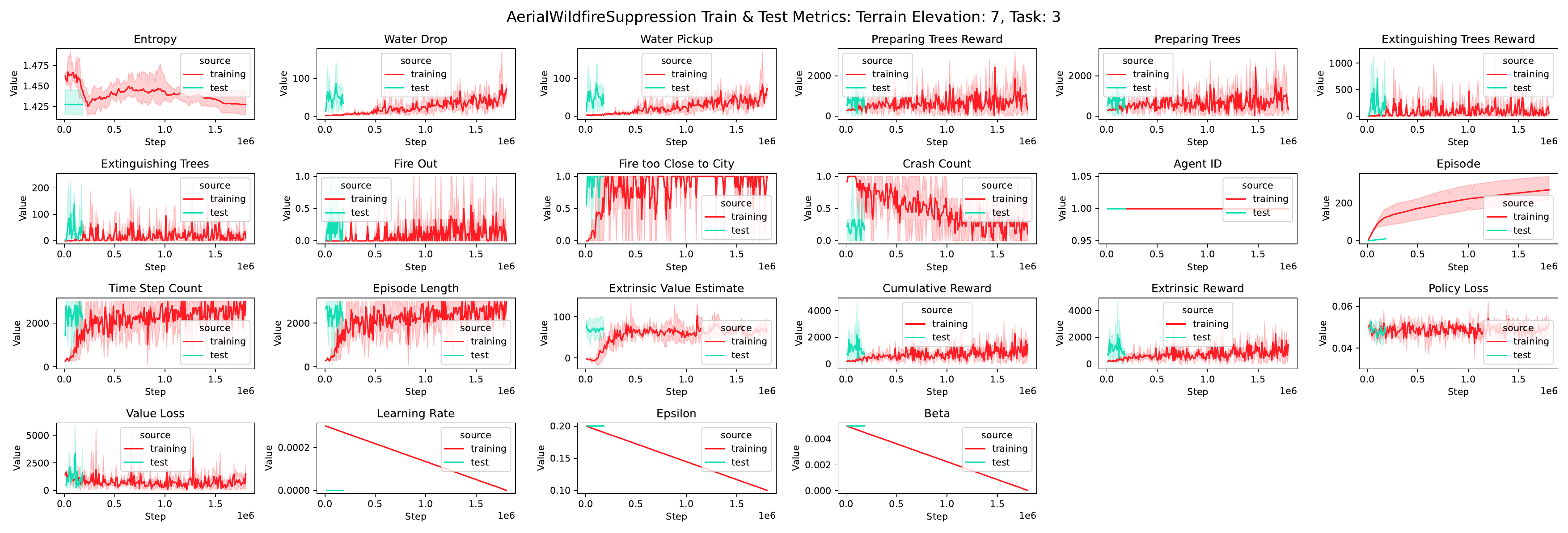}
\vspace{-0.6cm}
\caption{Aerial Wildfire Suppression: Train \& Test Metrics: Terrain Elevation 7, Task 3.}
\end{figure}

\begin{figure}[h!]
\centering
\includegraphics[width=\linewidth]{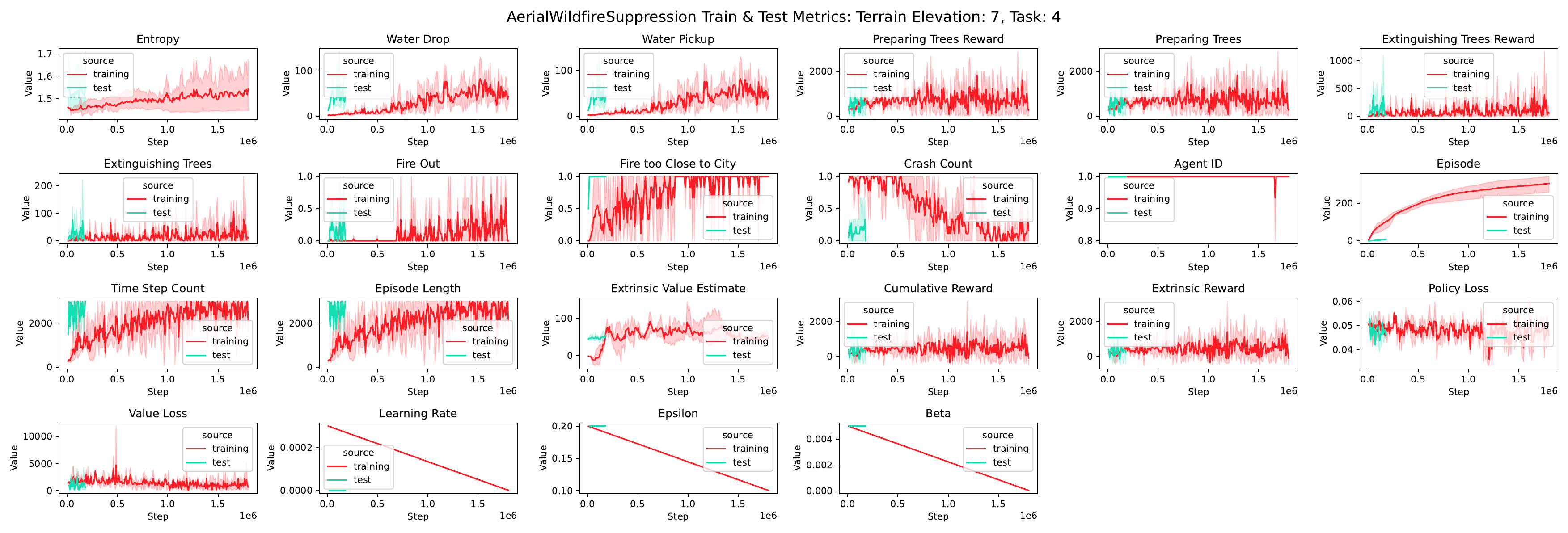}
\vspace{-0.6cm}
\caption{Aerial Wildfire Suppression: Train \& Test Metrics: Terrain Elevation 7, Task 4.}
\end{figure}

\begin{figure}[h!]
\centering
\includegraphics[width=\linewidth]{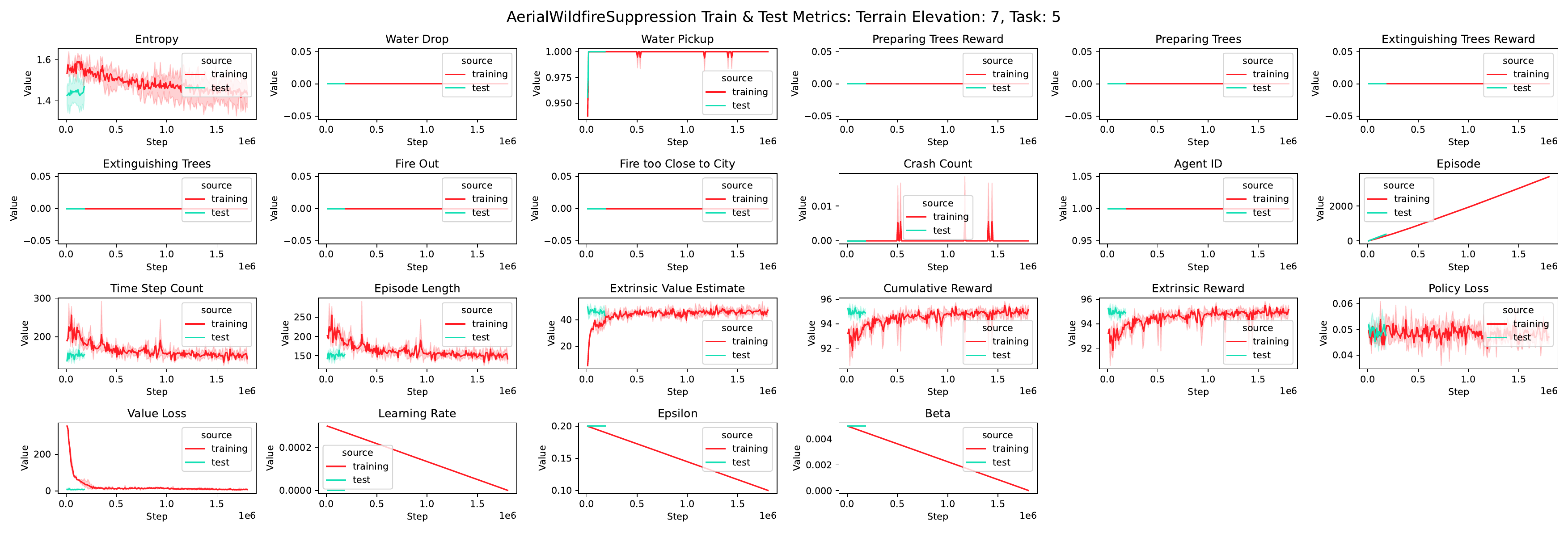}
\vspace{-0.6cm}
\caption{Aerial Wildfire Suppression: Train \& Test Metrics: Terrain Elevation 7, Task 5.}
\end{figure}

\clearpage

\begin{figure}[h!]
\centering
\includegraphics[width=\linewidth]{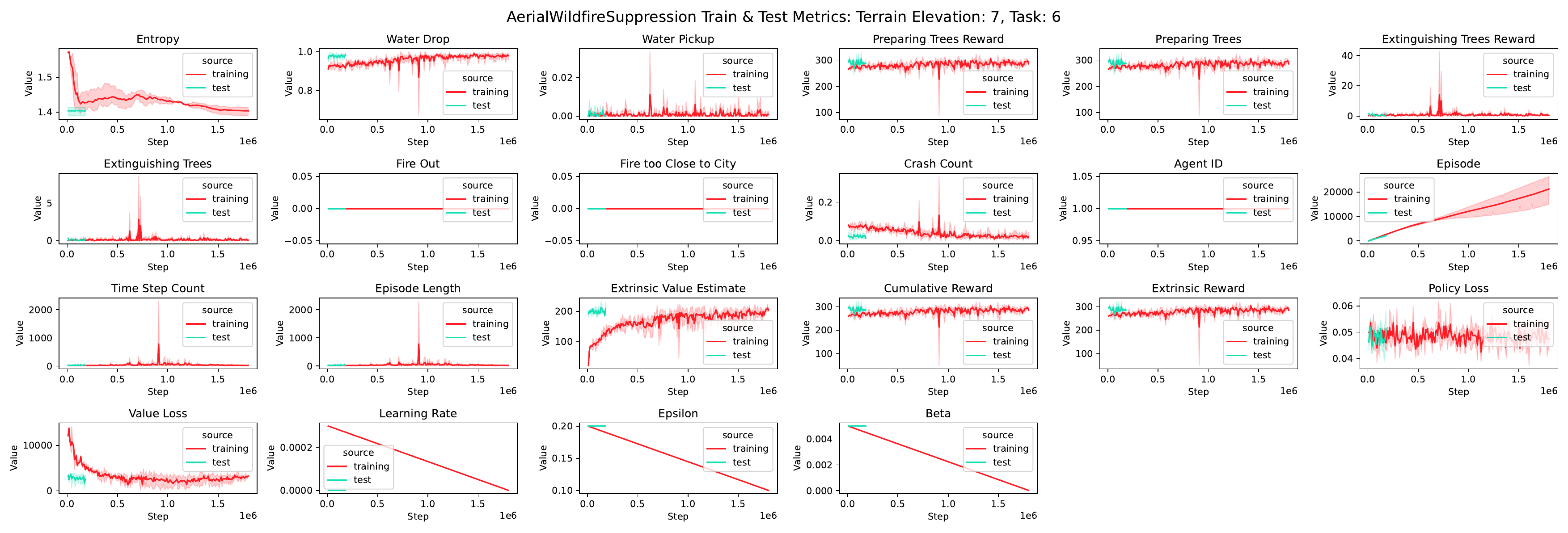}
\vspace{-0.6cm}
\caption{Aerial Wildfire Suppression: Train \& Test Metrics: Terrain Elevation 7, Task 6.}
\end{figure}

\begin{figure}[h!]
\centering
\includegraphics[width=\linewidth]{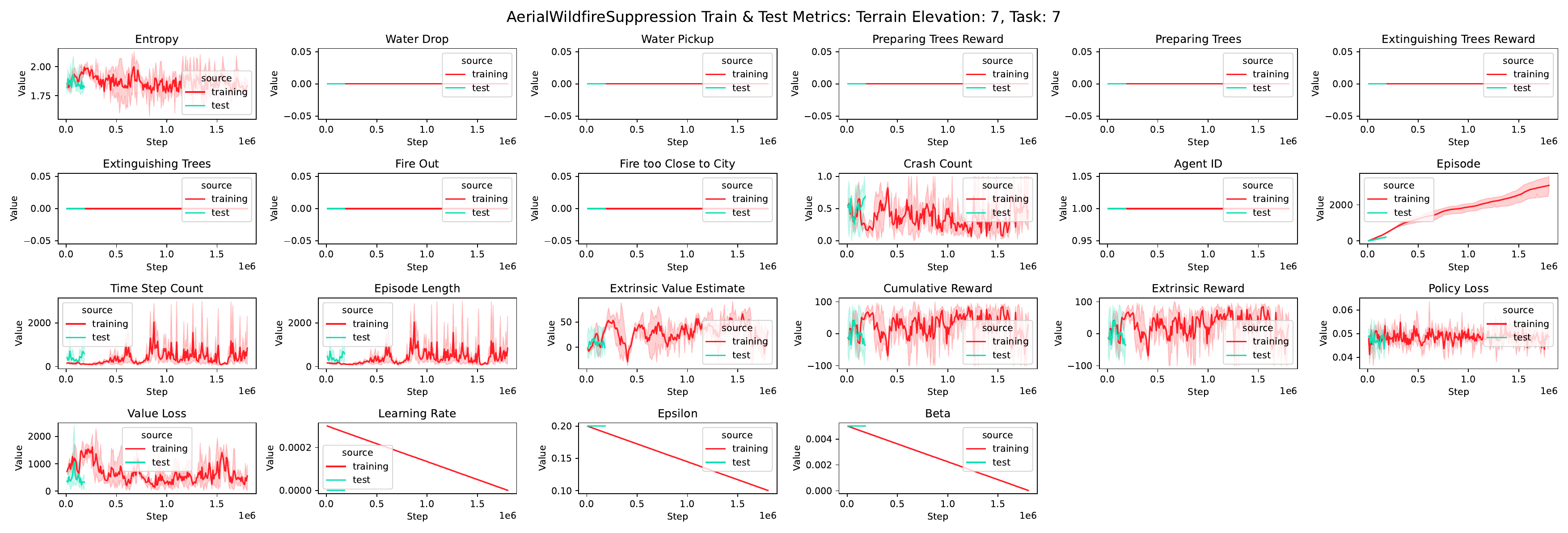}
\vspace{-0.6cm}
\caption{Aerial Wildfire Suppression: Train \& Test Metrics: Terrain Elevation 7, Task 7.}
\end{figure}

\begin{figure}[h!]
\centering
\includegraphics[width=\linewidth]{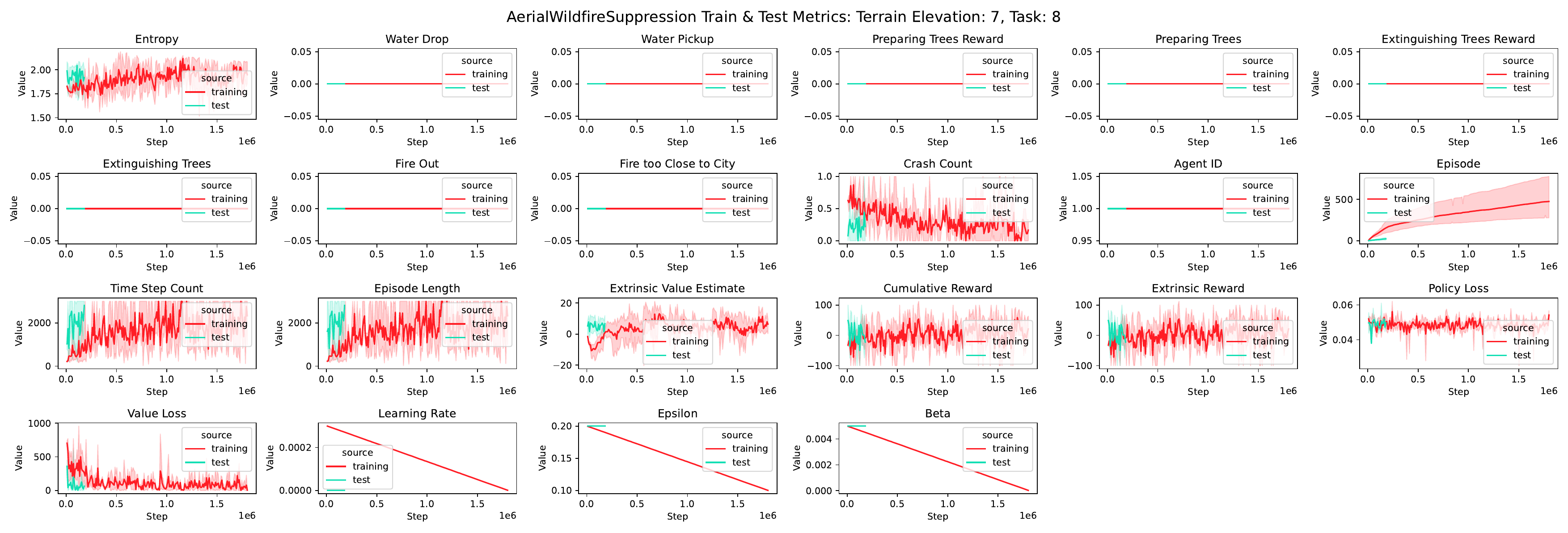}
\vspace{-0.6cm}
\caption{Aerial Wildfire Suppression: Train \& Test Metrics: Terrain Elevation 7, Task 8.}
\end{figure}

\clearpage

\begin{figure}[h!]
\centering
\includegraphics[width=\linewidth]{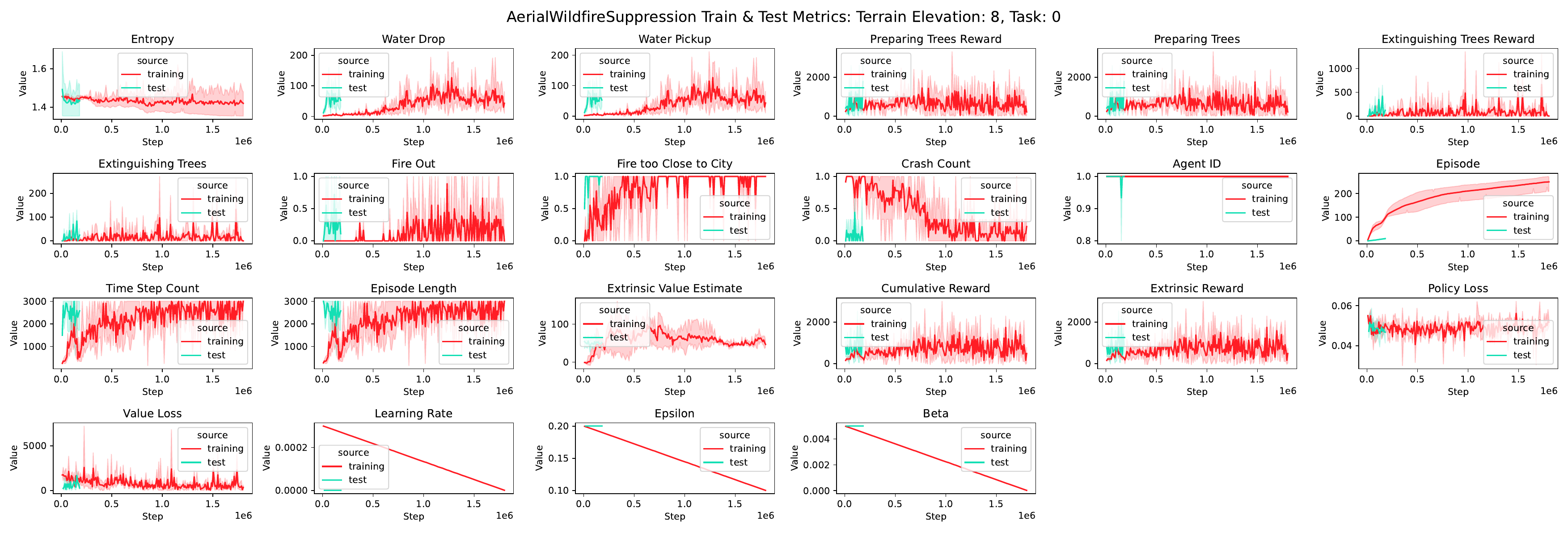}
\vspace{-0.6cm}
\caption{Aerial Wildfire Suppression: Train \& Test Metrics: Terrain Elevation 8, Task 0.}
\end{figure}

\begin{figure}[h!]
\centering
\includegraphics[width=\linewidth]{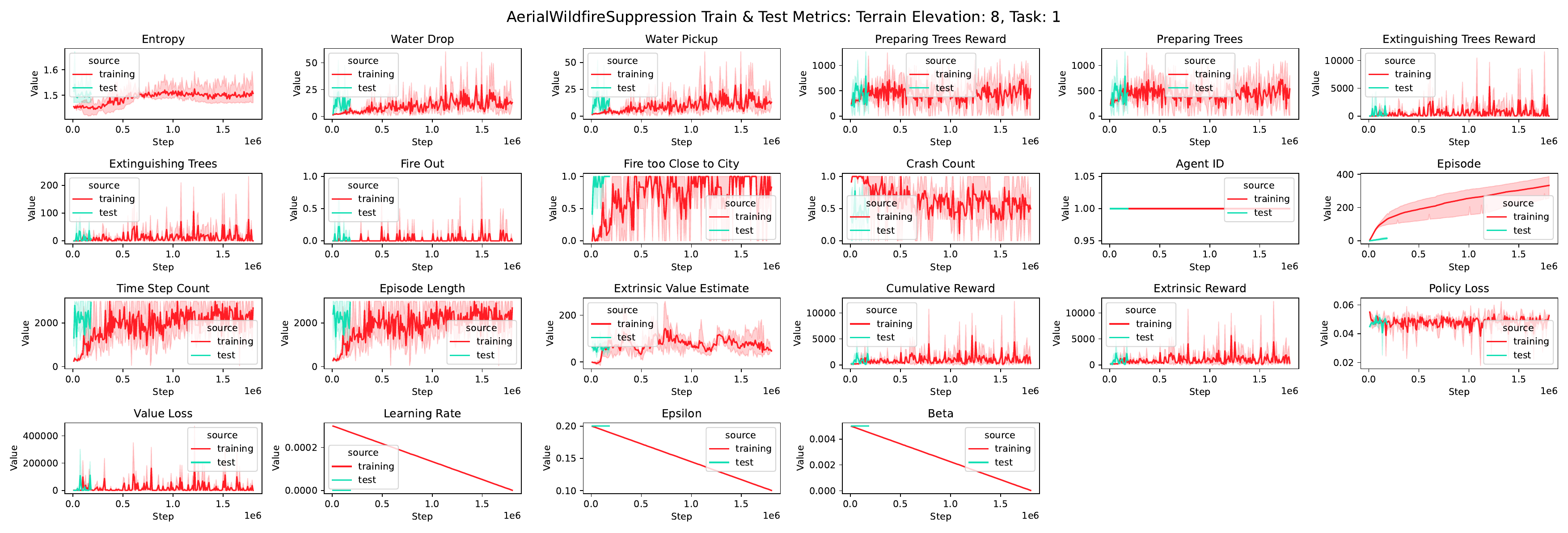}
\vspace{-0.6cm}
\caption{Aerial Wildfire Suppression: Train \& Test Metrics: Terrain Elevation 8, Task 1.}
\end{figure}

\begin{figure}[h!]
\centering
\includegraphics[width=\linewidth]{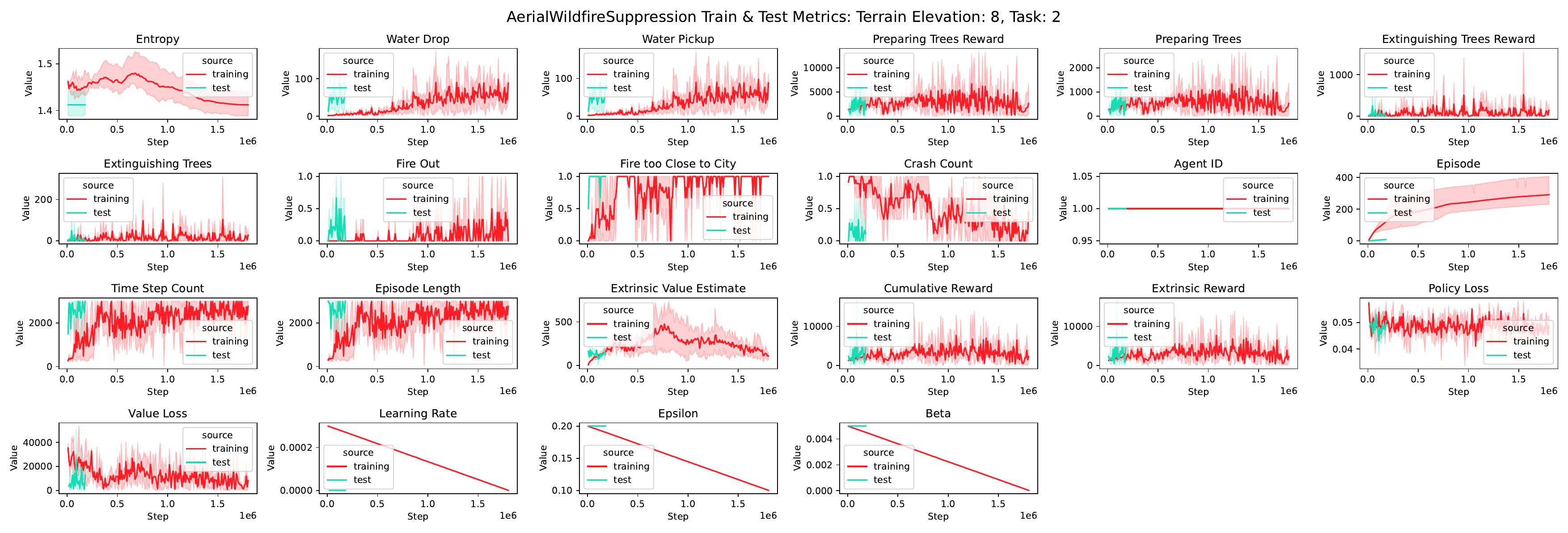}
\vspace{-0.6cm}
\caption{Aerial Wildfire Suppression: Train \& Test Metrics: Terrain Elevation 8, Task 2.}
\end{figure}

\clearpage

\begin{figure}[h!]
\centering
\includegraphics[width=\linewidth]{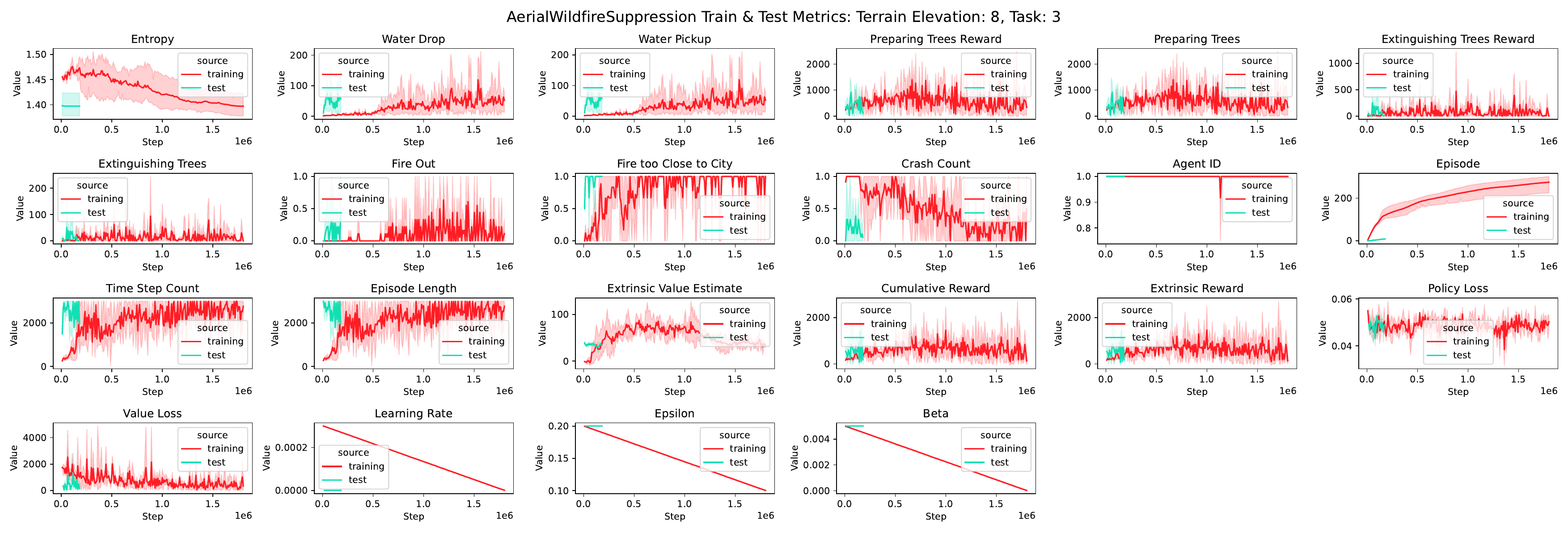}
\vspace{-0.6cm}
\caption{Aerial Wildfire Suppression: Train \& Test Metrics: Terrain Elevation 8, Task 3.}
\end{figure}

\begin{figure}[h!]
\centering
\includegraphics[width=\linewidth]{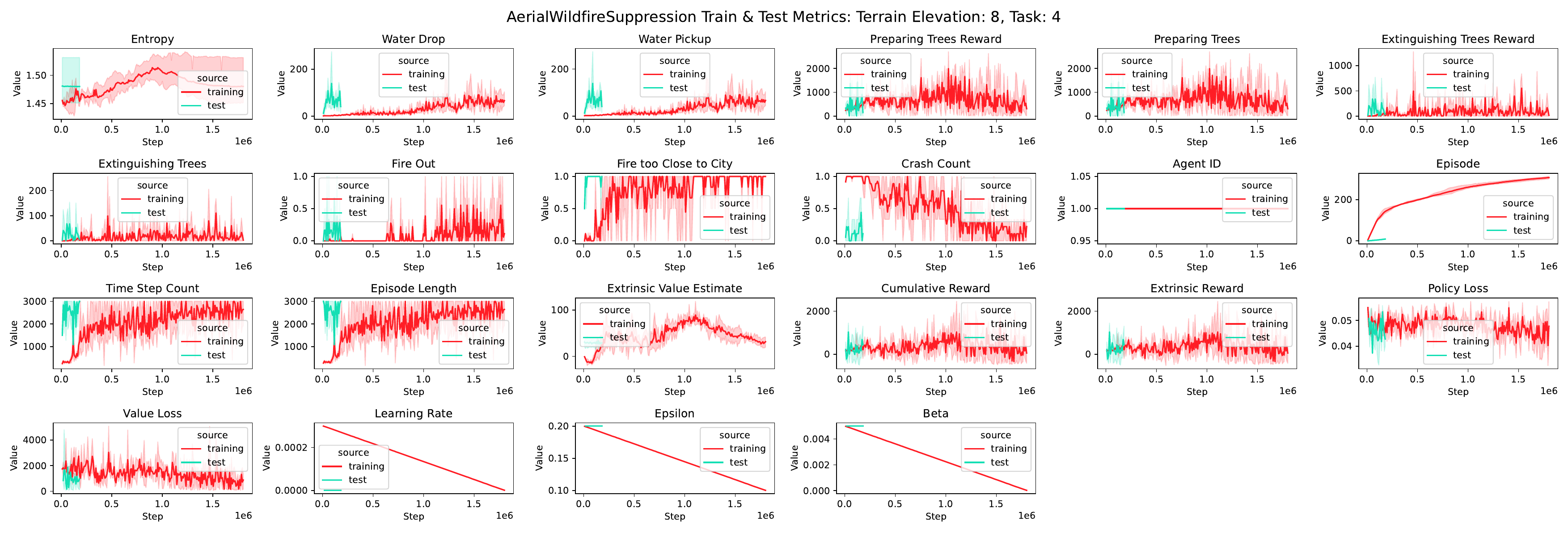}
\vspace{-0.6cm}
\caption{Aerial Wildfire Suppression: Train \& Test Metrics: Terrain Elevation 8, Task 4.}
\end{figure}

\begin{figure}[h!]
\centering
\includegraphics[width=\linewidth]{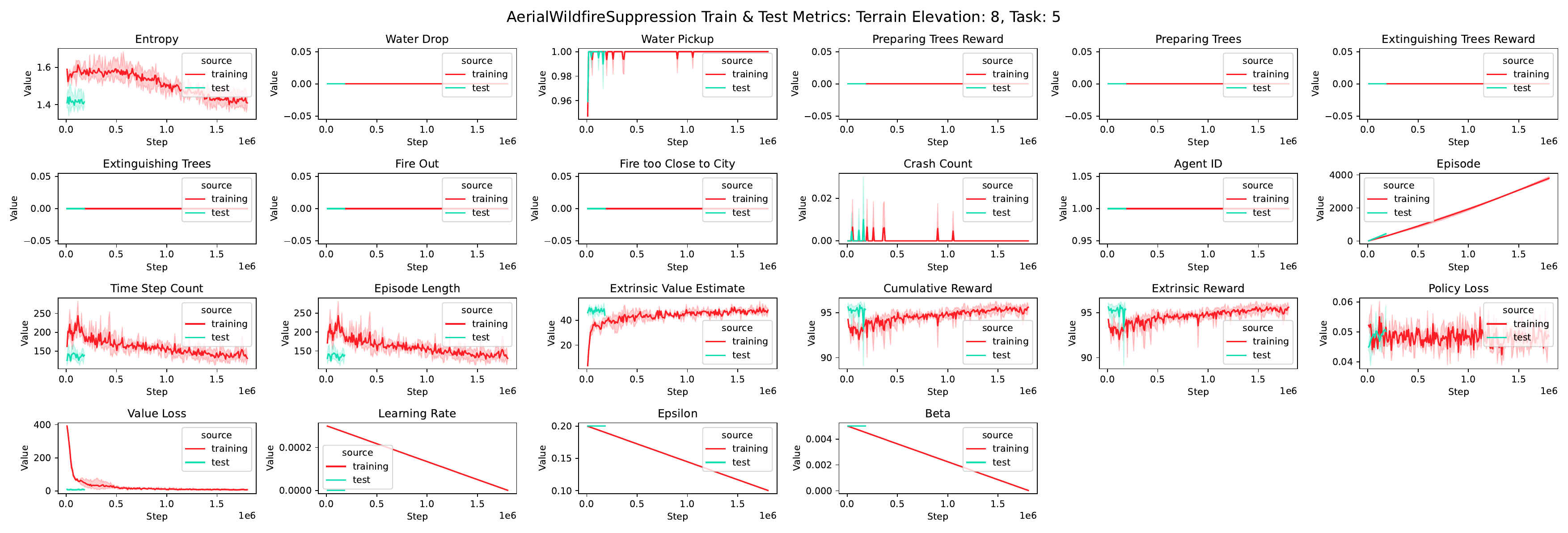}
\vspace{-0.6cm}
\caption{Aerial Wildfire Suppression: Train \& Test Metrics: Terrain Elevation 8, Task 5.}
\end{figure}

\clearpage

\begin{figure}[h!]
\centering
\includegraphics[width=\linewidth]{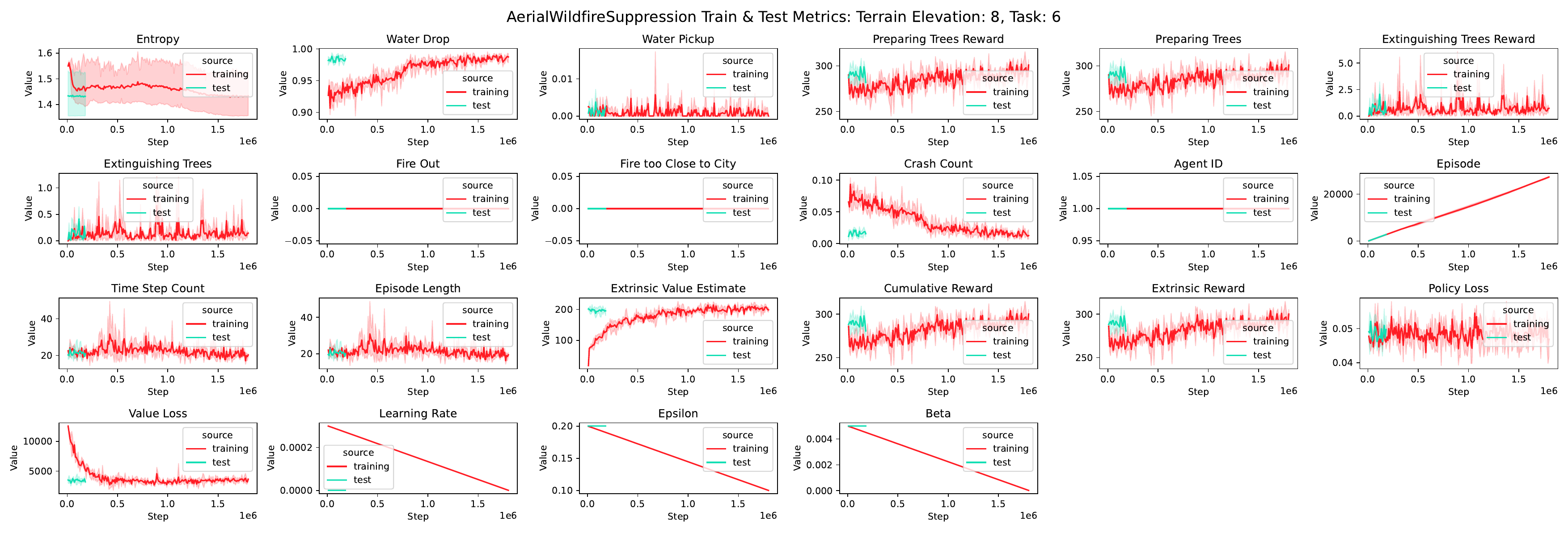}
\vspace{-0.6cm}
\caption{Aerial Wildfire Suppression: Train \& Test Metrics: Terrain Elevation 8, Task 6.}
\end{figure}

\begin{figure}[h!]
\centering
\includegraphics[width=\linewidth]{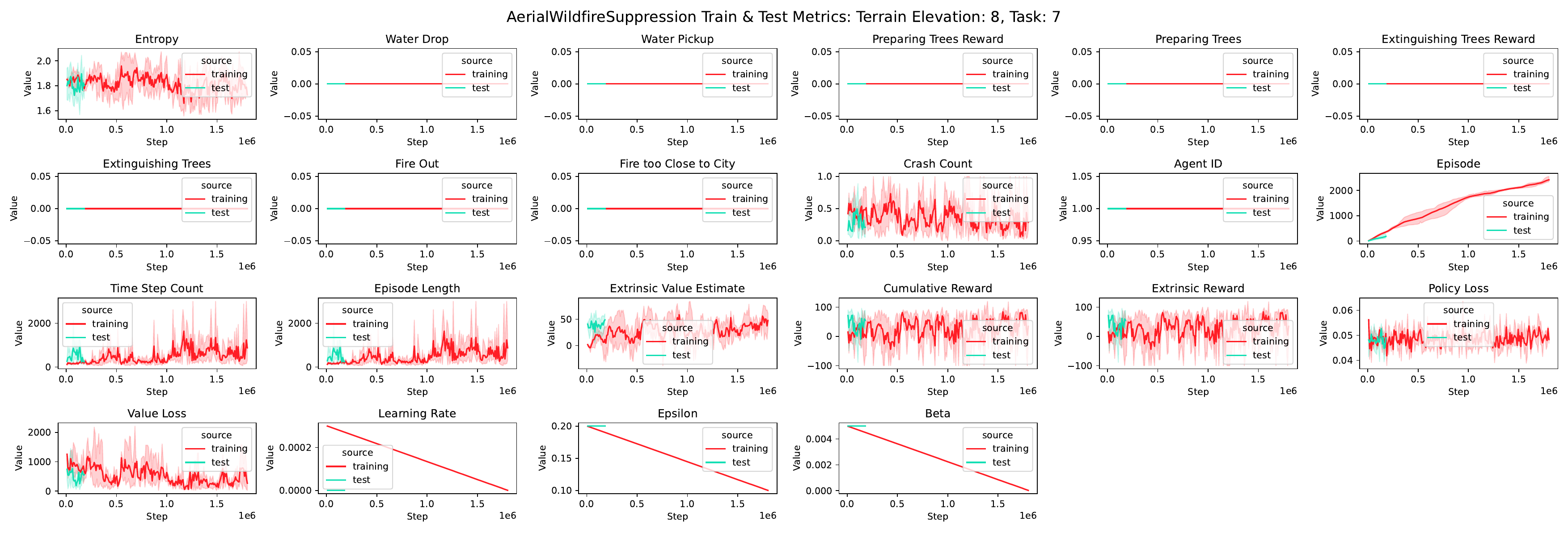}
\vspace{-0.6cm}
\caption{Aerial Wildfire Suppression: Train \& Test Metrics: Terrain Elevation 8, Task 7.}
\end{figure}

\begin{figure}[h!]
\centering
\includegraphics[width=\linewidth]{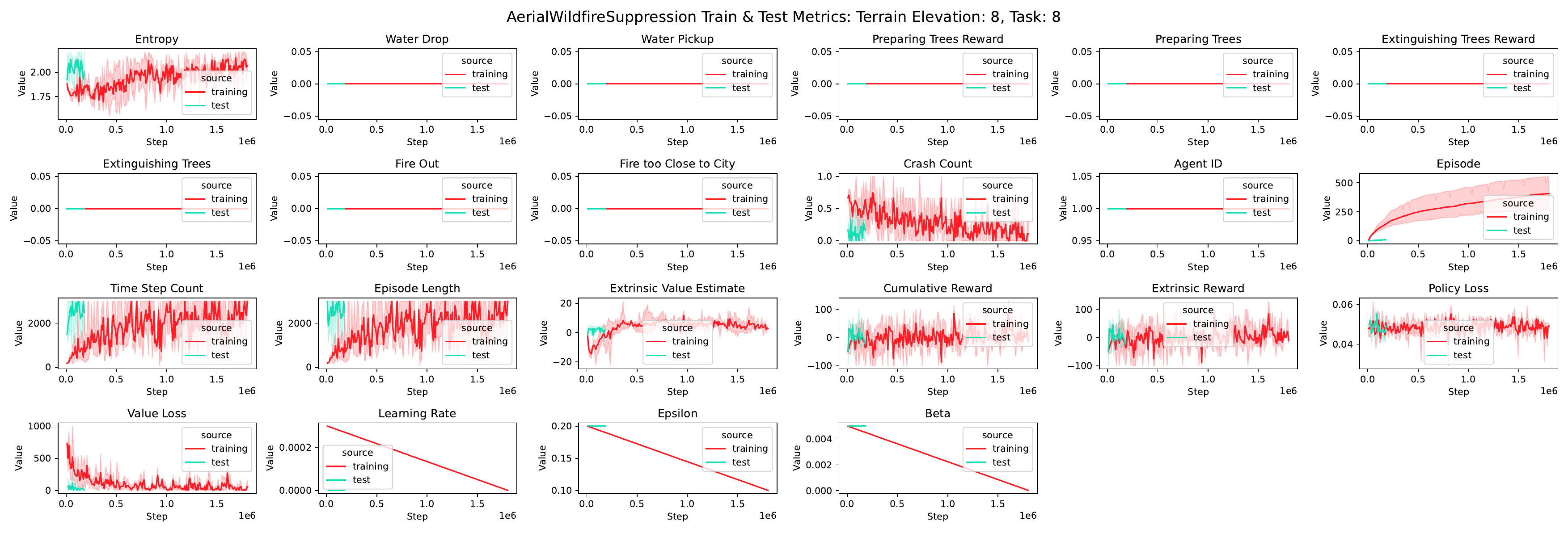}
\vspace{-0.6cm}
\caption{Aerial Wildfire Suppression: Train \& Test Metrics: Terrain Elevation 8, Task 8.}
\end{figure}

\clearpage

\begin{figure}[h!]
\centering
\includegraphics[width=\linewidth]{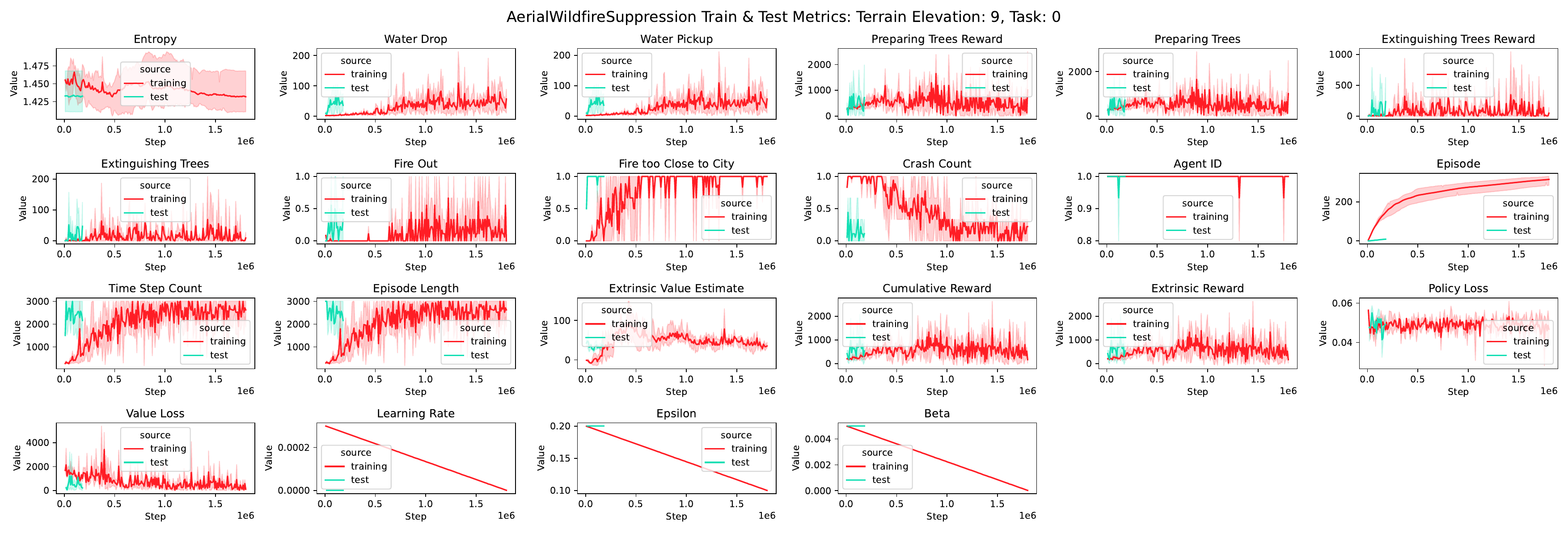}
\vspace{-0.6cm}
\caption{Aerial Wildfire Suppression: Train \& Test Metrics: Terrain Elevation 9, Task 0.}
\end{figure}

\begin{figure}[h!]
\centering
\includegraphics[width=\linewidth]{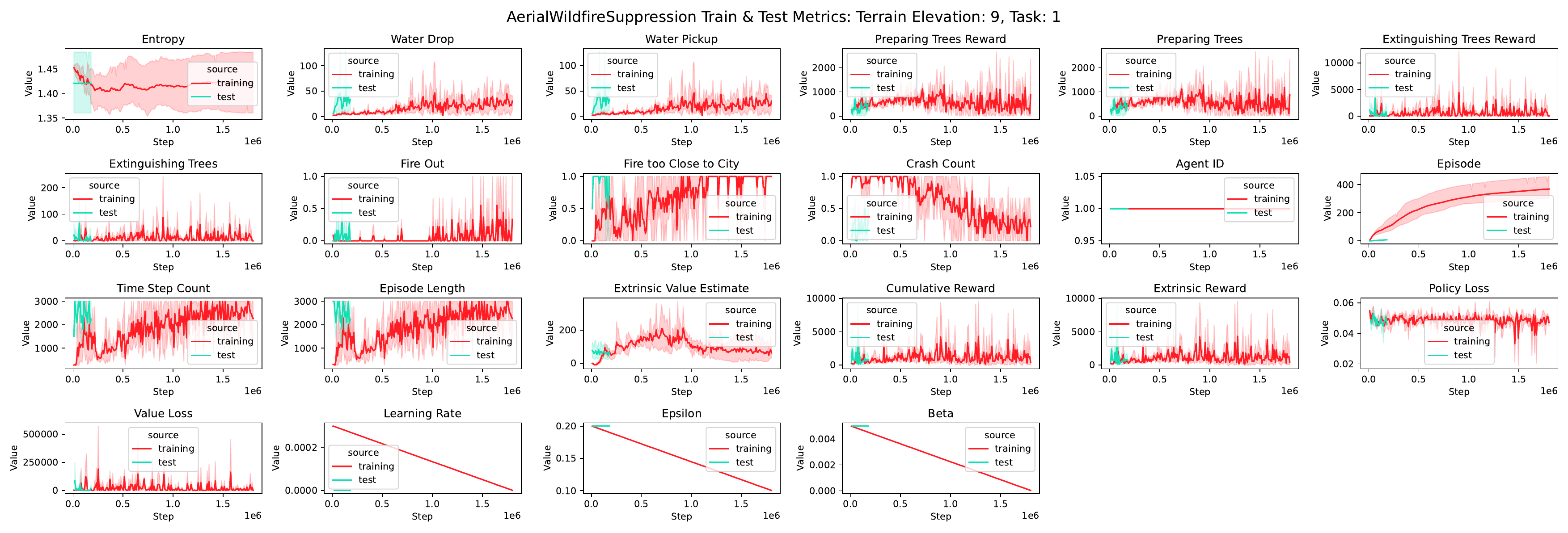}
\vspace{-0.6cm}
\caption{Aerial Wildfire Suppression: Train \& Test Metrics: Terrain Elevation 9, Task 1.}
\end{figure}

\begin{figure}[h!]
\centering
\includegraphics[width=\linewidth]{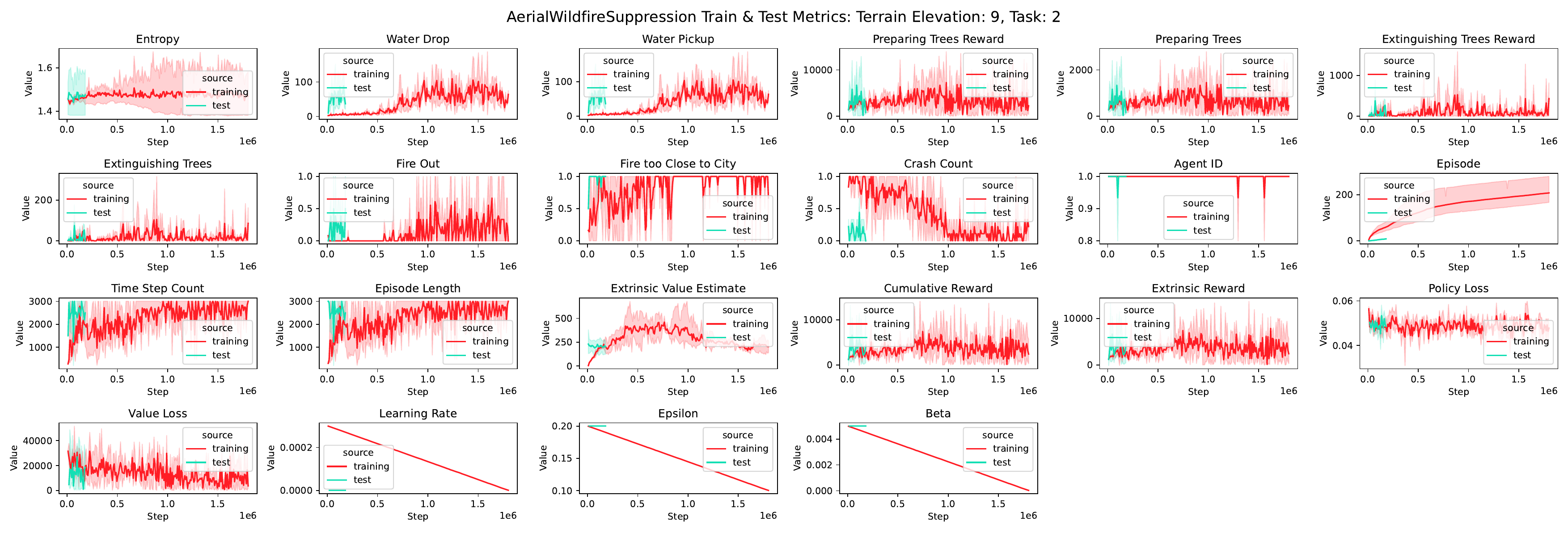}
\vspace{-0.6cm}
\caption{Aerial Wildfire Suppression: Train \& Test Metrics: Terrain Elevation 9, Task 2.}
\end{figure}

\clearpage

\begin{figure}[h!]
\centering
\includegraphics[width=\linewidth]{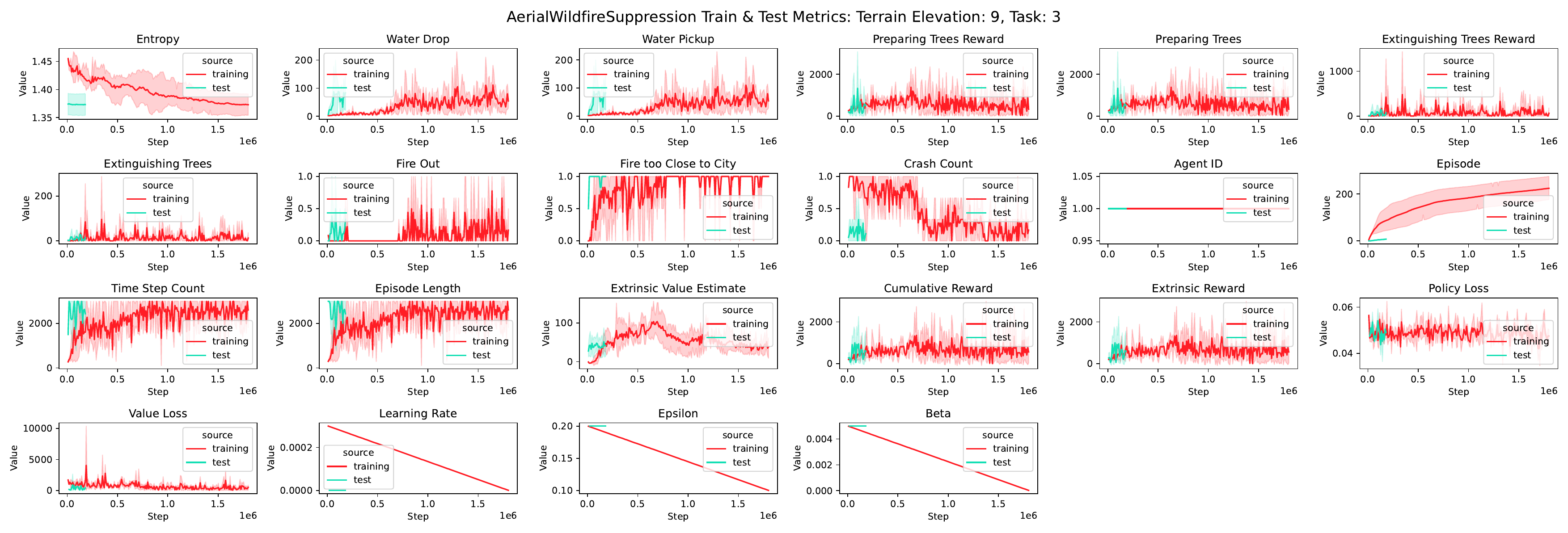}
\vspace{-0.6cm}
\caption{Aerial Wildfire Suppression: Train \& Test Metrics: Terrain Elevation 9, Task 3.}
\end{figure}

\begin{figure}[h!]
\centering
\includegraphics[width=\linewidth]{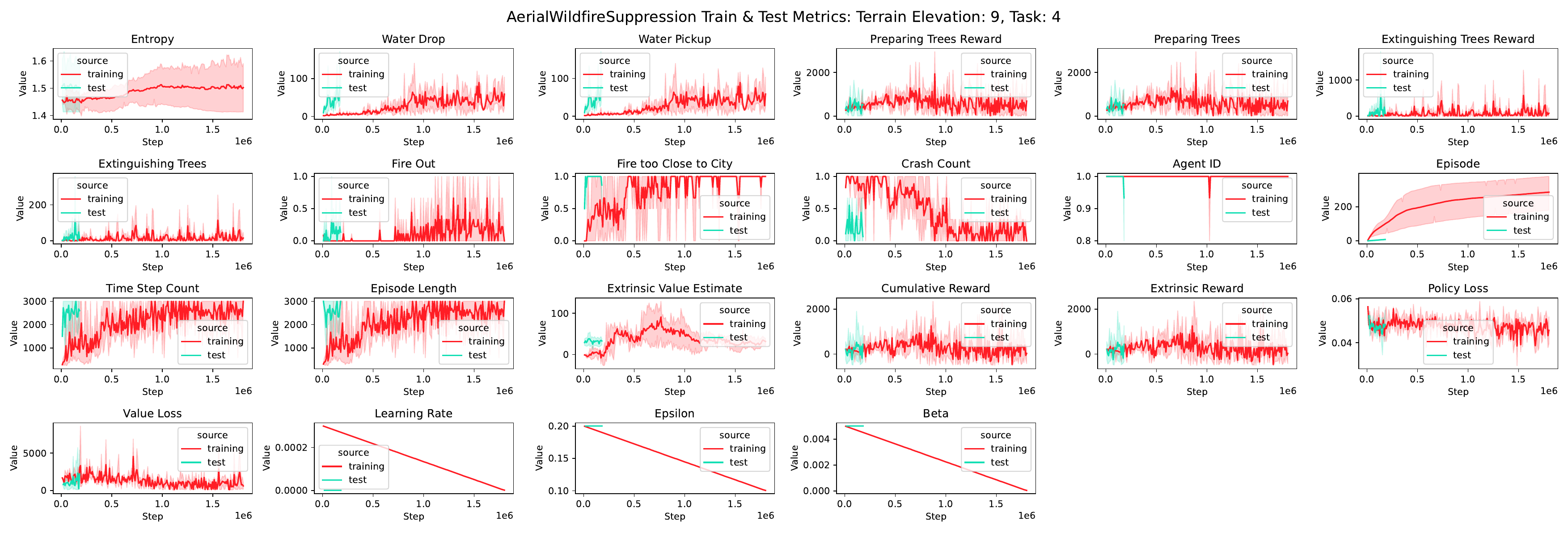}
\vspace{-0.6cm}
\caption{Aerial Wildfire Suppression: Train \& Test Metrics: Terrain Elevation 9, Task 4.}
\end{figure}

\begin{figure}[h!]
\centering
\includegraphics[width=\linewidth]{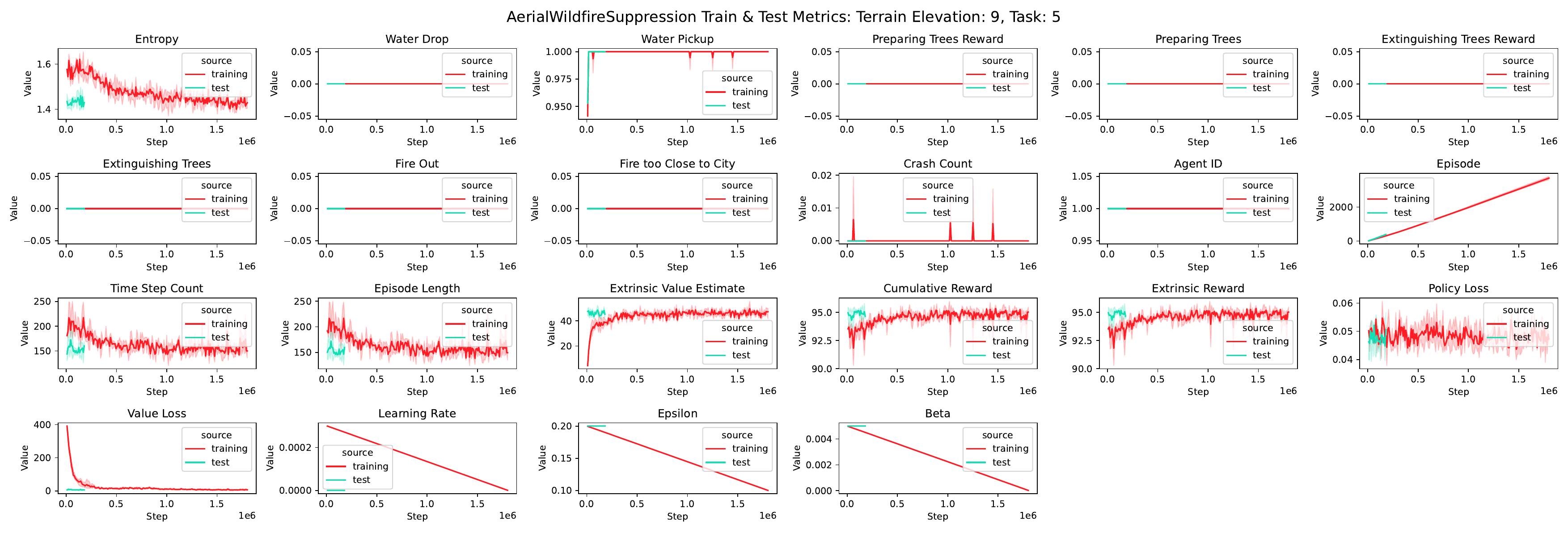}
\vspace{-0.6cm}
\caption{Aerial Wildfire Suppression: Train \& Test Metrics: Terrain Elevation 9, Task 5.}
\end{figure}

\clearpage

\begin{figure}[h!]
\centering
\includegraphics[width=\linewidth]{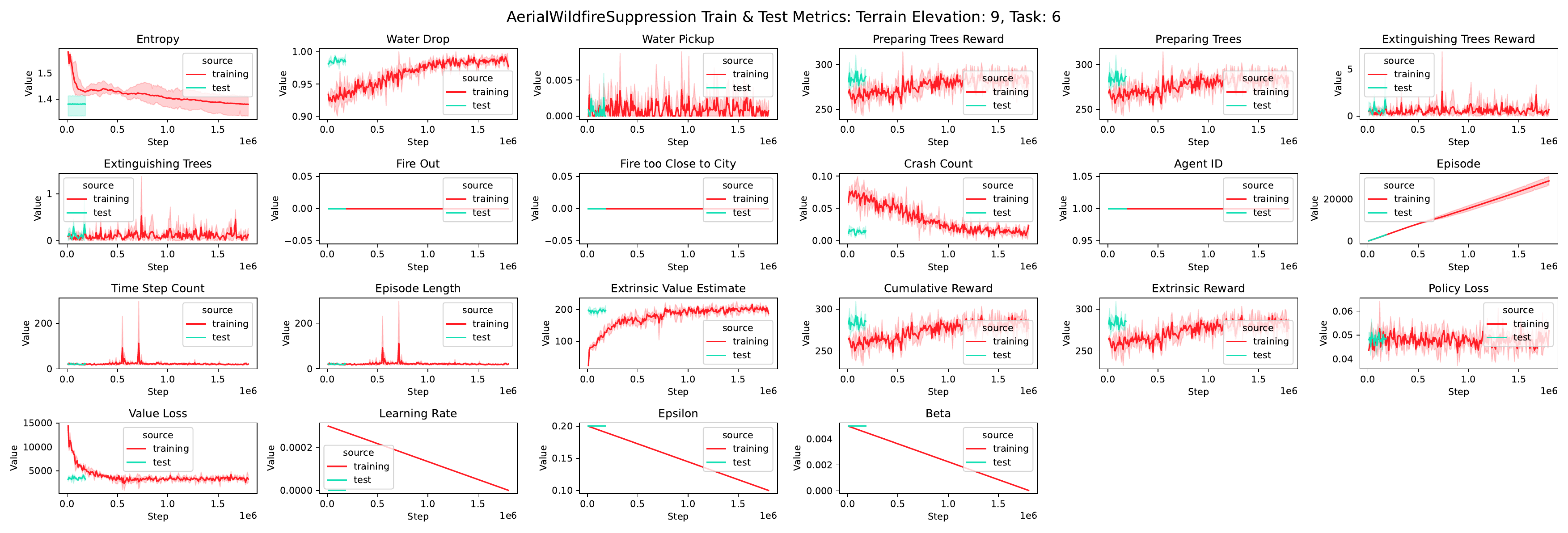}
\vspace{-0.6cm}
\caption{Aerial Wildfire Suppression: Train \& Test Metrics: Terrain Elevation 9, Task 6.}
\end{figure}

\begin{figure}[h!]
\centering
\includegraphics[width=\linewidth]{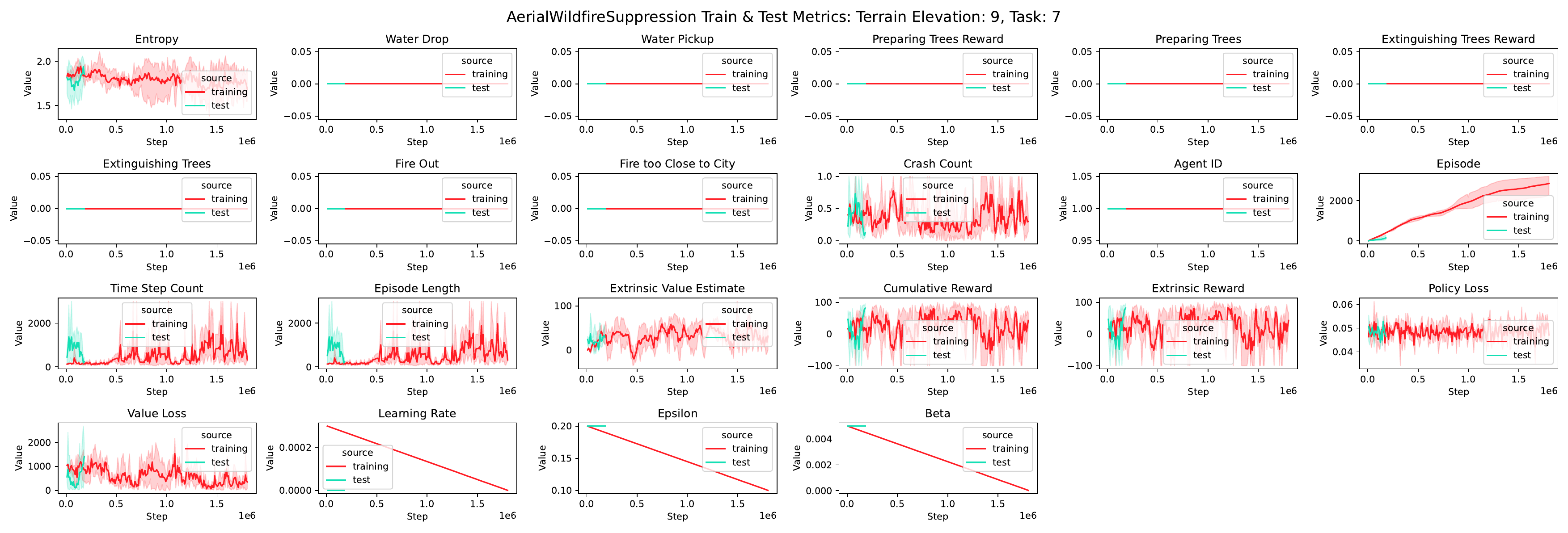}
\vspace{-0.6cm}
\caption{Aerial Wildfire Suppression: Train \& Test Metrics: Terrain Elevation 9, Task 7.}
\end{figure}

\begin{figure}[h!]
\centering
\includegraphics[width=\linewidth]{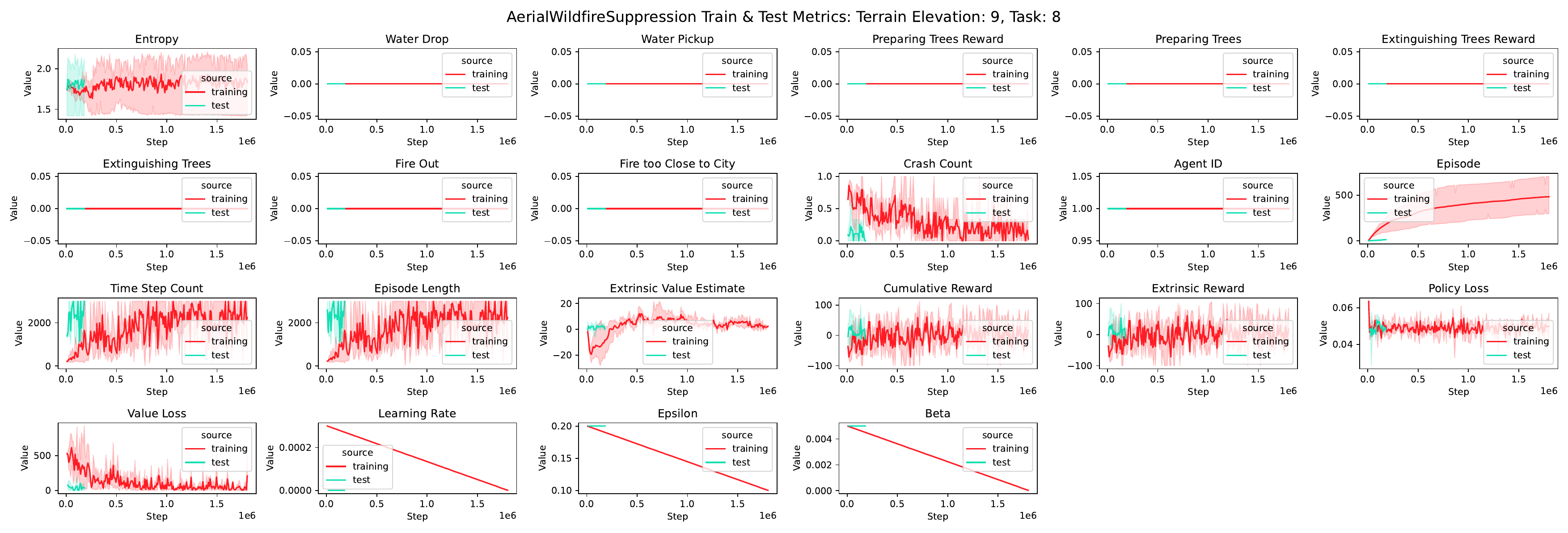}
\vspace{-0.6cm}
\caption{Aerial Wildfire Suppression: Train \& Test Metrics: Terrain Elevation 9, Task 8.}
\end{figure}

\clearpage

\begin{figure}[h!]
\centering
\includegraphics[width=\linewidth]{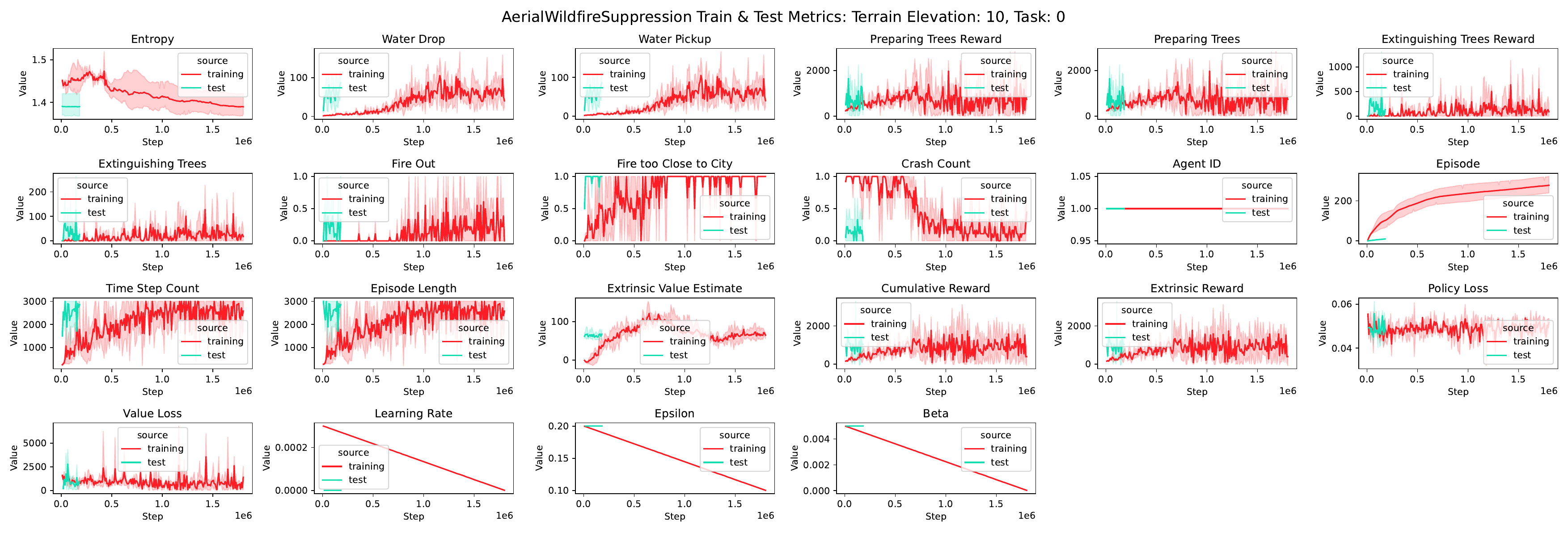}
\vspace{-0.6cm}
\caption{Aerial Wildfire Suppression: Train \& Test Metrics: Terrain Elevation 10, Task 0.}
\end{figure}

\begin{figure}[h!]
\centering
\includegraphics[width=\linewidth]{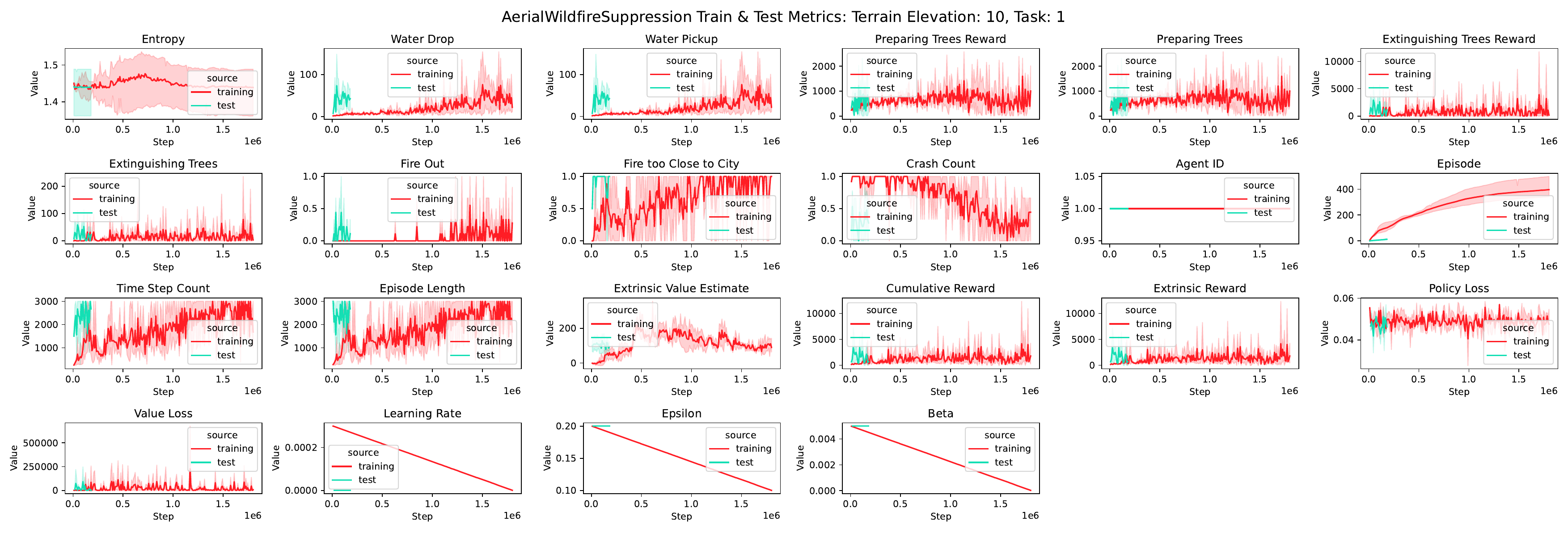}
\vspace{-0.6cm}
\caption{Aerial Wildfire Suppression: Train \& Test Metrics: Terrain Elevation 10, Task 1.}
\end{figure}

\begin{figure}[h!]
\centering
\includegraphics[width=\linewidth]{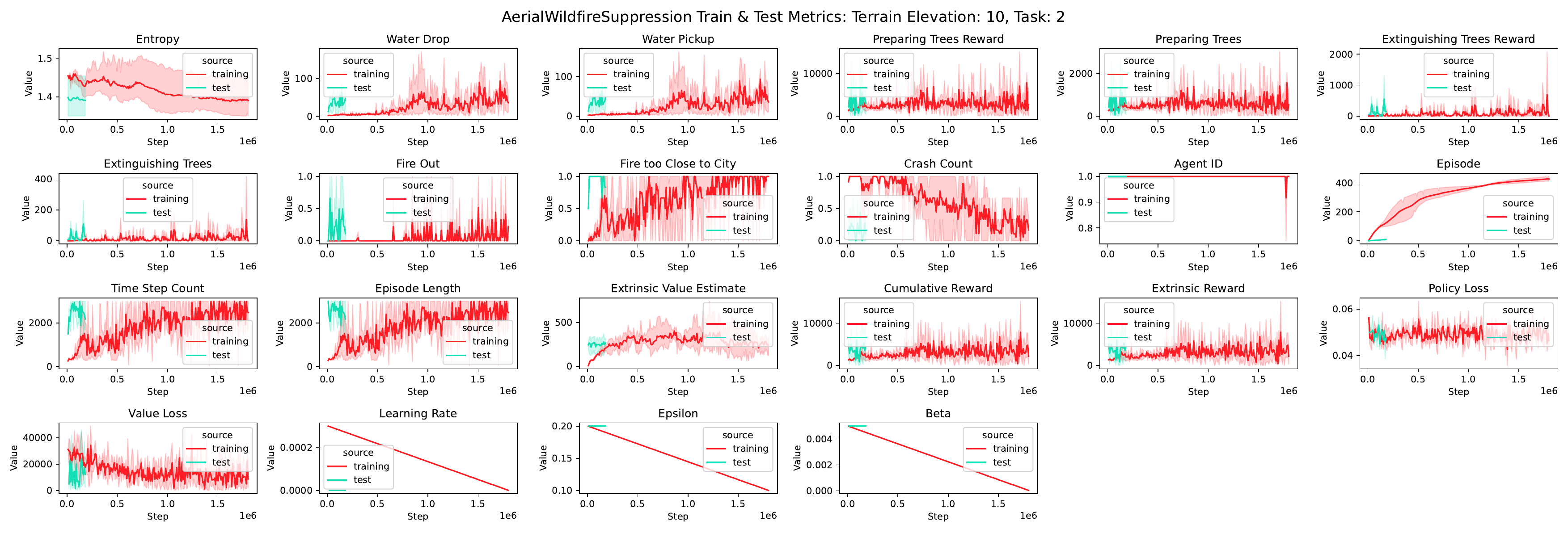}
\vspace{-0.6cm}
\caption{Aerial Wildfire Suppression: Train \& Test Metrics: Terrain Elevation 10, Task 2.}
\end{figure}

\clearpage

\begin{figure}[h!]
\centering
\includegraphics[width=\linewidth]{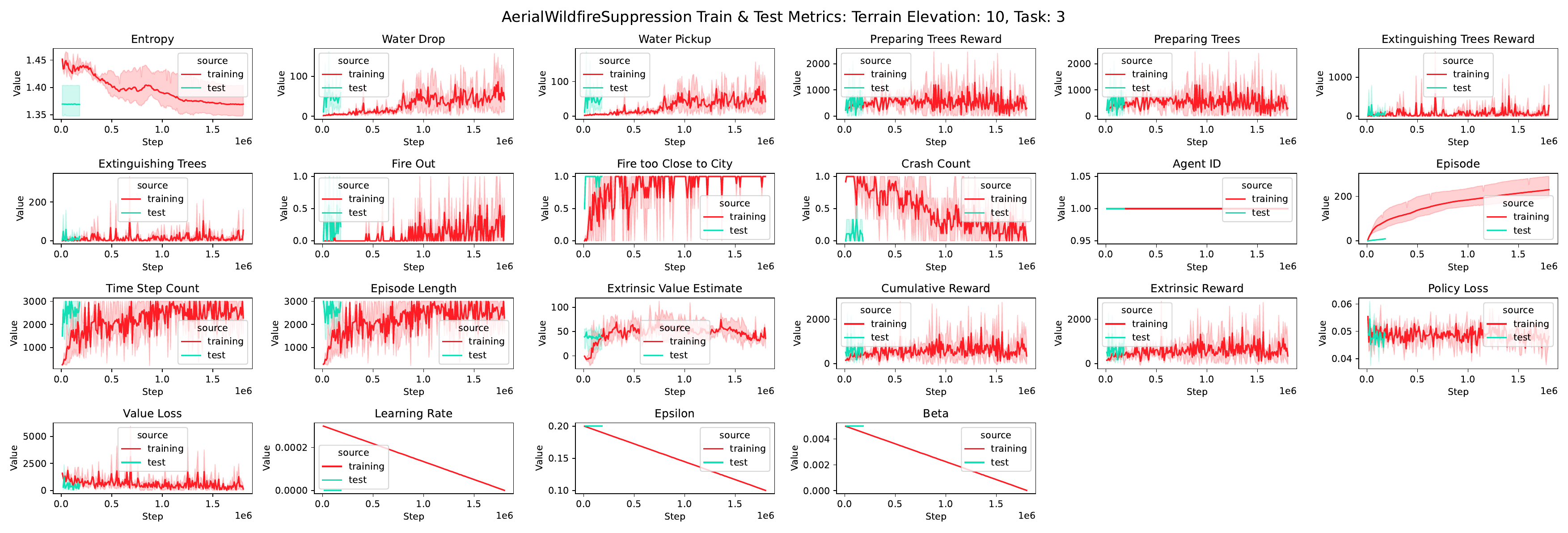}
\vspace{-0.6cm}
\caption{Aerial Wildfire Suppression: Train \& Test Metrics: Terrain Elevation 10, Task 3.}
\end{figure}

\begin{figure}[h!]
\centering
\includegraphics[width=\linewidth]{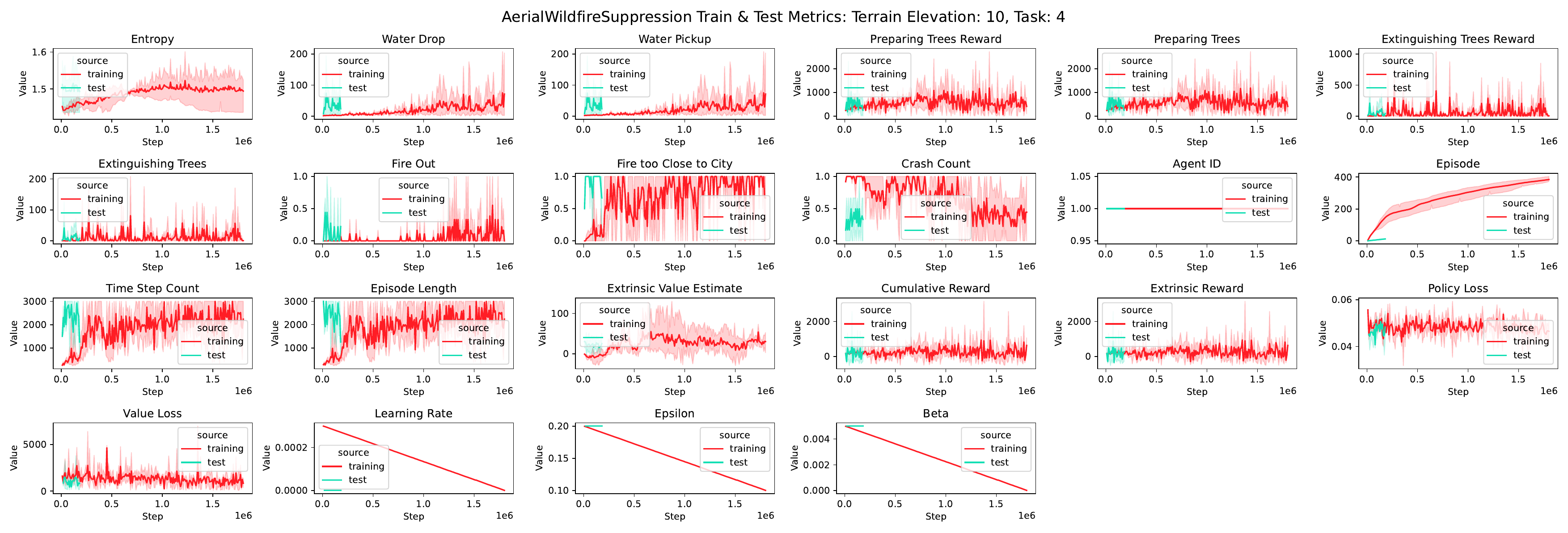}
\vspace{-0.6cm}
\caption{Aerial Wildfire Suppression: Train \& Test Metrics: Terrain Elevation 10, Task 4.}
\end{figure}

\begin{figure}[h!]
\centering
\includegraphics[width=\linewidth]{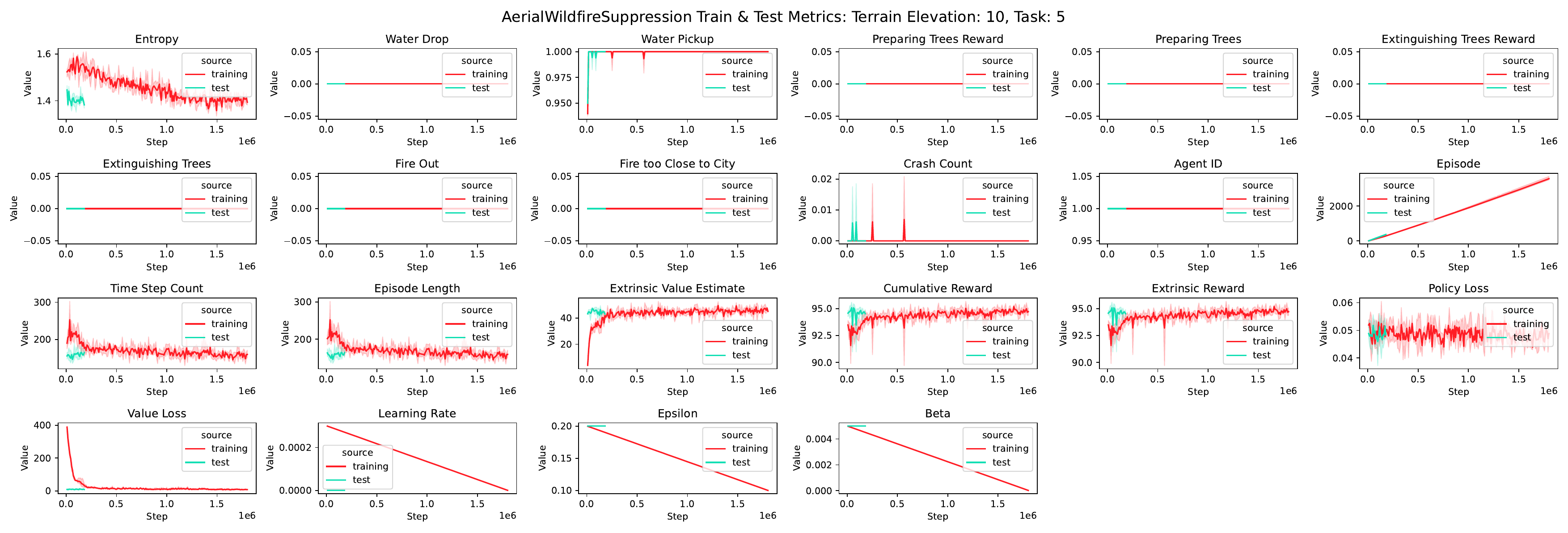}
\vspace{-0.6cm}
\caption{Aerial Wildfire Suppression: Train \& Test Metrics: Terrain Elevation 10, Task 5.}
\end{figure}

\clearpage

\begin{figure}[h!]
\centering
\includegraphics[width=\linewidth]{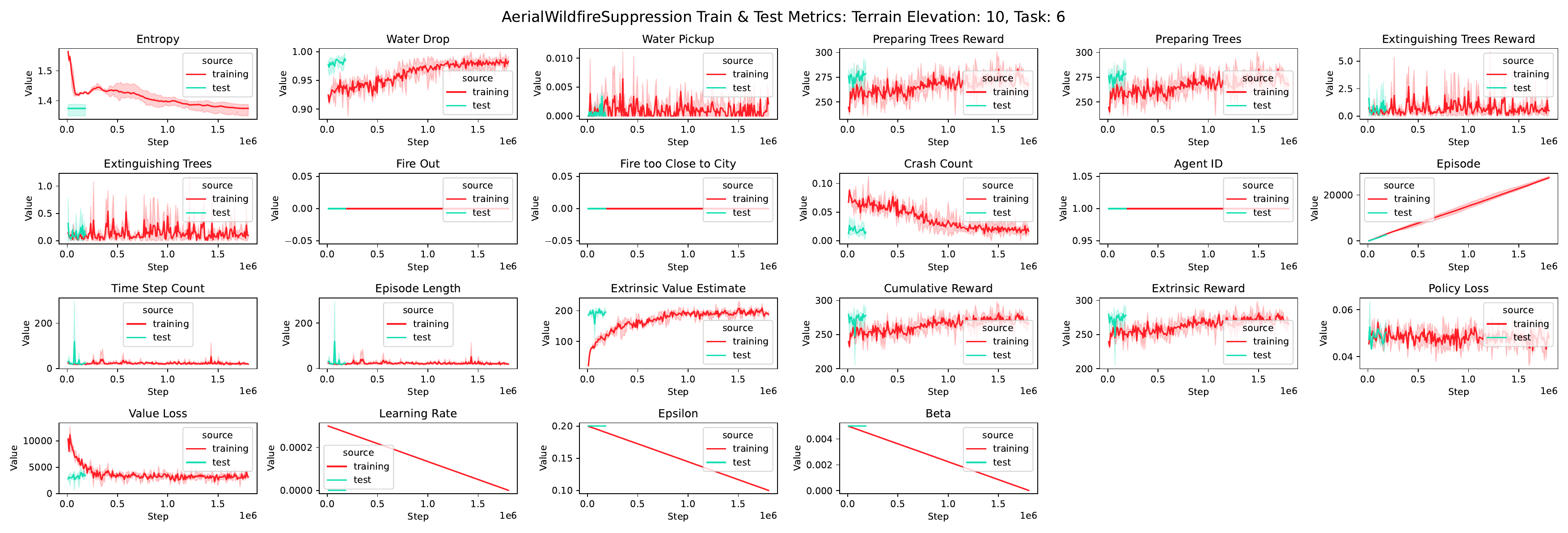}
\vspace{-0.6cm}
\caption{Aerial Wildfire Suppression: Train \& Test Metrics: Terrain Elevation 10, Task 6.}
\end{figure}

\begin{figure}[h!]
\centering
\includegraphics[width=\linewidth]{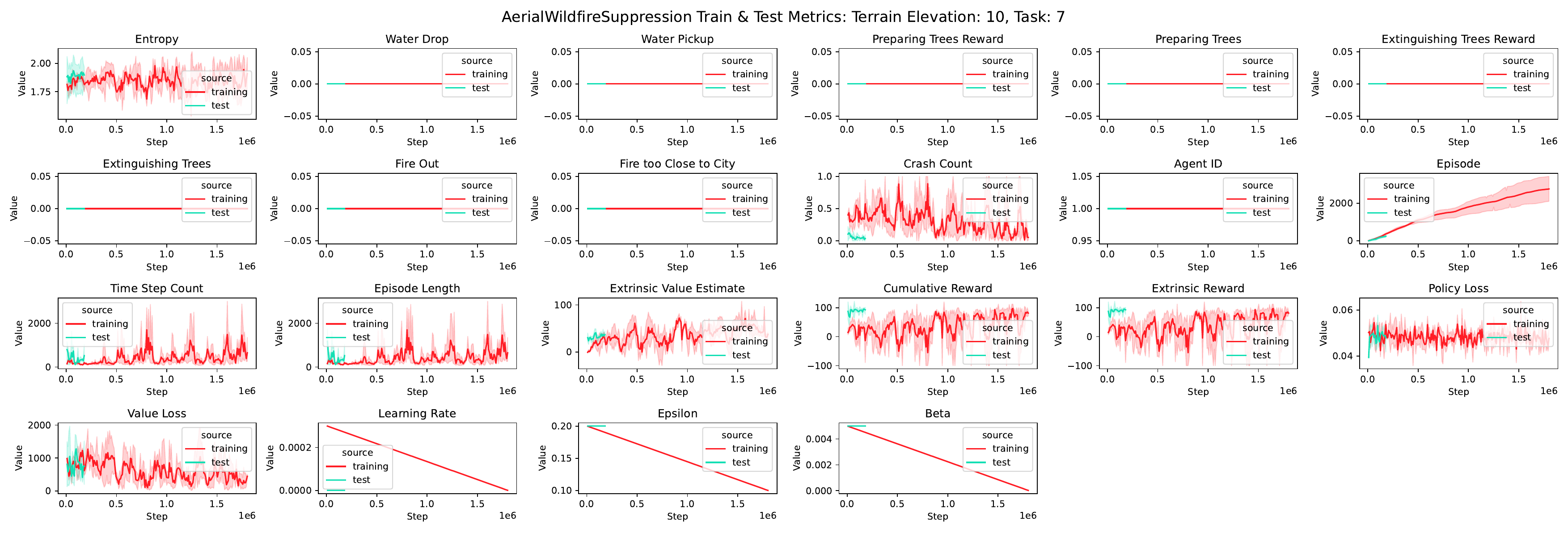}
\vspace{-0.6cm}
\caption{Aerial Wildfire Suppression: Train \& Test Metrics: Terrain Elevation 10, Task 7.}
\end{figure}

\begin{figure}[h!]
\centering
\includegraphics[width=\linewidth]{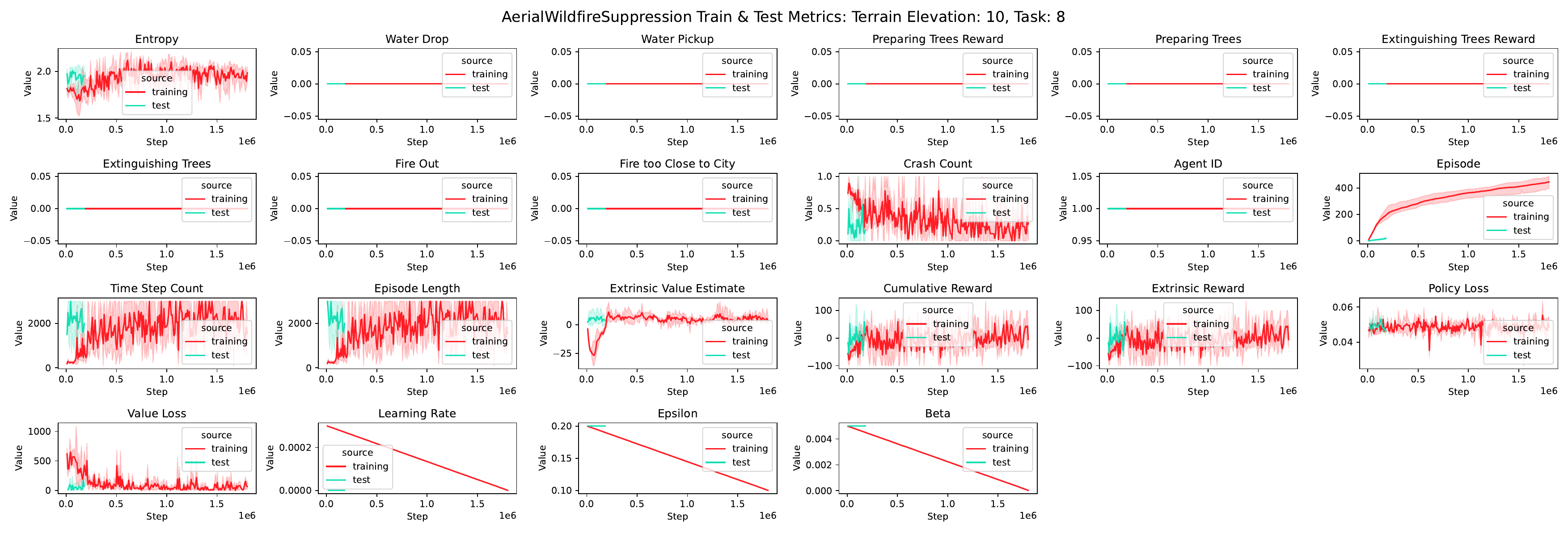}
\vspace{-0.6cm}
\caption{Aerial Wildfire Suppression: Train \& Test Metrics: Terrain Elevation 10, Task 8.}
\end{figure}

\clearpage

\subsubsection{Aerial Wildfire Suppression: Average Test Metric - Task VS Pattern}
\begin{figure}[h!]
\centering
\includegraphics[width=\linewidth]{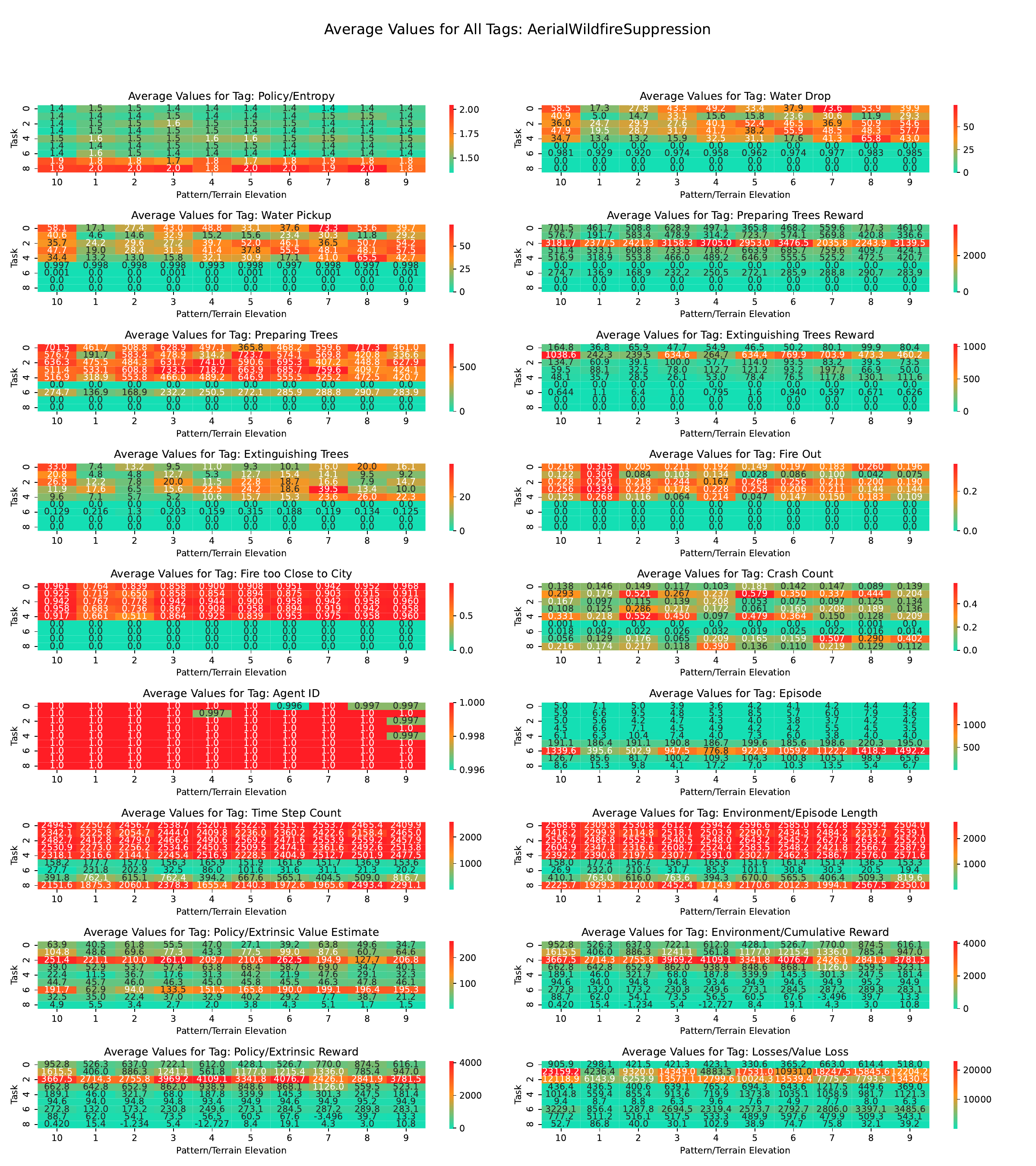}
\caption{Aerial Wildfire Suppression: Average Train \& Test Metrics.}
\end{figure}

\end{document}